\documentclass[prd,amssymb,amsmath,amsfonts,nofootinbib,reprint,showpacs]{revtex4-1}

\usepackage{graphicx}
\usepackage{lmodern}
\usepackage{amsmath,amssymb}
\usepackage{mathrsfs}
\usepackage{amsfonts}
\usepackage[utf8]{inputenc}
\usepackage{url}
\usepackage[colorlinks]{hyperref}
\usepackage[dvipsnames]{xcolor}
\usepackage{multirow}
\usepackage[normalem]{ulem}
\usepackage{lipsum}

\usepackage{marvosym}
\usepackage{enumerate}
\usepackage{color,soul}
\usepackage{acronym}

\usepackage{bm}
\usepackage{color}
\usepackage{commath}
\allowdisplaybreaks
\usepackage{multirow}

\usepackage{comment}

\def\lm{{\ell m}}

\newcommand{\abc}{\alpha, \beta, \gamma}
\newcommand{\bi}[1]{\textbf{\textit{#1}}}
\newcommand{\be}{\begin{equation}}
\newcommand{\ee}{\end{equation}}
\newcommand{\bea}{\begin{eqnarray}}
\newcommand{\eea}{\end{eqnarray}}
\newcommand{\bel}{\begin{align}}
\newcommand{\eel}{\end{align}}
\def\lm{{\ell m}}

\newacro{adm}[ADM]{Arnowitt-Deser-Misner}
\newacro{bbh}[BBH]{binary black hole}
\newacro{bh}[BH]{black hole}
\newacroplural{bh}[BHs]{black holes}
\newacro{bhns}[BHNS]{black hole-neutron star}
\newacro{bns}[BNS]{binary neutron star}
\newacro{bf}[BF]{Bayes' factor}
\newacro{cbc}[CBC]{compact binary coalescence}
\newacro{ce}[CE]{Cosmic Explorer}
\newacro{da}[DA]{data analysis}
\newacro{et}[ET]{Einstein Telescope}
\newacro{eob}[EOB]{Effective-One-Body}
\newacro{eom}[EOM]{equations of motion}
\newacro{fd}[FD]{frequency domain}
\newacro{fft}[FFT]{Fast Fourier transform}
\newacro{gw}[GW]{gravitational-wave}
\newacroplural{gw}[GWs]{Gravitational-waves}
\newacro{gr}[GR]{general relativity}
\newacro{grb}[GRB]{gamma-ray burst}
\newacro{grhd}[GRHD]{general-relativistic hydrodynamics}
\newacro{gwosc}[GWOSC]{Gravitational Wave Open Science Center}
\newacro{gwtc1}[GWTC-1]{the first gravitational-wave transients catalog}
\newacro{gsf}[GSF]{Gravitational Self Force}
\newacro{hm}[HM]{Higher mode}
\newacroplural{hm}[HMs]{Higher modes}
\newacro{ifo}[IFO]{interferometer}
\newacro{imr}[IMR]{inspiral-merger-ringdown}
\newacro{im}[IMR]{inspiral-to-merger}
\newacro{kagra}[KAGRA]{Kamioka Gravitational Wave Detector}
\newacro{ligo}[LIGO]{Laser Interferometer Gravitational-Wave Observatory}
\newacro{lso}[LSO]{Last Stable Orbit}
\newacro{lvc}[LVC]{LIGO-Virgo Collaboration}
\newacro{lvk}[LVK]{LIGO-Virgo-Kagra Collaboration}
\newacro{lo}[LO]{leading order}
\newacro{ns}[NS]{neutron star}
\newacroplural{ns}[NSs]{neutron stars}
\newacro{nr}[NR]{numerical relativity}
\newacro{nqc}[NQCs]{Next-to-quasicircular corrections}
\newacro{nlo}[NLO]{next-to-leading order}
\newacro{nnlo}[NNLO]{next-to-next-to-leading order}
\newacro{n3lo}[N3LO]{next-to-next-to-next-to-leading order}
\newacro{n4lo}[N3LO]{next-to-next-to-next-to-next-to-leading order}
\newacro{ode}[ODE]{Ordinary Differential Equation}
\newacroplural{ode}[ODEs]{Ordinary Differential Equations}
\newacro{pn}[PN]{post-Newtonian}
\newacro{pm}[PM]{post-Minkowskian}
\newacro{pe}[PE]{parameter estimation}
\newacro{psd}[PSD]{power spectral density}
\newacro{pa}[PA]{post-adiabatic}
\newacro{qnm}[QNM]{quasi-normal mode}
\newacro{qc}[QC]{quasi-circular}
\newacro{snr}[SNR]{signal-to-noise ratio}
\newacro{spa}[SPA]{stationary-phase approximation}
\newacro{sxs}[SXS]{Simulating eXtreme Spacetimes}
\newacro{td}[TD]{time domain}
\newacro{ng}[NG]{Nect Generation}
\def\Teukode{{\texttt{Teukode}}}

\definecolor{cyan}{rgb}{0,0.9,0.9}
\definecolor{orange}{rgb}{0.9,0.5,0}
\definecolor{magenta}{rgb}{1,0,1}
\definecolor{purple}{rgb}{0.8,0.4,0.8}
\definecolor{gray}{rgb}{0.8242,0.8242,0.8242}
\definecolor{dodgerblue}{rgb}{0.12, 0.56, 1.0}

\hypersetup{
    colorlinks=true,
    linkcolor=dodgerblue,
    citecolor=dodgerblue,
    urlcolor=dodgerblue
}

\newcommand{\rLSSO}{r_{\rm LSSO}}

\newcommand{\tLR}{t_{\rm LR}}
\newcommand{\thend}{\theta_\mathrm{end}}

\def\TEOBd{{\texttt{TEOBResumS-Dal\'i}}}
\def\TEOBg{{\texttt{TEOBResumS-GIOTTO}}}
\newcommand\SEOB{\texttt{SEOBNRv5PHM}}
\newcommand\SEOBE{\texttt{SEOBNRv5EPHM}}

\begin{document}

\title{Binary black hole merger in the extreme mass ratio limit:\\
a multipolar analysis of the inclined orbit case}
% Authorlist in random order atm
\author{Luca \surname{Nagni}${}^{1,2}$}
\author{Alessandro \surname{Nagar}${}^{1,3}$}
\author{Rossella \surname{Gamba}${}^{4,5}$}
\author{Simone \surname{Albanesi}${}^{1,6}$}
\author{Sebastiano \surname{Bernuzzi}${}^{1,6}$}

\affiliation{${}^{1}$ INFN sezione di Torino, Torino, 10125, Italy}
\affiliation{${}^{2}$ Dipartimento di Fisica, Università di Torino, Via Pietro Giuria 1, 10125, Torino Italy}
\affiliation{${}^{3}$ Institute des Hautes Etudes Scientifiques, 91440 Bures-sur-Yvette, France}
\affiliation{${}^{4}$ Institute for Gravitation \& the Cosmos, The Pennsylvania State University, University Park PA 16802, USA}
\affiliation{${}^{5}$ Department of Physics, University of California, Berkeley, CA 94720, USA}
\affiliation{${}^{6}$ Friedrich-Schiller-Universität Jena, Theoretisch-Physikalisches Institut, Max-Wien-Platz 1, 07743 Jena, Germany}

\begin{abstract}
We compute the gravitational waveform emitted during the transition from quasi-spherical inspiral to plunge, merger and ringdown 
for a system of two black holes in the extreme mass ratio limit, where the primary is spinning and  the secondary is represented by a 
nonspinning point-particle inspiralling along inclined orbits. The point-particle dynamics is described via a Hamiltonian formalism 
and the transition is driven by an effective-one-body like radiation reaction force. The gravitational waveform is obtained solving 
numerically, in the time-domain, the Teukolsky equation with a $\delta$-like source. The waveform is systematically characterized 
varying the black hole spin magnitude between $(0,0.9)$ and the inclination 
angle of the orbit between $(0,\pi)$. We consider all multipoles up to $\ell=4$ and compute the energy and angular momentum
losses during the plunge. The impact of the $m\neq \ell$  modes grows as the inclination angle is increased.
We also use our framework to quantify the accuracy of the approximate inspiral-merger-ringdown waveform for an inclined 
orbit that can be obtained by applying a suitable time-dependent rotation to a given spin-aligned waveform with approximately 
consistent (but constant) spin-orbit coupling. 
\end{abstract}

\date{\today}
\maketitle

% reset all acronyms
\acresetall

\section{Introduction}
\label{sec:intro}
Complete waveform models able to account for arbitrary non-planar orbits represent the fundamental theoretical goal
to help gravitational wave (GW) astronomy to properly analyze the GW transient signals 
detected by the LIGO-Virgo-KAGRA (LVK) collaboration~\cite{LIGOScientific:2016aoc,LIGOScientific:2018mvr,LIGOScientific:2020ibl,LIGOScientific:2021djp}.
Recently, a prescription in this direction based on the effective-one-body (EOB) 
approach~\cite{Buonanno:1998gg,Buonanno:2000ef,Damour:2000we,Damour:2001tu,Damour:2015isa} 
was proposed~\cite{Gamba:2024cvy, Albanesi:2025txj}, and the resulting \TEOBd{} waveform model allowed for the first 
ever inspiral-merger-ringdown analyses of the GW signals GW150914~\cite{Gamba:2025qfg} and GW200105~\cite{Jan:2025fps}
taking into account both eccentricity and spin-precession.
With the same model, also analysis for hyperbolic encounters proved possible~\cite{Gamba:2021gap,Henshaw:2025arb}.

The development of the highly accurate, Numerical-Relativity-informed, EOB waveform models that we 
have today, \TEOBg{}, \TEOBd{}~\cite{Nagar:2024oyk}, \SEOB\cite{Pompili:2023tna,Estelles:2025zah}, 
\SEOBE{}~\cite{Gamboa:2024hli} stemmed from the careful study and understanding 
of the structure of the waveform in the large mass ratio limit using black hole perturbation theory.
Starting from the seminal work of Davis, Ruffini, Press and Price~\cite{Davis:1971gg} and 
Davis, Ruffini and Tiomno~\cite{Davis:1972ud} (see also~\cite{Lousto:1996sx,Lousto:1997wf,Lousto:1997ge,Martel:2001yf}), 
black hole perturbation theory with a point-particle source proved to be an essential tool to 
compute waveforms of merging black holes in the large mass ratio limit for a variety of 
configurations, to be then used then as target waveform to develop EOB waveform or to complete the
knowledge of the parameter space in a region hardly reachable by NR 
simulations~\cite{Lousto:2020tnb,Lousto:2022hoq}.
The use of test-mass data to inform EOB models started with Refs.~\cite{Nagar:2006xv,Damour:2007xr},
that focused on the quasi-circular plunge using Regge-Wheeler-Zerilli perturbation theory with
a point-particle $\delta$-like source represented by a narrow Gaussian~\cite{Ruoff:2000et,Nagar:2004ns}
with a radiation-driven dynamics written in Hamiltonian form.
The method was progressively refined, in particular introducing a hyperboloidal layer approach
so to extract waveform at future null infinity~\cite{Bernuzzi:2010ty,Bernuzzi:2011aj,Bernuzzi:2012ku}.
The plunge on a spinning black hole on equatorial orbits was also examined in 
various works~\cite{Barausse:2011kb,Harms:2014dqa,Taracchini:2013wfa,Taracchini:2014zpa,Nagar:2014kha},
and recently extended to the case of eccentric orbits~\cite{Albanesi:2021rby,Albanesi:2023bgi,Faggioli:2024ugn,Faggioli:2025hff,Becker:2024xdi}.

The aim of this paper is to generalize the work of Ref.~\cite{Harms:2014dqa} about the plunge from equatorial orbits 
on a Kerr black hole to the case of {\it inclined orbits}. With this term we refer to orbits where the orbital angular momentum 
of the system (that is a vector orthogonal to the orbital plane) is initially rotated by a certain angle with respect to the 
spin of the black hole, which is chosen to coincide with the $z$-axis of an orthogonal reference frame. 
This orbital configuration yields an increased level of complexity of the waveform structure with respect to the equatorial case.

The gravitational waveform is decomposed in spin-weighted 
spherical harmonics ${}_{-2}Y_\lm(\Theta,\Phi)$ as
\be
\label{eq:strain}
D_L(h_+ - i h_\times) = \sum_{\ell =2}^{\infty}\sum_{m=-\ell}^{\ell} h_{\ell m}{}_{-2}Y_\lm(\Theta,\Phi)\ ,
\ee
where $D_L$ is the luminosity distance. Since, for inclined orbits, the invariance of the system under reflection across 
the orbital plane, usually identified as the (x,y) plane, is no longer satisfied, the relation $h_\lm=(-1)^\ell h^*_\lm$ does 
not hold anymore and it is needed to separately analyze modes with positive and negative values of $m$. It must be noted that
the transition from inspiral to plunge along inclined orbits on Kerr has already been studied applying a suitable generalization 
of the procedure developed by Ori and Thorne~\cite{Ori:2000zn} to model the dynamics. In this approach, one considers 
the transition from the adiabatic inspiral to plunge and merger as composed of three different stages 
(the adiabatic regime, the transition regime and the plunge regime), determined by the shape of the effective potential 
in which the test mass moves. The dynamics is described by different sets of equations of motion in each of the three stages and the solutions are matched to 
obtain a smooth transition between the early inspiral and the late plunge. The Ori-Thorne procedure was first applied to 
non-equatorial orbits in~\cite{Sundararajan:2008bw}; this work contains conceptual flaws that have been noticed and fixed in~\cite{Apte:2019txp}.
In a companion paper~\cite{Lim:2019xrb} 
the authors coupled this description of the dynamics to a solver for the Teukolsky equation to compute the waveforms and
study the QNM excitation versus the dynamical parameters. As they acknowledge, there remains an element of arbitrariness 
intrinsic to their setup, which inevitably follows from the necessity to choose when exactly the transition between the 
different phases begins: it is specifically this last choice, when exactly to end the transition phase and start the plunge, 
that non-negligibly affects the dynamics. In~\cite{Lim:2019xrb} it is argued that the ringdown modes obtained 
are independent of this choice as long as they are parameterized in terms of appropriate variables.
From a dynamical perspective, however, the existence of this arbitrariness is not fully satisfactory,
especially if one is interested in the study of the transition from inspiral to plunge as a whole.
On the other hand, within the Hamiltonian formalism the transition from the quasi-adiabatic inspiral to the geodesic 
plunge under the action of a radiation reaction force does not need special choices to relate the different stages 
of the dynamics as it is based on the continuous modification of the interaction potential due to gravitational wave losses. 
The only arbitrariness in this case may reside in the radiation reaction force, that depends on the specific analytic choice 
and thus impact the inspiral. However, the consistency of the flux and the waveform can be a-posteriori checked. 
Nonetheless, the nonadiabatic plunge is substantially insensitive to radiation reaction and
thus can be considered quasi-universal when the mass-ratio is sufficiently small
(say $\lesssim 10^{-3}$). Our aim here is to describe the particle dynamics within a suitable Hamiltonian
description (to be detailed below) complemented by a PN-resummed radiation reaction force and then compute
and characterize the waveforms by solving the Teukolsky equation in the time-domain.

This paper is organized as follows. Our Hamiltonian formalism (complemented by the radiation reaction force) is
extensively described in Sec.~\ref{sec:methods} and the numerical methods to compute the waveforms solving
the Teukolsky equation via {\tt Teukode}~\cite{Harms:2014dqa} are reminded as well. In Sec.~\ref{sec:schw} we 
extensively test the accuracy of our numerical approach by computing the waveforms emitted by a quasi-circular
plunge on Schwarzschild but along inclined orbits. The case of a Kerr black hole is discussed in Sec.~\ref{sec:insplunge},
and we also study the asymmetry between $+m$ and $-m$ waveform multipoles. 
Finally, in Sec.~\ref{sec:twist}
we demonstrate the quality of a spin-aligned, equatorial, waveform assumed to be in the coprecessing frame once 
rotated in the inertial frame, so to have a glimpse of the accuracy, in the small-mass-ratio regime, of a procedure that 
is routinely used (though with some subtleties) to generate waveforms for spin-precessing binaries within for 
example the EOB formalism~\cite{Pompili:2023tna,Khalil:2023kep,Gamba:2021ydi}.
If not otherwise stated, we use geometrized units $c=G=1$. 

\section{Dynamics of a point-particle on a Kerr background}
\label{sec:methods}
Consistently with the aligned-spin case~\cite{Harms:2014dqa}, the particle dynamics is
conveniently described via a Hamiltonian formalism with an EOB-resummed-type radiation reaction force. For equatorial
orbits, the Hamiltonian approach allows for a straightforward transition from the inspiral
to plunge and merger provided an expression of the radiation reaction is given. 
The same approach to the dynamics can be followed to describe a quasi-spherical\footnote{Or quasi-ellipsoidal, in analogy
with the planar, eccentric case.} inspiral in the sense introduced in Ref.~\cite{Damour:2001tu} within the EOB framework
and then used in  Ref.~\cite{Buonanno:2005xu} to compute the first EOB waveform for misaligned spins.
In the extreme mass ratio limit, with the nonspinning secondary, the Hamiltonian dynamics introduced in 
Refs.~\cite{Damour:2001tu,Buonanno:2005xu} describes the motion of inclined orbits around a Kerr black hole.
The main advantage of the Hamiltonian formalism is that the transition from inspiral to plunge is natural and there 
is no need of defining the end of the inspiral and the beginning
of the plunge. This idea of a {\it continuous transition} from inspiral to plunge was one of the main conceptual
new findings in the original EOB works~\cite{Buonanno:1998gg,Buonanno:2000ef}. As a side remark,
the plunge phase is considered to be quasi-geodesics, with a negligible dependence of both the dynamics and the waveform 
on radiation reaction. See for example Ref.~\cite{Bernuzzi:2011aj} (in particular Fig.~14 therein) 
for additional information on this point. 
In what follows we follow conceptually Refs.~\cite{Damour:2001tu,Buonanno:2005xu}, but we specify the dynamics 
to the limiting case of a point-particle on a Kerr background using the expression of the Hamiltonian 
introduced in Ref.~\cite{Balmelli:2015zsa}.
All technical details on the conservative and nonconservative description of the dynamics are reported next.

\subsection{Conservative dynamics}
\label{sec:Ham}
The Hamiltonian of Ref.~\cite{Balmelli:2015zsa} consists of a spin-orbit (odd-in-spin) 
term, $H_{\rm SO}$, and an orbital contribution (even-in-spin), $H_{\rm orb}$,
\be
H=H_{\rm SO}+H_{\rm orb} \ .
\ee
It is convenient to describe the dynamics in (Boyer-Lindquist-based) Cartesian-like coordinates
$\bi{r}=(x,y,z)$ with $x = r\sin\theta\cos\varphi$, $y = r\sin\theta\sin\varphi$ and $z=r\cos\theta$,
so that we have
\begin{widetext}
\begin{align}
    H_{\text{SO}} &= G_S(\bi{r},\bi{a}){\mathbf{L} \cdot \bi{a}} \ ,\label{eq:Hso}\\
    H_{\text{orb}} &= \sqrt{A^{\rm Kerr}(\bi{{r}},\bi{a})\left( \mu^2 + \frac{1}{1+\frac{(\bi{n}\cdot \bi{a})^2}{r^2}}\left[\bi{p}^2+\left( \frac{\Delta}{r^2}-1 \right) (\bi{n} \cdot \bi{p})^2-\frac{r^2+2r+(\bi{n}\cdot \bi{a})^2}{\mathcal{R}^4+\Delta (\bi{n}\cdot \bi{a})^2}((\bi{n}\times\bi{p})\cdot \bi{a})^2 \right]\right)} \ , \label{eq:Horb}
\end{align}
\end{widetext}
with $\bi{r}=r \bi{n}$, where $r$ is the Boyer-Lindquist radial coordinate, $\bi{n}$ the radial unit vector 
(see Fig.~\ref{fig:iotadef}), $\bi{a} = a\hat{\mathbf{z}}$ is the Kerr spin vector and
\begin{align}
    \Delta &= r^2 - 2Mr + a^2 \\
    \mathcal{R}^4 &= r^4 + r^2 a^2 +2Mr a^2 = r^2r_c^2 
    \label{eq:R4}  \ .
\end{align}
The quantity $r_c^2$ denotes the (squared) centrifugal radius defined by
\be
r_c^2 = r^2 + a^2 + \dfrac{2Ma^2}{r} \ .
\ee
The overall factor $A^{\rm Kerr}(\bi{r},\bi{a})$, 
\begin{align}
\label{eq:Apot}
A^{\rm Kerr}(\bi{r},\bi{a}) = \frac{\Delta (r^2+(\bi{n}\cdot\bi{a})^2)}{\mathcal{R}^4 + \Delta (\bi{n}\cdot\bi{a})^2} \ ,
\end{align}
is an anisotropic (spin-dependent) gravitational potential which generalizes the Schwarzschild (isotropic)
potential $1-2M/r$. Following~\cite{Balmelli:2015zsa}, it can be written as
\begin{equation}
    A^{\rm Kerr}(\bi{r},\bi{a}) = A^\mathrm{Kerr, eq}(r_c) \frac{1+\frac{(\bi{n}\cdot\bi{a})^2}{r^2}}{1+\frac{\Delta(\bi{n}\cdot\bi{a})^2}{r_c^2r^2}}.
\end{equation}
where $A^\mathrm{Kerr, eq}(r_c)$ denotes the equatorial Kerr radial potential~\cite{Damour:2014sva}
\begin{equation}
    A^\mathrm{Kerr, eq}(r_c) = \left( 1-\frac{2M}{r_c}\right)\frac{1+\frac{2M}{r_c}}{1+\frac{2M}{r}} \ .
\end{equation}
The gyrogravitomagnetic function $G_S$ reads
\begin{align}
    G_S(\bi{r},\bi{a}) &= \frac{2r}{\mathcal{R}^4 + \Delta (\bi{n}\cdot \bi{a})^2}\nonumber\\
                               &=\dfrac{2}{r r_c^2}\left(1+\dfrac{\Delta(\bi{n}\cdot \bi{a})^2}{r_c^2 r^2}\right)^{-1}.
\end{align}
In this form it is apparent that the equatorial expression is dressed by an anisotropic, 
spin-dependent, contribution. One can then write Hamilton's equations as
\begin{align}\label{eq:hameq}
    \dot{x}^i &= \partial_{p_i} H \\
    \dot{p}_i &= \partial_{x^i} H+ \mathcal{F}_i \ ,
\end{align}
where  $\mathcal{F}_i$ are the Cartesian components of the radiation reaction force discussed in Sec.~\ref{sec:RR} below. 

Let us finally note that, although the dynamics is solved in Cartesian coordinates, 
the source of {\tt Teukode} is written in Boyer-Lindquist polar-like coordinates, so 
that the $(x,y,z)$ dynamics must be transformed into the $(r,\theta,\varphi)$ in the
standard way. We write the results explicitly for completeness. We have
$r=\sqrt{x^2+y^2+z^2}$, $\theta = \mathrm{atan2}(\sqrt{x^2+y^2},z)$,  $\varphi = \mathrm{atan2}(y,x)$, 
while the relations between the momenta are\footnote{These are readily obtained from the standard relation $x = r\sin\theta\cos\varphi$, $y = r\sin\theta\sin\varphi$ and $z=r\cos\theta$ by recalling that, 
upon a coordinate transformation $x^i \rightarrow \tilde x^i$, contravariant vectors transform as $\tilde v^i = \tfrac{\partial \tilde x^i}{\partial x^j}v^j$, 
whereas covariant vectors obey $\tilde w_i = \tfrac{\partial x^j}{\partial \tilde x^i} w_j$.}
\begin{align}
p_r &= (p_x\cos\varphi + p_y\sin\varphi)\sin\theta +  p_z\cos\theta \ ,\\
p_\theta &= r\left[(p_x\cos\varphi + p_y\sin\varphi)\cos\theta -p_z\sin\theta\right] \ , \\
p_\varphi &= r\sin\theta(p_y\cos\varphi - p_x\sin\varphi) \ .
\end{align}
In these coordinates, the Hamiltonian reads:
\begin{widetext}
\begin{align}
    H_{\text{orb}} &= \sqrt{A(r,\theta)\left( \mu^2
    + \frac{r^2+a^2\cos^2\theta}{\mathcal{R}^4(r)+a^r\Delta(r)\cos^2\theta}\frac{p_\varphi^2}{\sin^2\theta}
    +\frac{p_\theta^2}{r^2+a^2\cos^2\theta}
    +\frac{\Delta(r)p_r^2}{r^2+a^2\cos^2\theta} \right)} \\
    \label{SO}
    H_{\text{SO}} &= G_S(r,\theta)a p_\varphi \, .    
\end{align}
\end{widetext}
Having the Hamiltonian written in this form might be useful to improve the characterization of the dynamics
according to the relative importance of the various term proportional to the momenta.

For the actual implementation it is more convenient to use dimensionless rescaled quantities as 
$\hat{H^{\rm K}}\equiv H^{\rm K}/\mu$, $\hat{r}\equiv r/M$, $\hat{t}\equiv t/M$, $\hat{p}_i=p/\mu$, $\hat{a}\equiv a/M$. 
However, in order to lighten the notation, we will omit to display the hats on the rescaled dynamical variables,
though they are intended to be so.

\subsection{Radiation reaction force}
\label{sec:RR}
As mentioned in the introductory paragraph, we adopt the expression for the three components of the radiation
reaction force given in Sec.~III of Ref.~\cite{Buonanno:2005xu}, Eq.~(3.27) therein, that is here specialized
to the extreme mass ratio limit with a nonspinning secondary. More precisely, the expression of 
Ref.~\cite{Buonanno:2005xu}, that depends on both masses $m_1$ and $m_2$, is evaluated in the
limit $m_2/m_1\ll 1$.
This explicitly yields the following formal expression
\begin{equation}
\label{eq:RRforce}
    \hat{\mathcal{F}}_i = -\frac{32}{5}\nu x^{7/2} \hat{f}\dfrac{p_i}{|\bf{L}|} + \frac{488}{15}  \frac{ \mathbf{p} \cdot \bi{a}}{L^2 r}\nu x^{4} {L_i} \ ,
\end{equation}
where $x\equiv (M\Omega)^{2/3}$ is the PN-expansion frequency variable that is expressed
in terms of $\Omega$, the modulus of the orbital frequency vector ${\bf \Omega}$.
This vector is defined as 
\be
{\bf \Omega} = \frac{\mathbf{r}\times\mathbf{v}}{r^2}
\ee
with $\mathbf{v}=\dot{\mathbf{r}}$.
Although Hamilton's equations are solved in Cartesian coordinates, it is convenient to
express this vector in polar coordinates, i.e. decompose it along the basis $(\mathbf{n},\mathbf{e_\theta},\mathbf{e_\varphi})$
\be
\label{eq:OmgVec}
{\bf \Omega} = \frac{\mathbf{r}\times\mathbf{v}}{r^2} 
 = \frac{1}{r}\left[-\left(\sin\theta\Omega_\varphi\right)\mathbf{e_\theta} + \left(\frac{1}{\sin\theta}\Omega_\theta\right)\mathbf{e_\varphi}\right]
\ee
where $\Omega_\theta\equiv \dot{\theta}$ is the {\it polar} frequency, $\Omega_\varphi\equiv \dot{\varphi}$
is the azimuthal frequency and $(\mathbf{e_\theta} ,\mathbf{e_\varphi} )$ are the unit vectors along 
the azimuthal and polar directions. The vector modulus, $\Omega\equiv |\mathbf{\Omega}|$, then reads
\be
\label{eq:Omg}
\Omega=\sqrt{\Omega_\theta^2 + \sin^2\theta\,\Omega_\varphi^2} \ ,
\ee
In Eq.~\eqref{eq:RRforce}, the quantity  $\mathbf{L} = \mathbf{x}\times\mathbf{p}$ is the orbital angular momentum 
while $\hat{f}$ is the reduced flux function, i.e. the relativistic energy flux along circular orbits divided by the
leading, quadrupolar, Newtonian contribution. 
For simplicity we use here $\hat{f}$ at the same PN-order of Ref.~\cite{Buonanno:2005xu}, i.e. at 3.5PN in
the orbital part and at 1.5PN in the spin-orbit part. In practice, we use the following PN-expanded function
\begin{equation}\label{eq:fhat}
    \begin{split}
        \hat{f} = 1 &+ f_2  x +\left( f_3+ f_{3\text{SO}} \right)  x^{3/2} + f_4  x^{2} + f_5  x^{5/2} \\
        &+ (f_6 + f_{\ell 6} \ln(4  x^{1/2}))  x^3 + f_7  x^{7/2} \ ,
    \end{split}
\end{equation}
where the $f_i$ coefficients are
\begin{align}
    f_2 &= -\frac{1247}{336} \ , \\
    f_3 &= 4\pi \ , \\
    \label{eq:f3SO}
    f_{3\mathrm{SO}} &= -\frac{11}{4}\frac{\mathbf{L}\cdot \bi{a}}{L} \ ,\\
    f_4 &= -\frac{44711}{9072} \ ,\\ 
    f_5 &= -\frac{8191}{672}\pi \ ,\\
    f_6 &= \frac{6643739519}{69854400} + \frac{16}{3}\pi - \frac{1712}{105 }\gamma_{\rm E} \ , \\
    f_{\ell 6} &= -\frac{1712}{105} \ ,\\
    f_7 & = -\frac{16285}{504}\pi \ ,
\end{align}
where $\gamma_E=0.57721 566\dots$ is the Euler constant. Following Ref.~\cite{Buonanno:2005xu}
the orbital part of $\hat{f}$ is resummed taking a $(3,4)$ Pad\'e approximant, while the spin-orbit term is added separately.
In Appendix~\ref{app:RR} we check the consistency of this expression of the flux against the waveform fluxes at infinity for 
a sample of spinning configurations at various inclinations.
Given the rather low PN-order used and the simplified resummed treatment 
(in a sense, analogous to the one originally proposed in Ref.~\cite{Damour:1997ub}) 
we expect this expression to break down 
as the particle approaches and crosses the last stable spherical orbit. 
As such, our flux is not intended to
be used to drive the long inspiral of an EMRI, whose dynamics is dominated by the radiation reaction. 
For this specific purpose,
various expressions for the resummed flux, at higher PN-order, were developed and can be found in the 
literature (see e.g. Refs.~\cite{Damour:2008gu,Fujita:2014eta,Albertini:2023aol}).

\subsection{Numerical Setup}
\begin{figure}
    \begin{center}
        \includegraphics[width=.4\textwidth]{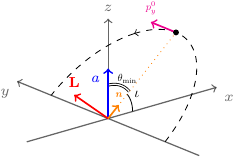}
    \end{center}
    \caption{\label{fig:iotadef}Relation between the inclination parameter 
    $\iota$, the angle $\theta_{\rm min} = \theta_0$ and the instataneous orbital plane 
    of the dynamics (identified by the dashed trajectory).
    For clarity we also include some relevant quantities that determine the initial conditions for the evolution
    of the particle (represented by a small dot) originating in the $(x,z)$ plane, as 
    explained thoroughly in the main text.}
\end{figure}

We compute the test particle dynamics by numerically solving Hamilton's equations~\eqref{eq:hameq}. 
Since we are interested in studying (quasi)-spherical orbits, we specify initial conditions by requiring that 
$\partial_r H = 0$; 
or simplicity and without loss of generality, we choose to position the particle 
in the $(x,z)$ plane and to orient the initial linear momentum along the positive $y$-direction. 
Since the spin of the central black hole is oriented along the positive $z$ axis, the resulting orbit 
will be prograde if $x_0$ is positive (or equivalently $\varphi_0=0$), retrograde if $x_0<0$ $(\varphi_0=\pi)$.
The distinction between prograde and retrograde orbits is conveniently expressed in terms of the inclination 
parameter $\iota$, which measures the inclination of the orbital plane with respect to equatorial one. 
Different definitions of $\iota$ exist in the literature which, albeit serving the same purpose, are not 
necessarily equivalent (see for instance Ref.~\cite{Warburton:2021kwk}). In this work we use the following formal 
definition 
\begin{equation}
    \cos \iota = \sin \left[ \theta_{\rm min} \mathrm{sign}(L_z) \right] \ ,
\end{equation}
where $\theta_{\rm min}$ is the smallest value taken by the coordinate $\theta$ during the time-evolution 
of the dynamics and $L_z$ is the $z$-component of the orbital angular momentum. 
This definition essentially guarantees that 
\begin{align}
    \iota + \theta_{\rm min} &= \pi/2 \qquad \text{if} \qquad \iota<\pi/2  \ ,\\
    \iota - \theta_{\rm min} &= \pi/2 \qquad \text{if} \qquad \iota>\pi/2\ ,
\end{align}
as exemplified in Fig.~\ref{fig:iotadef} (which is adapted from Fig.~1 of Ref~\cite{Warburton:2021kwk}):
hence $\cos\iota>0$ for prograde orbits and $\cos\iota<0$ for retrograde ones. This definition in terms 
of $\theta_{\rm min}$ is especially convenient since, as will be discussed in Sec.~\ref{sec:characterization_dynamics}, 
the range of $\theta$ remains essentially constant during the plunge, meaning that with our choice of initial 
conditions we simply have $\theta_{\rm min} = \theta_0$. Finally, note that by having the inclination parameter span 
the interval $(0,\pi)$ we can let the spin parameter $a$ be always positive, since anti-aligned configurations 
correspond to $\cos\iota<0$. 

We can then specify the initial position of the particle as $\mathbf{r_0} = (r_0,\theta_0,\varphi_0)$, and compute the 
momentum $\mathbf{p_0} = (0,p_{y}^0,0)$ corresponding to a spherical orbit of radius $r_0$ by solving
numerically the following equation
\begin{align}
    &\partial_r H \vert_{(\mathbf{r_0},\mathbf{p_0})} \nonumber\\
    &= \pm \sin\theta_0\partial_x H  + \cos\theta_0\partial_z H 
     \mp\frac{p_\varphi^0}{r_0^2\sin\theta_0}\partial_{p_y} H  = 0.
     \label{dHdr}
\end{align}
where the signs are determined depending on whether the orbit is prograde (top sign) or retrograde (bottom sign).
We solve for $p_y$, since $p_\varphi^0 = r_0\sin\theta_0p_y$, and indicate with $p_y^0$ this solution. 

%=============
% radius evolution - effect of including PA correction
%=============
\begin{figure}
    \begin{center}
    \includegraphics[width=.48\textwidth]{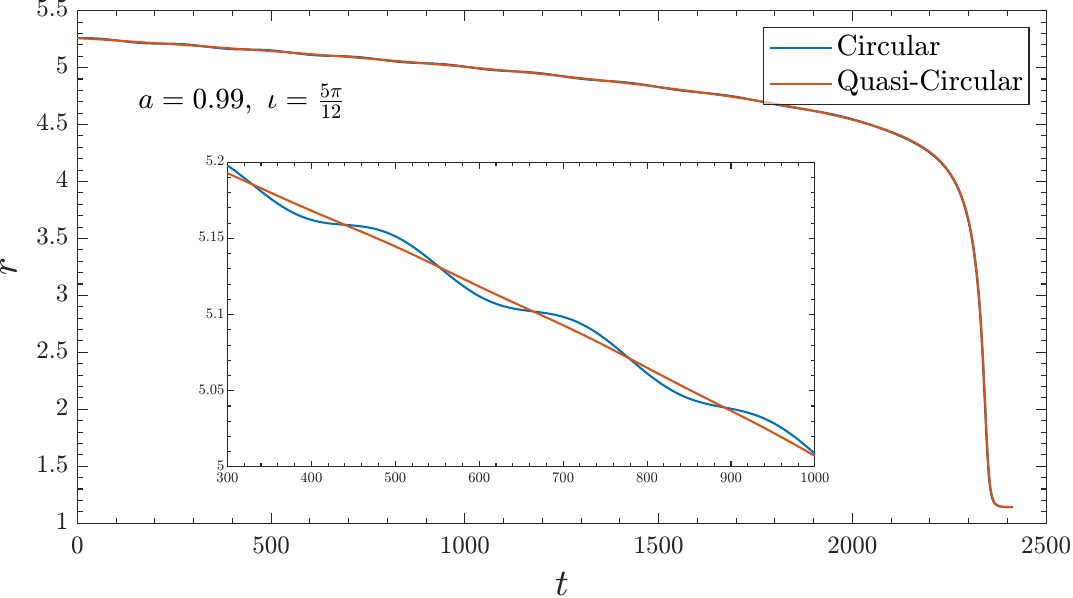}
    \caption{\label{fig:PAcorrection}From circular to post-circular initial data: reduction of initial eccentricity (i.e. oscillations in $r$) 
    including an initial radial momentum $p_r^0\neq 0$ obtained from Eq.~\eqref{eq:1PA}.}
\end{center}
\end{figure}
Analogously to the equatorial-orbit case, if we were to start the evolution with initial condition
$\bf {p}$ with the so-computed, circular $p_\varphi^0$ the system would develop spurious radial 
oscillations due to residual eccentricity coming from %not having used a nonzero value of the radial momentum. 
the condition $p_r^0 = 0$.
The procedure of initializing the relative dynamics in a consistent way beyond
the circular limit, known as {\it post-circular} initial data, was introduced as early as in Ref.~\cite{Buonanno:2000ef}
within the EOB framework for comparable mass binaries, generalized to the next order, dubbed {\it post-post-circular} 
in Ref.~\cite{Damour:2012ky} and to any post-circular (also called {\it post-adiabatic}) order in Ref.~\cite{Nagar:2018gnk} 
for spin-aligned configurations. For generic spins, an analogous, post-circular, procedure was presented
in Ref.~\cite{Buonanno:2005xu} using a simplified form of the Hamiltonian, while Ref.~\cite{Pan:2013rra} 
used an iterative, eccentricity-reduction, procedure considering the full Hamiltonian.
In our current setup, given our choice of initial conditions that guarantees $p_\theta=0$,
it is sufficient to obtain the initial value of $p_r$ using the equation valid for equatorial motion
\begin{equation}
    \frac{d p_\varphi}{dt} = \frac{d p_\varphi}{dr}\dot{r} = \frac{d p_\varphi}{dr}\frac{\partial H}{\partial p_r} = \mathcal{F}_\varphi.
    \label{eq:1PA}
\end{equation}
In practice, the procedure for initializing the dynamics consists of the following three steps:
(i) specify an initial separation and polar angle $(r_0,\theta_0)$, while $\varphi_0=0$;
(ii) compute the tangential momentum $p_y^0$ solving Eq.~\eqref{dHdr}; (iii) compute $p_r$
from Eq.~\eqref{eq:1PA}; (iv) express the results in Cartesian coordinates as
$(x_0,y_0,z_0)=(r_0\sin\theta_0,0,r_0\cos\theta_0)$ and
 $(p_x^0,p_y^0,p_z^0)=(p_{r}^0\sin\theta_0, p_{y}^0, p_{r}^0\cos\theta_0)$.
The reduction of the initial eccentricity moving from circular to post-circular initial data is
illustrated in Fig.~\ref{fig:PAcorrection}, that shows the time-evolution of the radial separation for 
$a=0.99$ and initial inclination $\iota = 5\pi/12$.

\subsection{Characterization of the dynamics}
\label{sec:characterization_dynamics}

\begin{figure}[t]
    \begin{center}
    \includegraphics[width=.38\textwidth]{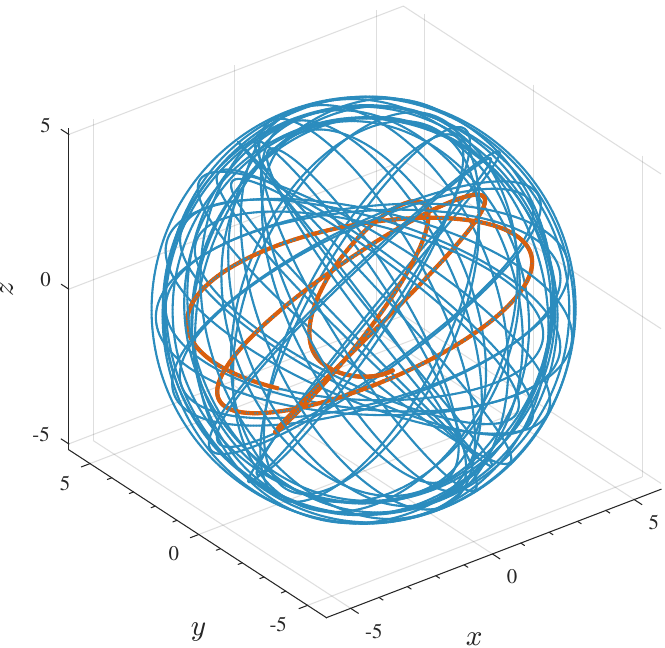} %\hspace{5mm}
    \vskip 2mm
    \includegraphics[width=.43\textwidth]{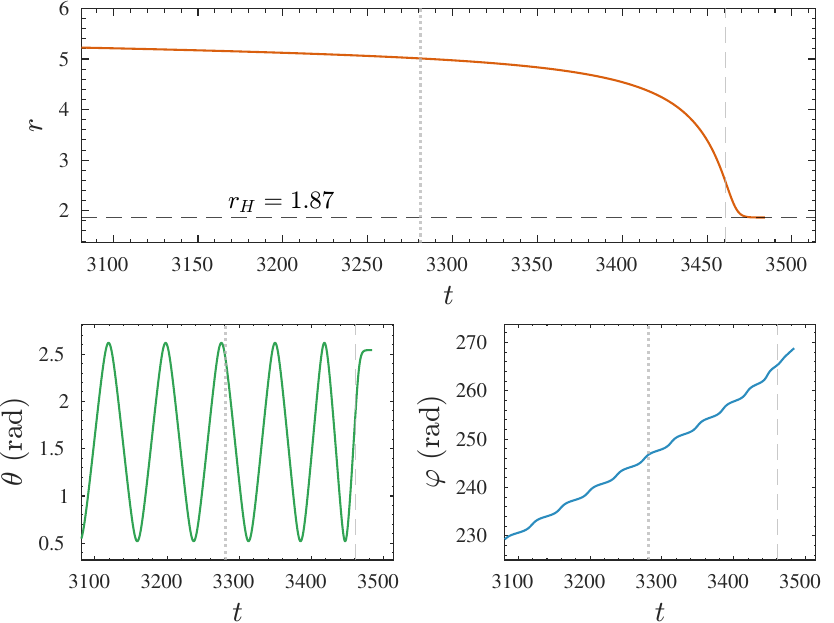}
    \caption{\label{fig:ExDynamics}Characterizing an inspiralling and plunging orbit: the parameters identifying 
    the trajectory are $a=0.5,\ \iota = \tfrac{\pi}{3},\ r_0 = 6$. Top panel: the full 3D trajectory, with the portion
    successive to the LSSO crossing highlighted in orange. The smaller plots on the bottom show the time evolution 
    of $(r,\theta,\varphi)$; the two vertical lines correspond to the crossing of the LSSO (dotted) and LR (dashed).
    Note that $\theta_{\rm min}=\pi/2-\pi/3\simeq 0.52$, according to the definitions of Fig.~\ref{fig:iotadef}.}
\end{center}
\end{figure}

\begin{figure}
    \begin{center}
        \includegraphics[width=0.23\textwidth]{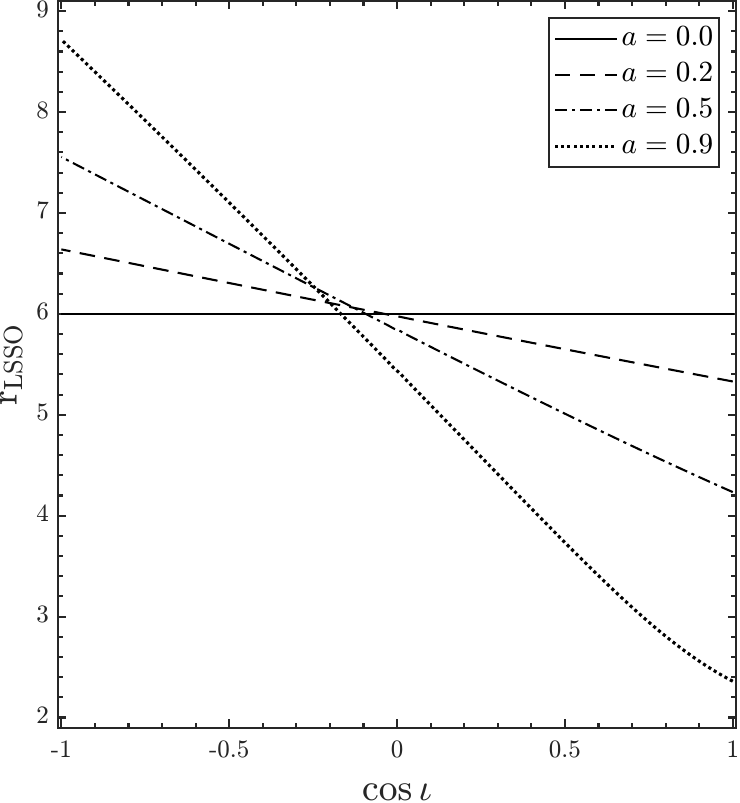}
        \hspace{1mm}
        \includegraphics[width=0.235\textwidth]{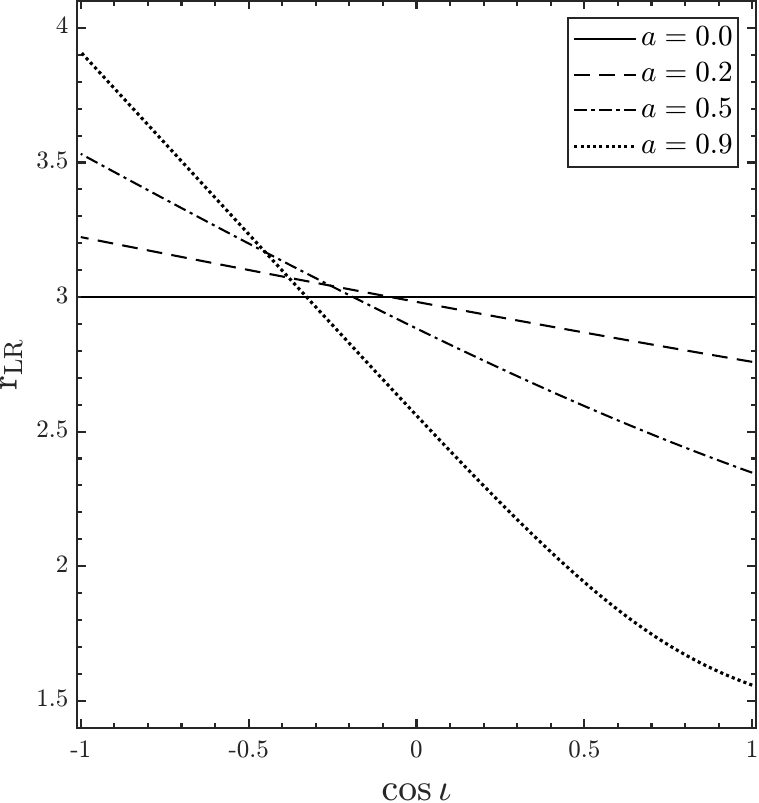}
    \end{center}
    \caption{\label{fig:LSSOLR}Dependence of the location last stable spherical orbit (left panel) and of the light ring (right panel) 
    on the inclination of the orbit $\iota$ for a range of values of black hole spin $a$.}
\end{figure}

Now that we have set up quasi-spherical (eccentricity free) initial data, let us move to a qualitative
discussion of the dynamics when varying spin and inclination. Each configuration we are going
to consider is labelled by the black hole spin $a$ and the inclination angle $\iota$. We define
a naming convention for each configuration as ${\tt aXiY}$ where ${\tt X}$ is the numerical value of $a$
and ${\tt Y}$ is the numerical value of $\iota$ expressed in degrees. For example, a configuration with
$a=0.5$ and $\iota=\pi/3=60^{\circ}$ is labelled as \texttt{a05i60}.
Figure~\ref{fig:ExDynamics} refers to this configuration with $\nu\equiv \mu/M=10^{-3}$ 
and $r_0=6$. The top panel of the figure illustrates the full quasi-spherical inspiralling trajectory.
In orange we highlight the portion corresponding to the nonadiabatic {\it plunge}, i.e. after the crossing 
of the Last Stable Spherical Orbit (LSSO)~\cite{Damour:2001tu,Apte:2019txp,Stein:2019buj},
$r<\rLSSO \simeq 5.01$. We note that, differently from the generalized Ori-Thorne procedure 
of Ref.~\cite{Apte:2019txp}, we do not need to know in advance the location of the LSSO 
in order to separate between an inspiral region, a transition region and a (geodesic) plunge,
but the transition occurs continuously under the action of radiation reaction as driven by the
corresponding evolution of the effective potential~\cite{Buonanno:2000ef,Damour:2001tu,Buonanno:2005xu}.
The plunge is thus quasi-geodesic, as it maintains a residual dependence on the radiation reaction
force through $\nu$. In order to minimize these effects, we consider $\nu=10^{-3}$, thus 
allowing for a robust characterization of the physics of the plunge~\cite{Bernuzzi:2011aj}.
The bottom panels of Fig.~\ref{fig:ExDynamics} show the time-evolution of the Boyer-Lindquist coordinates,
$(r,\theta,\phi)$. The radial separation decreases steadily and eventually settles on the value of the black hole
horizon, $r=r_H=1+\sqrt{1-a^2}$. The polar angle $\theta$ oscillates between $(\pi/6,5\pi/6)$, 
and as the particle approaches the horizon it freezes to  constant value $\theta_{\rm end}$; note that 
even as the particle loses energy the range of the polar oscillation does not change appreciably. 
Finally, the azimuthal angle $\varphi$ increases monotonically and exhibits visible oscillations throughout 
the inspiral and during most of the plunge. As the particle approaches the outer event horizon the 
oscillations disappear as the azimuthal frequency approaches its horizon value
\begin{equation}\label{eq:OmgHorizon}
    \Omega_H \equiv \frac{a}{2r_H} \ ,
\end{equation}
as we will explicitly see below. Note that our radiation-reaction-driven dynamics is 
in qualitative agreement with the one of Ref.~\cite{Apte:2019txp,Lim:2019xrb}, 
although our Hamiltonian approach is radically different. 
In Fig.~\ref{fig:ExDynamics} we also mark with a dotted vertical line the LSSO crossing and with
a dashed vertical line the light-ring (LR) crossing. The LSSO is computed numerically
following Ref.~\cite{Stein:2019buj} (that also includes eccentricity), while the LR is obtained
solving numerically the characteristic equation defining the null spherical 
geodesics~\cite{Chandrasekhar:1985kt}.
For completeness, in Fig.~\ref{fig:LSSOLR} we report the behavior of the LSSO and LR 
as function of $\cos\iota$ for a few values of the black hole spin.
%============
% Example of Dynamics figure
%============
%============
% LSO-LR figure
%============

For equatorial orbits, the time-evolution of the dynamics can be efficiently characterized 
using the azimuthal frequency $\Omega_\varphi$, that in this case coincides with the full $\Omega$.
In particular, the last maximum of $\Omega$ corresponds to the light-ring crossing and thus 
effectively marks the beginning of the ringdown phase of the waveform that is dominated
by quasi-normal modes. For this reason, within the EOB waveform modeling for nonprecessing 
binaries, the peak of the orbital frequency was conventionally used to identify a reference
merger time where the two-body description is continued with an effective description of 
postmerger and ringdown~\cite{Buonanno:2000ef,Damour:2007xr}. To be precise,
within the {\tt TEOBResumS} model~\cite{Damour:2014sva} it proved useful to use
the peak of the {\it pure} orbital frequency (i.e., without the spin-orbit contribution)
as reference point to switch to the postmerger-ringdown description. This choice was inspired
by studies in the test-mass limit that showed that the peak of the quadrupolar waveform might
be detached from the actual LR crossing as the spins gets large~\cite{Taracchini:2014zpa,Harms:2014dqa,Gralla:2016qfw}
whereas the peak of the {\it pure} orbital frequency is quite close to the quadrupolar waveform peak~\cite{Harms:2014dqa}.
For this reason, this pure orbital frequency is a pivotal quantity in the construction of any avatar 
of the  {\tt TEOBResumS} model~\cite{Damour:2014sva,Nagar:2024oyk}. Since one of the aims of 
this paper is to prepare the ground to the modeling of EOB-like waveform for inclined orbits on 
Kerr, so to generalize previous work for equatorial orbits~\cite{Albanesi:2023bgi} in the future, 
a detailed understanding of the properties of the orbital frequency is in order.

For inclined orbits the orbital frequency vector is given by Eq.~\eqref{eq:OmgVec} and its norm 
by Eq.~\eqref{eq:Omg}, i.e. a combination of the polar, $\Omega_\theta$ and azimuthal, 
$\Omega_\varphi$ frequencies. 
The structure of the full Hamiltonian, which separates in an orbital and a spin-orbit term~\eqref{eq:Horb}--\eqref{eq:Hso} term, 
allows us to write $\mathbf{\Omega}$ as
\begin{equation}\label{eq:OmgOrbSo}
    \mathbf{\Omega} =  \mathbf{\Omega^{\rm orb}} + \mathbf{\Omega^{\rm so}}  \ ,
\end{equation}
where the two terms on the right-hand side correspond to the purely orbital (even-in-spin only)
and spin-orbit (odd-in-spin only) frequency vectors. Their definition follows from that 
of $\mathbf{\Omega}$~\eqref{eq:OmgVec}, with $(\Omega_\varphi,\Omega_\theta)$ 
replaced by the corresponding {purely orbital}/{purely spin-orbit} frequencies which we define as 
\begin{align}
        \Omega_\varphi^{\rm orb/so}&\equiv \partial_{p_\varphi}H_{\rm orb/so} \\
        \Omega_\theta^{\rm orb/so}&\equiv \partial_{p_\theta}H_{\rm orb/so} \ .
\end{align}
%=============
% Figure of Omega,Omega^orb and Omega^so
%=============
\begin{figure}[t]
    \begin{center}
        \includegraphics[width=.43\textwidth]{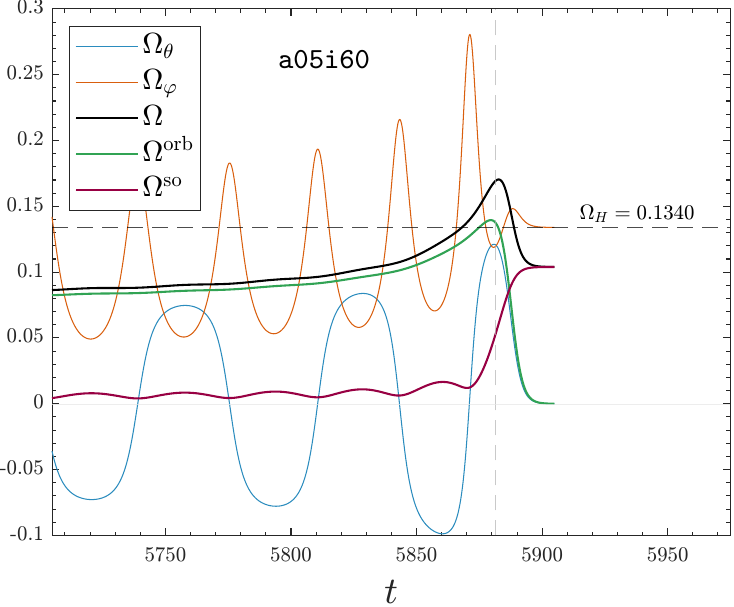}
        \caption{\label{fig:OmgOrbSo}Behavior of the total angular frequency $\Omega$ along with its components 
        $\Omega_\theta,\Omega_\varphi,\Omega^{\rm orb},\Omega^{\rm so}$ for the configuration {\tt a05i60} shown in Fig.~\ref{fig:ExDynamics}. 
        The vertical dashed line indicates the light ring crossing, which for this configuration is quite close to the positions of $\Omega_{\rm peak}$ 
        and $\Omega^{\rm orb}_{\rm peak}$. The horizontal dashed line denotes the horizon frequency $\Omega_H$ and the thin grey line 
        simply highlights the position of the zero on the $y$ axis. }
    \end{center}
\end{figure}
In BL coordinates these vectors read
\begin{align}
\label{eq:OmgOrb}
{\bf \Omega}^{\rm orb/so}= \frac{1}{r}\left[-\left(\sin\theta\Omega^{\rm orb/so}_\varphi\right)\mathbf{e_\theta} + \left(\frac{1}{\sin\theta}\Omega^{\rm orb/so}_\theta\right)\mathbf{e_\varphi}\right]
\end{align}
and their modulus $\Omega^{\rm orb/so}\equiv |{\bf \Omega}^{\rm orb/so}|$,
\begin{equation}
\label{eq:Omg_orbAmp}
    \Omega^\textrm{orb/so} = \sqrt{(\Omega_\theta^{\rm orb/so})^2+\sin^2\theta(\Omega_\varphi^{\rm orb/so})^2 }\ .
\end{equation}
We stress the fact that we have introduced two separate decompositions of the total angular frequency $\Omega$: on the one hand, we can think of $\Omega$ as 
the combination of the polar and azimuthal motion according to Eq.~\eqref{eq:Omg}. On the other hand, the form of the Hamiltonian suggests an 
interpretation in terms of pure orbital and spin-orbit frequency, Eq.~\eqref{eq:OmgOrbSo}, from which one sees that 
$\Omega^2 = (\Omega^{\rm orb})^2 + (\Omega^{\rm so})^2 + 2\mathbf{\Omega^{\rm orb}}\cdot\mathbf{\Omega^{\rm so}}$. 
Depending on our goal we will prefer one description or the other, as both can offer precious insights in the phenomenological features 
characterizing the dynamics. To illustrate this in Fig.~\ref{fig:OmgOrbSo} we plot the time evolution of
$(\Omega,\Omega_\theta,\Omega_\varphi, \Omega^{\rm orb},\Omega^{\rm so})$ for the 
configuration {\tt a05i60} of Fig.~\ref{fig:ExDynamics}. The vertical dashed line denotes the 
light ring crossing. We also highlight the horizon frequency of the black hole~\eqref{eq:OmgHorizon} 
by a black, horizontal dashed line and the position of the zero on the $y$ axis by a thin grey line. 
An inspection of the figure reveals the following facts:
(i) $\Omega_\varphi$ approaches the horizon frequency $\Omega_H$ for late times;
(ii) $\Omega_\varphi$ is largest when $\Omega_\theta$ is zero: this corresponds to the particle 
being at the maximum distance from the equatorial plane (either above or below it). Conversely, stationary points 
of $\Omega_\theta$ correspond to minima of $\Omega_\varphi$, when the particle is crossing the equator. 
This is indeed what one would expect from the type of motion shown in Fig.~\ref{fig:ExDynamics} and 
it explains  why the period of the oscillations for the polar frequency is twice the 
azimuthal one; 
(iii) the overall behavior of $\Omega$ is mostly determined by $\Omega^{\rm orb}$, with $\Omega^{\rm so}$ 
driving the oscillations on top\footnote{$\Omega^{\rm orb}$ also contains small oscillations throughout the inspiral 
when the dynamics is non-equatorial: these are barely visible in Fig.~\ref{fig:OmgOrbSo} but they become relevant for 
larger spins. Even then, however, their amplitude is much smaller than that of the oscillations in $\Omega^{\rm so}$, 
since the two grow comparably.}. Only in the late dynamics $(t\gtrsim\tLR)$, when $\Omega^{\rm orb}$ goes to zero, the 
behavior of $\Omega$ is determined by $\Omega^{\rm so}$;
(iv) the positions of the peaks of both $\Omega$ and $\Omega^{\rm orb}$ appear to be reasonably good 
indicators of the light ring crossing.
It should be noted, however, that the presence of a peak in $\Omega$ depends on the delicate interplay 
between $\Omega_\theta^2$  and $\sin^2\theta\,\Omega_\varphi^2$
during the final stages of the dynamics.
This might be illustrated in a simple way as follows. We consider the configuration {\tt a07i75} 
for a range of initial conditions, starting with $r_0=1.4993r_{\rm LSSO}$
and increasing the separation by $\delta r=2.5\cdot 10^{-4}$ until we reach $r_0 = 1.5r_{\rm LSSO}$.
The resulting $\Omega$'s are shown in Fig.~\ref{fig:omgpeak}, with all curves 
aligned at the light-ring crossing:
notice how, depending on the value of $\thend$, the shape $\Omega$ changes dramatically.
Because of this strong dependence on the choice of initial conditions, the peak of the orbital frequency 
cannot be considered a robust dynamical feature to characterize the merger. This is an important
qualitative difference with respect to the spin-aligned case. On the contrary, the pure orbital frequency 
$\Omega^{\rm orb}$ behaves consistently when $r_0$ is modified. 
%=============
% Figure of the peak of Omega disappearing
%=============
\begin{figure}[t]
    \includegraphics[width=.42\textwidth]{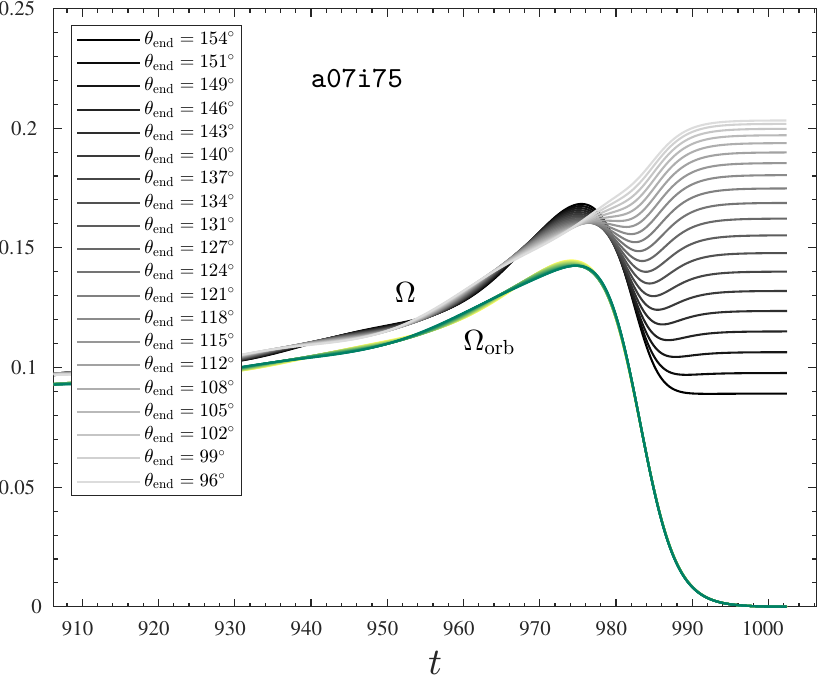}
    \caption{\label{fig:omgpeak}Small modifications in the initial condition $r_0$ (see text)  for the {\tt a07i75} 
    configuration (corresponding to $\theta_0=15^\circ$) yield large qualitative changes in $\Omega$ around merger 
    time due to the final value of $\theta$, $\thend$. In particular, the local maximum of $\Omega$ disappears 
    when $\thend$ becomes sufficiently small.}
\end{figure}
In Ref.~\cite{Harms:2014dqa} it was shown that, for equatorial dynamics, the peak of the purely orbital frequency 
is a good indicator of the light ring crossing for spins $a\lesssim0.8$; beyond this value the profile of $\Omega^{\rm orb}$ 
flattens drastically and the peak gets pushed back quite significantly. It turns out that for large values of the inclination 
parameter $\iota$ this is no longer the case and the peak of $\Omega^{\rm orb}$ is very close to the light ring crossing 
even for $a>0.8$, provided that $\iota$ is large enough. To illustrate this point we report in Table~\ref{tab:OmgOrbPeak} the 
time differences $\Delta t = \tLR - t_{\Omega^{\rm orb}}^{\rm peak}$ for a significative sample of the parameter space $(a,\iota)$. 
\begin{table*}[t]
    \begin{center}
        \begin{ruledtabular}
        \caption{\label{tab:OmgOrbPeak}The table contains time 
        differences $\Delta t = \tLR - t_{\Omega^{\rm orb}}^{\rm peak}$ for a significative sample of the parameter space $(a,\iota)$.
        Notice how delays that are relatively large when $\iota=0$ decrease progressively
        as $\iota$ is increased, up to the order of $M$ or below, i.e. $t_{\Omega^{\rm orb}}^{\rm peak}$ yields an
        increasingly good approximation of the light-ring crossing.}
            \begin{tabular}{c | c c c c c c c c } 
                $a$ & $\iota = 0$ & $\iota = 30^\circ$ & $\iota = 60^\circ$ & $\iota = 85^\circ$ & $\iota = 95$ & $\iota = 120^\circ$ & $\iota = 150^\circ$ & $\iota = 180^\circ$ \\ \hline
                $0.0$ & $0.00$ & $0.00$ & $0.00$ & $0.00$ & $0.00$ & $0.00$ & $0.00$ & $0.00$ \\ 
                $0.1$ & $0.69$ & $0.59$ & $0.34$ & $0.05$ & $-0.06$ & $-0.30$ & $-0.51$ & $-0.59$ \\ 
                $0.2$ & $1.51$ & $1.27$ & $0.72$ & $0.14$ & $-0.13$ & $-0.60$ & $-0.97$ & $-1.10$ \\ 
                $0.3$ & $2.52$ & $2.12$ & $1.12$ & $0.17$ & $-0.12$ & $-0.85$ & $-1.39$ & $-1.55$ \\ 
                $0.4$ & $3.81$ & $3.07$ & $1.57$ & $0.25$ & $-0.18$ & $-1.08$ & $-1.77$ & $-1.95$ \\ 
                $0.5$ & $5.58$ & $4.34$ & $2.06$ & $0.45$ & $-0.21$ & $-1.32$ & $-2.12$ & $-2.31$ \\ 
                $0.6$ & $8.23$ & $6.02$ & $2.89$ & $0.52$ & $-0.16$ & $-1.65$ & $-2.41$ & $-2.64$ \\ 
                $0.7$ & $12.84$ & $8.91$ & $3.06$ & $0.94$ & $-0.57$ & $-2.09$ & $-2.67$ & $-2.93$ \\ 
                $0.8$ & $23.86$ & $16.85$ & $7.05$ & $0.20$ & $-0.87$ & $-2.02$ & $-3.02$ & $-3.19$ \\ 
                $0.9$ & $101.83$ & $38.09$ & $5.64$ & $1.71$ & $0.09$ & $-2.63$ & $-3.35$ & $-3.43$ 
            \end{tabular}
        \end{ruledtabular}
    \end{center}
\end{table*}
It is interesting to note that time delays that are relatively large when $\iota=0$ decrease progressively
as $\iota$ is increased, up to the order of $M$ or below, i.e. $t_{\Omega^{\rm orb}}^{\rm peak}$ yields an
increasingly good approximation of the light-ring crossing.
As such, in view of the construction of an EOB waveform modeling for inclined orbits, it seems 
that $t_{\Omega^{\rm orb}}^{\rm peak}$ may offer a reasonable definition of merger-time 
encoded in the dynamics where the inspiral waveform is attached to some ringdown description.
This interesting subject will possibly be explored in future work.

\subsection{Waveform computation: Teukode summary}
\label{subsec:teuk}

To apply black-hole perturbation theory to our waveform calculation, we solve, in the time-domain, 
the Teukolsky equation~\cite{Teukolsky:1973ha} in the presence of a point-particle source, that is represented by
a $\delta$-function. We use {\Teukode}~\cite{Harms:2014dqa} to
solve the Teukolsky equation in the time-domain. In particular, the code uses horizon-penetrating and 
hyperboloidal coordinates that allow for the inclusion of the horizon and the future null infinity 
in the computational domain~\cite{Zenginoglu:2007jw,Zenginoglu:2009hd,Zenginoglu:2010cq}.
The 3+1 equation is decomposed exploiting the axisimmetry of the Kerr spacetime obtaining a 2+1 TD 
equation for each Fourier $m$-mode in the azimuthal direction. Then the wave equation is solved for 
gravitational perturbations, obtaining the Weyl scalar $\Psi_4$, i.e. the contraction of the Weyl scalar 
with a null-tetrad (the Hawking-Hartle tetrad in our case). Each waveform multipole is then obtained
by a double time integration of the Weyl scalar, since at infinity $\ddot{h}_\lm = \Psi_4^\lm$, 
where $\Psi_4^\lm$ is the multipolar decomposition  of $\Psi_4$. 
The formal Dirac $\delta$ functions present in the source term are approximated using 
narrow Gaussian functions~\cite{Nagar:2004ns,Nagar:2006xv}.
In our simulations, we use  horizon-penetrating, hyperboloidal coordinates with scri-fixing 
at $S=10$ and, typically, a resolution of $N_r\times N_\theta = 3601\times 321$, where $(N_r,N_\theta)$
are the number of points in the radial and angular directions, respectively. 
Only for highly-inclined orbits and some higher multipolar modes this resolution proves insufficient, as discussed in 
the next section: when this is the case it is increased accordingly. 

\section{Nonspinning case: waveform consistency between equatorial and inclined orbits}
\label{sec:schw}
%=============
% Figure of frames
%=============
\begin{figure}[t]
\centering
\includegraphics[width=.48\textwidth]{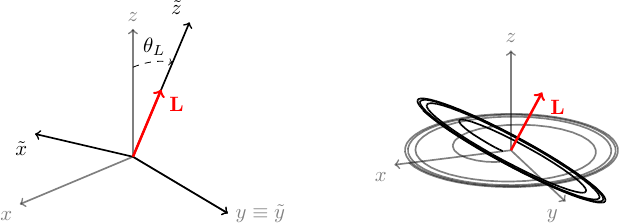}
\caption{\label{fig:axesrot}Comparing the original $(x,y,z)$ frame, in which the decomposition of the waveform in 
spin-weighted spherical harmonics is computed, with the rotated frame  $(\tilde{x},\tilde{y},\tilde{z})$ having the $\tilde{z}$
axis aligned with the direction of the orbital angular momentum $\mathbf{L} = \mathbf{r} \times \mathbf{p}$ of the 
inclined orbit. Right: illustrative example of two orbits inspiralling and plunging from $r_0=6.1$ on the equatorial 
plane (grey) and on a plane inclined by $\iota=\theta_L=\pi/6$ with respect to it (black).}
\end{figure}

%==========================================
% Figure of hpc in the two frames for different inclinations
%==========================================
\begin{figure*}[t]
\begin{center}
\includegraphics[width=.44\textwidth]{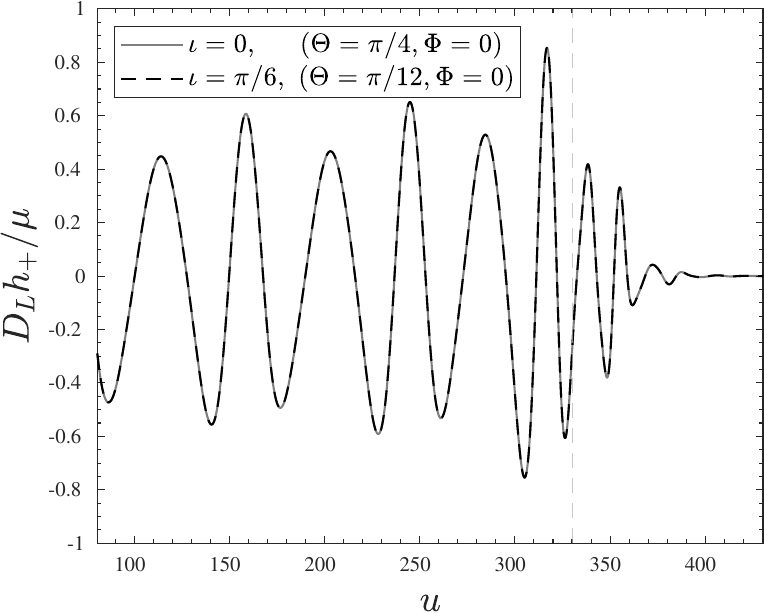}
\hspace{3mm}
\includegraphics[width=.465\textwidth]{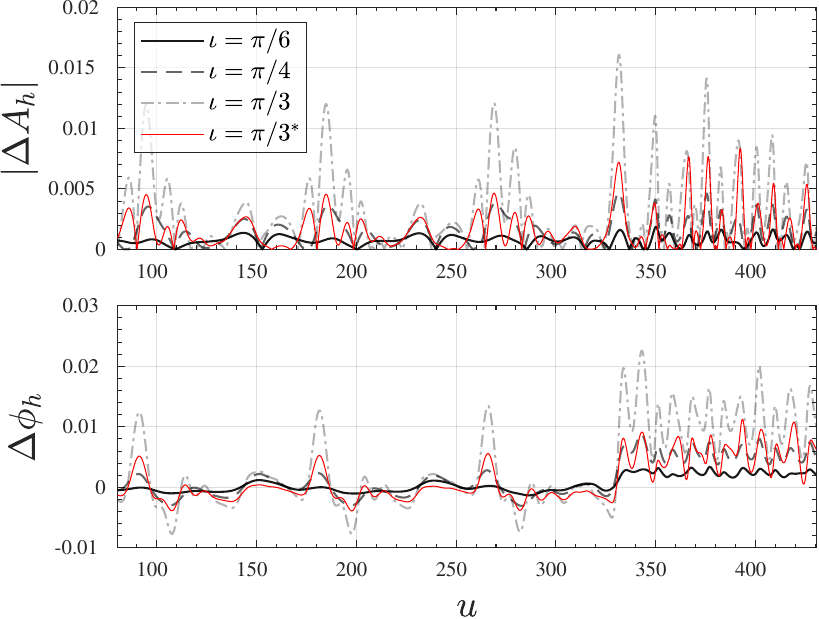}
\caption{\label{fig:schwarzschild_hpc}Plunge on a Schwarzschild black hole. Consistency between the waveform
in the two frames, $(x,y,z)$ and $(\tilde{x},\tilde{y},\tilde{z})$ of Fig.~\ref{fig:axesrot} for three inclination angles.
Left panel: the $D_L h_+/\mu$ polarization of the gravitational waveform including all multipoles up to $\ell=4$ 
for equatorial and inclined dynamics $(\iota = 0,\pi/6)$. We choose $(\Theta=\pi/4,\Phi=0)$ as the fiducial direction 
for the equatorial case and adjust the values of the observational angles such that $(\tilde\Theta=\pi/4,\tilde\Phi=0)$ 
for the inclined configuration; again coordinates identified by a tilde are measured in the $(\tilde x, \tilde y, \tilde z)$ 
frame (see Fig.~\ref{fig:axesrot}). Right panel: relative amplitude differences $|\Delta A_h| \equiv |\tilde A_h-A_h|/A_h$ 
and phase differences $\Delta\phi_h =  \tilde\phi_h -\phi_h$ between inclined and equatorial waveforms for $\iota = (\pi/6, \pi/4, \pi/3)$. 
The discrepancies increase with $\iota$, starting from $\sim 0.1\%$ for $\iota = \pi/6$ up to $\sim 2\%$ when $\iota=\pi/3$.
To show that these differences can be reduced by increasing the size of the numerical grid, we also show in red 
(and denote by an asterisk in the legend) the errors resulting from a finer 
$N_r\times N_\theta = 5401\times 481$ grid: notice how these are now much smaller and become comparable with the $\iota = \tfrac{\pi}{4}$ configuration.}
\end{center}
\end{figure*}

%=================
% FIg. test mode 22, 21
%=================
\begin{figure*}[t]
\begin{center}
    \includegraphics[width=.4\textwidth]{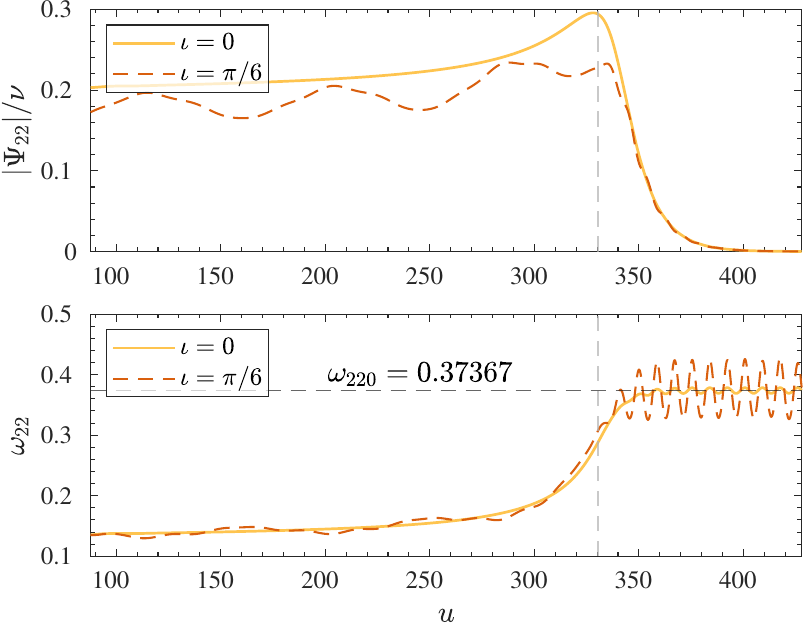} 
    \hspace{5mm}
    \includegraphics[width=.405\textwidth]{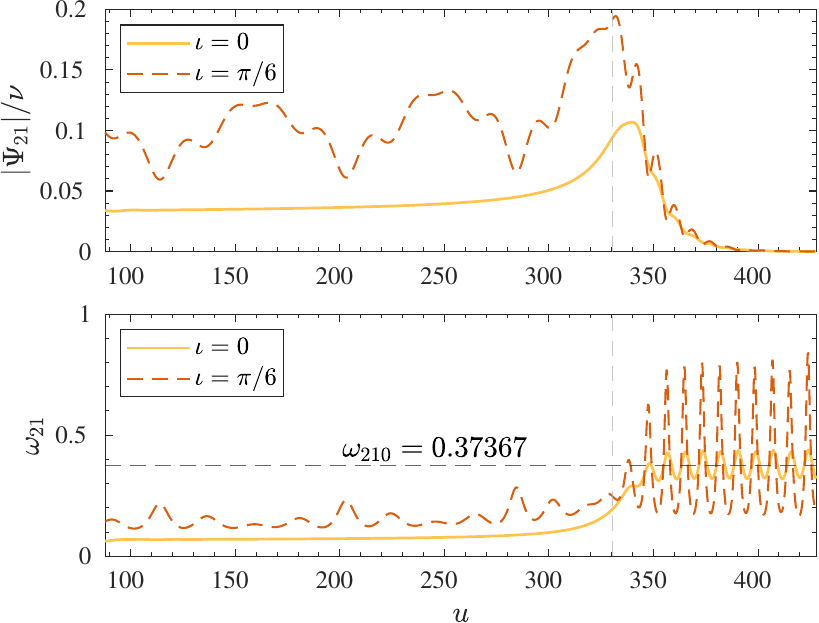}
    \caption{Plunge on a Schwarzschild black hole.  Amplitude and frequency of the $\ell=m=2$ mode (left panel)
    and $\ell=2$, $m=1$ mode for an equatorial configuration ($\iota = 0$) and an inclined one $(\iota = \pi/6)$. 
    For $\iota=\pi/6$, the amplitudes, $|\Psi_{\lm}|$, and frequencies $\omega_{\ell m}$ are modulated by the time-evolution of $\theta$ 
    during  the inspiral, while the well-known oscillations (in amplitude and frequency)~\cite{Damour:2007xr} during the ringdown 
    regime are amplified. Both effects are due to (different) mixing between modes with $\ell=2$ and different $m$, as detailed in the text.}
    \label{fig:schw_nonrot_AF}
\end{center}    
\end{figure*}

The gravitational waveform emitted by a small body of mass $\mu$ inspiralling and plunging on 
a Schwarzschild black hole of mass $M$ with $\nu \ll 1$ has been thoroughly 
studied in the literature, either in the case of quasi-circular
inspiral~\cite{Nagar:2006xv,Bernuzzi:2010xj,Bernuzzi:2011aj,Bernuzzi:2012ku,Nagar:2013sga,Harms:2014dqa,Nagar:2014kha} 
or eccentric inspiral~\cite{Albanesi:2021rby,Albanesi:2023bgi}, 
using Regge-Wheeler-Zerilli~\cite{Regge:1957td,Zerilli:1970se,Zerilli:1970wzz,Nagar:2005ea,Martel:2005ir} 
black-hole perturbation theory.
The waveform is decomposed in multipoles with respect to the natural coordinate frame of the system 
where the binary orbit lays in the $(x,y)$ plane (identified by the polar angle $\theta=\pi/2$) with the $z$ 
direction aligned with the orbital angular momentum, see the light grey plots in Fig.~\ref{fig:axesrot}.
Since our final goal is to describe the waveform emitted by a particle inspiralling and plunging on 
a Kerr black hole along inclined orbits, it is pedagogically useful to first study the features of such systems 
in the simpler non-spinning case:
an example of inclined orbit in Schwarzchild is 
illustrated by the black trajectory in the right panel of Fig.~\ref{fig:axesrot}. 
The motivation for this choice is two fold: (i) on the one hand, it allows us to test the numerical 
accuracy of {\tt Teukode} for inclined orbits, in particular the impact of the angular resolution;
(ii) on the other hand, it will give us a flavor of the complexity of the multipolar waveform as 
measured with respect to the $(x,y,z)$ frame, a complexity that will additionally increase once
the black hole is rotating. 

Non-equatorial orbits in Schwarzschild constitute an optimal testing ground for the accuracy of
the numerical computation. Since the Schwarzschild geometry is spherically
symmetric, all orbital configurations that differ only by the inclination  $\iota$ of the orbital plane are 
physically identical, and thus the waveform $(h_+,h_\times)$ observed along the same direction 
must coincide modulo numerical truncation errors. A simple test of the numerical accuracy is thus 
given by comparing the strain waveform generated by and equatorial plunge with an inclined 
one. Since we are solving numerically a 2D equation in $(r,\theta)$, the inclined result will allow 
us to test the quality of the polar resolution.
We consider three binaries with mass ratio $\nu=10^{-3}$, 
inclination angles $\iota=(\pi/6,\pi/4,\pi/3)$ and initial separation $r_0=6.1$. 
For each value of $\iota$ we solve the Teukolsky equation considering all multipoles up to $\ell=4$.
We work with $N_r=3601$ radial points and $N_\theta=321$ angular points. The strain is obtained from 
Eq.~\eqref{eq:strain} where we choose  $(\Theta=\pi/4,\Phi=0)$ as the fiducial direction for the equatorial 
case and adjust the values of the observational angles such that $(\tilde\Theta=\pi/4,\tilde\Phi=0)$ 
for the inclined configuration; coordinates identified by a tilde are measured in 
the $(\tilde x, \tilde y, \tilde z)$ frame (see Fig.~\ref{fig:axesrot}). 
The left panel of Fig.~\ref{fig:schwarzschild_hpc} compares the $h_{+}$ polarization 
for $\iota=0$ and $\iota=\pi/6$ as seen along the same direction with respect to the 
perpendicular to the orbital plane. In the $(x,y,z)$ frame, the direction $(\tilde\Theta=\pi/4,\tilde\Phi=0)$ 
corresponds to $(\Theta=\pi/12,\Phi=0)$.
The strain, Eq.~\eqref{eq:strain}, is decomposed in amplitude and phase as $h=A_h e^{-i\phi_h}$.
The right panel of Fig.~\ref{fig:schwarzschild_hpc} exhibits the relative amplitude and absolute 
phase differences with respect to the reference equatorial configuration. The phase difference
during ringdown is seen to increase by almost one order of magnitude from $\iota=\pi/6$
to $\iota=\pi/3$. This suggests that for high inclination it is advisable to increase the resolution 
further for production runs. We do so by setting $N_r\times N_\theta = 5401\times481$ and
recomputing the full waveform only for $\iota=\pi/3$, due to the increase in computational cost.
The result is depicted as a red line in the right panel of Fig.~\ref{fig:schwarzschild_hpc}, where
we see that the differences are now comparable to those for $\iota=\pi/4$.
%================
% Consistency for \pi/6
%================
\begin{figure}[t]
\begin{center}
\includegraphics[width=.23\textwidth]{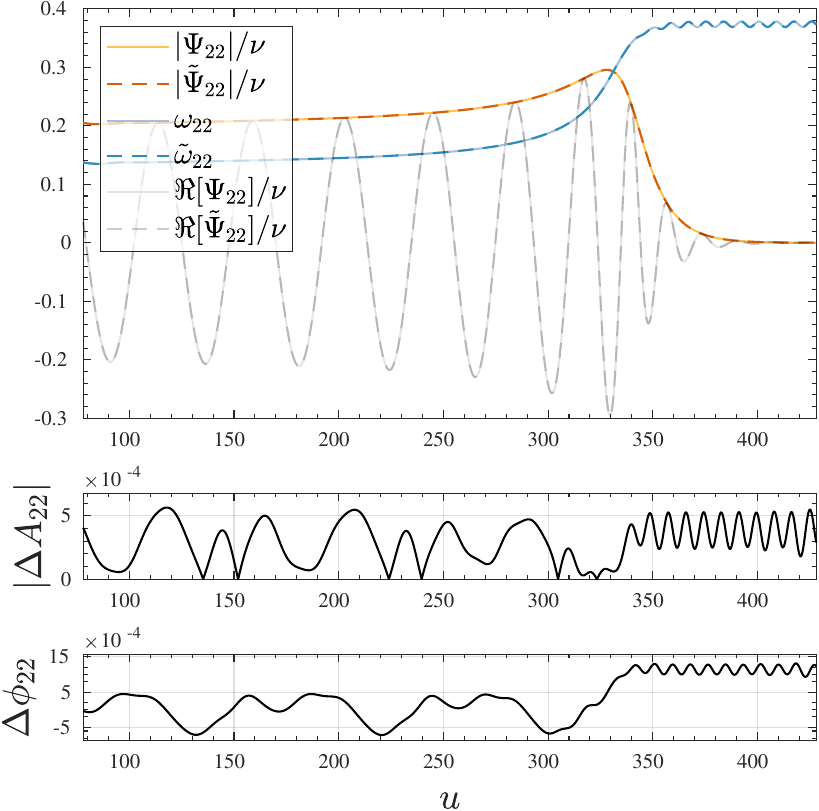} 
\includegraphics[width=.23\textwidth]{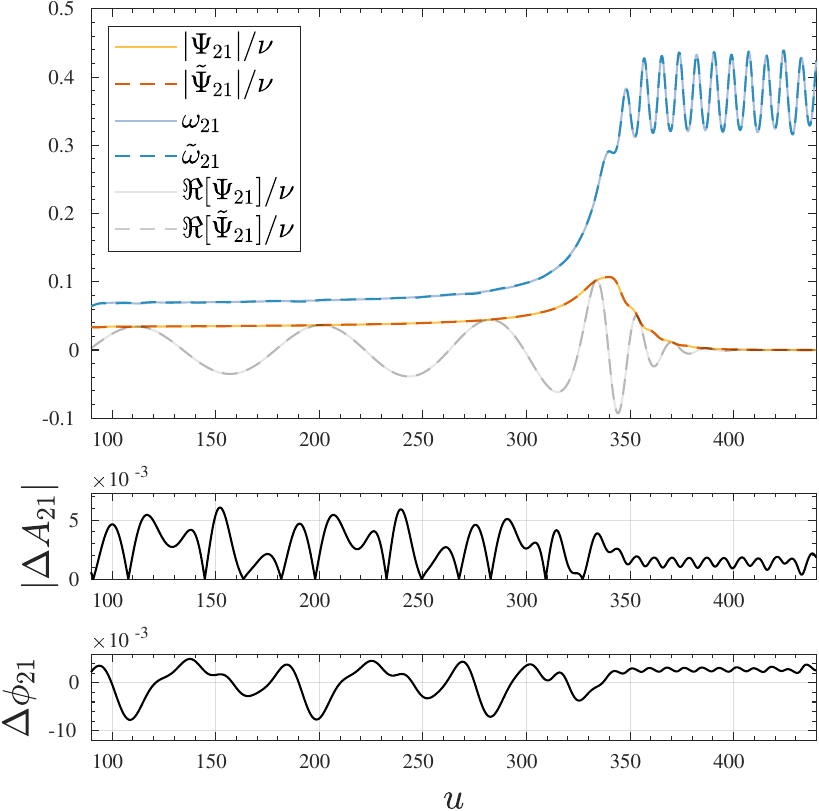}
\caption{Schwarzschild case: consistency between $\iota=0$ and $\iota=\pi/6$ waveforms upon rotation of the latter in the frame $(\tilde{x},\tilde{y},\tilde{z})$, 
where the $\tilde{z}$ axis is aligned with the direction of the orbital angular momentum (see Fig.~\ref{fig:axesrot}). For each multipole we report waveform amplitude, frequency, relative amplitude difference 
$|\Delta A_{2m}|\equiv|\tilde{A}_{2m}-A_{2m}|/A_{2m}$ and phase difference $\Delta\phi_{2m}\equiv \tilde{\phi}_{2m}-\phi_{2m}$.}     
 \label{carrellata1}
    \end{center}
\end{figure}
%
%==================
% Accuracy for pi/6 - pi/3
%==================
\begin{figure}[t]
\begin{center}
    \includegraphics[width=.425\textwidth]{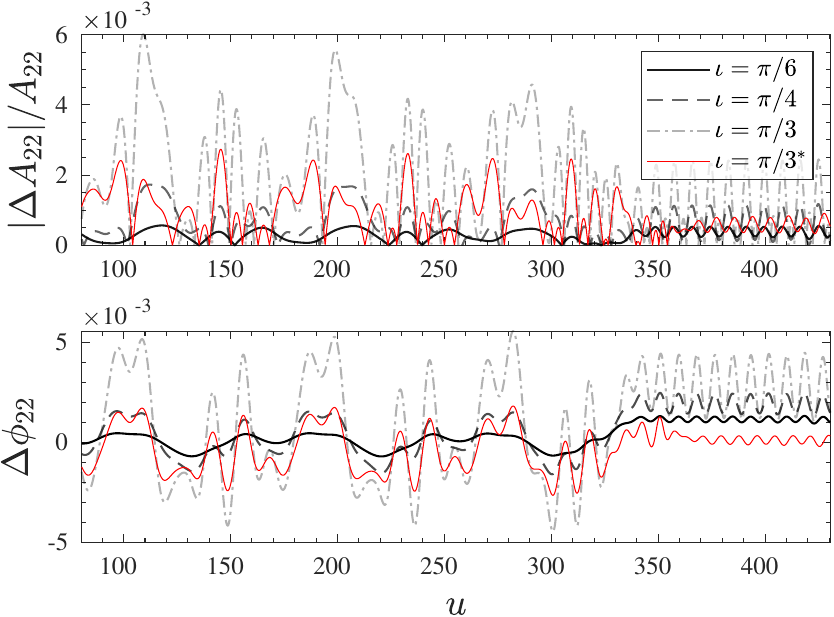}\\
    %\hspace{5mm}
    \includegraphics[width=.425\textwidth]{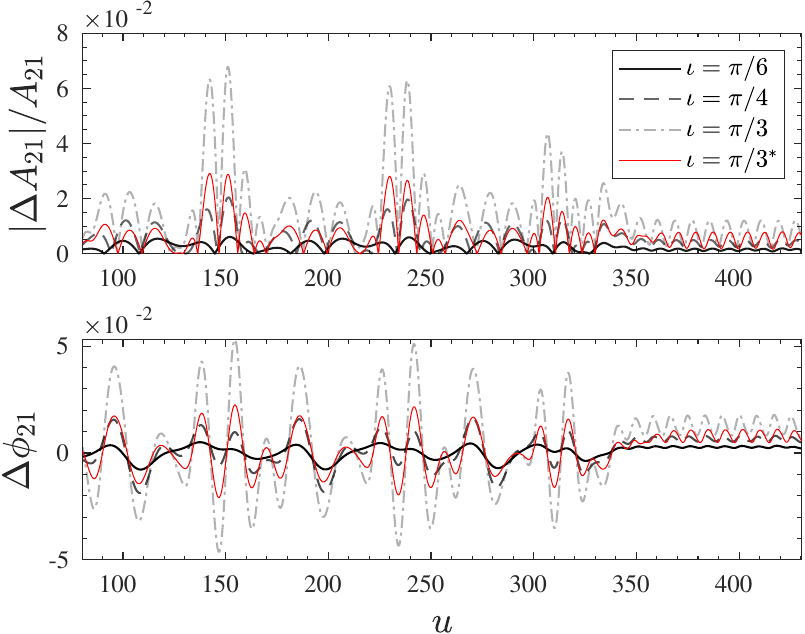}
    \caption{Schwarzschild case: comparing equatorial modes and three inclined configurations with $\iota=(\pi/6,\pi/4,\pi/3)$ after rotation.
    The comparison was performed after rotating the inclined modes in a reference frame where the $z-$axis is aligned with the orbital angular momentum 
    as was done for Fig.~\ref{carrellata1}. All configurations have initial separation $r_0 = 6.1$ and the numerical grid used for the extraction of the 
    waveforms is always $N_r\times N_\theta = 3601\times321$. The top panel shows relative amplitude differences and absolute phase difference 
    for the (2,2) mode, the bottom panel the same values for the (2,1) mode. The numerical errors grow with $\iota$: for moderate inclinations they 
    don't exceed $0.2\%$ and $2\%$ respectively, but as we reach $\iota = {\pi}/{3}$ they grow roughly threefold and start becoming 
    quite significant. We also show in red (and denote with an asterisk in the legend) the the errors obtained with a finer 
    $N_r\times N_\theta = 5401\times481$ grid for this highly-inclined configuration: these are noticeably much smaller and comparable 
    with the ${\pi}/{4}$ case.}
    \label{fig:Errors}
\end{center}
\end{figure}
%====================
% Table of merger quantities
%====================
\begin{table*}[t]
\caption{Comparison of values obtained from our data against those of Ref.~\cite{Harms:2014dqa}. All configurations for which $\iota \neq 0$ 
were rotated in a reference frame where the $z-$axis is aligned with the orbital angular momentum before computing 
the reported values. $\Delta t_{\ell m}$ is the time delay between the light ring crossing and the maximum of the 
mode amplitude $\Delta t_\lm = u_{{A}^\text{max}_\lm} - u_{\rm LR}$, $\hat{A}^{\max}_{\ell m}$ represents the maximum value 
of the amplitude and $\omega^{A^{\max}_\lm}_{\ell m}$ is the mode frequency at $u_{{A}^\text{max}_\lm}$. 
The data are obtained with $N_r=3601$ radial points and $N_\theta=321$ polar points.}
    \label{tab:enno}
    \centering
    \begin{ruledtabular}
    \setlength{\tabcolsep}{2pt}
    \renewcommand{\arraystretch}{1.1}
    \scalebox{1}{
    \begin{tabular}{ll|ccccc|ccccc|ccccc}
        %\hline
        \multicolumn{2}{c|}{} &
        \multicolumn{5}{c|}{\small $\Delta t_{\ell m}$} &
        \multicolumn{5}{c|}{\small $\hat{A}^{\max}_{\ell m}$} &
        \multicolumn{5}{c}{\small $\omega^{A^{\max}_\lm}_{\ell m}$} \\
        $\ell$ & $m$ &
        Ref.~\cite{Harms:2014dqa} & $\iota=0$ & $\pi/6$ & $\pi/4$ & $\pi/3$ &
        Ref.~\cite{Harms:2014dqa} & $\iota=0$ & $\pi/6$ & $\pi/4$ & $\pi/3$ &
        Ref.~\cite{Harms:2014dqa} & $\iota=0$ & $\pi/6$ & $\pi/4$ & $\pi/3$ \\
        \hline
        2 & 2 & $-2.38$ & $-2.39$   & $-2.39$   & $-2.27$   & \iffalse $-1.72$ \fi $-2.07$  & 0.2959    & 0.2957    & 0.2956    & 0.2957    & \iffalse 0.2959  \fi 0.2959   & 0.2733    & 0.2734    & 0.2735    & 0.2744    & \iffalse 0.277  \fi 0.2754\\
        2 & 1 & 9.41    & 9.40      & 9.39      & 9.44      & \iffalse 9.76 \fi 9.57        & 0.1069    & 0.1068    & 0.1070    & 0.1075    & \iffalse 0.1083  \fi 0.1076   & 0.2907    & 0.2907    & 0.2913    & 0.2922    & \iffalse 0.2931 \fi 0.2921\\
        3 & 3 & 1.11    & 1.10      & 1.10      & 1.13      & \iffalse 1.16 \fi 1.13        & 0.0517    & 0.0516    & 0.0515    & 0.0515    & \iffalse 0.0517  \fi 0.0517   & 0.4546    & 0.4547    & 0.4548    & 0.4554    & \iffalse 0.4559 \fi 0.4553\\
        3 & 2 & 6.85    & 6.84      & 6.81      & 6.76      & \iffalse 6.86 \fi 6.84        & 0.0182    & 0.0181    & 0.0181    & 0.0183    & \iffalse 0.0187  \fi 0.0184   & 0.4518    & 0.4518    & 0.4517    & 0.4519    & \iffalse 0.456  \fi 0.4536\\
        3 & 1 & 10.55   & 10.55     & 10.56     & 10.58     & \iffalse 10.59 \fi 10.57      & 0.00569   & 0.00566   & 0.00568   & 0.00574   & \iffalse 0.00597 \fi 0.00582  & 0.4118    & 0.4118    & 0.4118    & 0.4133    & \iffalse 0.4187 \fi 0.4150\\
        4 & 4 & 2.90    & 2.89      & 2.91      & 2.91      & \iffalse 2.92 \fi  2.90       & 0.0146    & 0.0145    & 0.0145    & 0.145     & \iffalse 0.0146  \fi 0.0146   & 0.635     & 0.635     & 0.636     & 0.637     & \iffalse 0.639  \fi 0.637 \\
        4 & 3 & 7.22    & 7.21      & 7.20      & 7.13      & \iffalse 6.97 \fi 7.07        & 0.00496   & 0.00494   & 0.00492   & 0.00495   & \iffalse 0.00509 \fi 0.00502  & 0.637     & 0.637     & 0.637     & 0.636     & \iffalse 0.636  \fi 0.636 \\
        4 & 2 & 9.54    & 9.54      & 9.49      & 9.34      & \iffalse 7.77 \fi 9.88        & 0.00165   & 0.00164   & 0.00163   & 0.00164   & \iffalse 0.00168 \fi 0.00166  & 0.626     & 0.626     & 0.625     & 0.623     & \iffalse 0.582  \fi 0.637 \\
    \end{tabular}
    }
    \end{ruledtabular}
\end{table*}
A complementary and more informative analysis is given by comparing the waveform multipole by multipole.
This requires additional technical steps because, since the multipolar decomposition in spin-weighted 
spherical harmonics of the numerical waveform is performed with respect  to the $(x,y,z)$ frame, 
the individual multipoles for equatorial and inclined orbits will differ. 
Nonetheless, once the multipoles for inclined orbits are expressed in the rotated frame $(\tilde{x},\tilde{y},\tilde{z})$,
defined by the condition that the $(\tilde x,\tilde y)$ plane is perpendicular to the orbital angular 
momentum $\mathbf{L}  = \mathbf{r}\times \mathbf{p}$ (see Fig.~\ref{fig:axesrot}),
they must coincide with those in the $(x,y,z)$ frame modulo numerical truncation errors. 
The rotated modes $\tilde h_{\ell m}$ are obtained from the original $h_{\ell m}$ via Wigner's D-matrices according to 
\begin{equation}\label{eq:new_hlms}
    \tilde h_{\ell m} = \sum_{m'=-\ell}^\ell \left[\mathfrak{D}^{(\ell)}_{m'm}(-\gamma,-\beta,-\alpha)\right]^* h_{\ell m'} \ ,
\end{equation}
where $(\alpha,\beta,\gamma)$ are the Euler angles relating the rotated frame to the original decomposition frame. 
We define the $\mathfrak{D}^{(\ell)}_{m'm}$'s and derive the above formula in Appendix~\ref{app:rotations}.
In the present case the new frame is obtained via a single clockwise rotation of 
$\beta=\theta_L$ about the $y$-axis, where $\theta_L$ is the polar angle identifying 
the direction of $\mathbf{L}$  in the original frame, as %precisely 
illustrated in\footnote{Note that this implies $\iota=\theta_L$. This simple relation no longer 
holds in the case of rotating black holes} Fig.~\ref{fig:axesrot}. 
Note that in this simple, nonspinning case, the angle $\theta_L$ coincides with the inclination $\iota$. 
This of course follows from the fact that the vector $\mathbf{L}$ is perpendicular to the orbital plane, 
and geometrically can be seen by comparing Fig.~\ref{fig:axesrot} with Fig.~\ref{fig:iotadef}.  

%==========================================
% l=2 modes for Schwarzschild with orbit inclined by pi/6
%==========================================
\begin{figure*}[t]
\begin{center}
    \includegraphics[width=.32\textwidth]{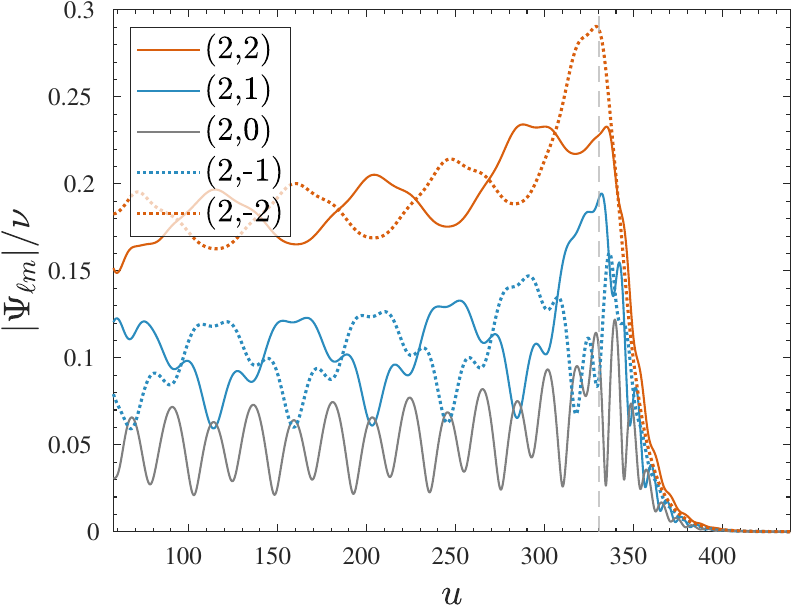}
    \includegraphics[width=.32\textwidth]{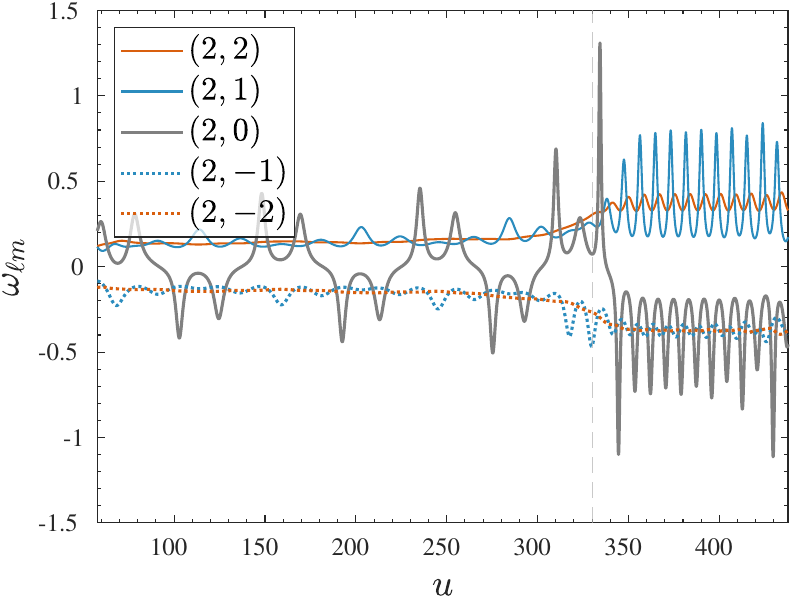}
    \includegraphics[width=.325\textwidth]{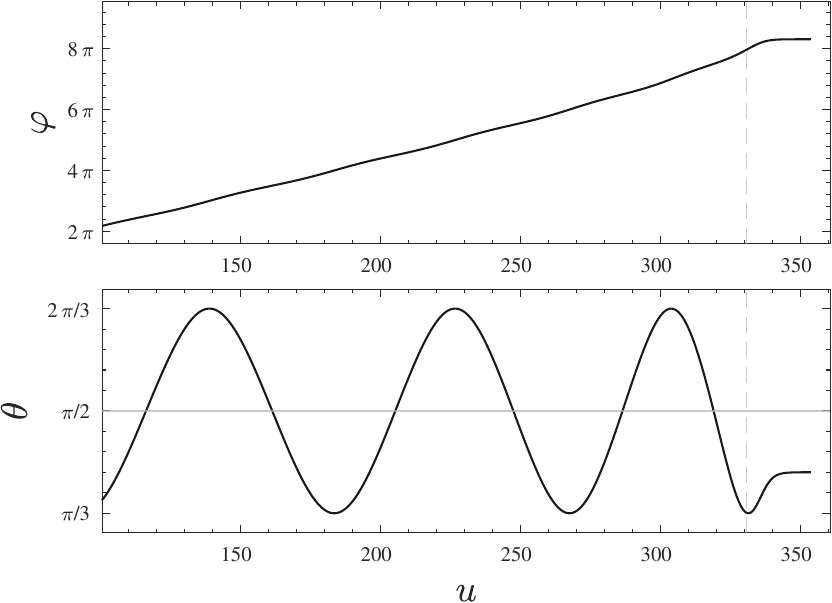}
    \caption{Schwarzschild case. Amplitude and frequency for the $\ell=2$ waveform modes for a test-mass plunging on an inclined orbit with $\iota=\pi/6$. 
    Note the asymmetry between $+m$ and $-m$ modes. The oscillations during the inspiral follow the time-evolution of $\theta$ (rightmost panel).
    The oscillations during the ringdown are due to the mixing of QNMs with positive and negative frequency. The magnitude of the mixing 
    depends on $m$. The off-equatorial location makes the $m=0$ mode complex. The vertical gray line marks the light ring crossing.}
    \label{fig:multipoles}
\end{center}
\end{figure*}
%
%===============
% Ringdown structure
%===============
\begin{figure*}[t]
    \begin{center}
    \includegraphics[width=.4\textwidth]{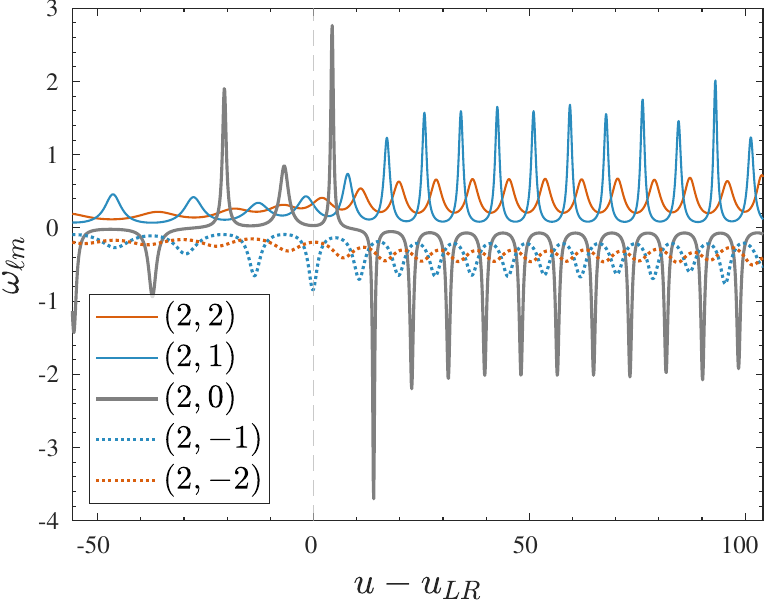}
    \hspace{5mm}
    \includegraphics[width=.415\textwidth]{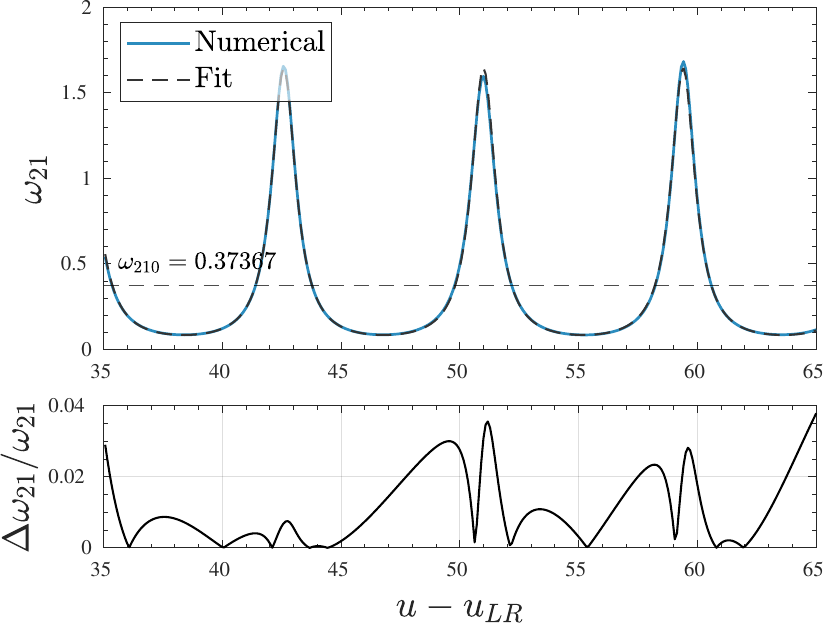}
    \caption{Schwarzschild case. Left: multipolar structure of the ringdown frequency for a highly-inclined dynamics, $\iota=\pi/3$. The hierarchy of the 
    various modes is very similar to that of Fig.~\ref{fig:multipoles}, but the amplitude of the oscillations is larger. Right: performance of 
    the fit of $\omega_{21}$ using the template of Eq.~\eqref{QNM_omegas} on the interval $35<t<65$ after the light ring crossing.}
    \end{center}
\end{figure*}
For consistency with previous literature, we adopt the following normalization
of the waveform multipoles (see Ref.~\cite{Nagar:2005ea}) 
$\Psi_{\ell m} = h_{\ell m}/\sqrt{(\ell+2)(\ell+1)\ell(\ell-1)}$. Each multipole is
decomposed in amplitude and phase as $\Psi_{\ell m}=A_{\ell m} e^{-i \phi_{\ell m}}$
and the corresponding frequency is defined as   $\omega_{\ell m}\equiv \dot{\phi}_{\ell m} = \Im\left(\dot\Psi_{\ell m}/\Psi_{\ell m}\right)$. 
Figure~\ref{fig:schw_nonrot_AF} compares amplitude and frequency of the $\ell=2,\ m=1,2$ 
modes of the equatorial configuration ($\iota = 0$) against the one with $\iota = \pi/6$.
Two qualitative differences are immediately apparent: oscillations in both $A_{\ell m}$ and $\omega_{\ell m}$ 
appear during the inspiral when $\iota=\pi/6$; in addition, the well-known frequency oscillations 
during ringdown~\cite{Damour:2007xr}  are amplified.
To express the waveform in the rotated frame we proceed as follows. We first need to determine 
the Euler angles $(\abc)$ relating the two frames, which allow us to compute the rotated 
modes according to Eq.~\eqref{eq:new_hlms}. 
In this example the rotation matrix entering Eq.~\eqref{eq:new_hlms} is
\begin{equation}
    R_y(\theta_L) = \begin{pmatrix} \cos\theta_L & 0 & \sin\theta_L \\ 0 & 1 & 0 \\ -\sin\theta_L & 0 & \cos\theta_L \end{pmatrix} = R(0,-\theta_L,0) \ ,
\end{equation}
and the full rotation formula for the (2,2) mode reads
\begin{align}
\label{eq:modemixing}
    \tilde{h}_{22} &= d_{22}^{(2)}(\theta_L) h_{22} + d_{12}^{(2)}(\theta_L) h_{21} + d_{02}^{(2)}(\theta_L) h_{20} \nonumber\\
    & + d_{-12}^{(2)}(\theta_L) h_{2-1} + d_{-22}^{(2)}(\theta_L) h_{2-2},
\end{align}
with the coefficients $d^{(\ell)}_{m'm}(\beta)$ being the elements of Wigner's $d-$matrices defined in Eq.~\eqref{eq:Sakurai's_d}. 
Equation~\eqref{eq:modemixing} explicitly shows how, when changing reference system, 
all modes with the same $\ell$ mix and contribute to the resulting one. This mode mixing is the responsible of the 
complicated structure of the waveforms from inclined orbits seen in the $(x,y,z)$ frame.
The $\ell=2,m=1,2$ multipolar waveform modes in the rotated frame are compared to the equatorial ones in Fig.~\ref{carrellata1}.
For each multipole, the bottom panels show the relative amplitude difference and phase difference with the equatorial modes: these are 
of the order of $10^{-4}$ for $m=2$ and grow one order of magnitude for $m=1$.
For the other values of $\iota$ we repeat the same procedure but only show the differences in amplitude and phase (Fig.~\ref{fig:Errors}):
the equatorial/inclined differences are nonegligible when $\iota = \tfrac{\pi}{3}$ especially for the $(2,1)$ mode, 
reaching values of almost $10\%$ in the amplitude.
As we have done in Fig.~\ref{fig:schwarzschild_hpc}, we show here in red the the differences obtained with 
a finer $N_r\times N_\theta = 5401\times481$ grid for this last configuration to show that they become comparable with the 
$\iota=\tfrac{\pi}{4}$ case. 

As a final quantitative analysis of the quality of the inclined waveform we compute three quantities that are useful to
characterize the merger wavefeform~\cite{Albanesi:2023bgi}. These are: (i) the maximum value of the amplitude of each mode, 
$\hat{A}^{\rm max}_{\lm}\equiv A^\mathrm{max}_{\ell m}/\nu$;
(ii) the corresponding value of the (multipolar) gravitational wave frequency $\omega_{\ell m}^{A^\mathrm{max}_{\lm}}$; 
(iii) the time delay between the time where each multipole peak and the light ring crossing, 
$\Delta t_{\ell m} = u_{A^\mathrm{max}_\lm} - u_{\rm LR}$.
The results for the rotated waveforms at all inclinations are reported in Table~\ref{tab:enno}. These numbers are highly 
consistent with those obtained in Ref.~\cite{Harms:2014dqa}, although we see that for $\Delta t_\lm$, and in particular
for $\Delta t_{22}$, the agreement degrades progressively as $\iota$ is increased; again, all values corresponding to 
the highest inclination $(\iota=\tfrac{\pi}{3})$ are obtained using the $N_r\times N_\theta=5401\times 481$ grid.

\subsection{Characterizing the multipolar waveform from inclined Schwarzschild orbits}
\label{sec:schwarzschild_results}
We now turn to a more detailed discussion of the phenomenology of the waveform multipoles emitted by inclined orbits on Schwarzschild. 
Even though, as discussed in the previous section, there is no new physics, many of the effects (specifically due to mode mixing) that shape 
the waveform for inclined orbits on a Kerr spacetime are already present in this simplified scenario: it is thus pedagogically useful to analyze
them in detail so to pave the way to the more complex spinning case.

When discussing the differences between equatorial and inclined waveforms in Fig.~\ref{fig:schw_nonrot_AF} we commented on the effects of 
mode mixing and how it disrupts the well know hierarchy of modes, highlighting that the amplitudes of the (2,2) and (2,1) modes are comparable. 
We also noted qualitatively that the modes coming from inclined orbits exhibit characteristic amplitude and frequency oscillations modulated
by the time-evolution of the polar coordinate $\theta$.
We now complete the analysis by expanding on these observations and inspecting the behavior of other modes: 
in Fig.~\ref{fig:multipoles} we show the amplitudes (left panel) and frequencies (middle panel) of all $\ell=2$ multipoles 
for an inclined configuration with $\iota=\pi/6$. 
The rightmost panel of the same figure shows the time evolution of $(\varphi,\theta)$; the first is monotonically 
increasing, with some small oscillations due to inclination, while the value of $\theta$ oscillates in the range $\pi/3<\theta<2\pi/3$.
Focusing first on the leftmost panel of Fig.~\ref{fig:multipoles} we see rather evident qualitative differences between  
the amplitudes of $(\ell,m)$ and $(\ell,-m)$: when the dynamics takes place in the $(x,y)$ plane, reflection symmetry guarantees 
that\footnote{More precisely, reflection symmetry guarantees that $h_{\ell, -m} = (-1)^\ell \bar{h}_{\ell, m}$, from which the above relation follows. For a detailed discussion on the symmetry-related properties of the multipolar modes, see \cite{Boyle:2014ioa}, Appendix C.} $|h_{\ell m}| = |h_{\ell -m}|$. As the inclination angle 
is increased, this relation is not satisfied anymore, the various modes mix according to Eq.~\eqref{eq:modemixing}, and this results in different 
amplitude oscillations between modes of opposite $m$ on the timescale of the $\theta$ motion. These oscillations can have significant effects 
on the waveform structure not only during the inspiral but also during plunge, merger and ringdown: notice for example how in this scenario 
the peak of the $(2,2)$ mode is reduced with respect to the equatorial case (which will of course affect the distribution of energy 
and angular momentum fluxes between the various modes). 
For the same reason, the $(2,0)$ mode amplitude does not show the usual monotonic behavior, but oscillates 
following the same $\theta$-evolution as it is shaped by the mode-mixing effects.
Such phenomenology is also evident (and even more readable) at the level of the frequency (middle panel). In this case, it is interesting
to focus in particular on the ringdown part. To ease the interpretation of Fig.~\ref{fig:multipoles} let us recall the  basic features of the 
stationary part of the ringdown waveform as explained in Ref.~\cite{Damour:2007xr}. It was pointed out there that black holes can 
be viewed as resonant objects that are non-negligibly excited only during the last part of the plunge, when the source varies 
non-adiabatically (in this respect see also Refs.~\cite{Albanesi:2023bgi,DeAmicis:2024eoy}).
After a transient of about $~20M$ where the presence of an external source term drives the quasi-normal-modes (QNMs) 
excitation, with the low-frequency ones excited first, the ringdown can be described as the 
linear superposition of QNMs with constant coefficients. The specific value of these coefficients (also called excitation
coefficients) depends on the dynamics of the binary. In particular, in Ref.~\cite{Nagar:2006xv,Damour:2007xr} pointed out 
that the gravitational wave frequency of each mode does not simply saturate to the constant value $\omega_{\ell 0}$ of
the fundamental model, but rather oscillates around this value with an amplitude depending on $m$. This oscillation 
comes from the mixing between QNMs with positive and negative frequency.
More precisely, the ringdown waveform can be written as 
\begin{equation}
    \Psi_{\ell m}^{\rm QNMs} = \sum_nC^+_{\ell m n}e^{-\sigma^+_{\ell n}t} + \sum_nC^-_{\ell m n}e^{-\sigma^-_{\ell n}t} \ ,
    \label{SchwRingdown}
\end{equation}
where $n=0,1,2,3,\dots$ indicates the order of the mode and 
$\sigma^{\pm}_{\ell n} = \alpha_{\ell n} \pm i\omega_{\ell n}$ are the complex QNM frequencies 
(where $\alpha_{\ell n}=1/\tau_{\ell n}$ is the inverse damping time and $\omega_{\ell n}$ its oscillatory frequency). 
The $C^{\pm}_{\ell mn}$ are  the corresponding  complex amplitudes (or excitation coefficients) that preserve 
a dependence on $m$, i.e. a dependence on the source ad thus on the dynamics. 
When we consider only the fundamental mode (i.e. during the ringdown part corresponding to the final frequency plateau-like phase) 
from Eq.~\eqref{SchwRingdown} one obtains the following expression for the multipolar gravitational wave frequency 
\begin{equation}
    \omega_{\ell m}^{\rm QNM} = \frac{(1-a_{\ell m 0}^2)\omega_{\ell 0}}{1 + a_{\ell m 0}^2 + 2a_{\ell m 0}\cos(2\omega_{\ell 0}t + \vartheta_{\ell m 0})} \ ,
    \label{QNM_omegas}
\end{equation}
where $a_{\ell m n}e^{i\vartheta_{\ell m n}}\equiv C^-_{\ell m n}/C^+_{\ell m n}$. In the most general case, this formula tells us that $\omega_{\ell m}$ 
oscillates around the value of $\omega_{\ell 0}$, the amplitude of these oscillations determined by $a_{\ell m 0}$: 
in particular if one of the two modes in Eq.~\eqref{SchwRingdown} is much more excited than the other $(a_{\ell m n} \rightarrow 0,\infty)$, 
then $\omega_{\ell m} \rightarrow \pm\omega_{\ell0}$ and the GW frequency plateaus with minimal oscillations. 
On the other hand, when the two modes are equally excited $(a_{\ell m n} \rightarrow 1)$ the instantaneous frequency goes to zero everywhere 
except when $2\omega_{\ell 0}t + \vartheta_{\ell m 0} = (2k+1)\pi, \ k = 0,1,2...$, where the denominator of Eq.~\eqref{QNM_omegas} vanishes.
This allows us to understand the behavior of the frequencies during the ringdown, see the middle panel of Fig.~\ref{fig:multipoles}: 
for modes such as the $(2,1)$ and $(2,0)$ mixing causes the positive and negative ringdown frequencies to be excited comparably, 
thus the amplitudes of the oscillations are greatly increased with respect 
to the equatorial case. The opposite happens for the $(2,-1)$ and the $(2,-2)$ modes.

\begin{table}[t]
    \centering
    \begin{ruledtabular}
     \caption{\label{tab:coeffs}Relative excitation coefficient of the positive and negative 
     frequency fundamental QNM versus inclination angle $\iota$.
     They are obtained fitting the ringdown part of the waveform frequency using Eq.~\eqref{QNM_omegas}.
     As $\iota$ is increased the excitation of the $\sigma_{20}^{-}$ mode becomes stronger for $m>0$
     modes and weaker for $m<0$ modes. This explains quantitatively the oscillations seen in Fig.~\ref{fig:multipoles}.}
    \begin{tabular}{c|cccc}
        & $\iota = 0$ & $\iota = \pi/6$ & $\iota = \pi/4$ & $\iota = \pi/3$ \\
        \hline
        \hline
        $a_{220}$  & 0.00601 & 0.06329 & 0.14091 & 0.28264 \\
        $a_{210}$  & 0.07554 & 0.35658 & 0.48950 & 0.62986 \\
        $a_{200}$  & $\dots$  & 0.32526 & 0.52151 & 0.68778 \\
        $a_{2-10}$ & 0.07554 & 0.04508 & 0.12328 & 0.27234 \\
        $a_{2-20}$ & 0.00601 & 0.00460 & 0.02599 & 0.10074    
    \end{tabular}
    \end{ruledtabular}
\end{table}

%====================
 % Fig for J flux
 %====================
As a final, global, test of our numerical setup we compare the gravitational wave emission of energy and angular momentum from the
LSO and from the light ring in both the equatorial case ($\iota=0$) and inclined case $\iota=\pi/6$. For this particular calculation 
the initial radial separation is $r_0=6.5$. While in the equatorial configuration the angular momentum in gravitational waves is 
emitted entirely along the $z$-direction, in the most general inclined case one expects emission in all directions. 
Specifically, since in our configurations the angular momentum lies entirely in the $(x,z)$ plane, we have 
$\dot{J}_x\neq 0$, $\dot{J}_y=0$ and $\dot{J}_z\neq 0$. 
Nonetheless, the magnitude of the angular momentum flux vector must be identical between the 
two configurations, since the physics described is the same. 
In Table~\ref{tab:Sfluxes} we report the values of energy and total angular momentum emitted in the two configurations, 
both from the LSSO and light ring summing all multipoles up to $\ell=4$. The numbers are consistent at the third digit
despite the additional complexity of the calculation (in terms of sum over modes) yielded by the inclined configuration.

%=================
% Schwarzschild fluxes
%=================
\begin{table}[t]
\begin{center}
\begin{ruledtabular}
\caption{\label{tab:Sfluxes}Consistency test for the Schwarzschild case. Total energy and angular momentum emitted from either the  
LSSO or the LR crossing for either the equatorial configuration, $\iota=0$, or the inclined one, $\iota = \pi/6$. The initial separation
is $r_0=6.5$. As expected, the numbers are consistent although for the $\iota=\pi/6$ configuration the calculation involves more modes.}
    \begin{tabular}{c|c c|c c}
        $\iota$   & $M\Delta E^{\rm LR}/\mu^2$ & $M\Delta E^{\rm LSSO}/\mu^2$ & $\Delta J^{\rm LR}/\mu^2$ & $\Delta J^{\rm LSSO}/\mu^2$ \\ 
        \hline\hline
        $0$       & 0.2181 & 0.6590 & 1.2019 & 6.2364 \\
        $\pi/6$   & 0.2178 & 0.6584 & 1.2006 & 6.2317\\                
    \end{tabular}    
    \end{ruledtabular}
\end{center}
\end{table}

To connect our results with the previous literature we mention that we have checked the agreement with the energy and angular momentum fluxes 
reported in Tab I of Ref.~\cite{Bernuzzi:2010ty}. The authors there identify the quasigeodesic part of the plunge by the condition 
$M\omega_{22}\geq0.167$, which is verified after the LSO but before the LR crossing. 
By implementing the same condition 
with our setup we get, for an equatorial orbit, $M\Delta E^{\rm LR}/\mu^2 = 0.43890$ and $\Delta J^{\rm LR}/\mu^2 = 3.22803$, whereas 
the original authors proposed the values $M\Delta E^{\rm LR}/\mu^2 = 0.47688$ and $\Delta J^{\rm LR}/\mu^2 = 3.48918$ for the same 
mass ratio. The small difference between the values is unsurprising given that the condition $M\omega_{22}\geq0.167$ 
is somewhat vague and that the authors of~\cite{Bernuzzi:2010ty} used an older numerical solver with extraction at finite distance.
We regard the fact that the fluxes agree to the first significant digit as sufficient evidence that our procedure is robust. 

%=========
% Kerr fluxes
%=========
\begin{table}[t]
    \begin{center}
    \caption{\label{tab:Kfluxes}Kerr case. Total emitted energy and angular momentum from either the 
    LSSO crossing or the LR crossing. All waveforms modes up to $\ell=4$ are considered. 
    A general trend emerges of decreasing emission as the inclination increases}
    \begin{ruledtabular}
    \begin{tabular}{l|c|c c|c l}
       $\#$& Name           & $M\Delta E^{\rm LR}/\mu^2$ & $M\Delta E^{\rm LSSO}/\mu^2$ & $\Delta J^{\rm LR}/\mu^2$ & $\Delta J^{\rm LSSO}/\mu^2$\\ 
       \hline\hline
        1 & {\tt a02i0} & $0.2533$ & $0.8370$ & $1.3147$ & $7.0940$ \\ 
        2 & {\tt a02i30} & $0.1971$ & $0.8076$ & $0.9752$ & $6.8963$ \\ 
        3 & {\tt a02i45} & $0.2469$ & $0.7840$ & $1.2999$ & $6.8321$ \\ 
        4 & {\tt a02i60} & $0.2410$ & $0.7523$ & $1.2796$ & $6.7034$ \\ 
        5 & {\tt a02i120} & $0.2115$ & $0.6042$ & $1.1812$ & $5.9379$ \\ 
        6 & {\tt a02i135} & $0.2033$ & $0.5756$ & $1.1495$ & $5.7833$ \\ 
        7 & {\tt a02i150} & $0.1999$ & $0.5569$ & $1.1390$ & $5.6717$ \\ 
        8 & {\tt a02i180} & $0.1951$ & $0.5476$ & $1.1199$ & $5.6983$ \\ \hline
        9 & {\tt a05i0} & $0.3487$ & $1.3322$ & $1.5817$ & $9.0226$ \\ 
        10 & {\tt a05i30} & $0.3469$ & $1.2431$ & $1.5970$ & $8.7234$ \\ 
        11 & {\tt a05i45} & $0.3408$ & $1.1425$ & $1.6159$ & $8.3789$ \\ 
        12 & {\tt a05i60} & $0.3169$ & $1.0092$ & $1.5213$ & $7.8376$ \\ 
        13 & {\tt a05i120} & $0.2272$ & $0.5719$ & $1.2313$ & $5.6972$ \\ 
        14 & {\tt a05i135} & $0.2050$ & $0.5064$ & $1.1452$ & $5.3199$ \\ 
        15 & {\tt a05i150} & $0.1887$ & $0.4615$ & $1.0834$ & $5.0524$ \\ 
        16 & {\tt a05i180} & $0.1766$ & $0.4280$ & $1.0376$ & $4.8469$ \\ \hline 
        17 & {\tt a09i0} & $0.7558$ & $4.1985$ & $2.3377$ & $16.2620$ \\ 
        18 & {\tt a09i30} & $0.8405$ & $3.5991$ & $2.6678$ & $14.9434$ \\ 
        19 & {\tt a09i45} & $0.8407$ & $3.0289$ & $2.8555$ & $14.0220$ \\ 
        20 & {\tt a09i60} & $1.0336$ & $2.5788$ & $3.7659$ & $13.3298$ \\ 
        21 & {\tt a09i120} & $0.5017$ & $0.8160$ & $2.4227$ & $6.7031$ \\ 
        22 & {\tt a09i135} & $0.4450$ & $0.6868$ & $1.9106$ & $5.6446$ \\ 
        23 & {\tt a09i150} & $0.2865$ & $0.4883$ & $1.3032$ & $4.7046$ \\ 
        24 & {\tt a09i180} & $0.1890$ & $0.3636$ & $0.9537$ & $4.1087$ \\ 
    \end{tabular}
    \end{ruledtabular}
    \end{center}
\end{table}

%===================
% a=0.5, two inclinations
%===================
\begin{figure*}
\begin{center}
    \includegraphics[height=38mm]{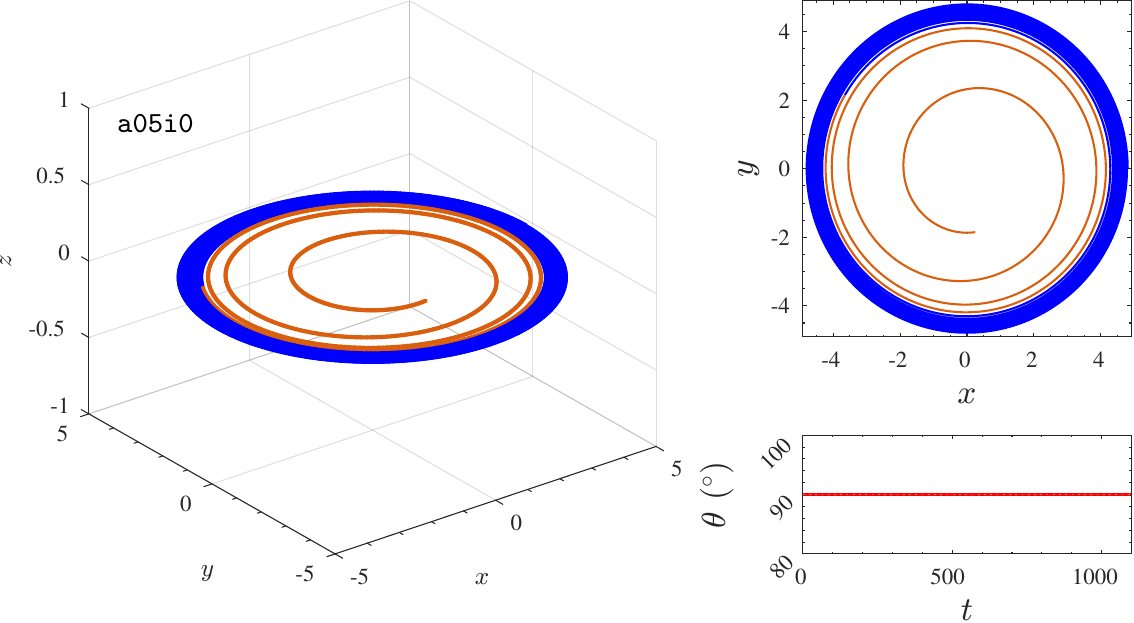} \hspace{1mm}
    \includegraphics[height=39mm]{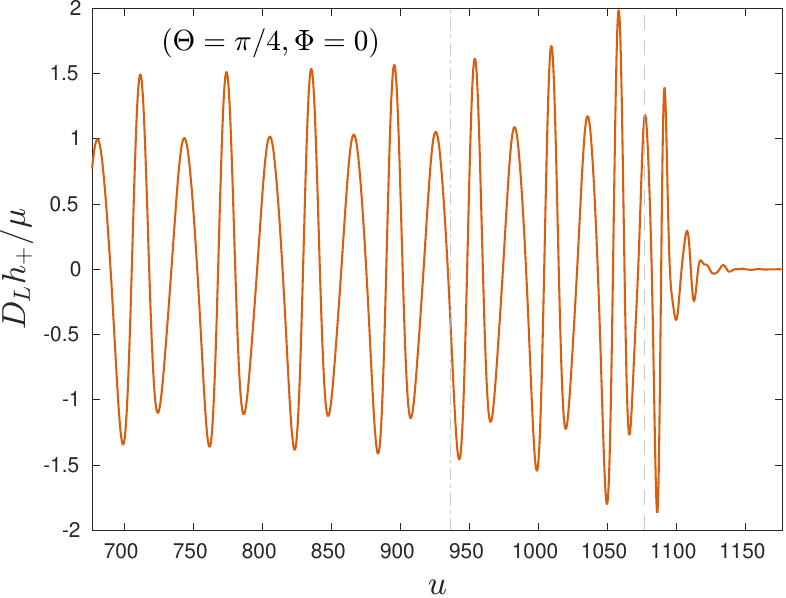} \hspace{2mm}
    \includegraphics[height=38mm]{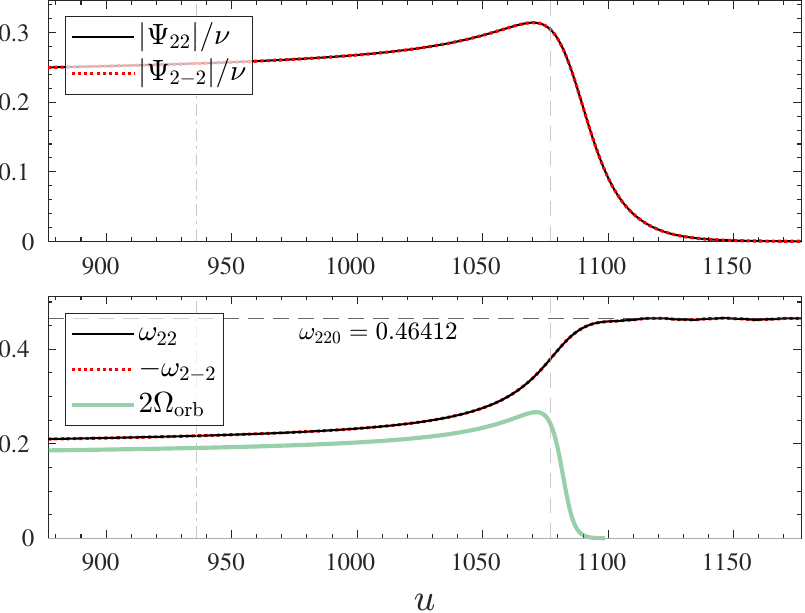} \\
    \vspace{.5cm}
    \includegraphics[height=38mm]{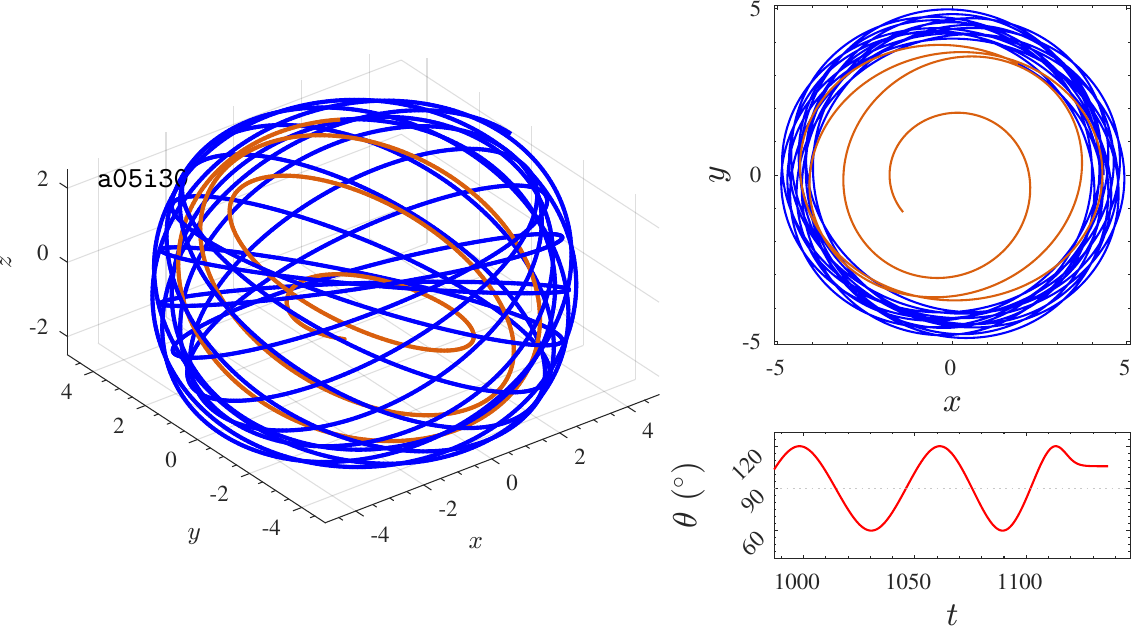} \hspace{1mm}
    \includegraphics[height=39mm]{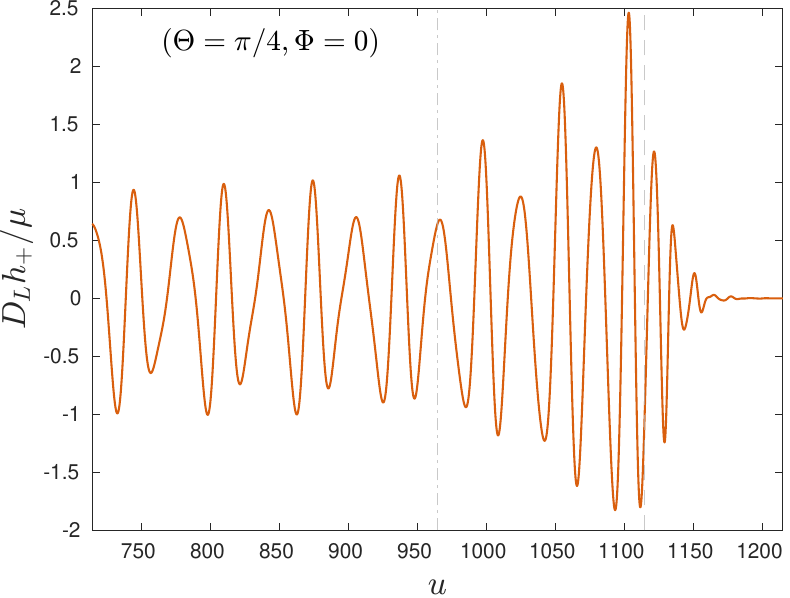} \hspace{2mm}
    \includegraphics[height=38mm]{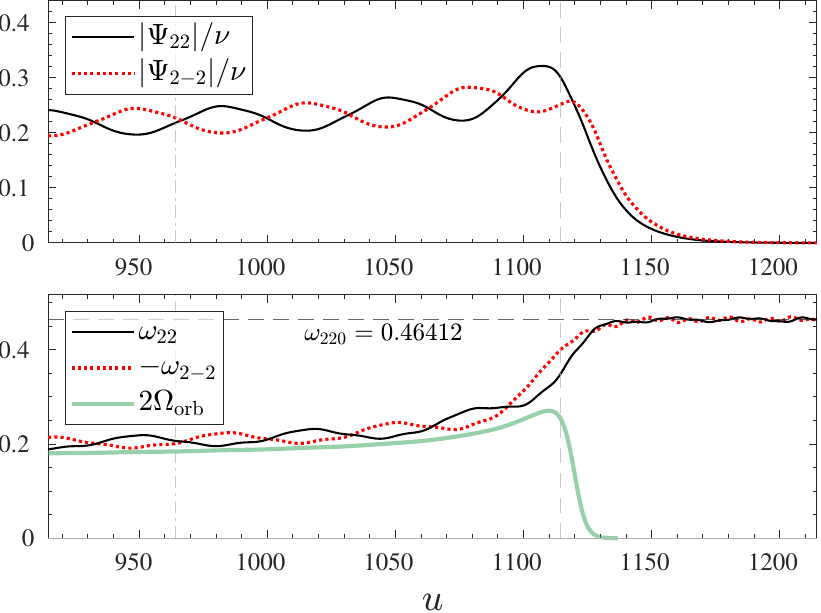} \\
    \end{center}
\caption{\label{fig:a05_comparison}Characterization of the dynamics and main features of the waveform for two 
spinning configurations with $a=0.5$, one equatorial (top) and one inclined by $\iota=\pi/6$ (bottom). 
From left to right we report: (i) the full 3D trajectory and its projection on the $(x,y)$ plane, with the portion after 
the LSSO crossing highlighted in red, as well as the last stages of the $\theta$ evolution; (ii) the real part of 
the gravitational wave strain $h_+$, obtained summing all modes up to $\ell=4$, as observed from the 
fiducial direction $(\Theta=\pi/4,\Phi=0)$: the two vertical lines represent the LSSO crossing (dot-dashed) 
and the light ring crossing (dashed); (iii) the amplitude and frequency of the $\ell=2, m=\pm 2$ modes 
of each configuration. A horizontal dashed line denotes the value of the fundamental QNM ringdown 
frequency, while the light ring and LSSO crossing (when visible) again identified by vertical gray lines. 
The green curve in the frequency plot represents twice the orbital frequency of the binary.}
\end{figure*}

\section{Transition from inspiral to plunge: black hole spin and orbital inclination}
\label{sec:insplunge}
Let us move now to discussing the case of plunge from inclined orbits on a spinning black hole. 
The binary parameter space is covered as follows. We consider three values of the (positive) 
spin $a=(0.2,0.5,0.9)$ and inclination angles varying from \hbox{$0$°$\leq \iota \leq 180$°}. 
With this convention, the cases with a component of the black hole spin anti-aligned with the orbital
angular momentum correspond to $\iota>90$°. In total we simulated 24 different configurations 
labeled as reported in the first column of Table~\ref{tab:Kfluxes}. 
For each case we consider all multipoles up to $\ell=4$. As an illustrative example of the waveform 
phenomenology induced by orbital inclination, in Fig.~\ref{fig:a05_comparison} we contrast 
the equatorial configuration with $a=0.5$ with one with $\iota = \pi/6$. The qualitative differences 
shown in this case remain substantially the same either increasing $\iota$ or $a$, so that this
analysis can be considered general. The dynamics and waveforms for  all other configurations 
of Table~\ref{tab:Kfluxes} are shown in Appendix~\ref{app:catalog}, Figs.~\ref{fig:a02prog}--\ref{fig:a09retr}, 
to which we address the reader for a comprehensive vision of our results.

\begin{figure*}
\begin{center}
    \includegraphics[width=.31\textwidth]{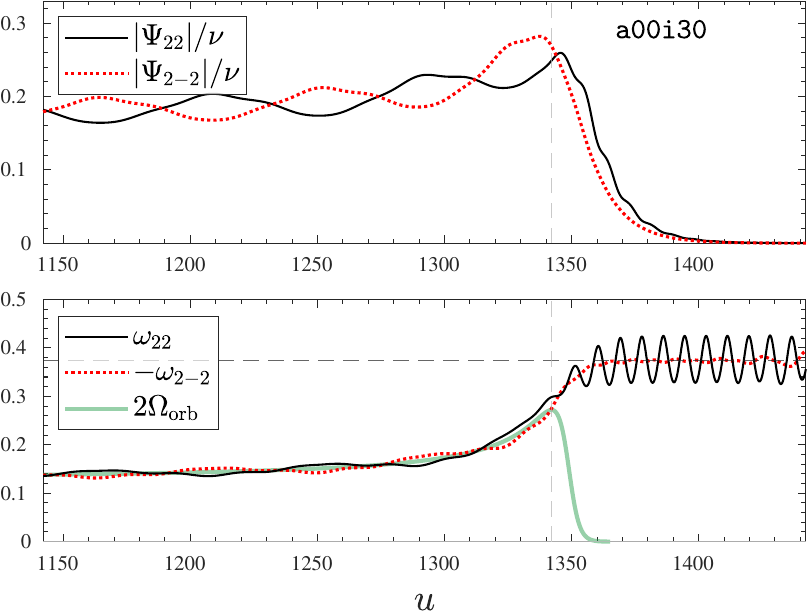}
    \hspace{2mm}
    \includegraphics[width=.31\textwidth]{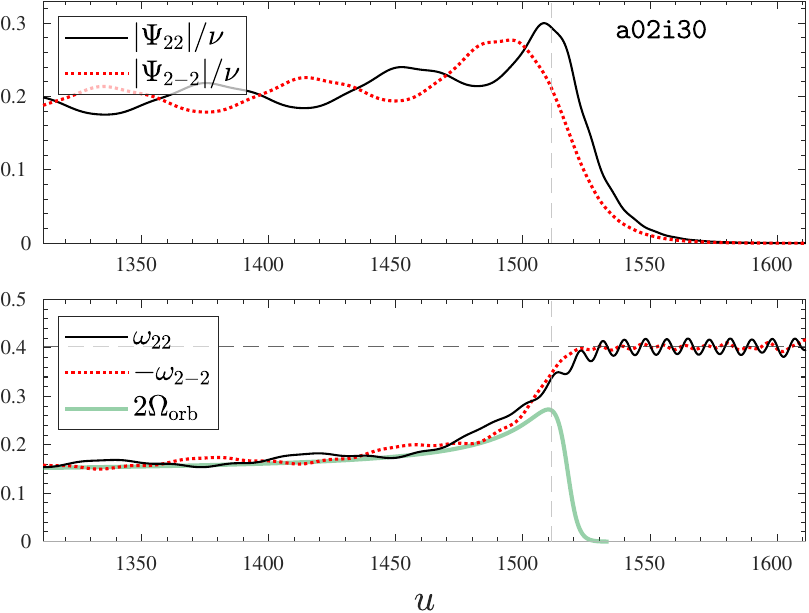} 
     \hspace{2mm}
    \includegraphics[width=.32\textwidth]{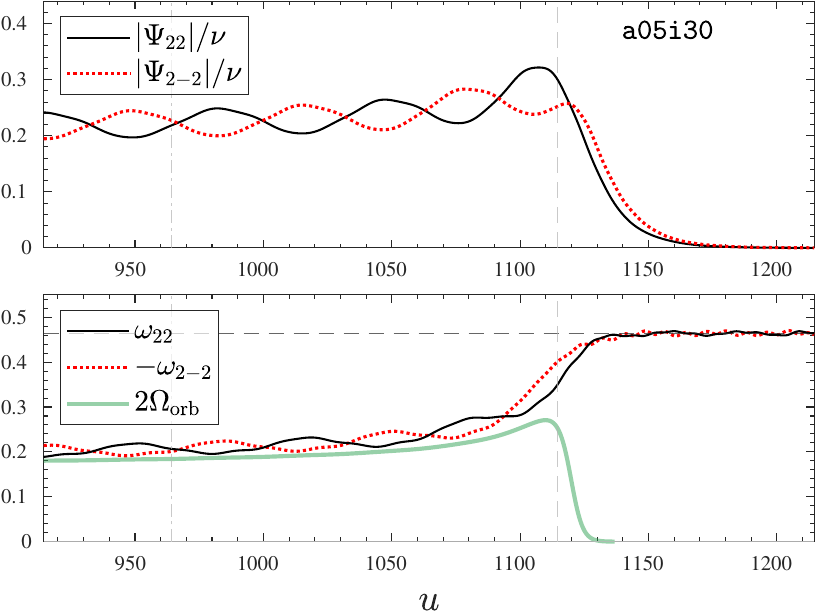}
\end{center}
\caption{\label{fig:i30_comparison}
    Amplitude and frequency of the $(2,\pm 2)$ modes of the waveform emitted 
    by a range of configurations, all with the same inclination $\iota=\pi/6$ but with increasing spin $a\in(0,0.9)$.
    A horizontal dashed line denotes the value of the fundamental QNM ringdown frequency, while the light ring and LSSO crossing 
    (when visible) are identified by a vertical gray line. The green curve in the frequency plot represents twice the modulus of the 
    pure orbital frequency, Eq.~\eqref{eq:Omg_orbAmp}, that always peaks close to the light-ring crossing.
    The modulations in amplitude and frequency are visually very similar for small values of the spin parameter $a\lesssim 0.5$, 
    but become much more pronounced for high spin values: as the amplitude of these modulations increases, 
    the asymmetry between the positive- and negative-$m$ modes consequently becomes larger. }
\end{figure*}

The left panels of Fig.~\ref{fig:a05_comparison} show the full 3D trajectory and its projection on 
the $(x,y)$ plane, with the portion after the LSSO crossing highlighted in red. The last stages 
of the time-evolution of $\theta$ are also shown. In the middle panel we report the $h_+$ polarization
of the gravitational wave strain. This is obtained summing all modes up to $\ell=4$, and is observed 
from the fiducial direction $(\Theta=\pi/4,\Phi=0)$. 
The two vertical lines indicate the LSSO crossing (dot-dashed) and the light ring crossing (dashed). 
Finally, the rightmost panels illustrate the amplitude and frequency of the $\ell=2, m=\pm 2$ modes 
of each configuration: a horizontal dashed line denotes the value of the fundamental QNM ringdown frequency, 
while the light ring and LSSO crossing (when visible) are again identified by vertical gray lines. 
The green curve in the frequency plot represents twice the amplitude of the pure  orbital frequency 
vector of the binary, as defined in Eq.~\eqref{eq:OmgOrb}. Focusing on the rightmost plots, 
we see how the main difference between equatorial and non-equatorial orbits are the modulations in amplitude 
and frequency: these are visually very similar to the ones shown in Fig.~\ref{fig:schw_nonrot_AF}, which we recall 
are related to the inclination of the orbital plane with respect to the observer. When the central black hole 
is spinning we expect these modulations to be amplified proportionally to (i) the value of $a$ and (ii) 
the angle between $\bi{a}$ and $\mathbf{L}$~\cite{Kidder:1995zr}.
In order to illustrate this point, in Fig.~\ref{fig:i30_comparison} we compare amplitude and frequency of the $(2,\pm 2)$ modes of the 
waveform emitted a few configurations with different values of $a$ but with the same inclination $\iota=\pi/6$.
One notes that the frequency and amplitude modulations, that are driven by the time-evolution of $\theta$, are rather
similar for small values of the spin parameter $a\lesssim 0.5$. They become progressively more pronounced for
higher spin values (see in particular Figs.~\ref{fig:a09prog}-\ref{fig:a09retr} in Appendix~\ref{app:catalog} that refer to the $a=0.9$ case). 
In addition, as the amplitude of these modulation increases, the asymmetry  between 
the positive- and negative-$m$ modes becomes larger. 
%
%===================
% i=30, four spins
%===================

The asymmetry between $+m$ and $-m$ modes is the genuine characteristic of the dynamics on inclined orbits around
a Kerr black hole and can be quantified rigorously~\cite{Boyle:2014ioa}. In particular, although the waveform phenomenology
looks close to the one of the Schwarzschild waveform, the absence of the background spherical symmetry due to spin precession
entails that the oscillations cannot be completely removed via a rotation of the reference frame.
As pointed out in great detail in Ref.~\cite{Boyle:2014ioa}, their physical origin is that, for precessing systems, 
there is no symmetry related to reflection across the orbital plane (as there is for spin-aligned binaries). As a consequence we can't relate 
modes with positive and negative value of $m$ in any reference frame (not even in the coprecessing frame) and the asymmetry between $(\ell,\pm m)$ 
modes gives rise to a residual oscillation.

\begin{figure}
    \begin{center}
        \includegraphics[width=.232\textwidth]{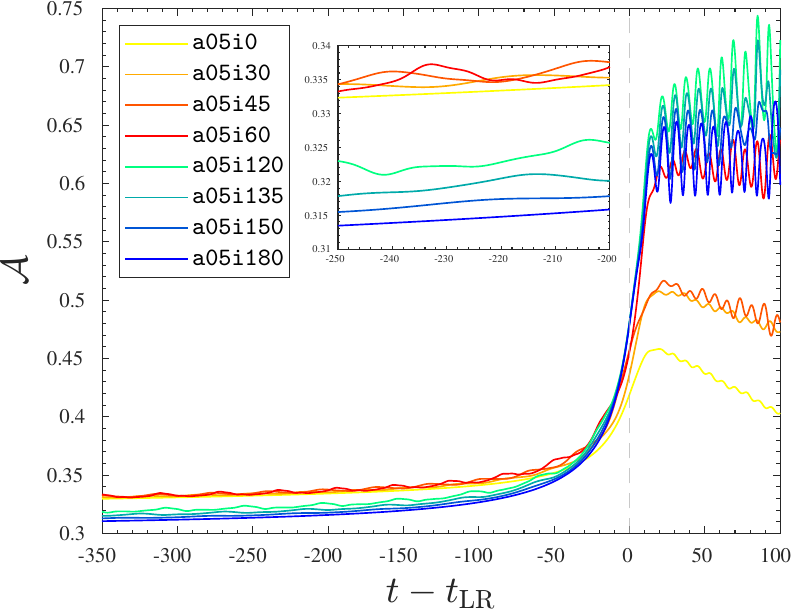}
        \includegraphics[width=.23\textwidth]{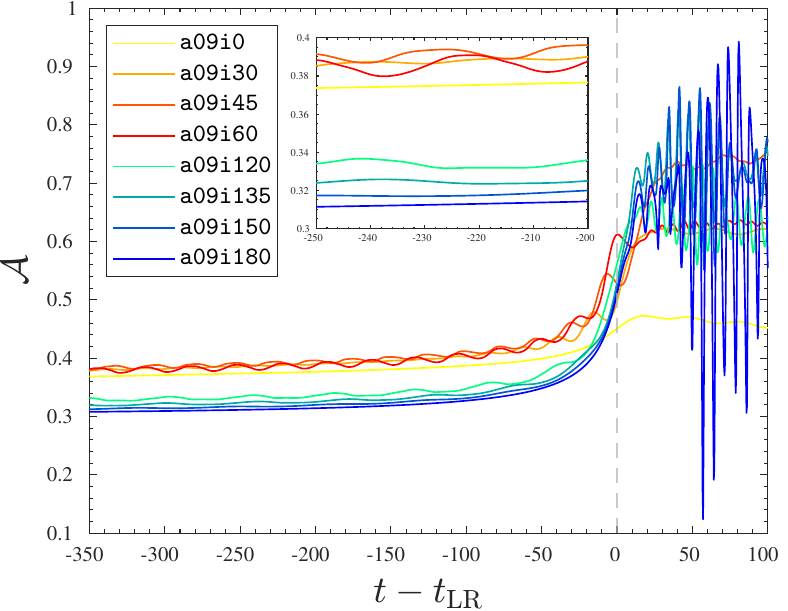}
    \end{center}
    \caption{\label{fig:asym}Time evolution of the asymmetry ${\cal A}$, Eq.~\eqref{eq:asymmetry}, for a range of configurations with spin $a=0.5$ (left) and $a=0.9$ (right), 
    including multipolar modes up to $\ell=4$. We use warm colors (red through yellow) if the underlying orbit is prograde and cold colors (blue through green) 
    if it is retrograde: during the inspiral we can visually distinguish the two, as the prograde configurations tend to have slightly higher asymmetry. 
    The insets show that, in each case, the equatorial configuration is the one with lowest $\mathcal{A}$, which then increases with $\iota$. 
    }
\end{figure}

This was proven rigorously in Ref.~\cite{Boyle:2014ioa}: the authors there define the following {\it asymmetry}
\begin{equation}
\label{eq:asymmetry}
    {\cal A} \equiv \sqrt{\frac{\sum_{\ell,m} \left| h_{\ell,m} - (-)^{\ell+m} {h^*}_{\ell,-m} \right|^2}{4 \sum_{\ell,m} \left| h_{\ell,m} \right|^2}},
\end{equation}
which is a rotationally invariant measure of the difference between $\pm m$ modes
\footnote{The fact that $(-)^{\ell + m}$ appears in Eq.~\eqref{eq:asymmetry} as opposed to $(-)^\ell$ is essential to have invariance under rotations.}. 
This quantity is nonzero for precessing systems and for precessing systems if the masses or spins of the black holes are unequal.
However, the fact that it is rotationally invariant justifies the assertion (for inclined orbits) that no 
frame exists in which modes with opposite $m$ can be related. Note that this does \textit{not} necessarily mean that the oscillations in 
a single mode cannot be removed: it just means that even if one were to find some complicated reference frame in which, say, 
the $(2,2)$ mode exhibits no oscillations, we'd observe them twice as large in the $(2,-2)$ mode.
In Fig.~\ref{fig:asym} we show the time-evolution of ${\cal A}$ for a range of configurations with spins $a=0.5$ and $a=0.9$. All curves
are aligned at the light-ring crossing. The inset focus on the inspiral part and highlights that ${\cal A}$ increases with the inclination angle 
for both prograde and retrograde orbits. Similarly, the magnitude of the asymmetry grows with the spin magnitude.

\section{Approximate waveforms from time-dependent rotations}
\label{sec:twist}

Now that we have analyzed the phenomenological features of the waveforms from inclined orbits, we explore the possibility of approximating them 
by rotating the modes of a given spin-aligned configuration. To do this we define two reference frames: an \textit{inertial} frame, which corresponds 
to the fixed $(x,y,z)$ frame in which the multipolar decomposition is originally performed, and a \textit{coprecessing frame}, which 
has the advantage of smoothing out many of the complex features characterizing the precessing waveforms. 
This frame will be rigorously defined shortly. We will then use the Euler angles relating the two frames to rotate 
the multipolar modes of one configuration and relate them to those of the other. 
This procedure is conceptually well known and considered standard, as it is routinely 
used to compute semi-analytical waveforms for precessing systems in the comparable-mass 
case~\cite{Akcay:2020qrj,Gamba:2021ydi,Ramos-Buades:2023ehm,Khalil:2023kep,Gamba:2024cvy,Hamilton:2025xru}. 
It stems from the observation that the waveform modulations due to spin-orbit coupling, as well as the related change of orientation
of the orbital plane with respect to an observer, can be reduced (though not totally eliminated, because of the intrinsic absence of
equatorial asymmetry, as mentioned above) by performing a change of frame~\cite{Kidder:1995zr} (see also~\cite{Boyle:2011gg})

\begin{figure}[t]
    \begin{center}
    \includegraphics[width=.45\textwidth]{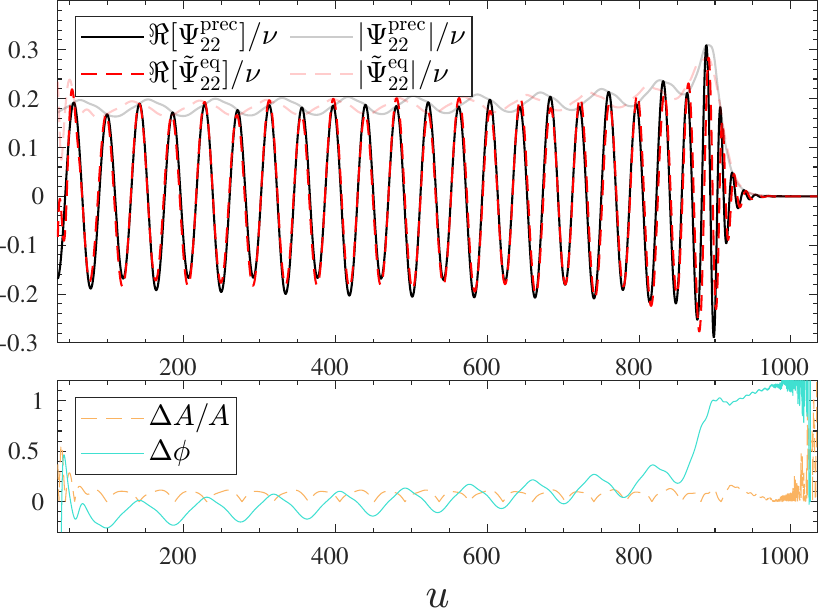} 
\end{center}
    \caption{\label{fig:twist02}Comparison between the exact $\ell=m=2$ waveform produced by an 
    inclined dynamics ({\tt a023i30}) and the corresponding approximate one, obtained upon rotation 
    of the spin-aligned modes ({\tt a02i0}). The two waveforms are aligned in the early inspiral, 
    in the interval $[u_L,u_R]=[100,500]$ fixing a relative time and phase shift 
    (see Ref.~\cite{Baiotti:2011am}) by minimizing the phase difference on that interval.
    The phase difference $\Delta\phi\equiv \tilde{\phi}^{\rm eq}-\phi^{\rm prec}$ 
    increases steadily versus time throughout the inspiral, reaching $0.5$~rad just before 
    merger: this is not surprising since the two underlying relative dynamics are different. 
    By contrast, note that this procedure entails that the relative amplitude difference hovers around $10\%$.
    }
\end{figure}

\begin{figure}[t]
    \begin{center}
    \includegraphics[width=.45\textwidth]{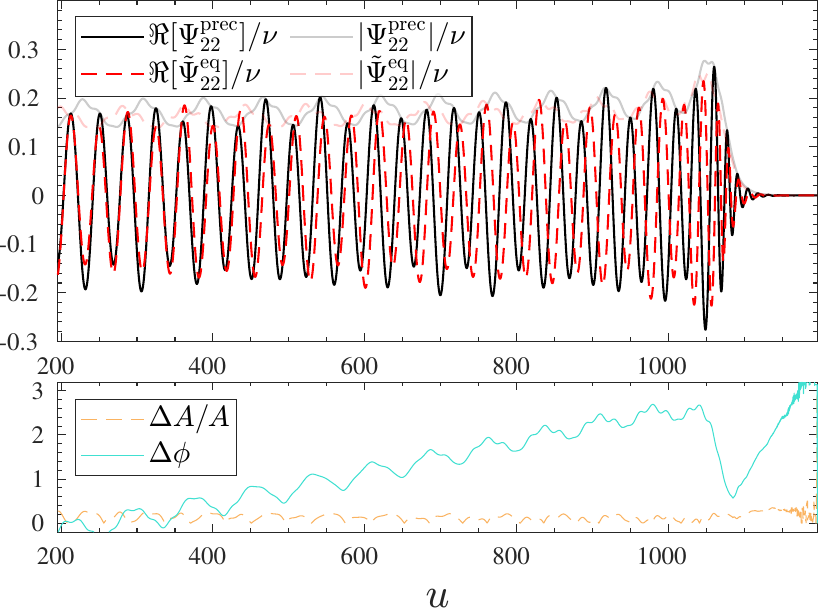} 
\end{center}
    \caption{\label{fig:twist05}Comparison between the exact waveform produced by an inclined dynamics ({\tt a05i45}) 
    and the approximate one, obtained upon rotation of the spin-aligned modes ({\tt a035i0}). The waveforms are aligned 
    in the interval $[u_L,u_R]=[100,500]$. We notice much greater differences both in phase 
    (which reaches almost 3 radians before merger) and amplitude (constantly around $25\%$) with respect to Fig.~\ref{fig:twist02}.
    This is a priori expected since both inclination and the spin are larger here.}
\end{figure}

The method to obtain approximate spin-precessing waveforms from spin-aligned ones was introduced long ago and 
extensively explored in the context of PN-theory, Numerical Relativity or semi-analytical models~\cite{Buonanno:2002fy,Schmidt:2010it,Schmidt:2012rh,Pan:2013rra,Estelles:2025zah}. 
However, for the case of a test-mass inspiralling and plunging on a Kerr black hole it is discussed here for the first time.
Although the current state-of-the-art spin precessing EOB models are rather sophisticated,
a look to the performance of the procedure in the test-mass limit might be useful to give a fresh cut
to the problem in a context where the dynamics and waveforms are consistent among themselves.
Various prescriptions for the choice of coprecessing frame exist in the literature.
For example, Refs.~\cite{Schmidt:2010it,Schmidt:2012rh} define it by tracking the direction of maximum emission 
of energy of the $\ell=2,m=\pm2$ modes of the binary: in this frame the system is essentially observed ``head-on'' throughout its evolution 
and consequently the hierarchy of modes characteristic of planar orbits is restored. A similar, though more generic, 
approach is the one proposed by O'Shaughnessy \textit{et al.}~\cite{OShaughnessy:2011pmr}. 
The features of these two approaches are analyzed in great detail in~\cite{Boyle:2011gg}, 
where the authors further introduce a \textit{minimal rotation condition} in order to remove the arbitrariness of choosing the orientation of 
the coprecessing frame. The idea of tracking the direction of maximum emission is clearly well-adapted to the study of numerical 
waveforms, since in that case the underlying dynamics is not necessarily known.
In the context of semi-analytical models, and in particular the EOB approach, Pan {\it et al.}~\cite{Pan:2013rra} 
suggested that a physically-reasonable choice of coprecessing frame is obtained by aligning the $z$ axis with the 
Newtonian angular momentum vector $\mathbf{L_N}\propto \mathbf{r}\times \dot{\mathbf{r}}$ at every instant of time, 
the underlying reason being that $\mathbf{L_N}$ identifies the instantaneous orbital plane of the binary. 
The authors there also mention that using $\mathbf{L}\propto \mathbf{r}\times {\mathbf{p}}$ 
as opposed to $\mathbf{L_N}$ to perform the rotation yields very similar results, but it is 
impossible to know \textit{a priori} which frame better encapsulates the effects of 
precession\footnote{We briefly investigate this point in Appendix~\ref{app:antitwist_geod}.}.
In this final section we follow the reasoning outlined in Ref.~\cite{Pan:2013rra}: our coprecessing frame is 
identified by the (instantaneous) direction of the angular momentum of the precessing binary $\mathbf{L}$ along 
with the minimal-rotation condition of Ref.~\cite{Boyle:2011gg} $\dot \gamma = -\dot \alpha \cos\beta$. 
(here $\{\alpha,\beta,\gamma\}$ are the Euler-angles that specify the rotation 
relating the original frame and the co-precessing frame). 
For the reasons outlined at the beginning of this section, we can consider the spin-aligned waveform as being produced by a precessing dynamics 
observed in the coprecessing frame: consequently, we can produce an approximate waveform 
by applying an inverse rotation $\{-\gamma,-\beta,-\alpha\}$ to the spin-aligned modes using the transformation 
rule~\eqref{eq:new_hlms} derived in Appendix~\ref{app:rotations}.
The accuracy of the procedure is then evaluated by comparing the resulting modes with the exact ones obtained from the 
precessing dynamics in the inertial frame\footnote{In Appendix~\ref{app:antitwist_geod} we also explore the accuracy of the inverse transformation, 
i.e. the non-equatorial waveform is rotated into the coprecessing frame and compare the resulting
modes with those produced by an equatorial dynamics.}.
The precessing dynamics and the spin-aligned dynamics (used to generate the waveform assumed to be in the coprecessing
frame) have to be generated consistently as follows. Since the precession of the orbital plane is mainly driven by the spin-orbit 
coupling in Eq.~\eqref{eq:Hso}, which contains the product $\bi{a}\cdot \mathbf{L}$, we require it to be the same at $t=0$
for both the inclined and equatorial configuration.
In practice, the equatorial and inclined configurations are connected by the condition
\be
\label{eq:aeq_aprec}
[a]^{\rm equatorial}_{t=0}=[a\cos\theta_L]^{\rm inclined}_{t=0} \ ,
\ee
where all other initial parameters are kept identical. Let us remind that $\theta_L$ is the initial polar angle identifying 
the direction of ${\bf L}$ and we have $\theta_L=\iota$ at $t=0$. With this relation, an inclined orbit configuration with 
$a=0.23$ and $\iota=\pi/6$ (dubbed {\tt a023i30}) should be (approximately) reproduced by rotating an equatorial waveform with
$a=0.2$ (dubbed {\tt a02i0}) . Similarly, an inclined configuration with $a=0.5$ and $\iota=\pi/4$ 
(dubbed {\tt a05i45}) should be obtained rotating an equatorial configuration with $a=0.35$ (dubbed {\tt a035i0}).
These two configurations should be considered explorative of the various effects involved increasing the 
spin and the inclination. A more comprehensive and systematic analysis is postponed to future work.
The comparison between equatorial-rotated and inclined waveforms for the parameters mentioned above
is displayed in Fig.~\ref{fig:twist02} for {\tt a023i30} versus {\tt a02i0} and in Fig.~\ref{fig:twist05} for
{\tt a05i45} versus {\tt a035i0}.
The top panel compares the real parts and amplitudes of the $\ell=m=2$ modes, where we indicate with $\Psi_{22}^{\rm prec}$ 
the waveform obtained from the precessing dynamics and with $\tilde{\Psi}_{22}^{\rm eq}$ the equatorial one
after rotation. Since the dynamical evolution is different (despite the initial conditions being the same) the two
waveforms are aligned during the inspiral in the interval $[u_L,u_R]=[100,500]$  by fixing an appropriate relative 
time and phase shift by minimizing the root-mean-square of the phase difference over this interval.
See e.g.~Ref.~\cite{Baiotti:2011am}, Eqs.~(30)-(31) therein for the details of the procedure.
Both waveforms are decomposed in amplitude and phase as $\Psi_{22}^{\rm prec}=A e^{-i\phi}$ and
as $\tilde{\Psi}_{22}^{\rm eq} =\tilde{A}e^{-i\tilde{\phi}}$, where the tilde always indicates equatorial
quantities rotated. The phase difference $\Delta\phi\equiv \tilde{\phi}^{\rm eq}-\phi^{\rm prec}$ 
and the relative amplitude difference $\Delta A/A \equiv |\tilde{A}^{\rm eq}-A^{\rm prec}|/A^{\rm prec}$ between
the rotated and the precessing waveform is reported in the bottom panels of each figure.
Figure~\ref{fig:twist02} shows that, when the spin and inclination are sufficiently small, the straightforward
application of the rotation procedure to the corresponding equatorial waveform is such to yield a rather
good approximation of the full precessing waveform, including also merger and ringdown.
By contrast, as the spin and inclination are increased, the quantitative agreement worsens.
This is a priori not surprising considering that (i) the effect of spin-spin coupling becomes stronger; (ii) the
larger variations of $\theta_L$ during the evolution imply that the identification~\eqref{eq:aeq_aprec} 
is less and less accurate as time evolves, affecting both the conservative Hamiltonian dynamics and the radiation reaction (
the latter through the spin-orbit term $f_{\rm 3SO}$ in Eq.~\eqref{eq:f3SO}).
A careful understanding of all these effects requires a dedicated study where each spin-dependent 
element of the Hamiltonian and radiation reaction is switched on and its effect analyzed. This would be
in particular the case of spin-spin effects, that could be analyzed progressively by PN-expanding 
$H_{\rm SO}$. Since this would require dedicated {\tt Teukode} simulations, it will be explored in future work.
The current explorative analysis seems however to suggest that the application of the rotation procedure
in the test-mass plunge, though promising, might be tricky and it success all over the parameter space
is not a priori evident.

%===========
% Twist: a=0.23
%===========

%==========
% twist:a=0.5
%==========

\section{Conclusions}
\label{sec:conclusions}
We have thoroughly studied the gravitational waveform emitted during the transition from quasi-spherical inspiral to plunge, merger 
and ringdown of a point-particle plunging on a Kerr black hole from inclined orbits. We do so using black hole perturbation
theory (notably solving the Teukolsky equation in the time-domain using {\tt Teukode}~\cite{Harms:2014dqa}) with 
a point-particle source whose inspiral is driven by an analytical, EOB-resummed-type, radiation reaction force. 
The current study can be considered as the latest development of the approach started 
with Refs.~\cite{Nagar:2006xv,Damour:2007xr,Bernuzzi:2010ty,Bernuzzi:2010xj,Bernuzzi:2011aj,Bernuzzi:2012ku}
for the quasi-circular plunge on a Schwarzschild black hole and then prosecuted with several studies including 
spin aligned with the orbital angular momentum~\cite{Harms:2014dqa,Nagar:2014kha} 
and eccentricity~\cite{Albanesi:2021rby,Albanesi:2023bgi,DeAmicis:2024not}.
The dynamics and waveforms from inclined orbits was recently studied also in Refs.~\cite{Lim:2019xrb,Hughes:2019zmt,Lim:2022veo} 
with special emphasis on the QNM content of the ringdown. The approach to evolve the dynamics implemented in Ref.~\cite{Lim:2019xrb} 
is based on a generalized Ori-Thorne~\cite{Ori:2000zn} algorithm (that requires a certain number of ad-hoc prescriptions),
while here we adopt the standard Hamiltonian approach popular within the EOB literature. 
This allows for a natural transition from the quasi-spherical inspiral to plunge and merger that 
is (i) substantially independent of the choice of radiation reaction force because of its intrinsic 
nonadiabatic character and (ii) does not need to fix additional parameters\footnote{It must be noticed that 
Ref.~\cite{Lim:2019xrb} is careful to stress that the main results of the paper, i.e. the QNMs excitation coefficients, are unaffected
by the details of the generalized Ori-Thorne algorithm.} as in Refs.~\cite{Lim:2019xrb}.
Our results can be summarized as follows.
\begin{itemize}
\item[(i)]We have implemented the transition from quasi-spherical inspiral to plunge from inclined orbits using the Hamiltonian 
of Ref.~\cite{Balmelli:2015zsa} and a (resummed) version of the radiation reaction of Ref.~\cite{Buonanno:2005xu}. This radiation
reaction is not such to give an accurate description of the phasing during the inspiral, but it should just be seen as a tool to 
drive the transition from the quasi-spherical inspiral to plunge and merger. Since our focus on this paper is on characterizing the
waveform during the plunge from inclined orbits, our results are in fact insensitive to the specific choice of radiation reaction.
\item[(ii)]We thoroughly tested the accuracy of our approach in the Schwarzschild case, considering orbits inclined by $(\pi/6,\pi/4,\pi/3)$ 
with respect to a given direction and evaluating in detail the impact of the resolution on the waveform accuracy. 
We did so mainly performing a rotation and transforming the multipoles in the equatorial frame into the inclined, frame and comparing 
them with the waveform multipoles obtained for equatorial orbits. This eventually allowed us to have uncertainties in the 
waveform below $1\%$ in amplitude and phase for any inclination angle considered.
The study of the plunge on Schwarzschild from inclined orbits allowed us to perform several other quantitative tests 
(e.g. computation of the losses and characterization of the merger) and to explore the complex phenomenology of the 
multipoles, like mode mixing effects. This pedagogical exercise proved useful to gain a deeper understanding of the 
waveform structure for inclined orbits on Kerr.
\item[(iii)]We have computed the waveform for a sample of configurations with black hole spin $(0.2, 0.5, 0.9)$ 
and inclination angles between $(0,\pi)$, extracting all waveform multipoles up to $\ell=4$. We have computed the losses
through the plunge phase (i.e., after the crossing of the LSSO) and quantified that as $\iota$ grows the contribution of the modes 
with $m\neq \ell$ progressively becomes dominant with respect to the $m= \ell$ ones. 
\end{itemize}
These results prove the accuracy and robustness of our approach to obtaining waveforms from
test-masses inspiralling and plunging on a Kerr black hole along quasi-spherical orbits.
The current study should be considered as a first step towards a systematic exploration
of the modeling of the radiation reaction and waveform for strongly precessing systems.

In future works, we plan to improve the consistency of the fluxes used to drive the dynamics,
and extend our study to a larger sample of configurations, including non-spherical ones.
We will produce a large catalog of waveforms, that will be made publicly available,
and explore the possibility of using them to inform semi-analytical models of precessing binaries
in the comparable-mass case.

\begin{acknowledgements}
We are grateful to P.~Rettegno for developing the first version of the numerical code for the 
time-evolution of the 3D dynamics. This original code was developed at IHES, that we thank 
for hospitality. We also thank G.~Faggioli for discussions.
S.~B. aknowledges support by the EU Horizon under ERC Consolidator Grant, 
no. InspiReM-101043372. S.~A. and S.~B. acknowledge support from the Deutsche
Forschungsgemeinschaft (DFG) project ``GROOVHY'' (BE 6301/5-1
Projektnummer: 523180871).

\end{acknowledgements}

\appendix
\section{Consistency of radiation reaction force}
\label{app:RR}
%===========================
% Check mechanical loss vs infty flux
%===========================
\begin{figure*}[t]
    \includegraphics[width=.30\textwidth]{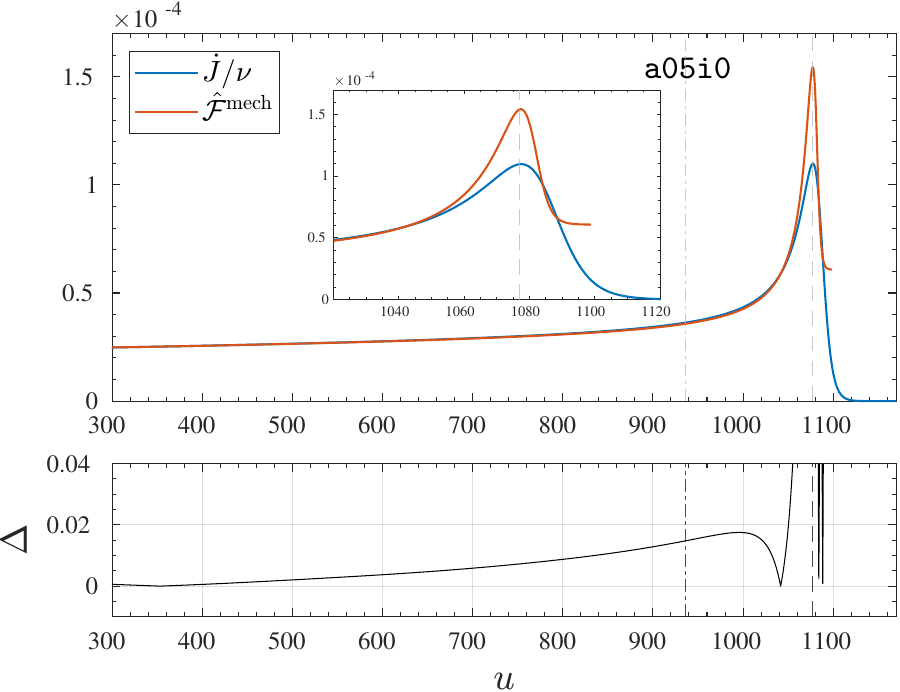}
    \hspace{2mm}
    \includegraphics[width=.30\textwidth]{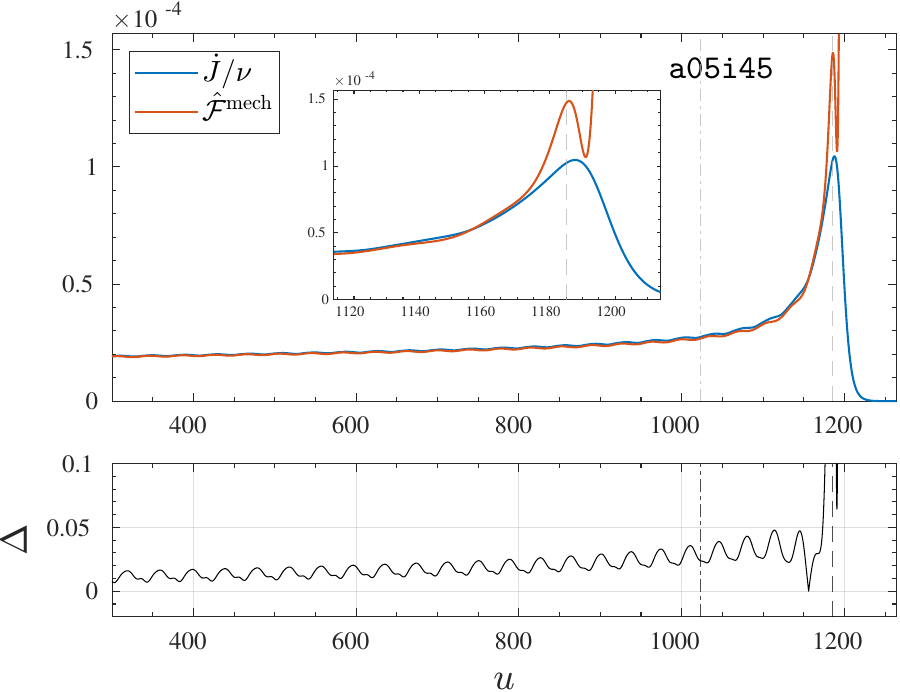}
    \hspace{2mm}
    \includegraphics[width=.30\textwidth]{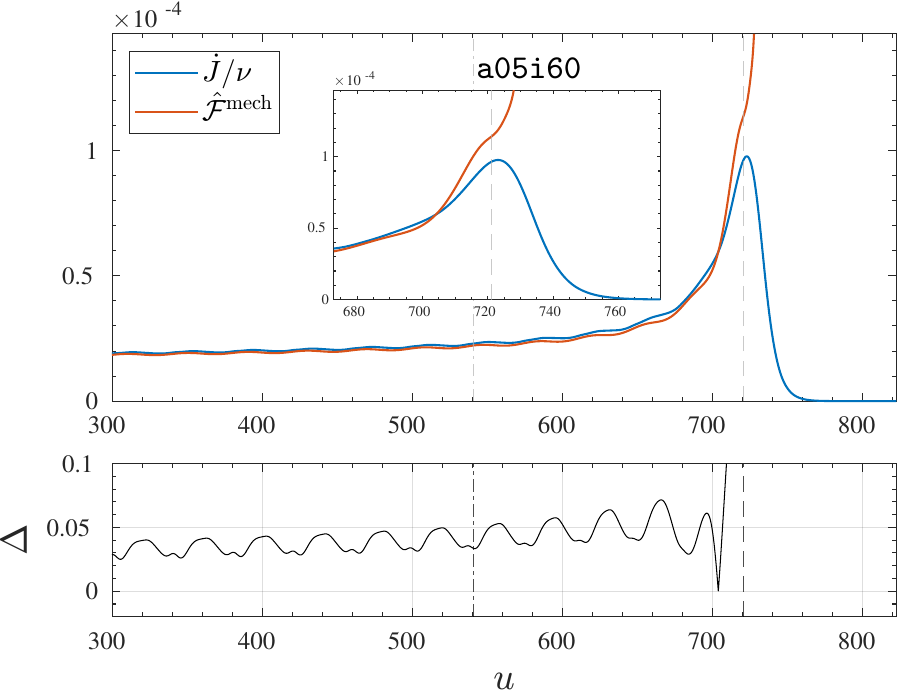}
    \caption{\label{fig:mech-exact-fluxes}Comparison between the mechanical loss of angular momentum 
    yielded by the radiation reaction force vector, Eq.~\eqref{eq:RRforce}, and the flux of angular momentum 
    computed from the waveform multipole extracted at future null infinity. Here $\hat{F}^{\rm mech}$ addresses
    the modulus of the vector in Eq.~\eqref{eq:RRforce}. Three illustrative configurations with spin $a=0.5$ 
    but different inclinations are considered. 
    In the bottom row we show the relative difference between the two fluxes, with
    $\Delta \equiv |\hat{\mathcal{F}}^{\rm mech} - \dot J/\nu|/(\dot J/\nu)$. 
    Although the discrepancies are larger with non-zero inclination, they never exceed $5\%$ at lightring crossing. 
    In the examples above the mechanical fluxes diverge when the inclination is non-zero: 
    this is caused by the second term in Eq.~\eqref{eq:RRforce}. Nonetheless, the behavior 
    of the flux $\dot J/\nu$ behaves as expected, since the plunge is largely independent of radiation reaction.}
\end{figure*}

In this Appendix we check the consistency between the mechanical angular momentum loss $\hat{\mathcal{F}}^{\rm mech}$ 
driven by the analytical radiation reaction force, Eq.~\eqref{eq:RRforce}, and the gravitational wave flux $\dot J/\nu$ computed 
summing up the multipoles extracted with {\tt Teukode} up to $\ell=4$ (as explained in section~\ref{subsec:teuk}). 
This is a useful consistency test which allows us to investigate the accuracy of our radiation reaction prescription, 
as pointed out in Refs.~\cite{Damour:2007xr,Bernuzzi:2010ty}, which present a similar analysis for the nonspinning case. 
Since for nonequatorial orbits the angular momentum fluxes are vectorial quantities with components in the three Cartesian directions $(x,y,z)$, 
we consider the modulus $\hat{\cal F}^{\rm mech}$ of the total flux to perform the comparison.
This is shown in the top row of Fig.~\ref{fig:mech-exact-fluxes} for a range of inclinations
$(\iota = 0^\circ, 45^\circ, 60^\circ)$ at fixes spin $a=0.5$. 
We denote the LSSO crossing by a vertical dotted line and the light ring crossing by a vertical dashed line. 

We find an agreement of a few percents between the two quantities early in the inspiral and during most of the plunging 
phase, as exemplified by the behavior of the relative flux difference $\Delta \equiv |\hat{\mathcal{F}}^{\rm mech} - \dot J/\nu|/(\dot J/\nu)$.
Around the LR crossing, because of the strong-field behavior of the second term in Eq.~\eqref{eq:RRforce},
the mechanical flux shows a divergence.
This behavior is consistent with the one observed in Refs.~\cite{Damour:2007xr,Bernuzzi:2010ty}. 
Two novel features emerge in the non-equatorial case which are worth noting. The first are oscillations during the 
inspiral which are more pronounced for large values of $\iota$. It is interesting to note that, even though we have taken 
into account the time shift necessary to relate the retarded time $u$ with the mechanical time $t$ when performing the 
comparison, there is a small residual difference in the positions of the peaks of the fluxes, particularly evident 
in the {\tt a05i20} configuration. Secondly, when $\iota\neq 0$ the mechanical fluxes tend to diverge close to merger. 
This does not represent a drawback at the moment since at this stage the motion is largely independent of the 
radiation reaction, but in future work it would be desirable to address this issue, possibly with the implementation 
of tortoise coordinates $(\mathbf{r_*},\mathbf{p_*})$ as defined for instance in Ref.~\cite{Pan:2009wj}.

\section{Rotating the multipole modes}\label{app:rotations}
We summarize here the definitions used throughout the article of Wigner's matrices $\mathfrak{D}^{(\ell)}_{m'm}$ and of the Spin-Weighted Spherical Harmonics (SWSH). The $\mathfrak{D}^{(\ell)}_{m'm}$'s are defined following Wigner \cite{WignerGroups} whereas the definition of the SWSH is the one found in \cite{Brown:2007jx}, 
appropriately modified to fit our conventions on the definition of the $d^{(\ell)}_{m'm}$'s.
We have
\begin{widetext}
  \begin{align}
    &\mathfrak{D}^{(\ell)}_{m'm}(\alpha, \beta, \gamma) = e^{im'\alpha}d^{(\ell)}_{m'm}(\beta)e^{im\gamma} \label{eq:WigD}\ , \\
    &d^{\ell}_{m',m}(\beta) = \sum_{k_i}^{k_f} (-1)^{k} \frac{\sqrt{(\ell+m)!(\ell-m)!(\ell+m')!(\ell-m')!}}{k!(\ell+m-k)!(\ell-k-m')!(k-m+m')!}
    \left[\cos\frac{\beta}{2}\right]^{2\ell-2k+m-m'} \left[\sin\frac{\beta}{2}\right]^{2k-m+m'} \label{eq:Sakurai's_d} \ ,\\
    &_sY^{\ell m}(\theta,\varphi) = (-1)^{-s}\sqrt{\frac{2\ell +1}{4 \pi}}\mathfrak{D}^{(\ell)}_{-s,m}(0, \theta, \varphi) \ .
    \label{eq:SWSH}
\end{align}  
\end{widetext}
Spin-weighted spherical harmonics are a generalization of the standard spherical harmonics defined on the sphere, to which they reduce when $s=0$. 
Recall that, under a generic rotation $R\in SO(3)$ identified by Euler angles $\{\abc\}$, the transformation law for the spherical harmonics is~\cite{WignerGroups}
\begin{equation}
    Y_{\ell m}(\theta', \varphi') = \sum_{m'=-\ell}^\ell \mathfrak{D}^{(\ell)}_{m'm}(\alpha, \beta, \gamma)Y_{\ell m'}(\theta, \varphi).
\end{equation}
Spin-weighted spherical harmonics inherit the transformation property of the $Y_{\ell m}$'s, but, being functions of spin-weight $s$, 
they have the additional property of transforming under rotation of the frame tangent to the sphere. 
Under such rotation a function of spin-weight $s$ transforms as 
\begin{equation}
    _sf\  \rightarrow\  _sfe^{is\chi},
    \label{SW_function}
\end{equation}
where $\chi$ is the rotation angle. Hence the most general transformation rule for SWSH reads
\begin{equation}
    _sY_{\ell m}(\theta',\varphi') = \sum_{m' = -\ell }^\ell  e^{is\chi}\mathfrak{D}_{m'm}^{(\ell) }(R) _sY_{\ell m'}(\theta, \varphi) \ .
    \label{SWSH transf}
\end{equation}
As already mentioned in the introduction, the gravitational wave strain $h \equiv h_+ -i h_\times $ is a function of 
spin-weight $-2$, and as such it can be decomposed in SWSH:
\begin{equation}
    h_+ - ih_\times = \frac{1}{D_L}\sum_{\ell, m} h_{\ell m}\ _{-2}Y^{\ell m}(\theta, \varphi) \ ,
    \label{GWstrain}
\end{equation}
where $D_L$ is the luminosity distance. The $h_{\ell m}$'s appearing here are the \textit{multipolar modes} 
and their transformation law under rotations of the coordinates can be derived directly from the one 
for the SWSH's. Considering a generic rotation and using the property\footnote{
This follows from the unitarity of the $\mathfrak{D}^{(\ell)}$'s and from the fact that the inverse of the rotation 
$R_{\{\abc\}}$ is $R_{\{-\gamma,-\beta,-\alpha\}}$.} $\left[\mathfrak{D}^{(\ell)}_{m'm}(R)\right]^\ast = \mathfrak{D}_{mm'}^{(\ell)}(R^{-1})$, 
it is straightforward to show that the transformation law for the modes reads
\begin{equation}\label{new_hlms}
    \tilde{h}_{\ell m} = \sum_{m'}h_{\ell m} e^{im'\gamma} d_{m'm}^{(\ell)}(-\beta)e^{im\alpha} \ ,
\end{equation}
where $(\abc)$ are the Euler angles identifying the rotation which relates the two frames. Consistently with the notation of section~\ref{sec:schw} 
we have used a tilde to denote the rotated modes $\tilde h_{\ell m}$ obtained from some original $h_{\ell m}$.

%====================
% Figure: anti-twist geodetic
%====================
\begin{figure*}[t]
    \begin{center}
        \includegraphics[width=.32\textwidth]{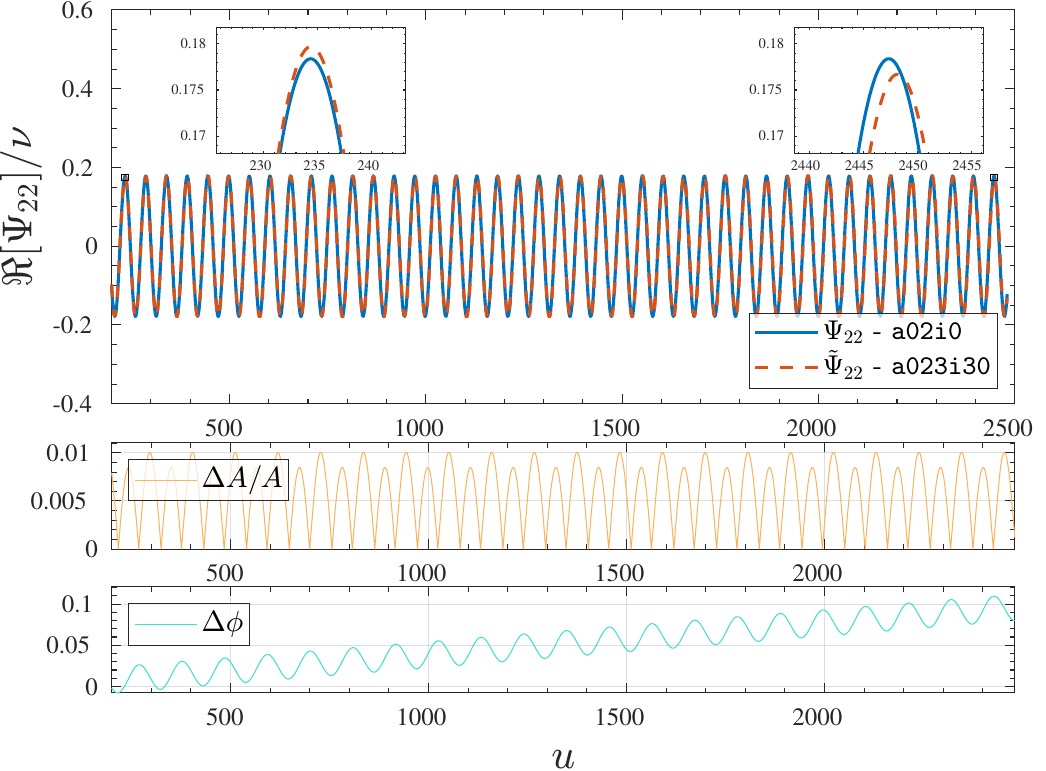} 
        \includegraphics[width=.32\textwidth]{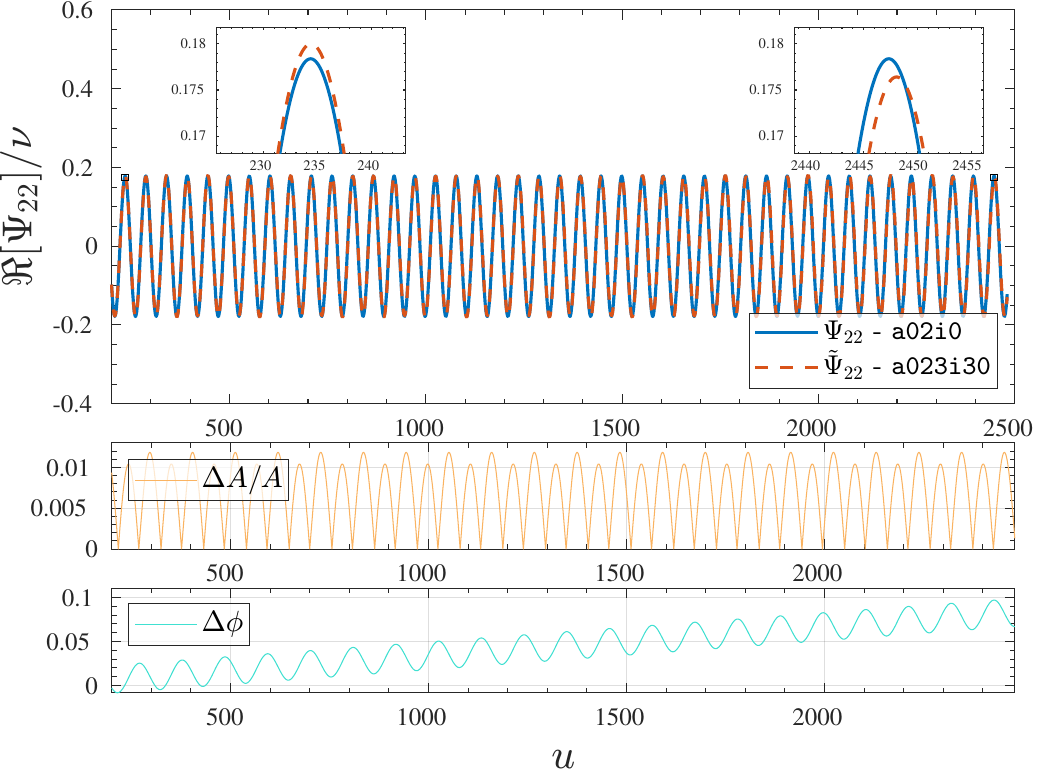} 
        \includegraphics[width=.32\textwidth]{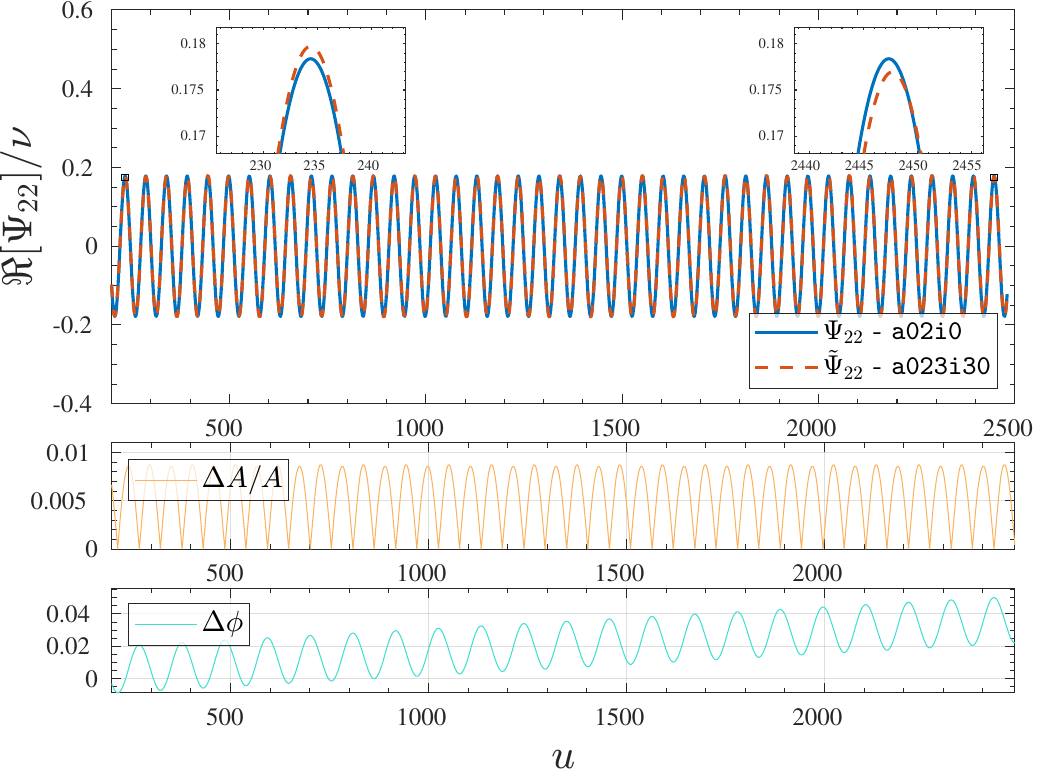}
    \end{center}
    \caption{\label{fig:geod_twist}Comparison between an exact spin-aligned waveform (blue) and a precessing 
    waveform as observed in the co-precessing frame (dashed red) 
    as we tweak the rotation procedure. In all panels the precessing configuration is chosen such that the spin-orbit coupling is the 
    same as for the spin-aligned one, see Sec.~\ref{sec:twist}. Left panel: the coprecessing frame is 
    identified by tracking the direction of $\mathbf{L} = \mathbf{r}\times \mathbf{p}$. Middle panel: we instead track 
    $\mathbf{L}_N \propto \mathbf{r}\times \mathbf{\dot r}$, as is commonly done in the relevant literature~\cite{Pan:2013rra,Gamba:2021ydi}. 
    The relative amplitude difference is slightly smaller when using $\mathbf{L}$, which justifies our choice of Sec.~\ref{sec:twist} to 
    use this quantity to define our coprecessing frames. In both panels we note a cumulative dephasing, stemming from the fact that 
    the underlying dynamics is different between the spin-aligned and precessing configurations. To address this issue 
    we consider a different precessing configuration, where the dynamics is governed by a modified Hamiltonian where 
    we manually set the spin-orbit {\it and} the spin-spin coupling to be the same: the resulting waveform in shown in 
    the rightmost panel, appropriately rotated in the coprecessing frame identified by $\mathbf{L}$. 
    The relative amplitude difference is similar, but the cumulative dephasing is halved. This is due to the fact that the underlying dynamics
    is still different, since we have not taken into account the fact that the orbital angular momentum vector undergoes 
    slight changes in modulus and direction during  the evolution since $L$ is not conserved for non-equatorial orbits in Kerr.}
\end{figure*}

\section{Time-Dependent Rotations: from the inertial frame to the co-precessing frame} 
\label{sec:twist_test}

In the main text, Sec.~\ref{sec:twist}, we investigated the possibility of approximating the waveform profile generated by 
an inclined plunging orbit by applying a suitable rotation to the multipolar modes of a spin-aligned wave. 
This was done to test, in a consistent framework, the accuracy of a procedure that is ubiquitously applied to 
compute spin-precessing waveforms rotating spin-aligned ones with a certain evolution of the Euler angles.
In this Appendix we intend to give more background to this approach: to this end it is convenient to perform a 
rotation of the modes produced by the precessing dynamics into the coprecessing frame defined in Sec.~\ref{sec:twist} (as opposed 
to applying the inverse rotation to the spin-aligned modes, as done in Sec.~\ref{sec:twist}).
This procedure, as largely documented in the literature (see e.g. Refs.~\cite{Kidder:1995zr,Boyle:2011gg}), 
allows one to reduce the oscillations in the waveform due to the precession of the orbital plane.
We first study the case of geodesic quasi-spherical orbits, Sec.~\ref{app:antitwist_geod}, and then briefly investigate the transition from inspiral 
to plunge, Sec.~\ref{sec:antitwist_mrg}. In particular, in this second case we will see that the waveform rotated in the coprecessing
frame has some qualitative and quantitative features that might be useful to characterize the merger and that are not accessible otherwise.

\subsection{Geodesic dynamics}
\label{app:antitwist_geod}

For consistency with our discussion of Sec.~\ref{sec:twist}, we begin again by considering configurations \texttt{a02i0} and \texttt{a023i30}. 
Since we are now considering geodesic orbits We set $r_0 = 6.5$ and evolve the system for $t=2500$, that yields roughly 23 full orbits. 
The Euler angles used to rotate the modes are again determined by tracking the direction of the orbital angular momentum 
$\mathbf{L} = \mathbf{r}\times\mathbf{p}$ along with the minimal rotation condition~\cite{Boyle:2011gg}.
Since we have full control of the (relatively simple) dynamics underlying the signal, we can investigate the effects of slightly 
tweaking the rotation procedure. As an illustrative example, we compare the $(2,2)$ mode of the spin-aligned waveform with the same mode 
generated by the precessing configuration rotated in the coprecessing frame via two different prescriptions: first (left panel of 
Fig.~\ref{fig:geod_twist}) we identify the coprecessing frame by tracking the direction of $\mathbf{L} = \mathbf{r}\times \mathbf{p}$, as was done in 
the main text. We then perform the same comparison by tracking the direction of the Newtonian angular momentum, $\mathbf{L}_N \propto \mathbf{r}\times \mathbf{\dot r}$, 
instead (middle panel of Fig.~\ref{fig:geod_twist}). From the discussion of Ref.~\cite{Pan:2013rra} we expect the outcome to be quite similar; 
the novelty of our setup is that it allows us to quantitatively establish which of the two approaches yields a waveform that is closer to the exact one. 
The top plot on each panel of Fig.~\ref{fig:geod_twist} shows the real part of the two waveforms (the rotated one is denoted with a tilde in the legend) 
along with two insets with a detailed view of the first and last maxima. The bottom plots show the 
relative amplitude difference $\Delta A/A \equiv |A^{\rm eq} - \tilde{A}^{\rm prec}|/A^{\rm eq}$ 
and the phase difference $\Delta \phi \equiv \phi^{\rm eq} - \tilde{\phi}^{\rm prec}$ between the two.
By comparing the left and middle panels we see that the amplitude of the residual oscillations is slightly reduced 
if the coprecessing frame is defined by tracking the total orbital angular momentum $\mathbf{L}$ 
instead of the Newtonian one $\mathbf{L}_N$.

%========================================
% Untwisted waveform frequency and merger location
%========================================
\begin{figure*}[t]
    \begin{center}
        \includegraphics[width=.30\textwidth]{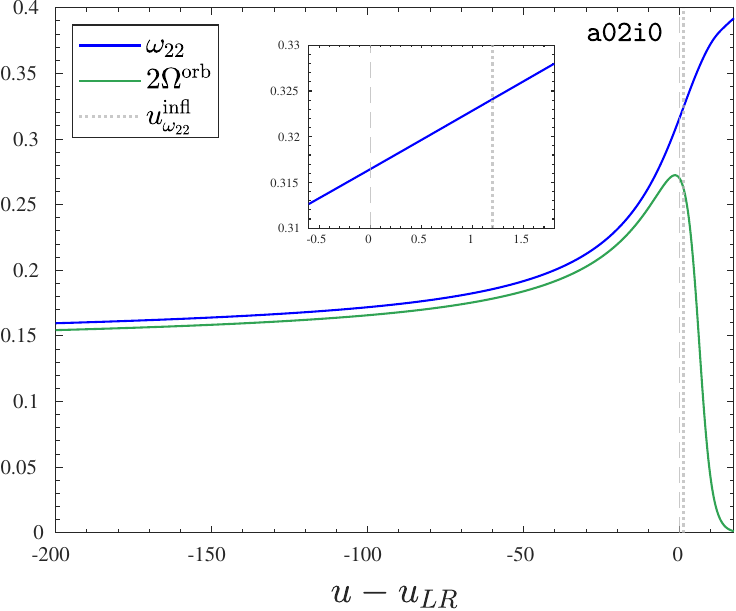}
        \hspace{2mm}
        \includegraphics[width=.30\textwidth]{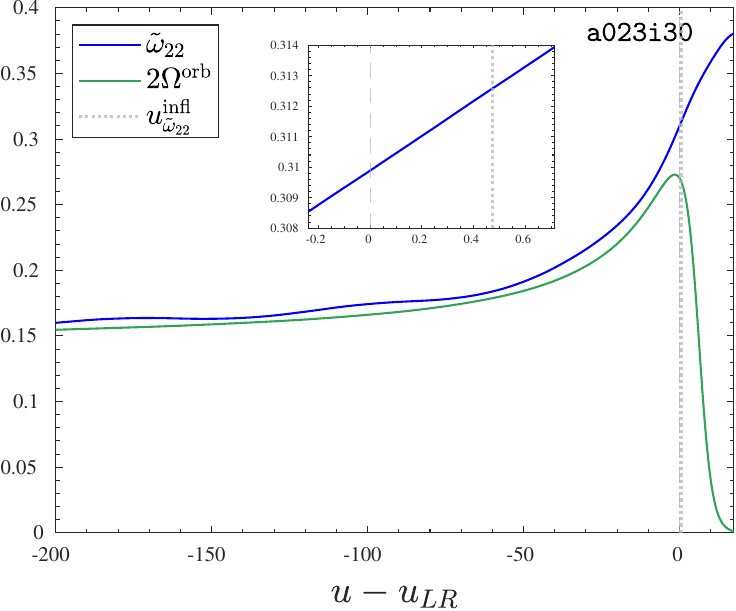}
        \hspace{2mm}
        \includegraphics[width=.30\textwidth]{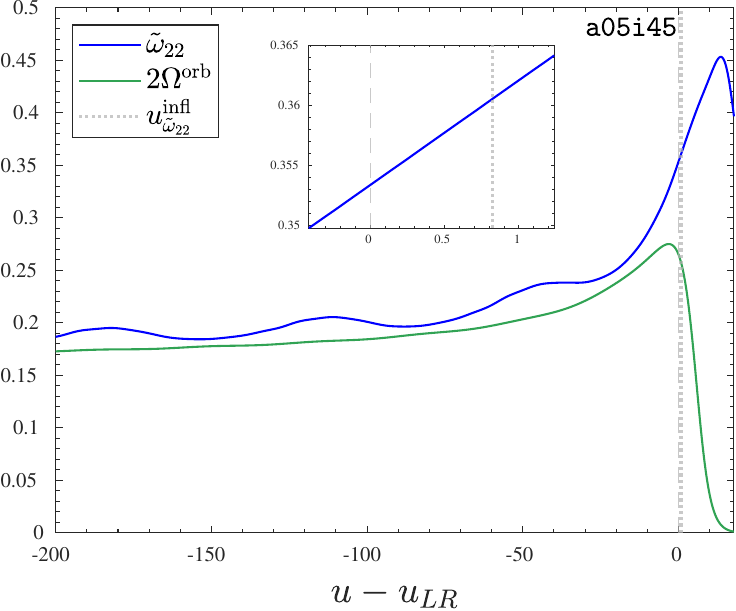}
    \end{center}
    \caption{\label{fig:twist_inflection}
    Illustrating the closeness between the light-ring crossing time ($u_{\rm LR}$ , vertical dashed line) 
    and the time location of the inflection point  of  the (rotated) gravitational wave frequency $\tilde{\omega}_{22}$
    once observed in the coprecessing frame ($u^{\rm infl}_{\tilde{\omega}_{22}}$, vertical dotted line). 
    The two inclined configurations, \texttt{a023i30} and \texttt{a05i45} are compared to the equatorial 
    configuration \texttt{a02i0}. The time delay alway lies within $\sim 1.5M$, showing that 
    $u^{\rm infl}_{\tilde \omega_{22}}$ can be considered to yield a good approximation of $u_{\rm LR}$.
    In each plot we also show twice the purely orbital frequency for completeness.}
\end{figure*}
The fact that we see a cumulative dephasing is not surprising since the underlying dynamics of the inclined configuration is substantially different: 
given that we have chosen the central spins such that the spin-orbit coupling is the same between the two (Eq.~\eqref{eq:aeq_aprec}), 
it is reasonable to make the hypothesis that any residual differences are due to the different spin-spin 
coupling (the $a^2$ terms in the Hamiltonian~\eqref{eq:Hso}--\eqref{eq:Horb}). This suggests to compute 
a ``hybrid'' dynamics obtained with a  modified Hamiltonian such that the spin-orbit \textit{and} the spin-spin terms 
are identical to those of the spin-aligned configuration. We can then compute the resulting waveform, that 
we expect to be closer to the exact one. The result is illustrated in the rightmost panel of 
Fig.~\ref{fig:geod_twist}: the relative amplitude difference is similar, but the cumulative dephasing 
is halved. It is interesting to note that, by enforcing the condition that the spin-spin coupling is 
identical between configurations, the amplitude oscillations are now exactly centered around the spin-aligned value. 
The fact that the phase difference increases overtime, however, mirrors the fact that the underlying dynamics 
is still different: in fact, in our discussions we have not taken into account the fact that the orbital angular 
momentum vector undergoes slight changes in modulus and direction during the evolution 
(this is because $L$ is not a constant of motion for non-equatorial orbits in Kerr). It is nonetheless 
interesting to note that the approximate waveform obtained in this way is much closer to the exact one.

\subsection{Transition from inspiral to plunge and merger}
\label{sec:antitwist_mrg}
In the main text we have discussed the use of the orbital frequency in the context of characterizing the dynamics and 
in doing so we pointed out that $\Omega_{\rm orb}$ always peaks at a time $t_{\Omega_{\rm orb}^{\rm peak}}$ 
close to the time the particle crosse the light ring, $t_{\rm LR}$. In the spin-aligned case (including the comparable-mass case) 
the peak of $\Omega_{\rm orb}$ identifies a crucial time in the dynamics that is used to determine 
the next-to-quasi-circular corrections and the transition to the postmerger-ringdown phase within a certain flavor 
of EOB waveform modeling~\cite{Harms:2014dqa,Damour:2014sva,Nagar:2024dzj,Nagar:2024oyk,Damour:2025uka}.
In practical terms $t_{\Omega_{\rm orb}^{\rm peak}}$ is instrumental to identify an approximate location 
of the time when the merger occurs that is useful {\it also} in the comparable mass case, 
where the underlying dynamics of the two black holes is in fact not accessible through Numerical Relativity (NR) simulations. 
In this respect, it would be important to have a method to identify the location of the merger, i.e. the moment where
the two black-hole dynamics is replaced by a single, ringing, black hole, that only relies on properties of the waveform
and does not need the knowledge of the dynamics. As such, it could be applied to NR waveforms.
Note that typically the merger is {\it conventionally assumed} to coincide with the peak of the $\ell=m=2$ 
waveform amplitude\footnote{This stems from the case of a test-mass on Schwarzschild, where the time when 
the $\ell=m=2$ mode peaks is very close to the light-ring crossing~\cite{Davis:1972ud,Nagar:2006xv,Damour:2007xr}.}. 
This is, generally, a good approximation but might need to be revisited when the black
hole are spinning or the inspiral is eccentric. Although a systematic investigation all over the binary 
black hole parameter space is missing, one has indications from the test-mass case that the 
commonly adopted conventional definition of the merger point might need to be revisited.
More precisely, in the presence of equatorial orbits on Kerr, it is well known that the quadrupolar
mode amplitude might peak at a time very displaced with respect to the light-ring crossing depending on the 
spin of the black hole~\cite{Taracchini:2013wfa,Harms:2014dqa,Taracchini:2014zpa,Gralla:2016qfw}
Similar results were found in the presence of eccentricity~\cite{Albanesi:2023bgi} and both spin and
eccentricity~\cite{Albanesi:2025prep,Faggioli:2025hff}, with even configurations where the waveform 
amplitude does not have a local maximum.
From this test-mass results it seems that the conventional definition of merger location should be revised.
As a matter of fact, Ref.~\cite{Damour:2012ky} had already observed long ago that, for a test-mass inspiralling
and plunging on a Schwarzschild black hole from a quasi-circular orbit, the time location of the {\it inflection point} 
of the  quadrupolar gravitational wave frequency gives a good approximation (within 1M) of the light-ring crossing time,
a proxy that is actually better than the one given by the location of the peak  of the $(2,2)$ 
mode (see Fig.~4 in Ref.~\cite{Damour:2012ky}). More recently, Ref.~\cite{Albanesi:2023bgi} (see Fig.~4 therein)
found that the same remains true for the case of eccentric plunge on a Schwarzschild black hole and
in the presence of spin as well~\cite{Faggioli:2025hff,Albanesi:2025prep}. All this information together seems 
to indicate that a revision of the definition of merger and postmerger in EOB models will be needed in the future.

In the case of inclined orbits, the problem is more intricate because of the $\theta$-related modulation of
the waveform, so that it is might be hard to identify a well-defined inflection point of a gravitational
wave frequency with oscillations more or less pronounced depending on the inclination angle.
This difficulty can be overcome by expressing the waveform in the coprecessing frame.
When this is done, the rotated frequency corresponding to the plunge becomes monotonic 
and it is possible to identify an inclination point point unambiguously (at least as long as the spin 
and inclination are not too extreme). 
We find that, when the inflection point of the frequency of the rotated mode is present, 
it is indeed a good indicator of the light ring crossing. Since the frame transformation can 
be performed without knowledge of the dynamics of the 
binary~\cite{Schmidt:2010it,OShaughnessy:2011pmr,Boyle:2011gg}, this observation 
can be a useful tool in better understanding the ringdown  of precessing systems. 
For consistency with the main text, Sec.~\ref{sec:twist}, we consider again the precessing 
configurations \texttt{a023i30} and \texttt{a05i45}, as well as the equatorial one \texttt{a02i0}. 
Figure~\ref{fig:twist_inflection} illustrates that the location of the inflection point,  
$u^{\rm infl}_{\omega_{22}}$ is within $1.5M$ from $u_{\rm LR}$ 
(dashed and dotted lines respectively in the plot).
We checked that for all configurations discussed in this paper up to spin $a=0.5$ 
the location of $u^{\rm infl}_{\omega_{22}}$ lies within $5M$ of the light 
ring crossing; for higher spins even in the co-precessing frame the frequency oscillations 
can be rather large and the identification of an inflection point is not trivial. 
A thorough analysis of the complications ensuing in the high-spin, high-inclination 
regime is postponed to future work.

\section{Waveform visualization for all configurations}
\label{app:catalog}
In this Appendix we report a list of figures where alle waveforms corresponding to
the configurations of Table~\ref{tab:Kfluxes} are reported. 
Each row of the figure reports; the full 3D trajectory and its projection on the $(x,y)$ plane, with 
the portion after the LSSO crossing (i.e., the actual plunging phase) highlighted in red, as well as the last stages 
of the $\theta$ evolution. In the middle we show show the real part of the gravitational wave strain $h_+$, 
obtained summing up all modes up to $\ell=4$, as observed from the fiducial direction $(\Theta=\pi/4,\Phi=0)$. 
The two vertical lines represent the LSSO crossing (dot-dashed) and the light ring crossing (dashed). 
Finally, the rightmost panels illustrate the amplitude and frequency of the $\ell=2, m=\pm 2$ modes 
of each configuration: an horizontal dashed line denotes the value of the fundamental QNM ringdown frequency, while the light ring and LSSO crossing 
(when visible) again identified by vertical grey lines. Comparing plots in successive rows we see how the differences between the positive and negative-$m$
modes become more pronounced as the inclination increases and the orbit becomes equatorial. We also notice how the ringdown frequency oscillations are much 
more pronounced when the orbit is retrograde than when it is prograde (Fig.~\ref{fig:a05prog}): 
this is of course a consequence of the source exciting the negative frequencies 
when the rotation direction changes in the late plunge. In the frequency plot of configuration {\tt a05i120} we also include the value of the 
negative QNM ringdown frequency $-\omega_{2-20}$ since it rapidly becomes the dominant mode for the $(2,2)$ multipole (notice how the oscillation is initially 
around $\omega_{220}$ and then immediately switches to $-\omega_{2-20}$).  
The green curve in the frequency plot represents twice the modulus of the pure orbital frequency of the 
binary,  $\Omega^{\rm orb}$, as defined in Eq.~\eqref{eq:Omg_orbAmp}: note that the peak 
of $\Omega^{\rm orb}$ always very close to the light ring crossing.

%================
% All waveform figures
%================
%=======
% Spin 0.2
%=======
\begin{figure*}[t]
\begin{center}
    \includegraphics[height=38mm]{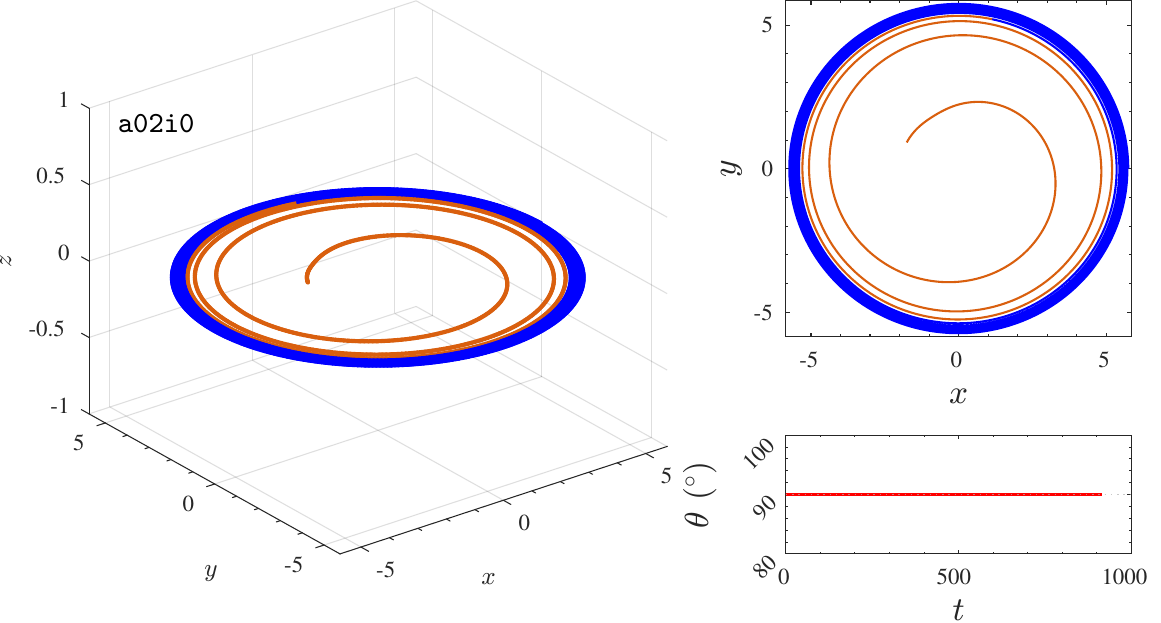} \hspace{1mm}
    \includegraphics[height=39mm]{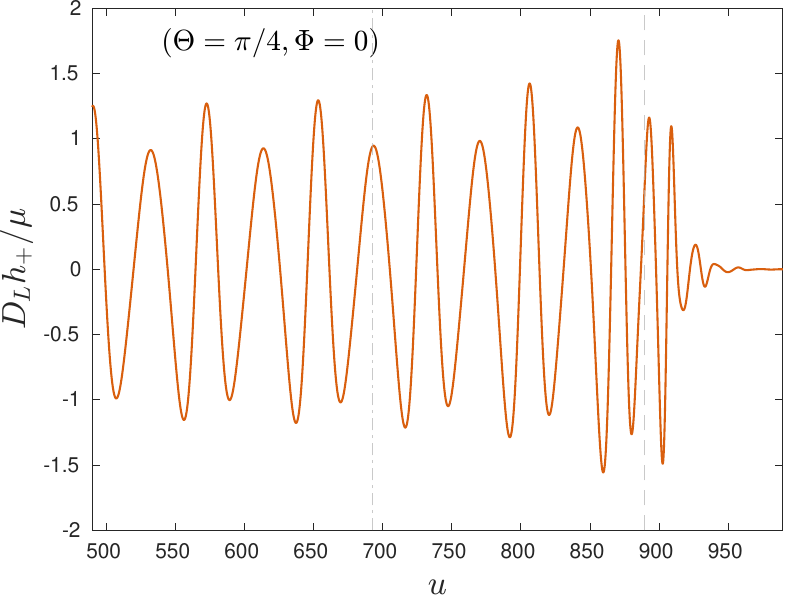} \hspace{2mm}
    \includegraphics[height=38mm]{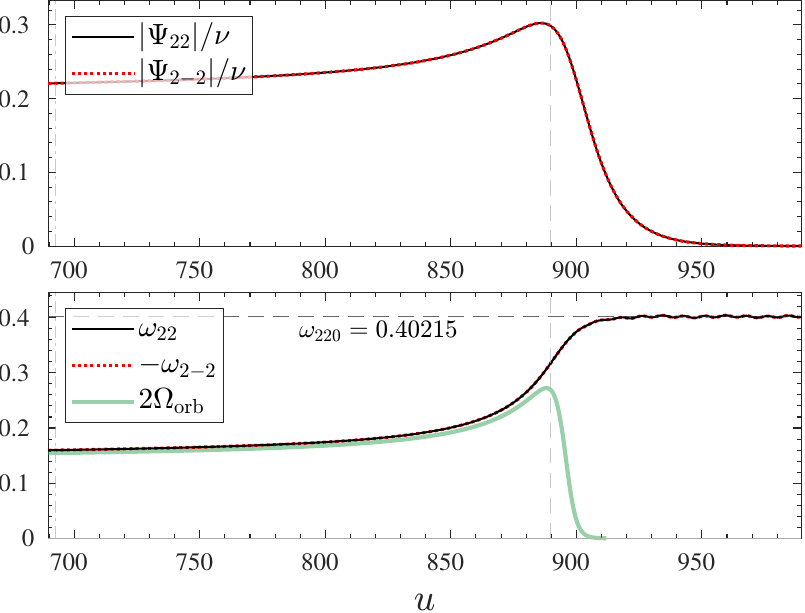} \\
    \vspace{.5cm}

    \includegraphics[height=38mm]{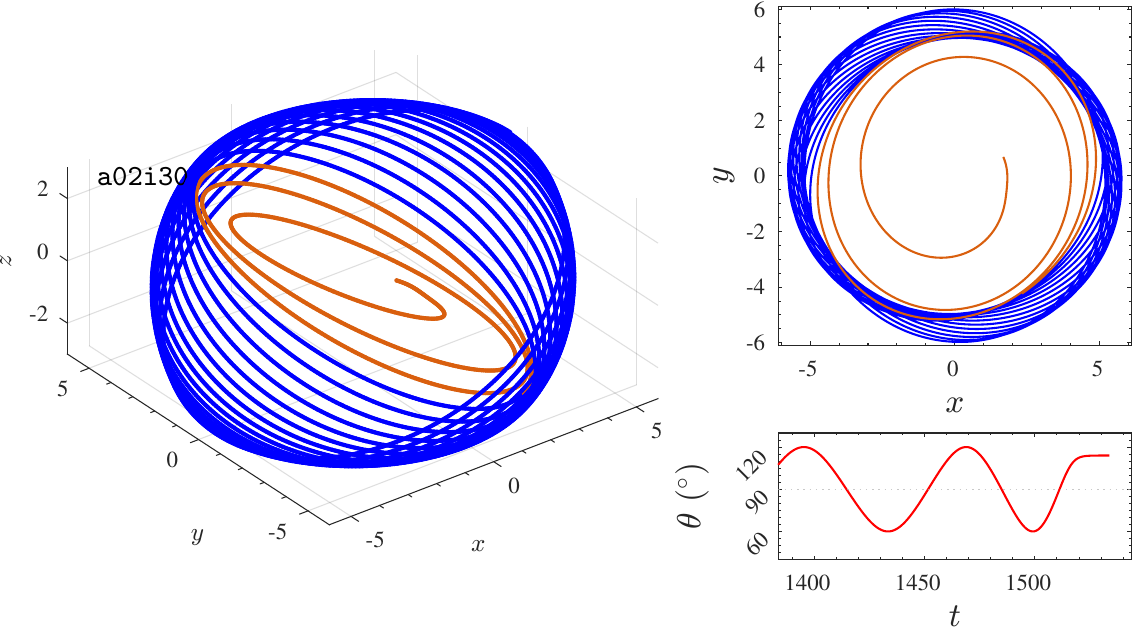} \hspace{1mm}
    \includegraphics[height=39mm]{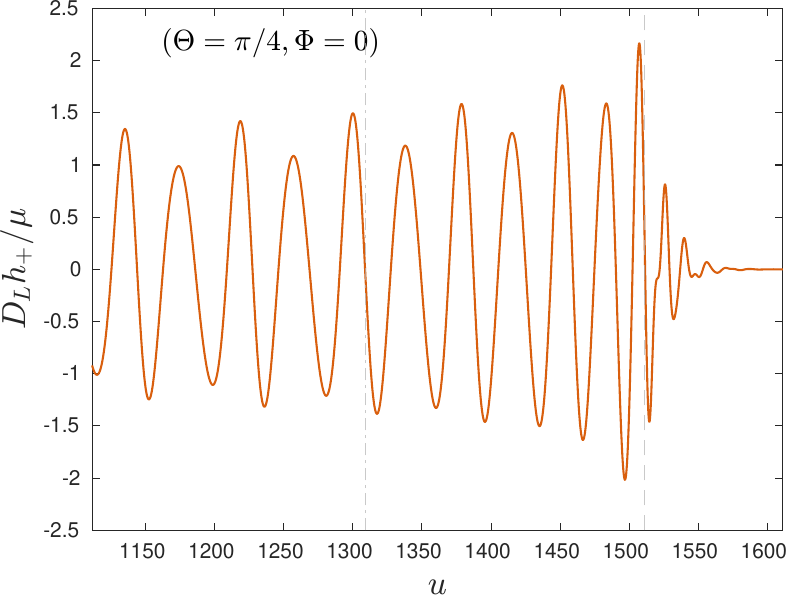} \hspace{2mm}
    \includegraphics[height=38mm]{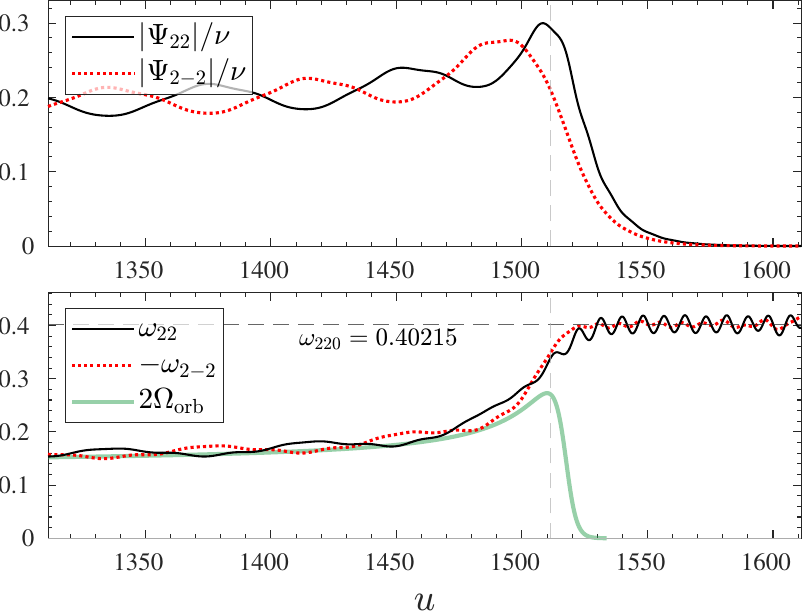} \\
    \vspace{.5cm}

    \includegraphics[height=38mm]{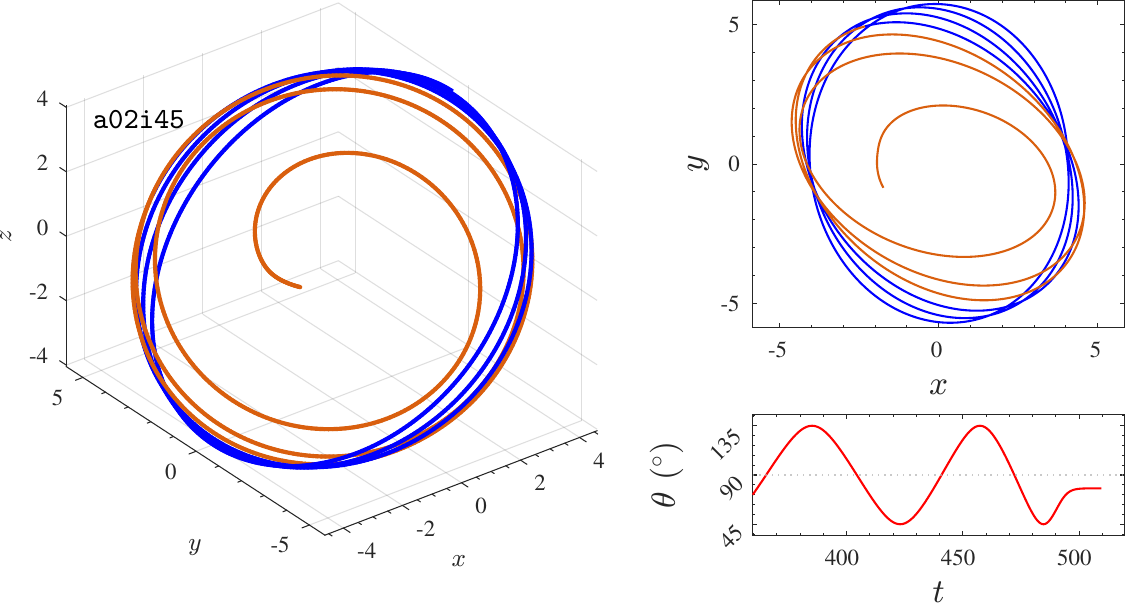} \hspace{1mm}
    \includegraphics[height=39mm]{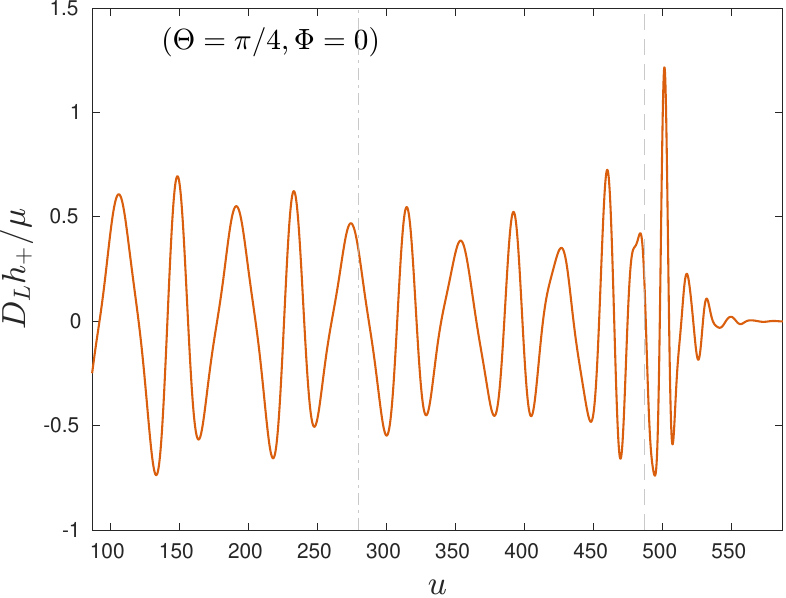} \hspace{2mm}
    \includegraphics[height=38mm]{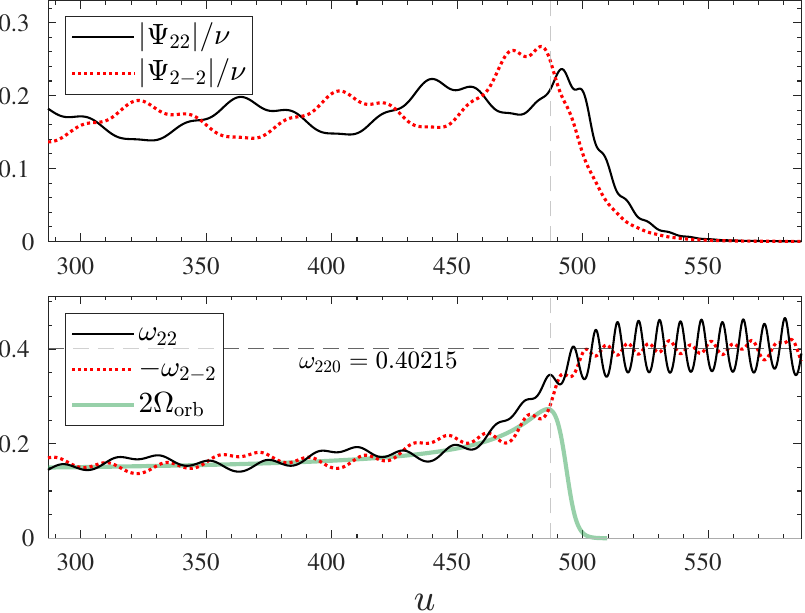} \\
    \vspace{.5cm}

    \includegraphics[height=38mm]{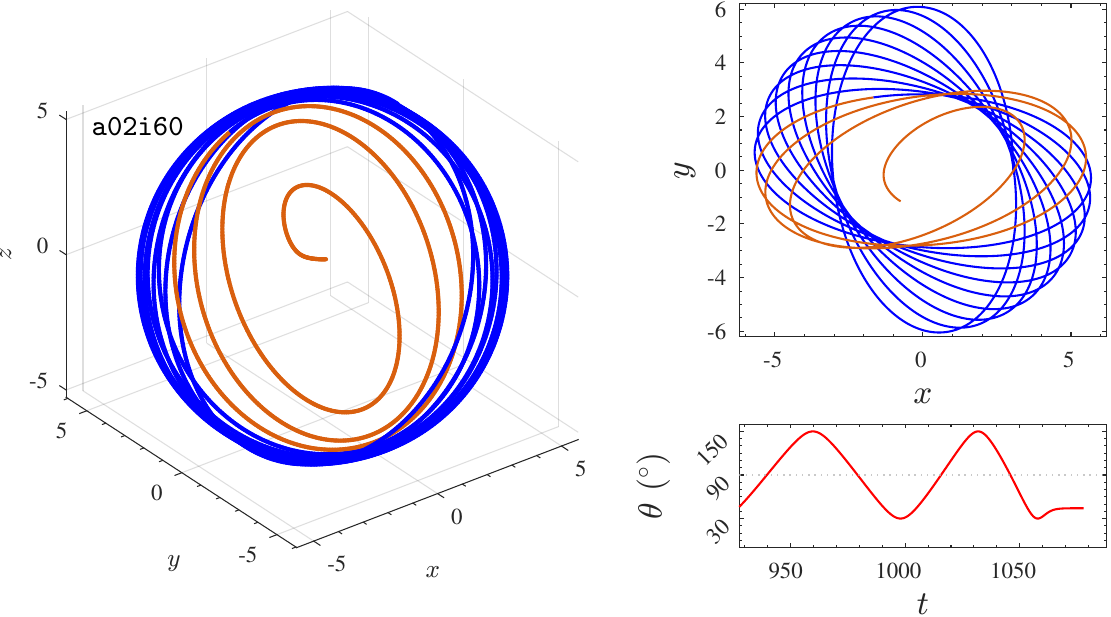} \hspace{1mm}
    \includegraphics[height=39mm]{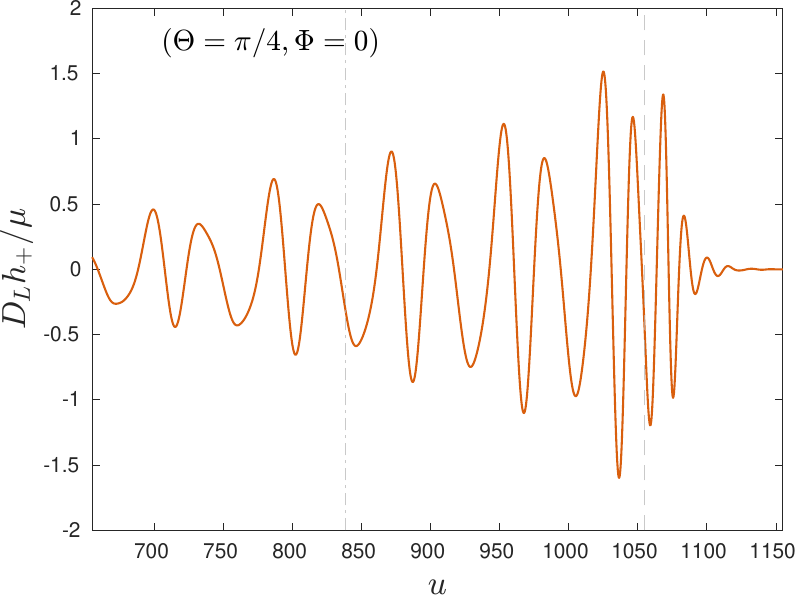} \hspace{2mm}
    \includegraphics[height=38mm]{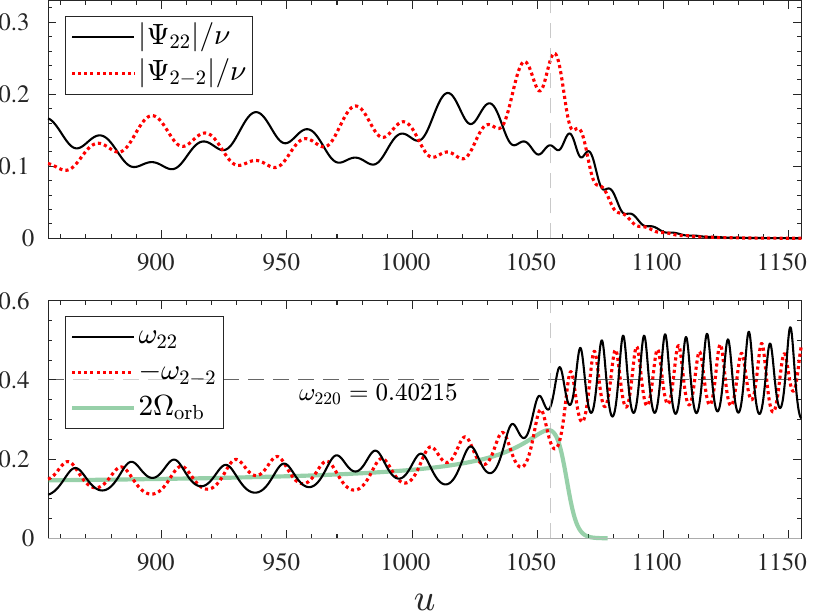} \\
    \end{center}
    \caption{\label{fig:a02prog}
    Waveform characterization for $a=0.2,\ \iota = 0,\tfrac{\pi}{6},\tfrac{\pi}{4},\tfrac{\pi}{3}$ (prograde orbits). In each row we have: on the left 
    the full 3D trajectory and its projection on the $(x,y)$ plane, with the portion 
    after LSSO crossing highlighted in red, as well as the last stages of the $\theta$ evolution. In the middle we show show the real part of the 
    gravitational wave strain $h_+$, including all multipoles up to $\ell=4$, as observed from the fiducial direction $(\Theta=\pi/4,\Phi=0)$. 
    The two vertical lines represent 
    the LSSO crossing (dot-dashed) and the light ring crossing (dashed). Finally, on the right we plot the amplitude and frequency of the $\ell=2, m=\pm 2$ modes 
    of each congifuration: an horizontal dashed line denotes the value of the fundamental QNM ringdown frequency, while the light ring and LSSO crossing 
    (when visible) again identified by vertical grey lines. Comparing plots in successive rows we see how the differences between the positive and negative-$m$
    modes become more pronounced as the inclination increases. The green curve in the frequency plot represents twice the modulus of the 
    pure orbital frequency, Eq.~\eqref{eq:Omg_orbAmp}, that always peaks close to the light-ring crossing.}
\end{figure*}

\begin{figure*}[t]
\begin{center}
    \includegraphics[height=38mm]{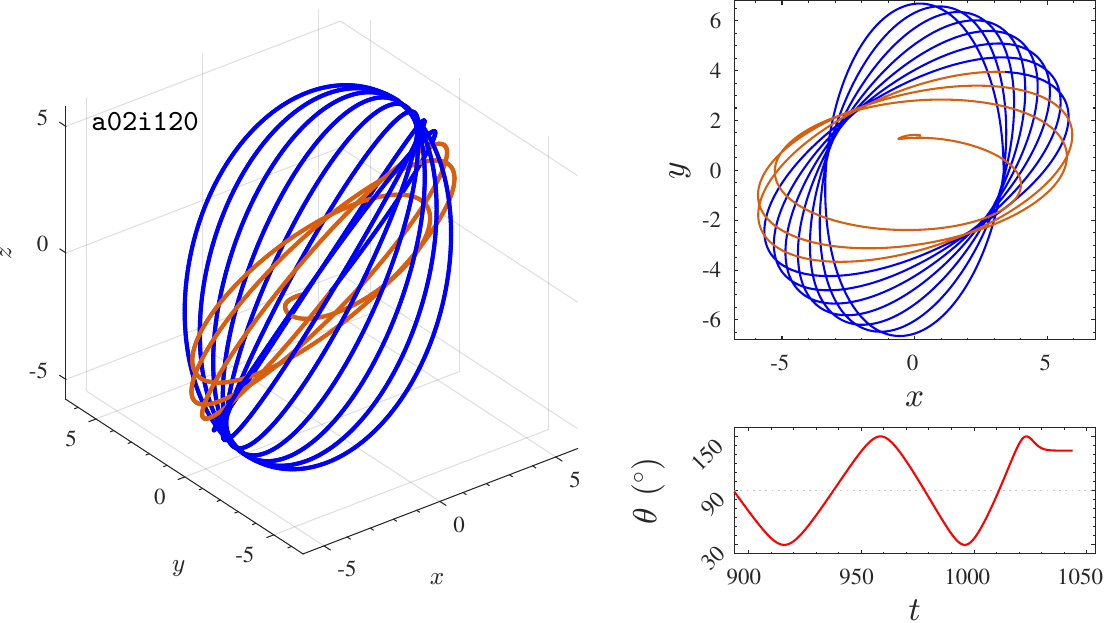} \hspace{1mm}
    \includegraphics[height=39mm]{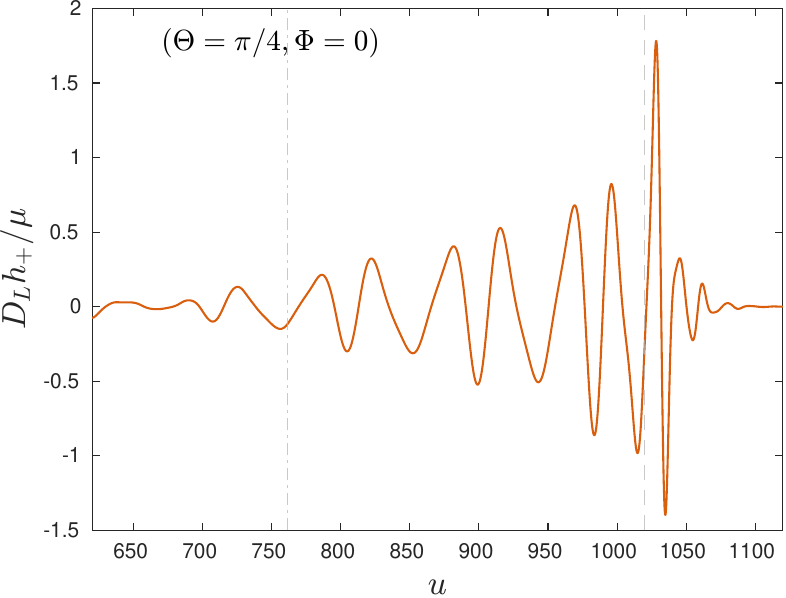} \hspace{2mm}
    \includegraphics[height=38mm]{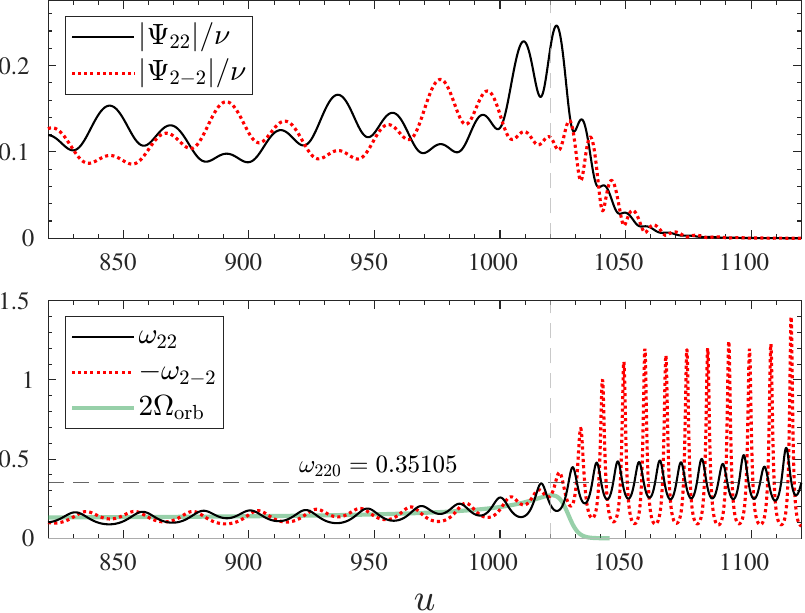} \\
    \vspace{.5cm}
    
    \includegraphics[height=38mm]{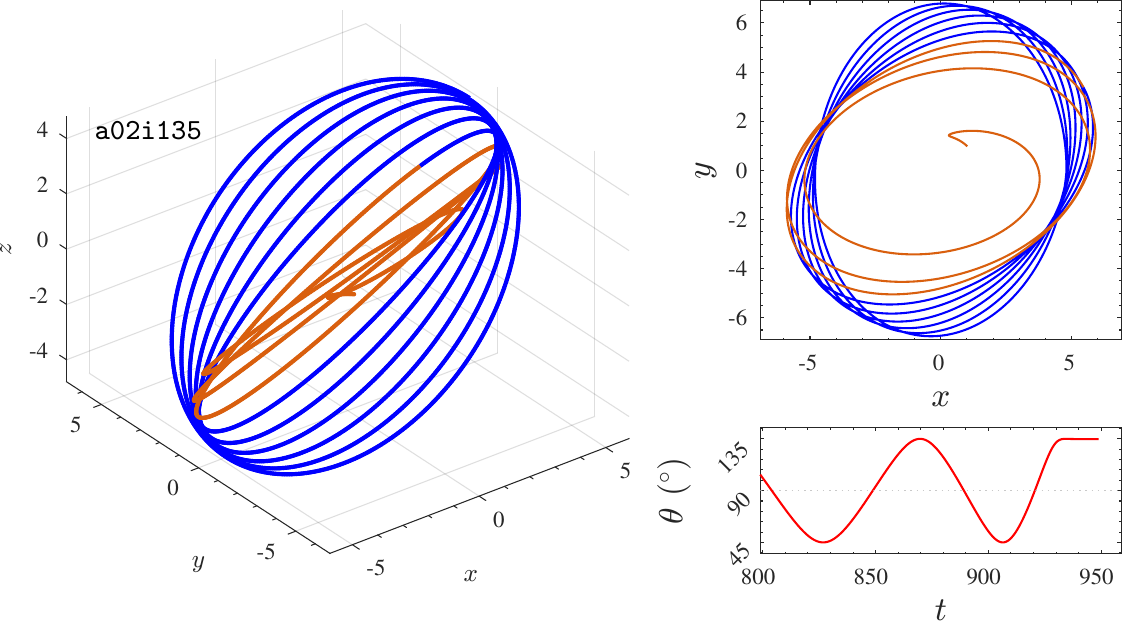} \hspace{1mm}
    \includegraphics[height=39mm]{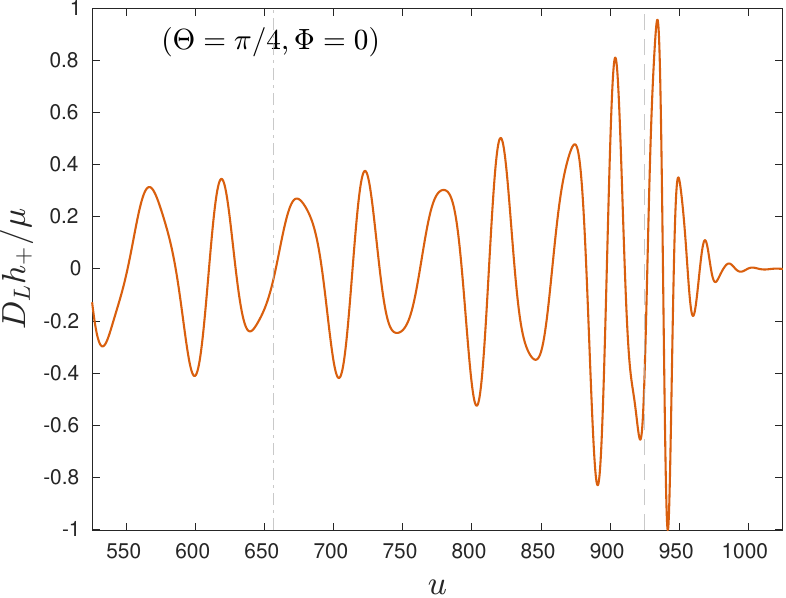} \hspace{2mm}
    \includegraphics[height=38mm]{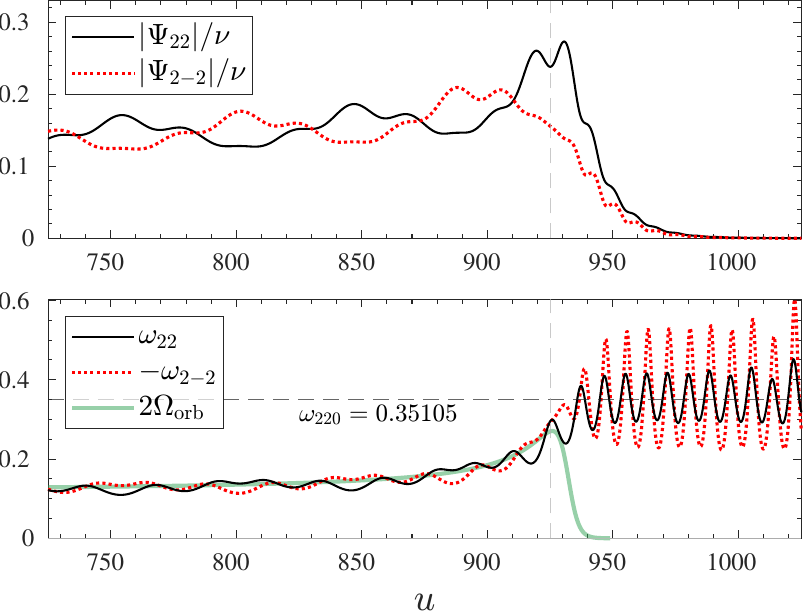} \\
    \vspace{.5cm}

    \includegraphics[height=38mm]{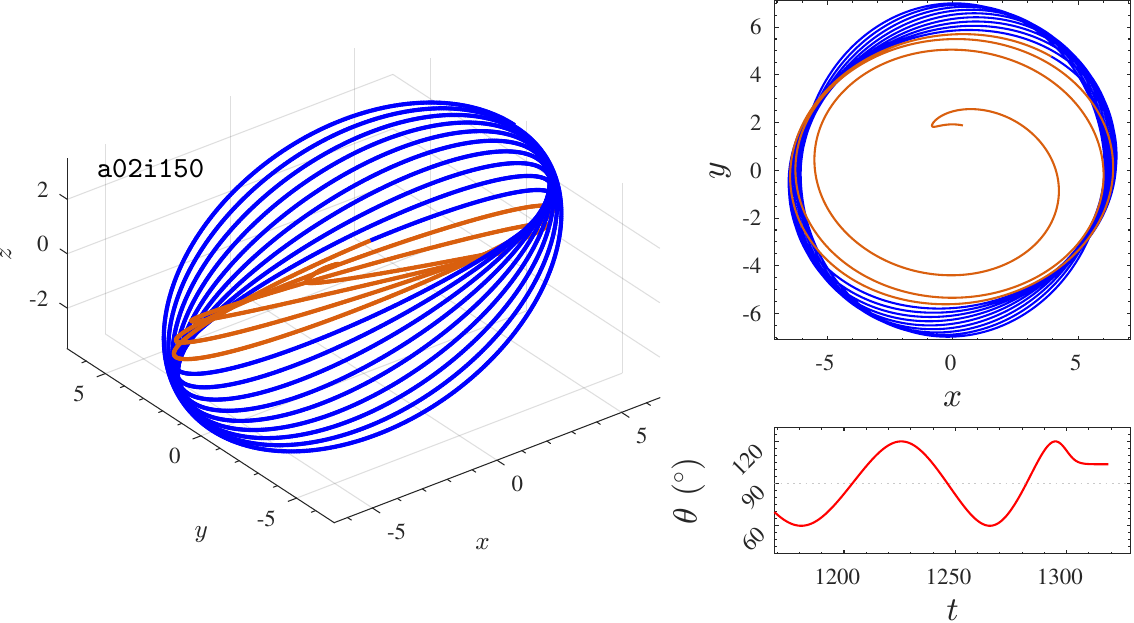} \hspace{1mm}
    \includegraphics[height=39mm]{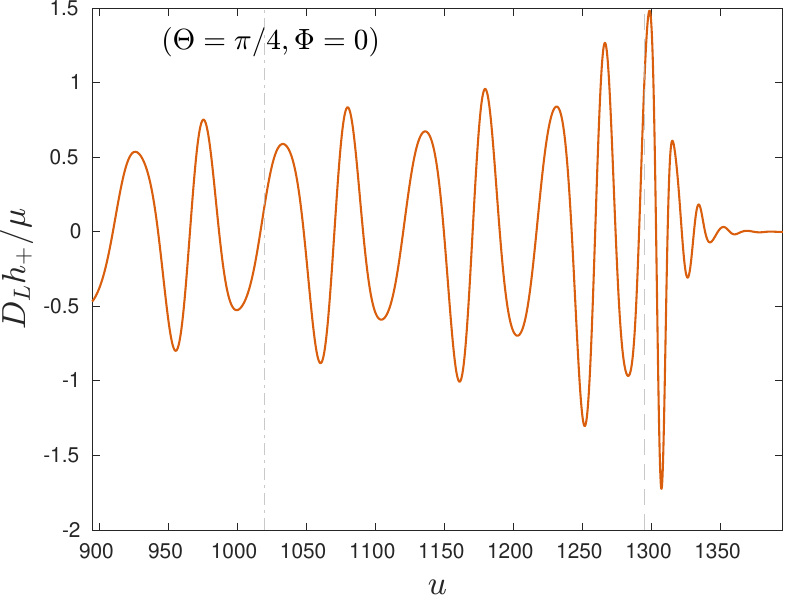} \hspace{2mm}
    \includegraphics[height=38mm]{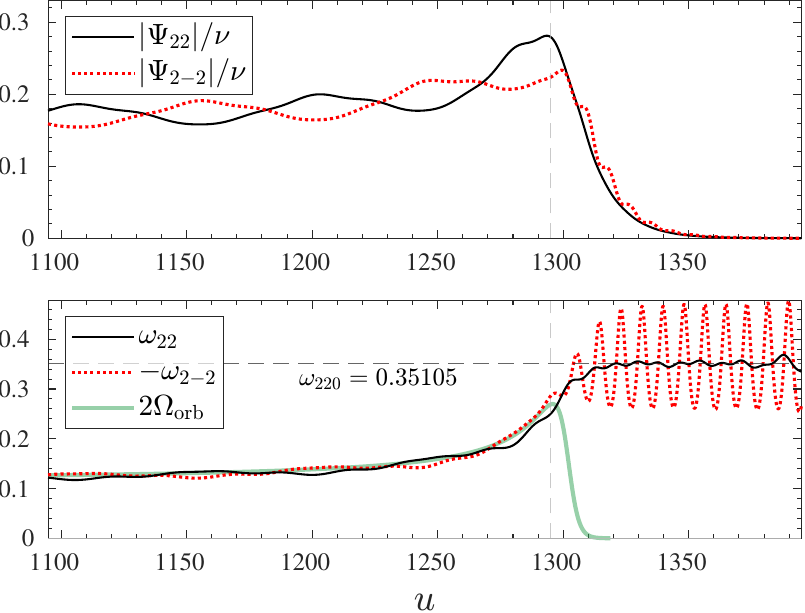} \\
    \vspace{.5cm}

    \includegraphics[height=38mm]{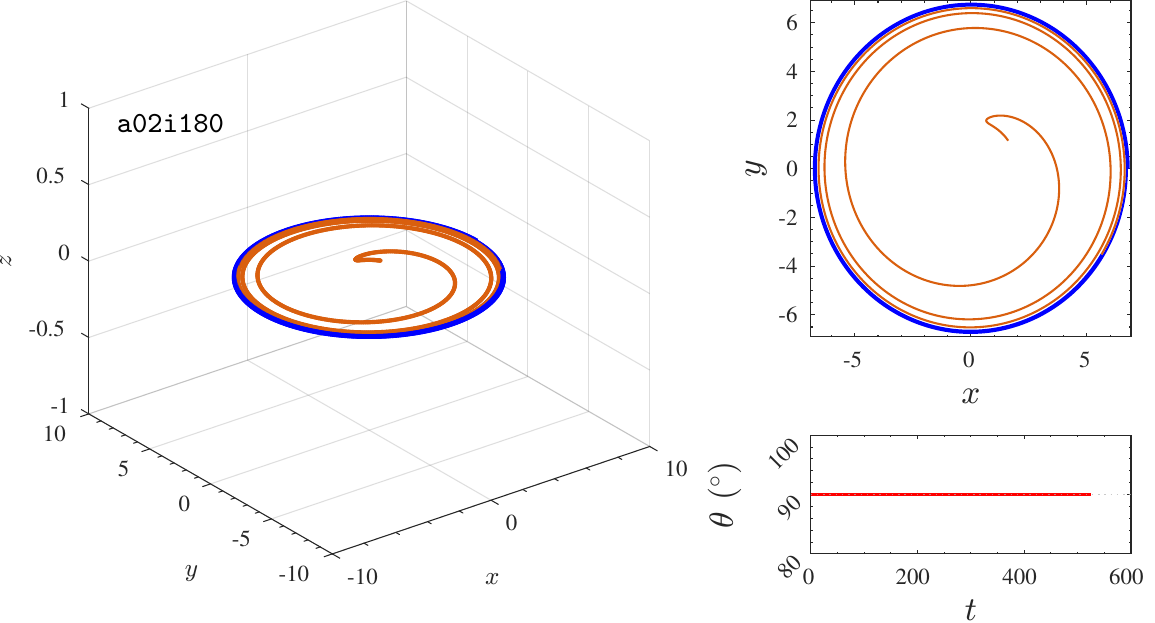} \hspace{1mm}
    \includegraphics[height=39mm]{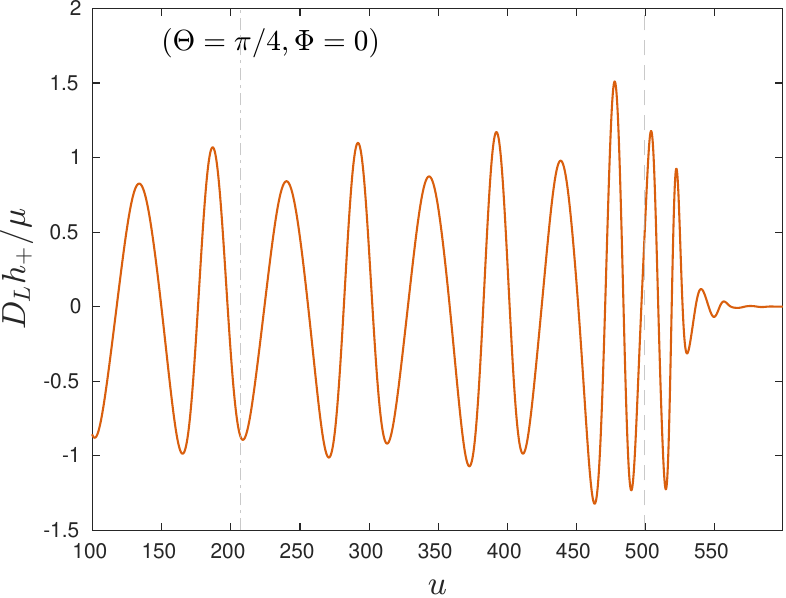} \hspace{2mm}
    \includegraphics[height=38mm]{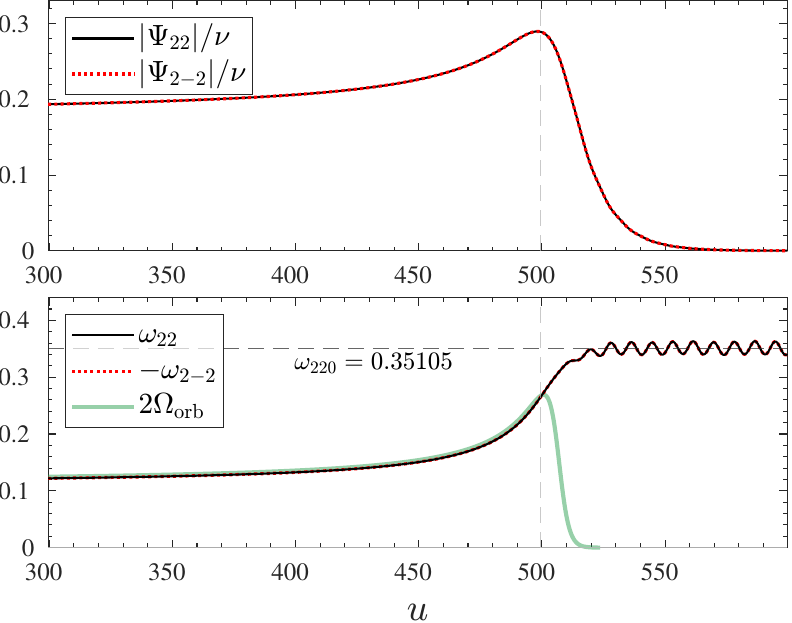} \\
\end{center}
\caption{\label{fig:a02retr}
Waveform characterization for $a=0.2,\ \iota = 0,\tfrac{2\pi}{3},\tfrac{3\pi}{4},\tfrac{5\pi}{6}$ (retrograde orbits). In each row we have: on the left 
    the full 3D trajectory and its projection on the $(x,y)$ plane, with the portion 
    after LSSO crossing highlighted in red, as well as the last stages of the $\theta$ evolution. In the middle we show show the real part of the 
    gravitational wave strain $h_+$, including all multipoles up to $\ell=4$, as observed from the fiducial direction $(\Theta=\pi/4,\Phi=0)$. 
    The two vertical lines represent 
    the LSSO crossing (dot-dashed) and the light ring crossing (dashed). Finally, on the right we plot the amplitude and frequency of the $\ell=2, m=\pm 2$ modes 
    of each congifuration: an horizontal dashed line denotes the value of the fundamental QNM ringdown frequency, while the light ring and LSSO crossing 
    (when visible) again identified by vertical grey lines. Comparing plots in successive rows we see how the differences between the positive and negative-$m$
    modes become more pronounced as the inclination increases and the orbit becomes equatorial. We also notice how the ringdown frequency oscillations are much 
    more pronounced when the orbit is retrograde than when it is prograde (Fig.~\ref{fig:a02prog}): 
    this is of course a consequence of the source exciting the negative frequencies 
    when the rotation direction changes in the late plunge. 
    The green curve in the frequency plot represents twice the modulus of the 
    pure orbital frequency, Eq.~\eqref{eq:Omg_orbAmp}, that always peaks close to the light-ring crossing.}
\end{figure*}

%=======
% Spin 0.5
%=======
\begin{figure*}[t]
\begin{center}
    \includegraphics[height=38mm]{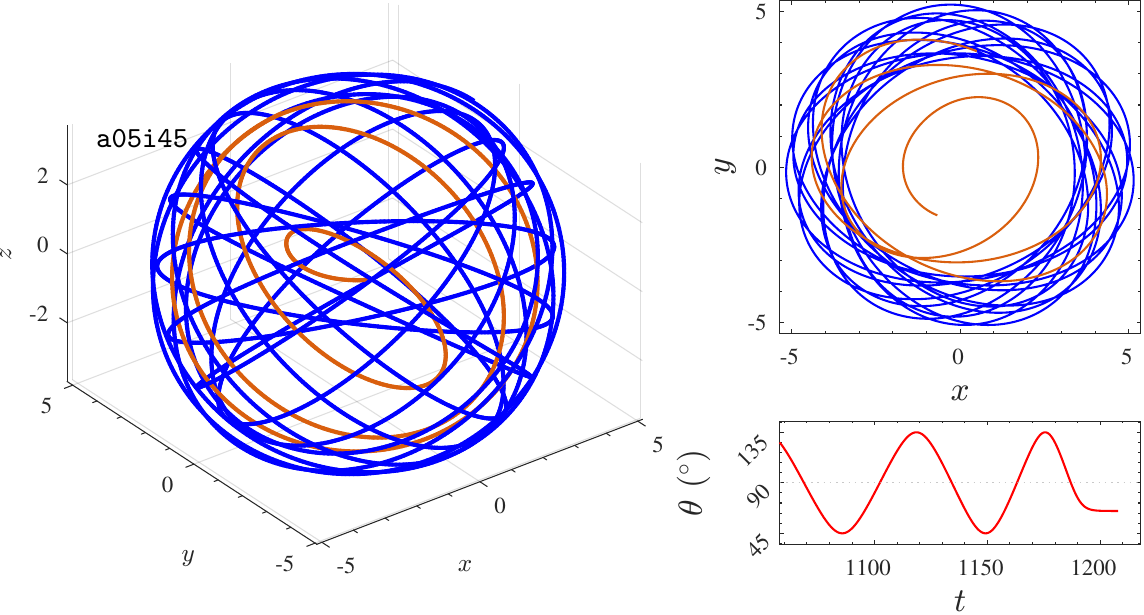} \hspace{1mm}
    \includegraphics[height=39mm]{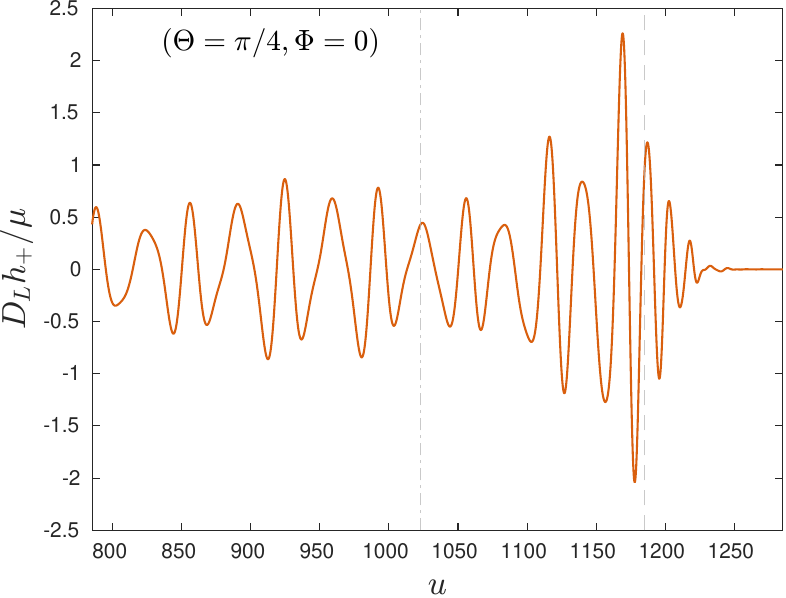} \hspace{2mm}
    \includegraphics[height=38mm]{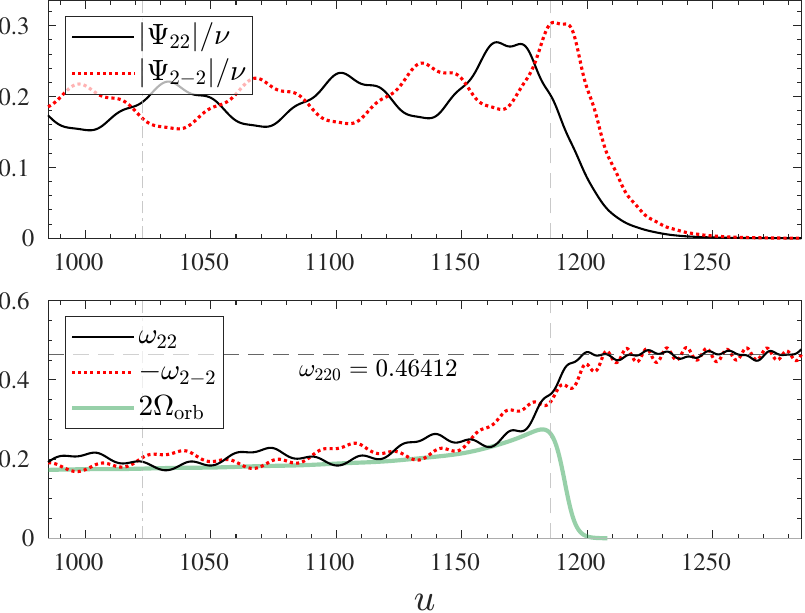} \\
    \vspace{.5cm}
    \includegraphics[height=38mm]{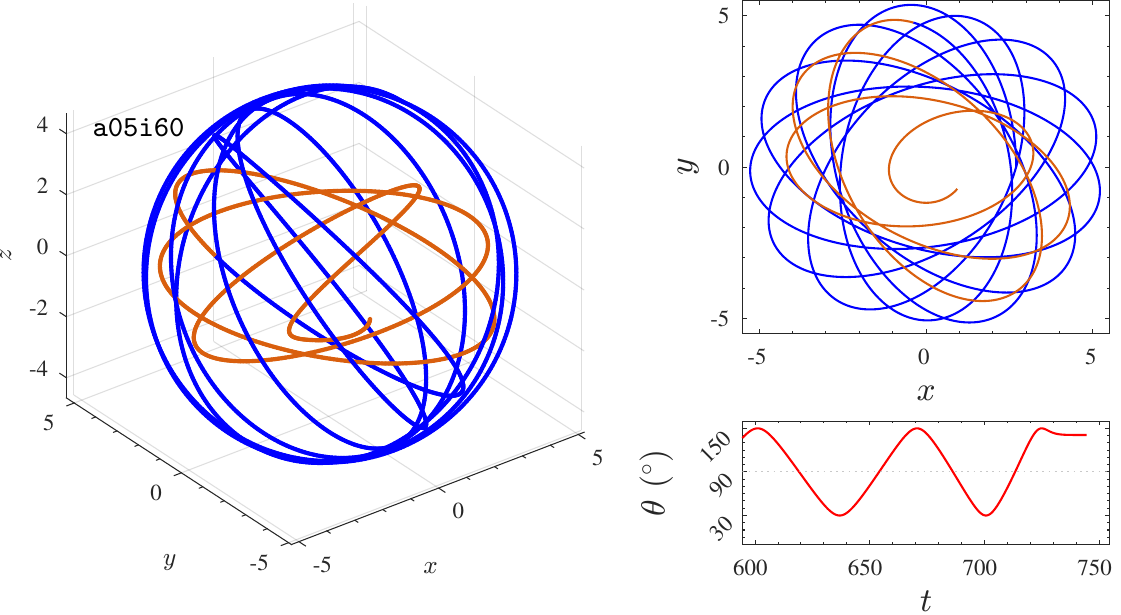} \hspace{1mm}
    \includegraphics[height=39mm]{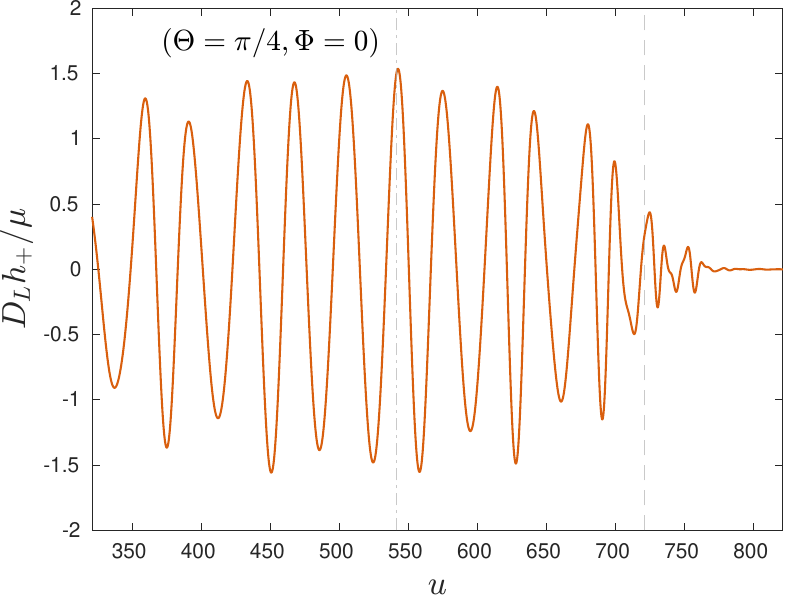} \hspace{2mm}
    \includegraphics[height=38mm]{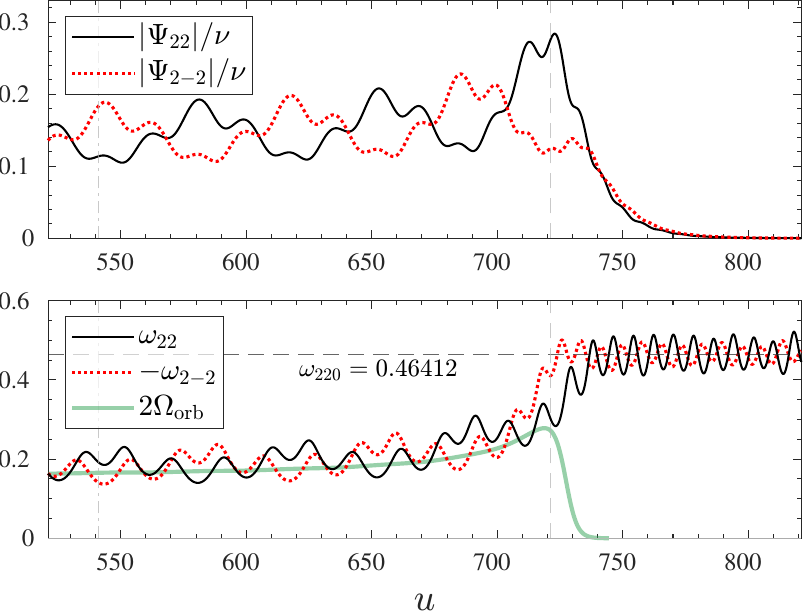} \\
    \end{center}
    \caption{\label{fig:a05prog}
    Waveform characterization for $a=0.5,\tfrac{\pi}{4},\tfrac{\pi}{3}$ (prograde orbits). In each row we have: on the left 
    the full 3D trajectory and its projection on the $(x,y)$ plane, with the portion 
    after LSSO crossing highlighted in red, as well as the last stages of the $\theta$ evolution. In the middle we show show the real part of the 
    gravitational wave strain $h_+$, including all multipoles up to $\ell=4$, as observed from the fiducial direction $(\Theta=\pi/4,\Phi=0)$. 
    The two vertical lines represent 
    the LSSO crossing (dot-dashed) and the light ring crossing (dashed). Finally, on the right we plot the amplitude and frequency of the $\ell=2, m=\pm 2$ modes 
    of each congifuration: an horizontal dashed line denotes the value of the fundamental QNM ringdown frequency, while the light ring and LSSO crossing 
    (when visible) again identified by vertical grey lines. Comparing plots in successive rows we see how the differences between the positive and negative-$m$
    modes become more pronounced as the inclination increases. The green curve in the frequency plot represents twice the modulus of the 
    pure orbital frequency, Eq.~\eqref{eq:Omg_orbAmp}, that always peaks close to the light-ring crossing. 
    We also mention that, as the spin of the black hole increases, the value of $2\Omega_{\rm orb}$ starts to differ noticeably from that of $\omega_{22}$ during the 
    early inspiral; this is especially noticeable for low inclinations.}
\end{figure*}
%\vfill

\begin{figure*}[t]
\begin{center}
    \includegraphics[height=38mm]{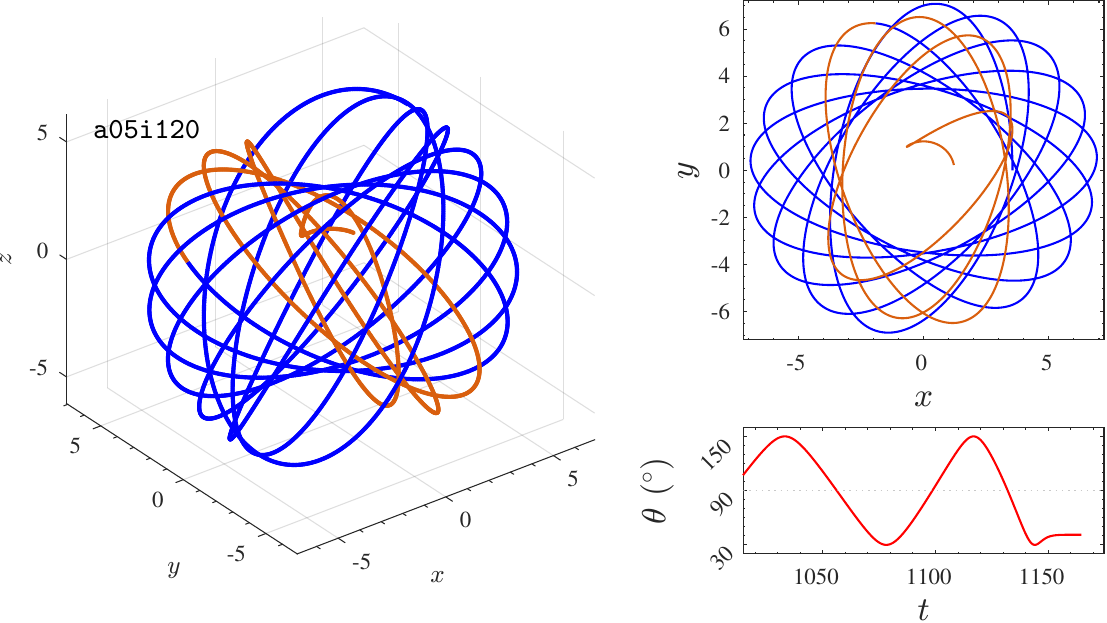} \hspace{1mm}
    \includegraphics[height=39mm]{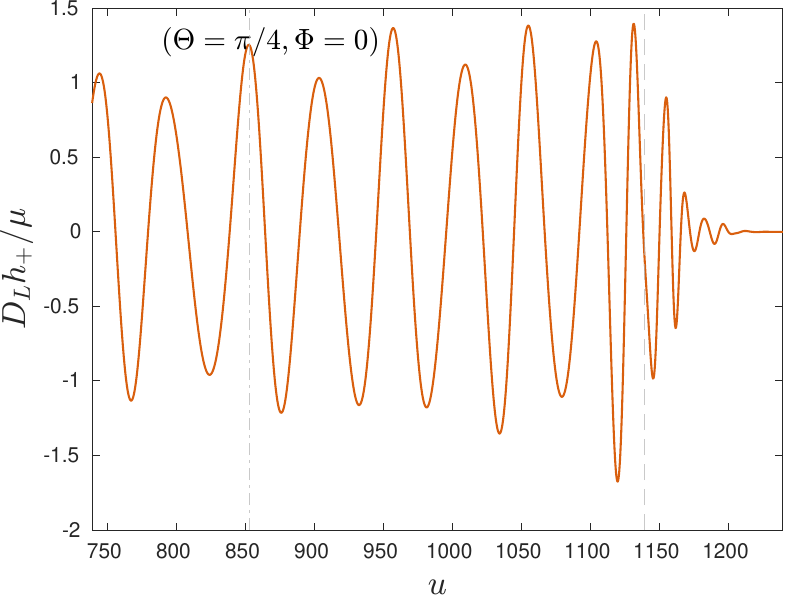} \hspace{2mm}
    \includegraphics[height=38mm]{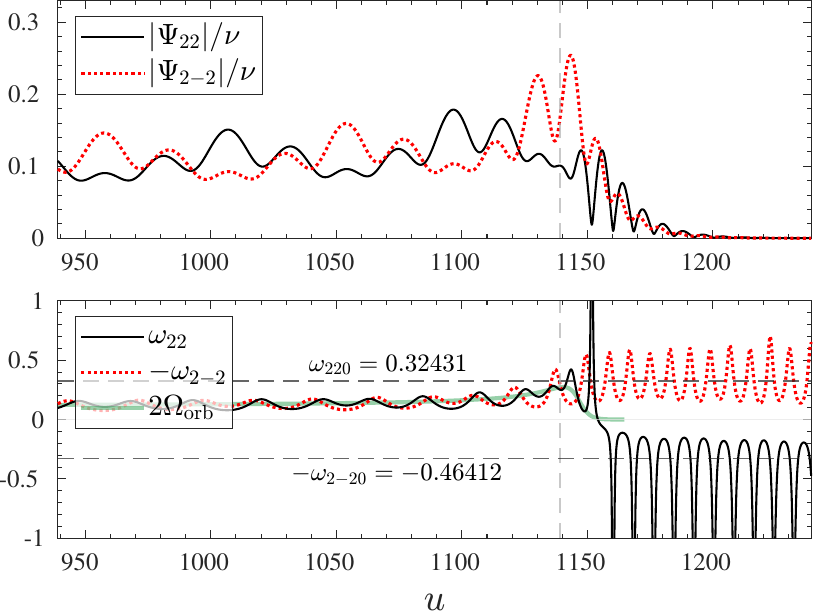} \\
    \vspace{.5cm}

    \includegraphics[height=38mm]{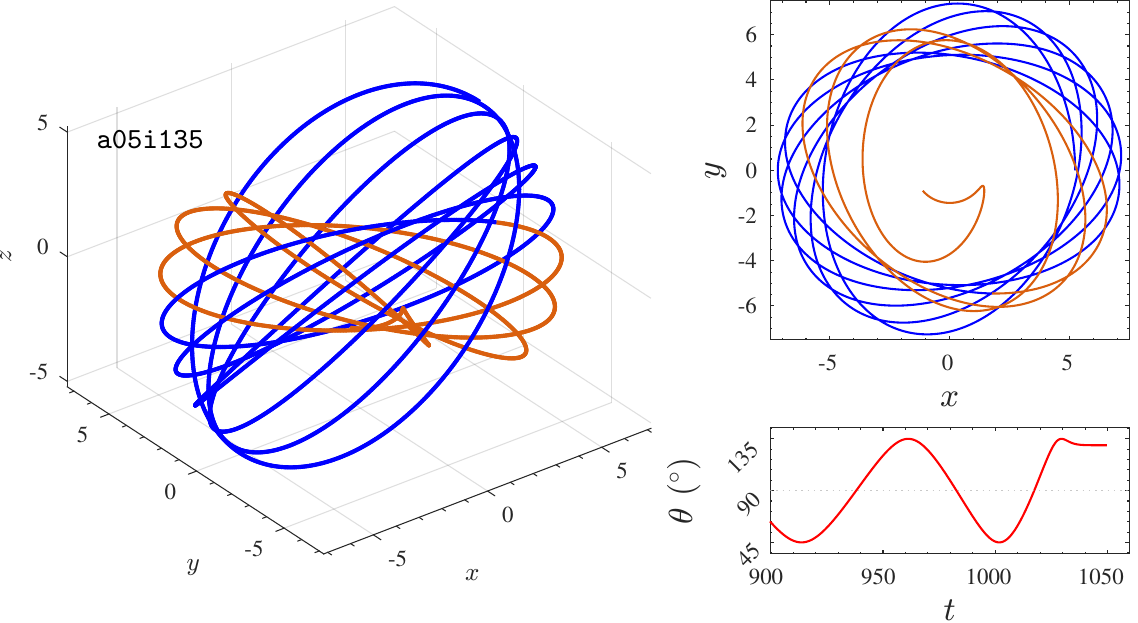} \hspace{1mm}
    \includegraphics[height=39mm]{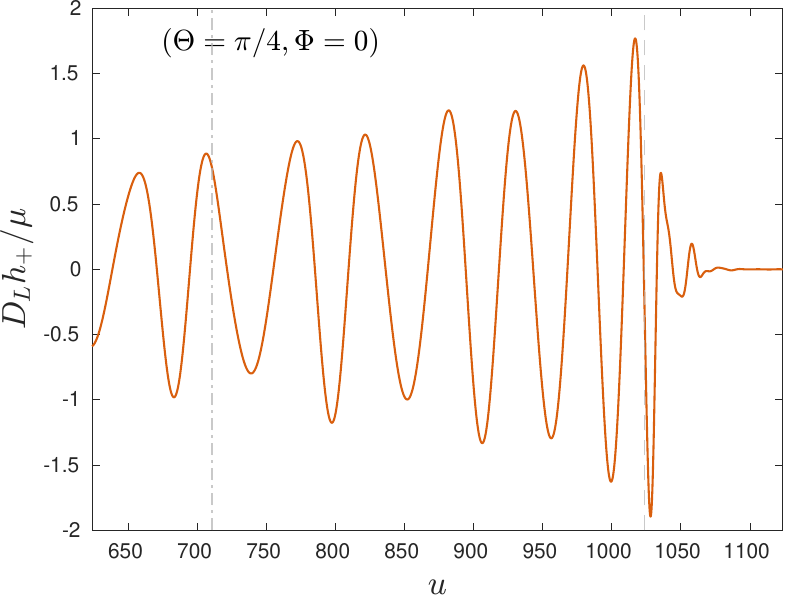} \hspace{2mm}
    \includegraphics[height=38mm]{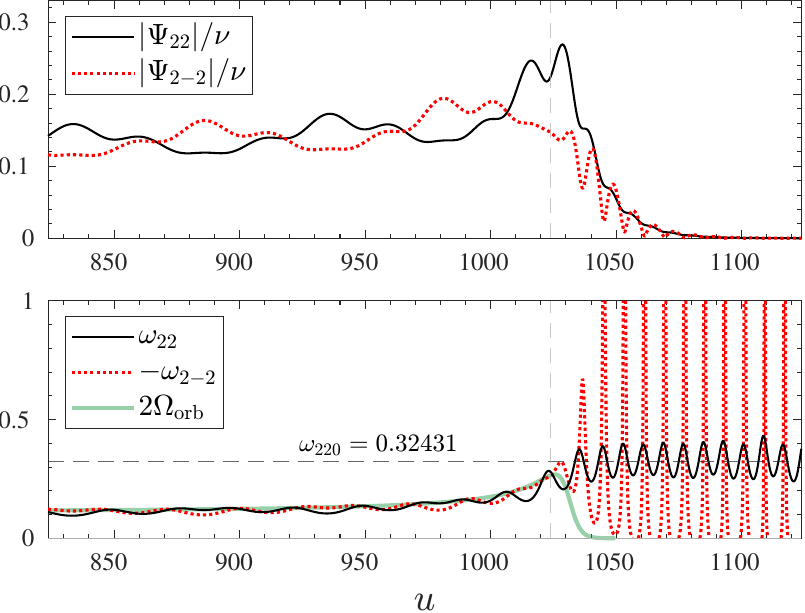} \\
    \vspace{.5cm}

    \includegraphics[height=38mm]{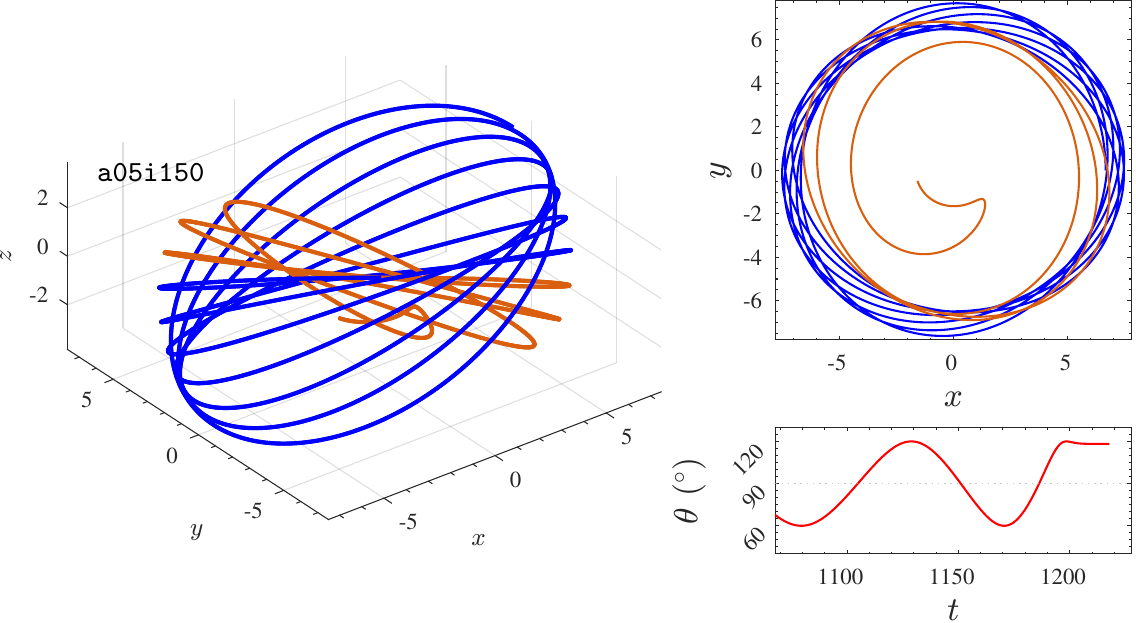} \hspace{1mm}
    \includegraphics[height=39mm]{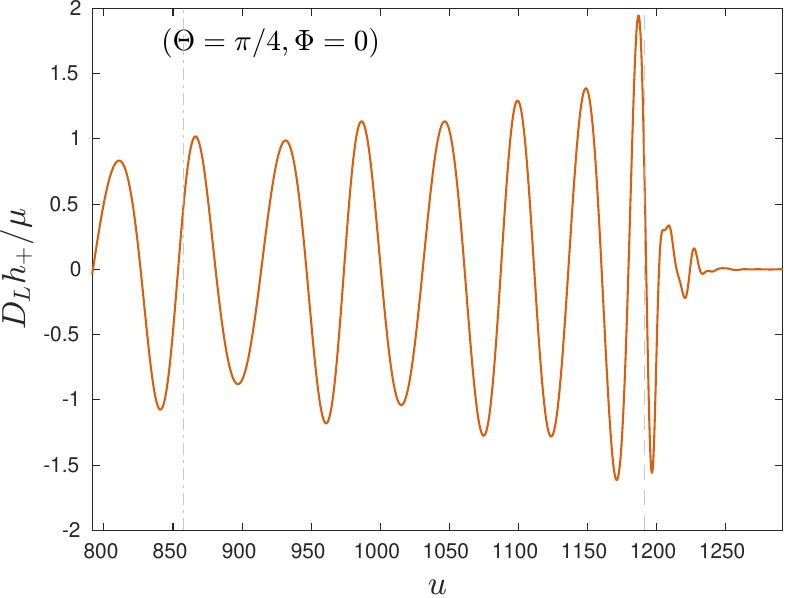} \hspace{2mm}
    \includegraphics[height=38mm]{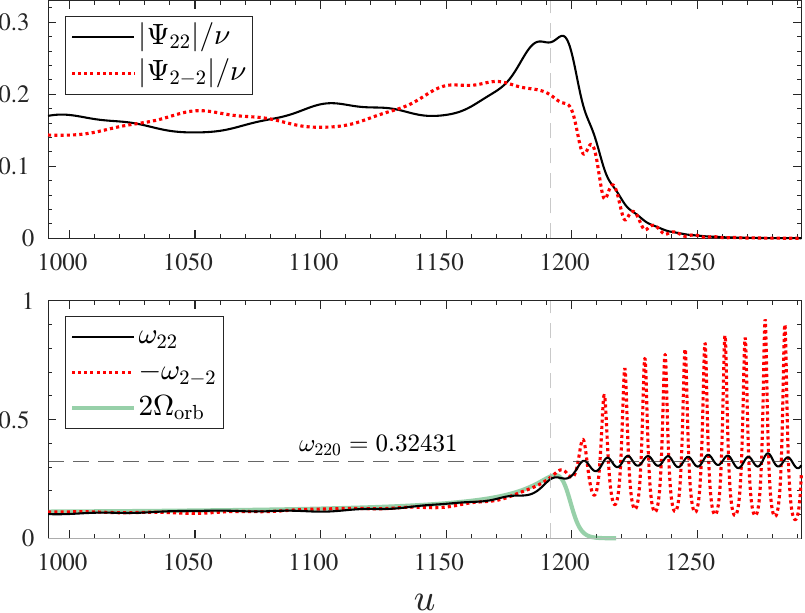} \\
    \vspace{.5cm}

    \includegraphics[height=38mm]{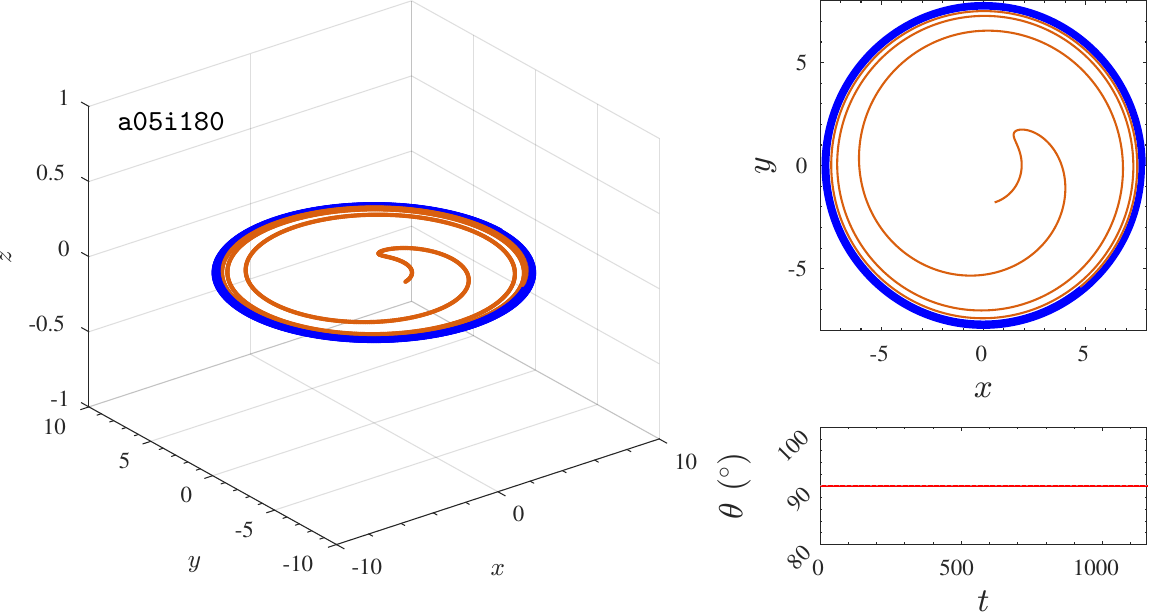} \hspace{1mm}
    \includegraphics[height=39mm]{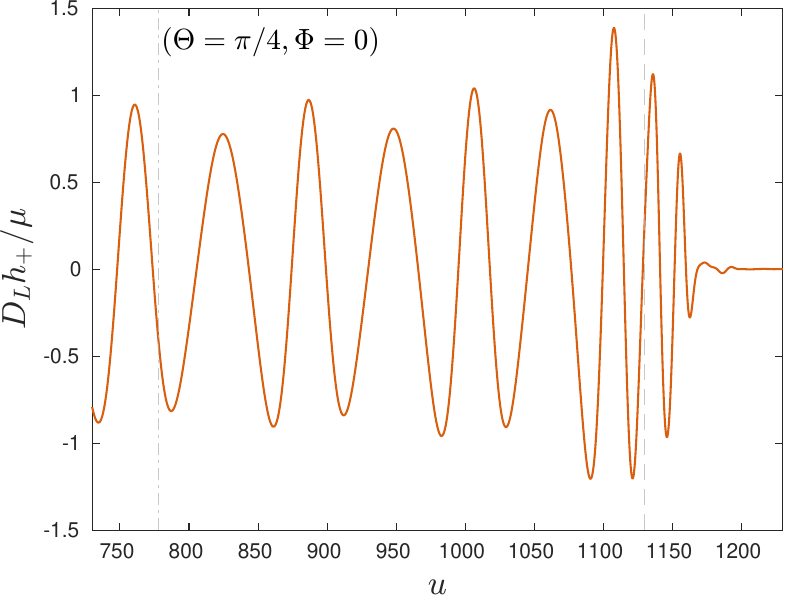} \hspace{2mm}
    \includegraphics[height=38mm]{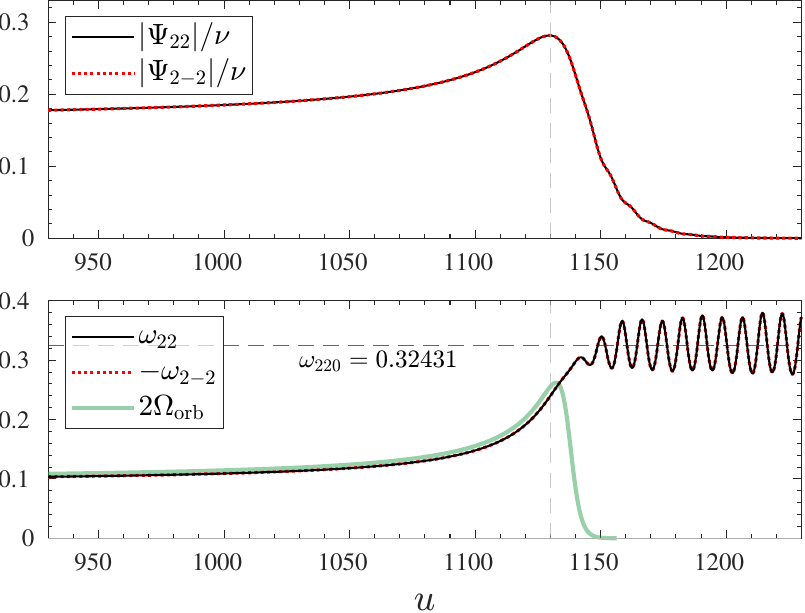} \\
\end{center}
\caption{\label{fig:a05retr}
Waveform characterization for $a=0.5,\ \iota = 0,\tfrac{2\pi}{3},\tfrac{3\pi}{4},\tfrac{5\pi}{6}$ (retrograde orbits). In each row we have: on the left 
    the full 3D trajectory and its projection on the $(x,y)$ plane, with the portion 
    after LSSO crossing highlighted in red, as well as the last stages of the $\theta$ evolution. In the middle we show show the real part of the 
    gravitational wave strain $h_+$, including all multipoles up to $\ell=4$, as observed from the fiducial direction $(\Theta=\pi/4,\Phi=0)$. 
    The two vertical lines represent 
    the LSSO crossing (dot-dashed) and the light ring crossing (dashed). Finally, on the right we plot the amplitude and frequency of the $\ell=2, m=\pm 2$ modes 
    of each congifuration: an horizontal dashed line denotes the value of the fundamental QNM ringdown frequency, while the light ring and LSSO crossing 
    (when visible) again identified by vertical grey lines. Comparing plots in successive rows we see how the differences between the positive and negative-$m$
    modes become more pronounced as the inclination increases and the orbit becomes equatorial. We also notice how the ringdown frequency oscillations are much 
    more pronounced when the orbit is retrograde than when it is prograde (Fig.~\ref{fig:a05prog}): 
    this is of course a consequence of the source exciting the negative frequencies 
    when the rotation direction changes in the late plunge. In the frequency plot of configuration {\tt a05i120} we also include the value of the 
    negative QNM ringdown frequency $-\omega_{2-20}$ since it rapidly becomes the dominant mode for the $(2,2)$ multipole (notice how the oscillation is initially 
    around $\omega_{220}$ and then immediately switches to $-\omega_{2-20}$).  
    The green curve in the frequency plot represents twice the modulus of the 
    pure orbital frequency, Eq.~\eqref{eq:Omg_orbAmp}, that always peaks close to the light-ring crossing.}
\end{figure*}
%\vfill

\begin{figure*}[t]
\begin{center}
    \includegraphics[height=37mm]{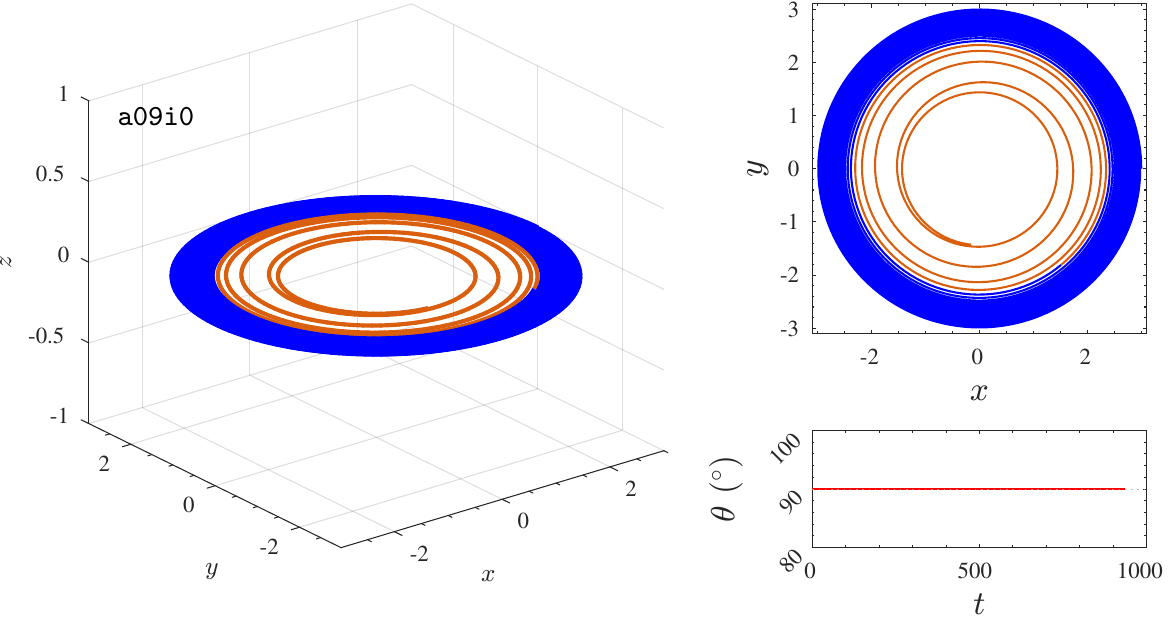} \hspace{1mm}
    \includegraphics[height=38mm]{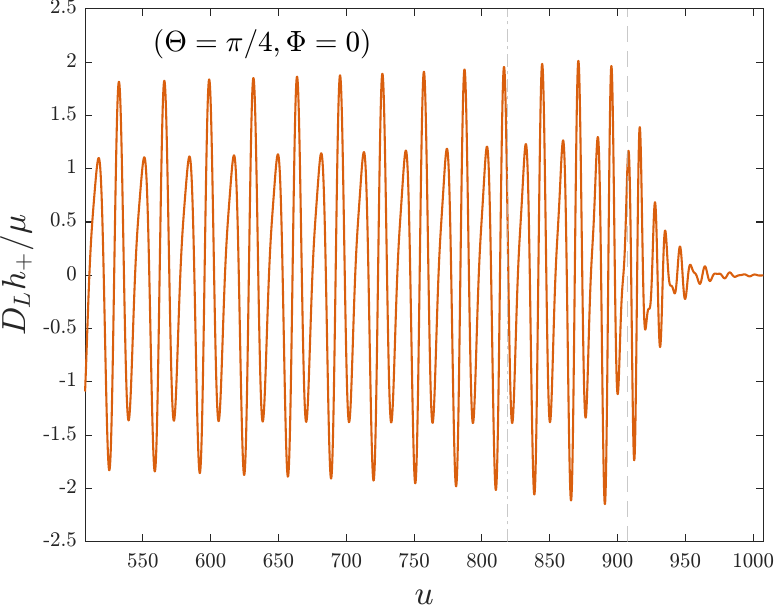} \hspace{2mm}
    \includegraphics[height=37mm]{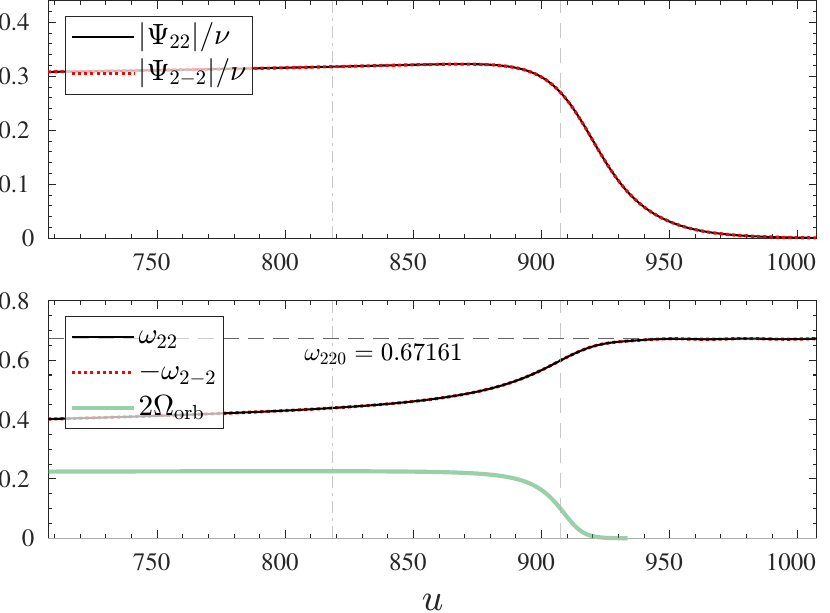} \\
    \vspace{.5cm}

    \includegraphics[height=38mm]{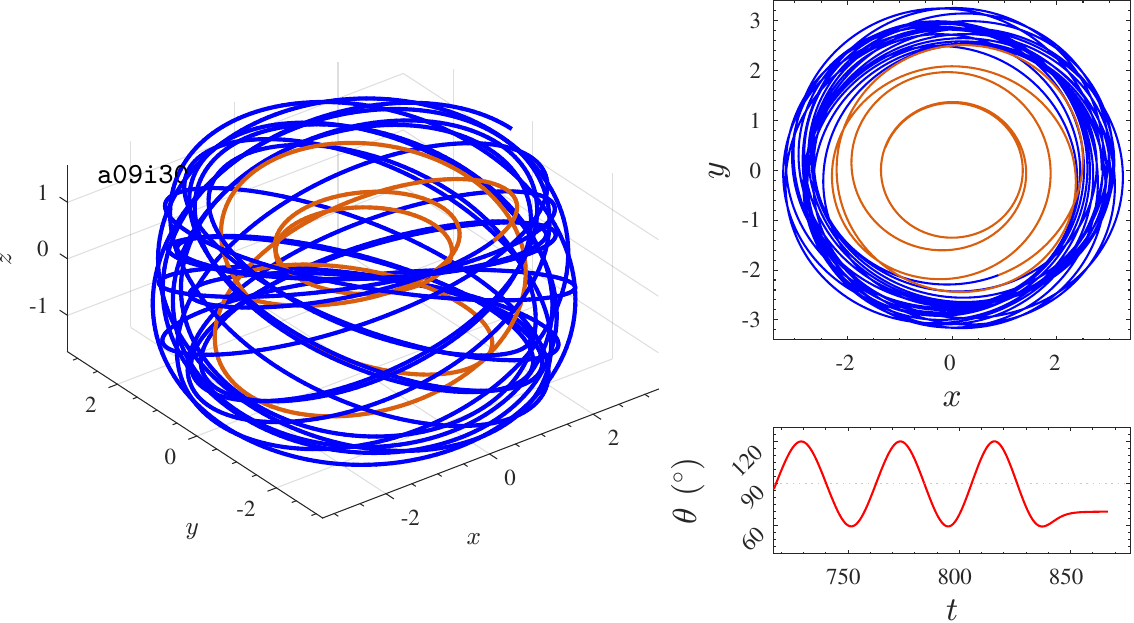} \hspace{1mm}
    \includegraphics[height=38mm]{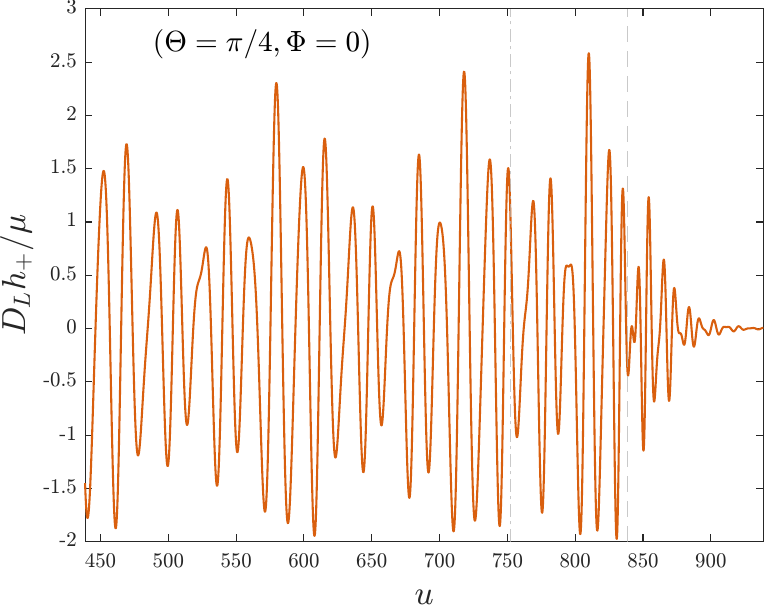} \hspace{2mm}
    \includegraphics[height=37mm]{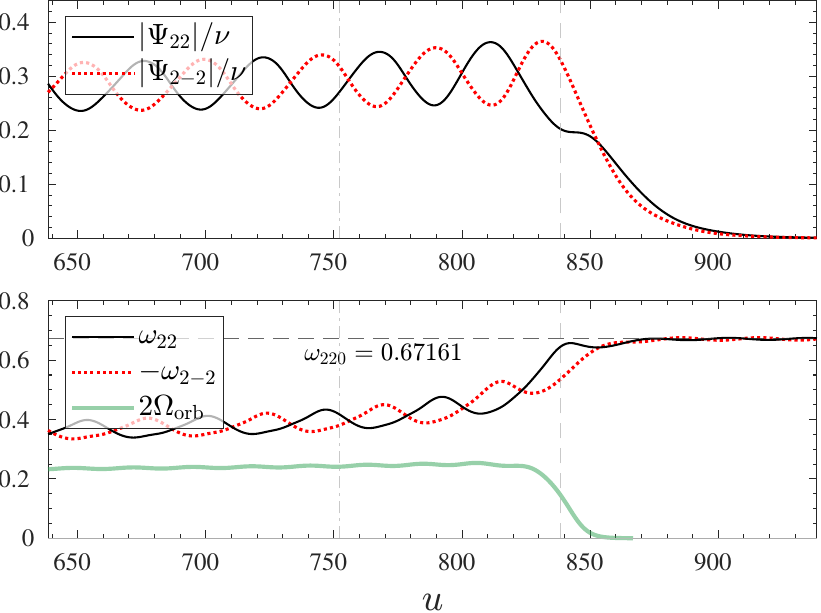} \\
    \vspace{.5cm}

    \includegraphics[height=38mm]{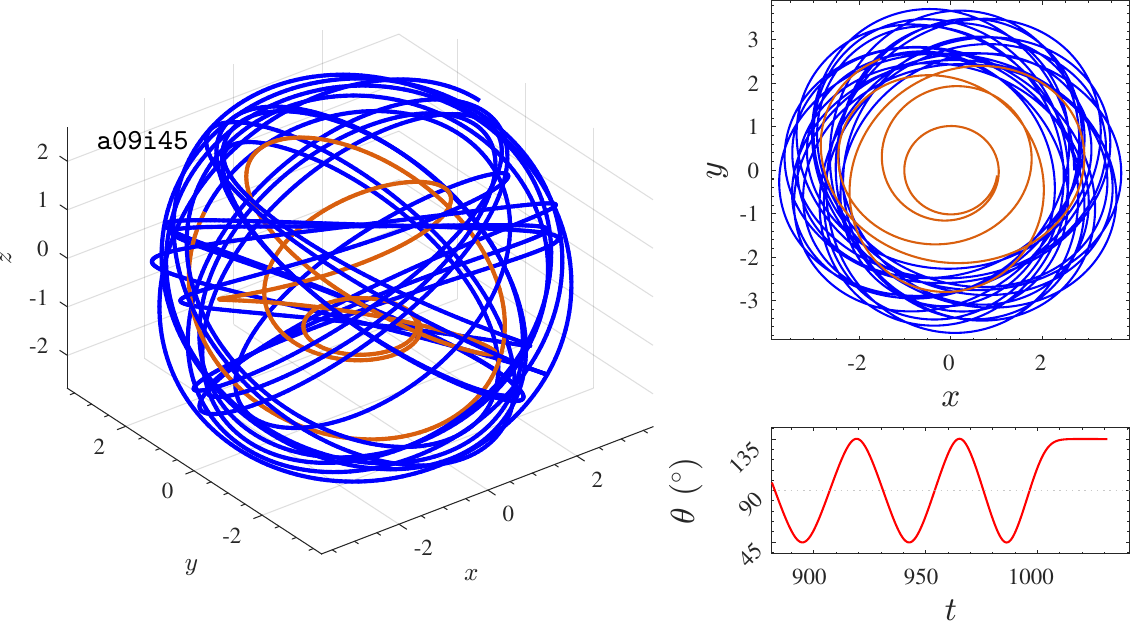} \hspace{1mm}
    \includegraphics[height=38mm]{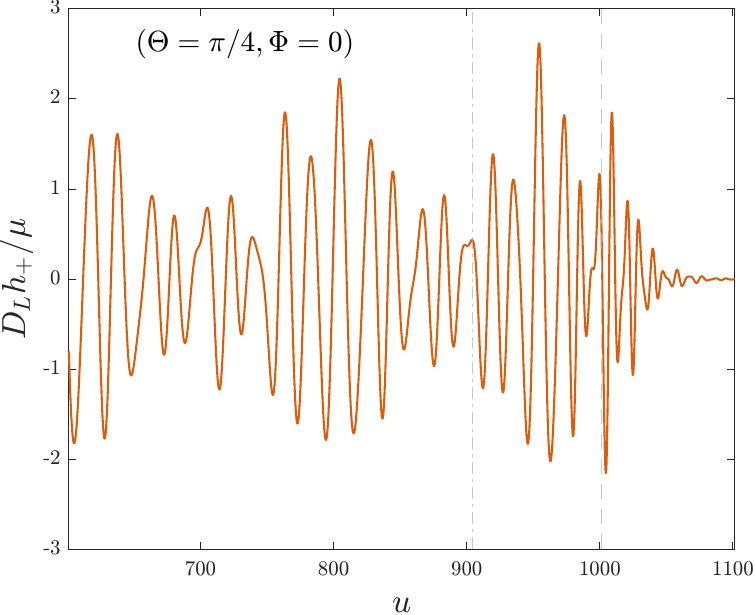} \hspace{2mm}
    \includegraphics[height=37mm]{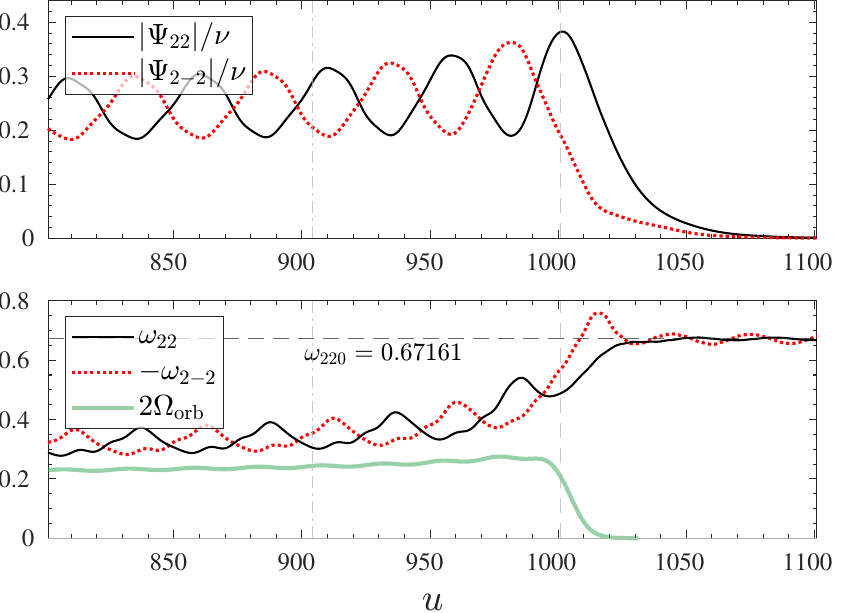} \\
    \vspace{.5cm}

    \includegraphics[height=38mm]{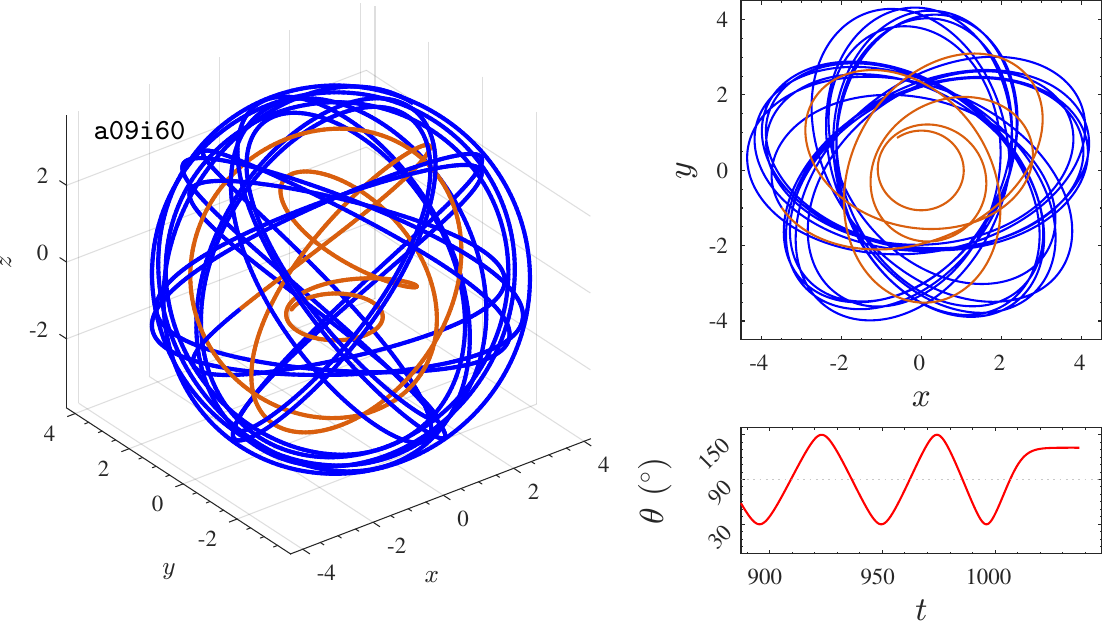} \hspace{1mm}
    \includegraphics[height=38mm]{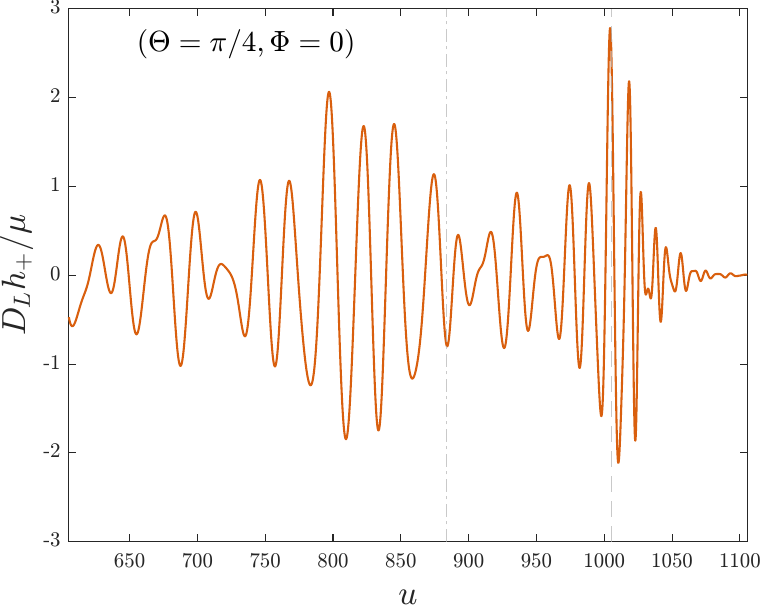} \hspace{2mm}
    \includegraphics[height=37mm]{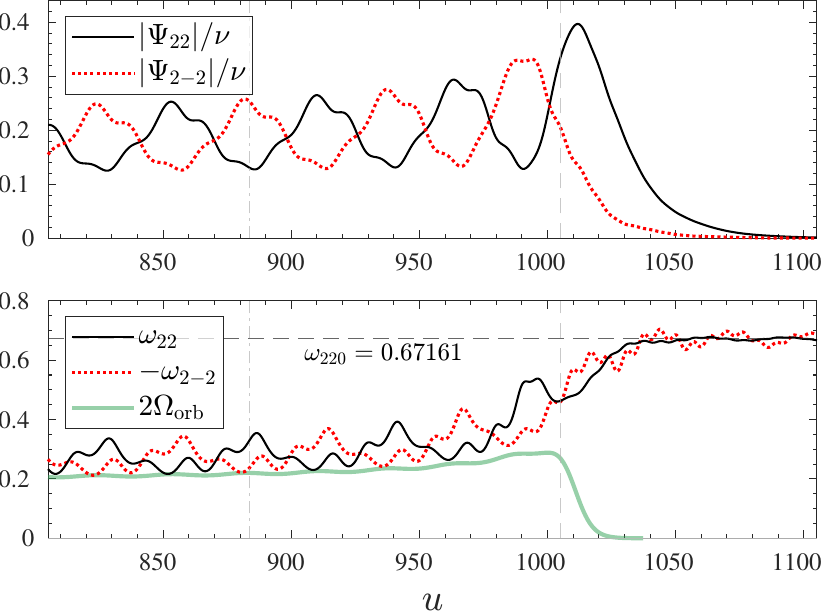} \\
\end{center}
\caption{\label{fig:a09prog}
    Waveform characterization for $a=0.9,\ \iota = 0,\tfrac{\pi}{6},\tfrac{\pi}{4},\tfrac{\pi}{3}$ (prograde orbits). In each row we have: on the left 
    the full 3D trajectory and its projection on the $(x,y)$ plane, with the portion 
    after LSSO crossing highlighted in red, as well as the last stages of the $\theta$ evolution. In the middle we show show the real part of the 
    gravitational wave strain $h_+$, including all multipoles up to $\ell=3$, as observed from the fiducial direction $(\Theta=\pi/4,\Phi=0)$. 
    The two vertical lines represent 
    the LSSO crossing (dot-dashed) and the light ring crossing (dashed). Finally, on the right we plot the amplitude and frequency of the $\ell=2, m=\pm 2$ modes 
    of each congifuration: an horizontal dashed line denotes the value of the fundamental QNM ringdown frequency, while the light ring and LSSO crossing 
    (when visible) again identified by vertical grey lines. The differences between the positive and negative-$m$ gets larger as the inclination increases. 
    Regardless of the inclination, there is very little excitation of the negative-frequency QNMs (on this respect see~\cite{Taracchini:2014zpa}). 
    The green curve in the frequency plot represents twice the modulus of the pure orbital frequency, Eq.~\eqref{eq:Omg_orbAmp}, that always peaks close to the light-ring crossing.
    As in Fig.~\ref{fig:a05prog} we notice again here that, with high values of the spin parameter $a$, $2\Omega_{\rm orb}$ starts to differ noticeably from that of $\omega_{22}$ during the early inspiral; this time this is quite evident even for large inclinations.}\end{figure*}
%\vfill

\begin{figure*}[t]
\begin{center}
    \includegraphics[height=37mm]{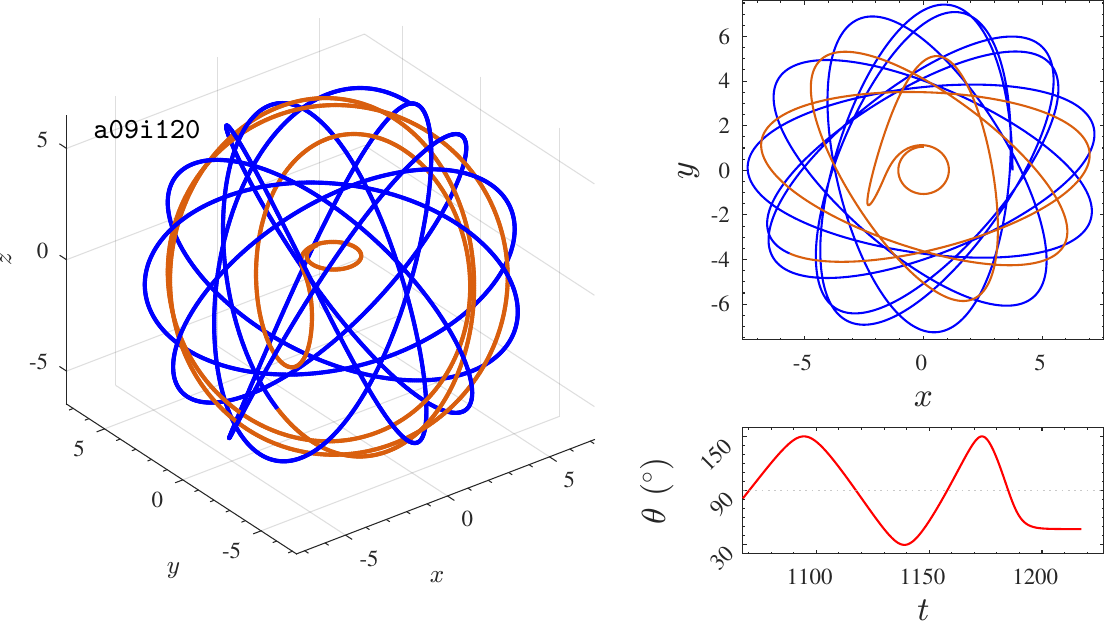} \hspace{1mm}
    \includegraphics[height=38mm]{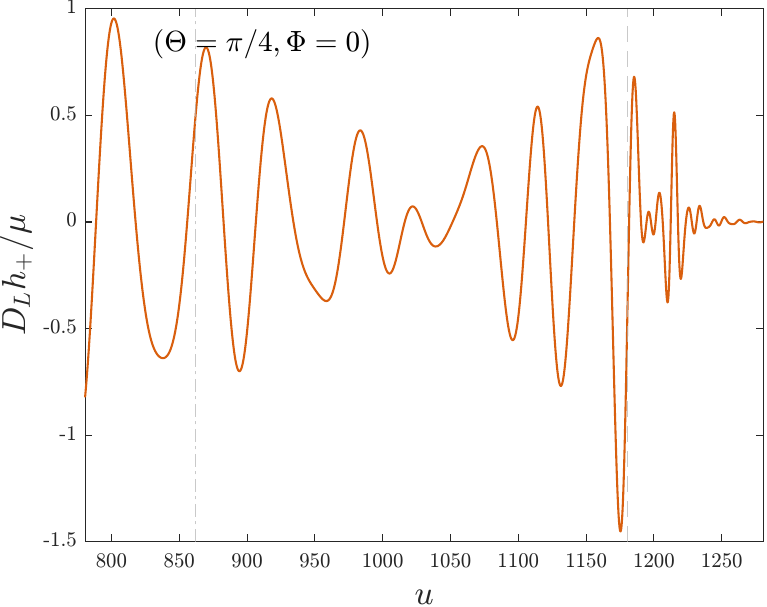} \hspace{2mm}
    \includegraphics[height=37mm]{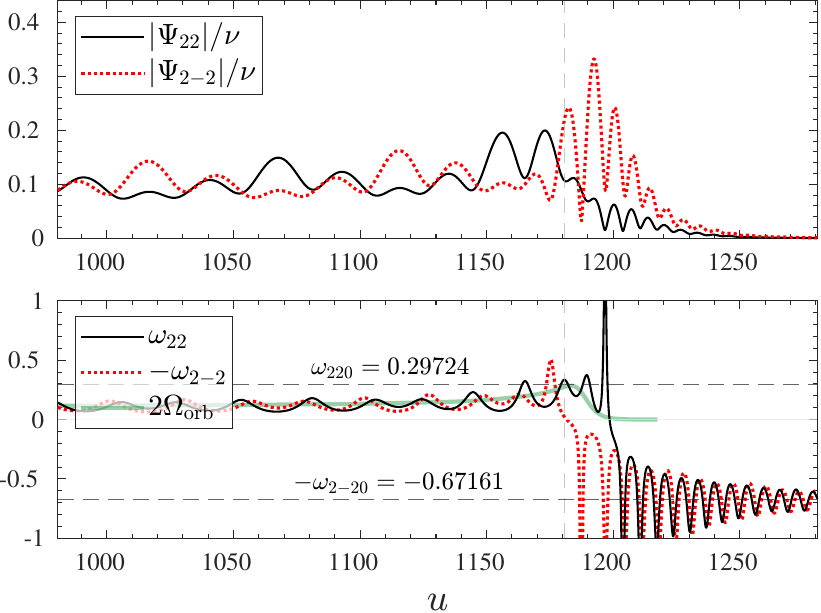} \\
    \vspace{.5cm}

    \includegraphics[height=37mm]{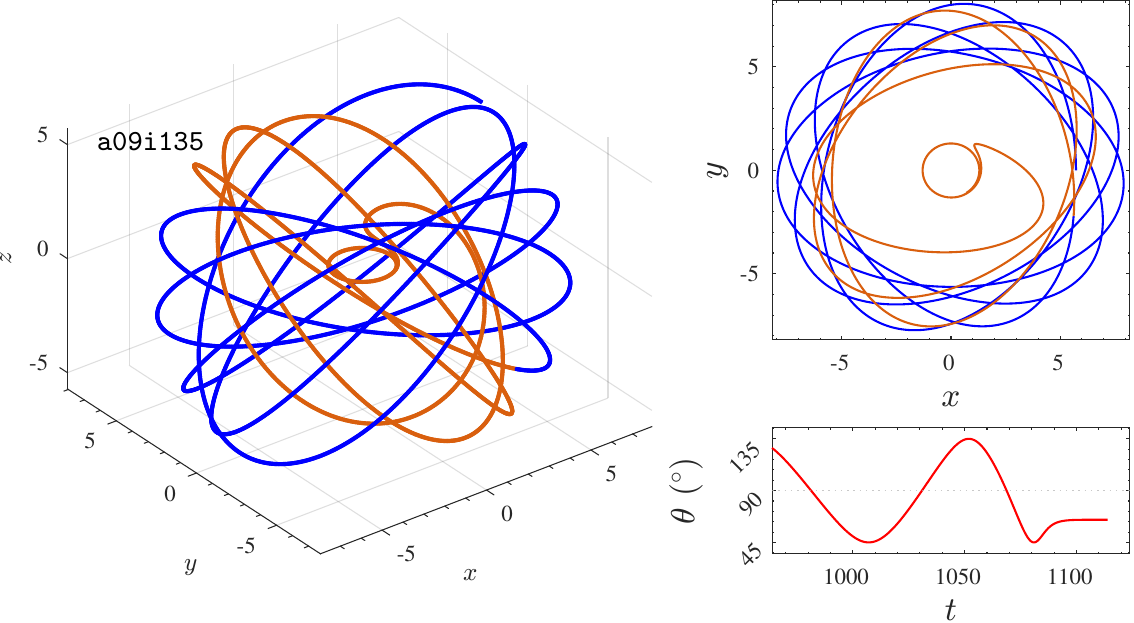} \hspace{1mm}
    \includegraphics[height=38mm]{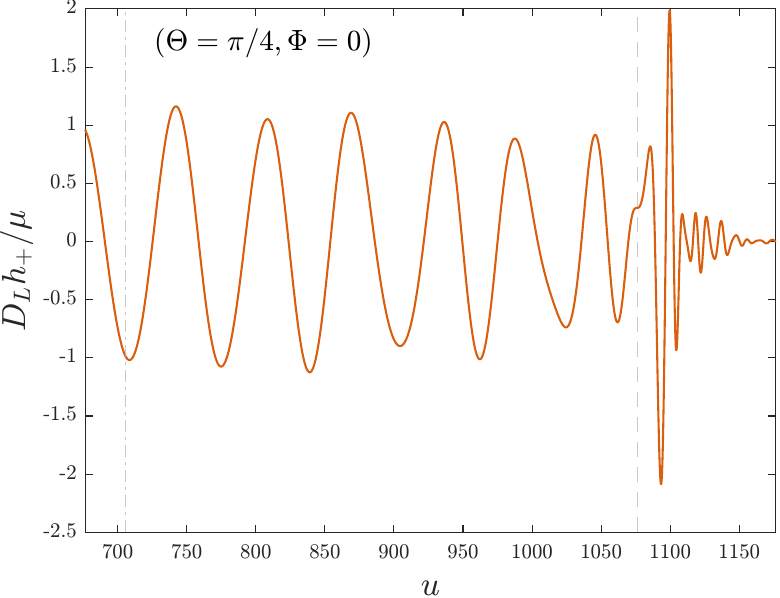} \hspace{2mm}
    \includegraphics[height=37mm]{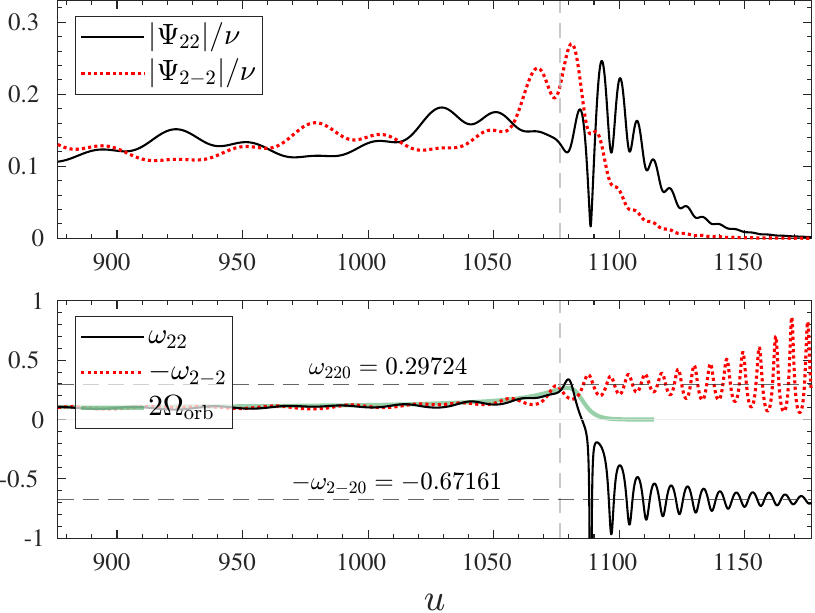} \\
    \vspace{.5cm}

    \includegraphics[height=37mm]{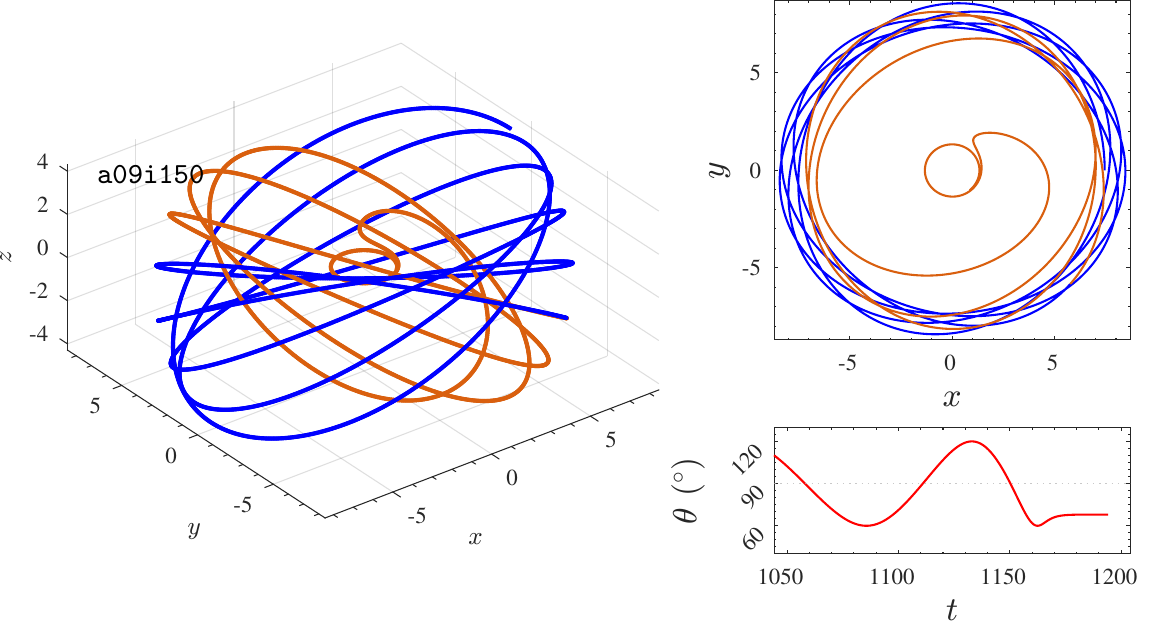} \hspace{1mm}
    \includegraphics[height=38mm]{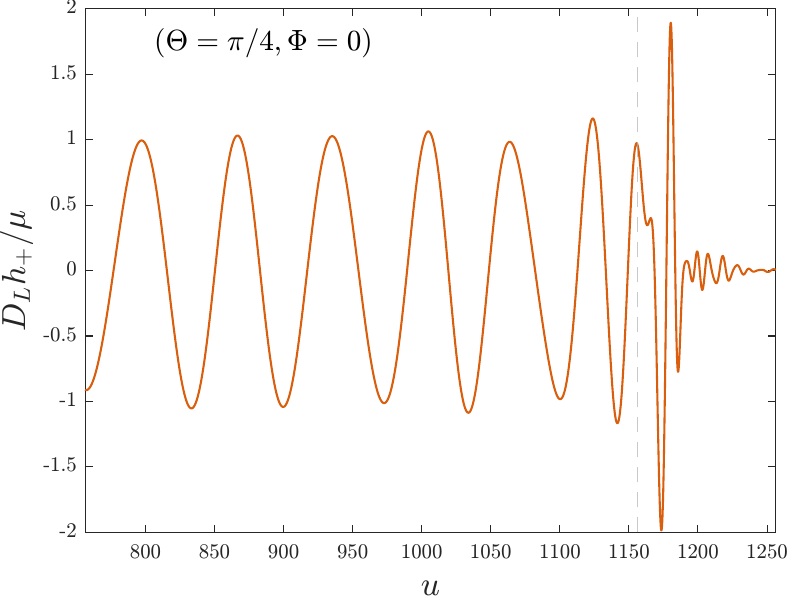} \hspace{2mm}
    \includegraphics[height=37mm]{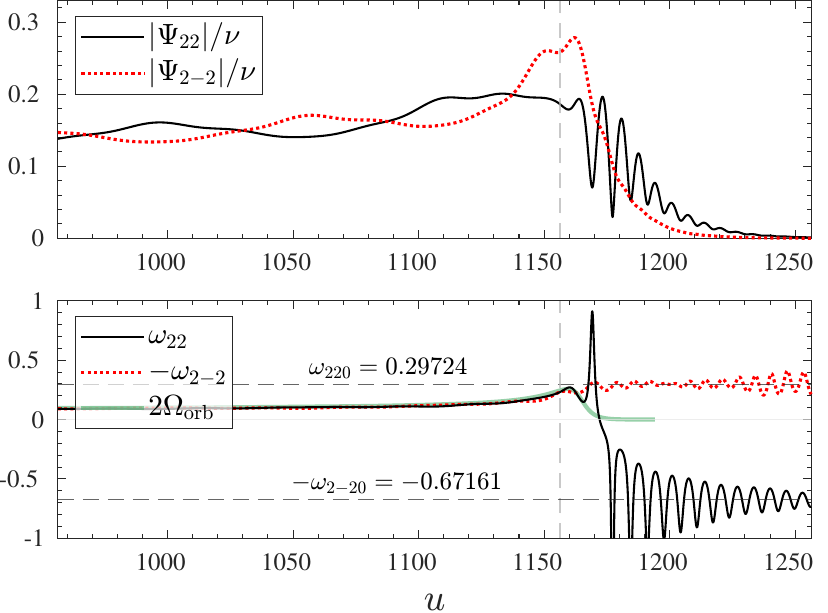} \\
    \vspace{.5cm}

    \includegraphics[height=37mm]{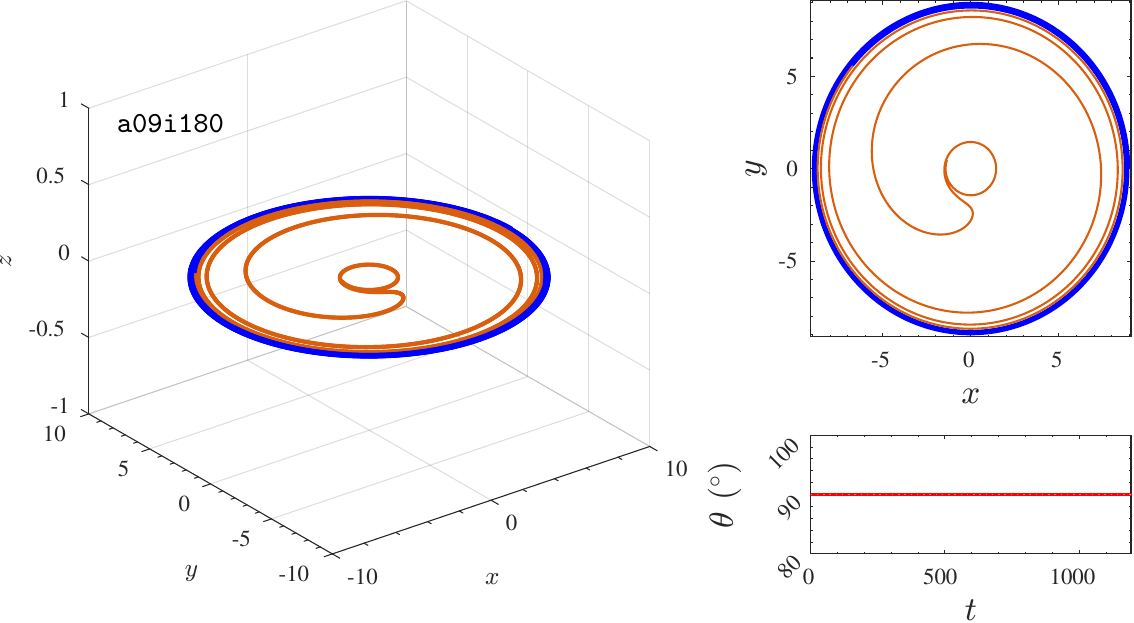} \hspace{1mm}
    \includegraphics[height=38mm]{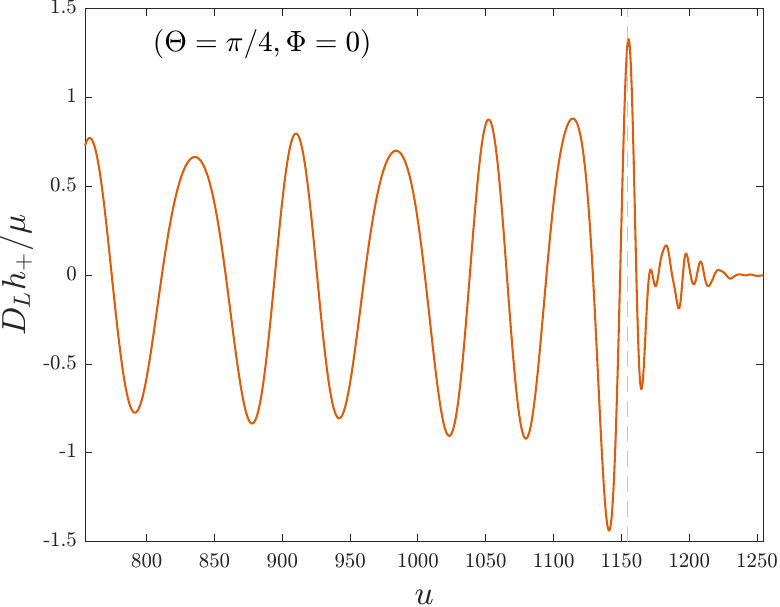} \hspace{2mm}
    \includegraphics[height=37mm]{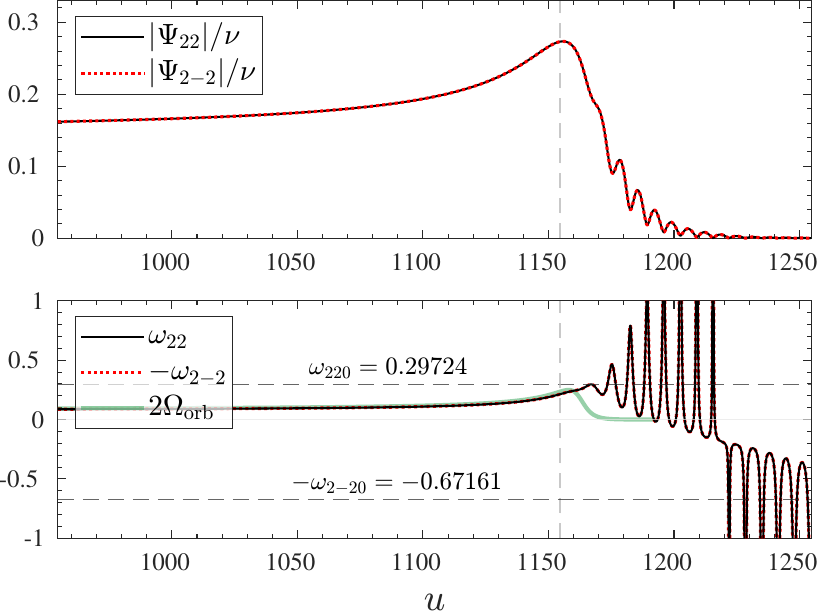} \\
\end{center}
\caption{\label{fig:a09retr}
Waveform characterization for $a=0.9,\ \iota = 0,\tfrac{2\pi}{3},\tfrac{3\pi}{4},\tfrac{5\pi}{6}$ (retrograde orbits). In each row we have: on the left 
    the full 3D trajectory and its projection on the $(x,y)$ plane, with the portion 
    after LSSO crossing highlighted in red, as well as the last stages of the $\theta$ evolution. In the middle we show show the real part of the 
    gravitational wave strain $h_+$, including all multipoles up to $\ell=3$, as observed from the fiducial direction $(\Theta=\pi/4,\Phi=0)$. 
    The two vertical lines represent 
    the LSSO crossing (dot-dashed) and the light ring crossing (dashed). Finally, on the right we plot the amplitude and frequency of the $\ell=2, m=\pm 2$ modes 
    of each congifuration: an horizontal dashed line denotes the value of the fundamental QNM ringdown frequency, while the light ring and LSSO crossing 
    (when visible) again identified by vertical grey lines.The differences between the positive and negative-$m$
    modes become more pronounced as the inclination increases and the orbit becomes equatorial. 
    We also notice how the ringdown frequency oscillations are much 
    more pronounced when the orbit is retrograde than when it is prograde (Fig.~\ref{fig:a09prog}): 
    this is of course a consequence of the source exciting the negative frequencies 
    when the rotation direction changes in the late plunge. 
    In all frequency plots we include the value of the 
    negative QNM ringdown frequency $-\omega_{2-20}$: this is either because, as was the case for {\tt a05i120} in 
    Fig.~\ref{fig:a05prog}, it becomes dominant during the ringdown ({\tt a09i120, a09i150, a09i180}) or simply because the $(2,2)$ and $(2,-2)$ 
    modes settle around different QNM frequencies ({\tt a09i135}).
    The green curve in the frequency plot represents twice the modulus of the pure orbital frequency, Eq.~\eqref{eq:Omg_orbAmp}, that always peaks close to the light-ring crossing.}
\end{figure*}
%\vfill

\newpage
\bibliography{refs, local}

\begin{thebibliography}{96}%
\makeatletter
\providecommand \@ifxundefined [1]{%
 \@ifx{#1\undefined}
}%
\providecommand \@ifnum [1]{%
 \ifnum #1\expandafter \@firstoftwo
 \else \expandafter \@secondoftwo
 \fi
}%
\providecommand \@ifx [1]{%
 \ifx #1\expandafter \@firstoftwo
 \else \expandafter \@secondoftwo
 \fi
}%
\providecommand \natexlab [1]{#1}%
\providecommand \enquote  [1]{``#1''}%
\providecommand \bibnamefont  [1]{#1}%
\providecommand \bibfnamefont [1]{#1}%
\providecommand \citenamefont [1]{#1}%
\providecommand \href@noop [0]{\@secondoftwo}%
\providecommand \href [0]{\begingroup \@sanitize@url \@href}%
\providecommand \@href[1]{\@@startlink{#1}\@@href}%
\providecommand \@@href[1]{\endgroup#1\@@endlink}%
\providecommand \@sanitize@url [0]{\catcode `\\12\catcode `\$12\catcode
  `\&12\catcode `\#12\catcode `\^12\catcode `\_12\catcode `\%12\relax}%
\providecommand \@@startlink[1]{}%
\providecommand \@@endlink[0]{}%
\providecommand \url  [0]{\begingroup\@sanitize@url \@url }%
\providecommand \@url [1]{\endgroup\@href {#1}{\urlprefix }}%
\providecommand \urlprefix  [0]{URL }%
\providecommand \Eprint [0]{\href }%
\providecommand \doibase [0]{http://dx.doi.org/}%
\providecommand \selectlanguage [0]{\@gobble}%
\providecommand \bibinfo  [0]{\@secondoftwo}%
\providecommand \bibfield  [0]{\@secondoftwo}%
\providecommand \translation [1]{[#1]}%
\providecommand \BibitemOpen [0]{}%
\providecommand \bibitemStop [0]{}%
\providecommand \bibitemNoStop [0]{.\EOS\space}%
\providecommand \EOS [0]{\spacefactor3000\relax}%
\providecommand \BibitemShut  [1]{\csname bibitem#1\endcsname}%
\let\auto@bib@innerbib\@empty
%</preamble>
\bibitem [{\citenamefont {Abbott}\ \emph {et~al.}(2016)\citenamefont {Abbott}
  \emph {et~al.}}]{LIGOScientific:2016aoc}%
  \BibitemOpen
  \bibfield  {author} {\bibinfo {author} {\bibfnamefont {B.~P.}\ \bibnamefont
  {Abbott}} \emph {et~al.} (\bibinfo {collaboration} {LIGO Scientific,
  Virgo}),\ }\href {\doibase 10.1103/PhysRevLett.116.061102} {\bibfield
  {journal} {\bibinfo  {journal} {Phys. Rev. Lett.}\ }\textbf {\bibinfo
  {volume} {116}},\ \bibinfo {pages} {061102} (\bibinfo {year} {2016})},\
  \Eprint {http://arxiv.org/abs/1602.03837} {arXiv:1602.03837 [gr-qc]}
  \BibitemShut {NoStop}%
\bibitem [{\citenamefont {Abbott}\ \emph {et~al.}(2019)\citenamefont {Abbott}
  \emph {et~al.}}]{LIGOScientific:2018mvr}%
  \BibitemOpen
  \bibfield  {author} {\bibinfo {author} {\bibfnamefont {B.~P.}\ \bibnamefont
  {Abbott}} \emph {et~al.} (\bibinfo {collaboration} {LIGO Scientific,
  Virgo}),\ }\href {\doibase 10.1103/PhysRevX.9.031040} {\bibfield  {journal}
  {\bibinfo  {journal} {Phys. Rev.}\ }\textbf {\bibinfo {volume} {X9}},\
  \bibinfo {pages} {031040} (\bibinfo {year} {2019})},\ \Eprint
  {http://arxiv.org/abs/1811.12907} {arXiv:1811.12907 [astro-ph.HE]}
  \BibitemShut {NoStop}%
%%CITATION = ARXIV:1811.12907;%%
\bibitem [{\citenamefont {Abbott}\ \emph
  {et~al.}(2021{\natexlab{a}})\citenamefont {Abbott} \emph
  {et~al.}}]{LIGOScientific:2020ibl}%
  \BibitemOpen
  \bibfield  {author} {\bibinfo {author} {\bibfnamefont {R.}~\bibnamefont
  {Abbott}} \emph {et~al.} (\bibinfo {collaboration} {LIGO Scientific,
  Virgo}),\ }\href {\doibase 10.1103/PhysRevX.11.021053} {\bibfield  {journal}
  {\bibinfo  {journal} {Phys. Rev. X}\ }\textbf {\bibinfo {volume} {11}},\
  \bibinfo {pages} {021053} (\bibinfo {year} {2021}{\natexlab{a}})},\ \Eprint
  {http://arxiv.org/abs/2010.14527} {arXiv:2010.14527 [gr-qc]} \BibitemShut
  {NoStop}%
\bibitem [{\citenamefont {Abbott}\ \emph
  {et~al.}(2021{\natexlab{b}})\citenamefont {Abbott} \emph
  {et~al.}}]{LIGOScientific:2021djp}%
  \BibitemOpen
  \bibfield  {author} {\bibinfo {author} {\bibfnamefont {R.}~\bibnamefont
  {Abbott}} \emph {et~al.} (\bibinfo {collaboration} {LIGO Scientific, VIRGO,
  KAGRA}),\ }\href@noop {} {\  (\bibinfo {year} {2021}{\natexlab{b}})},\
  \Eprint {http://arxiv.org/abs/2111.03606} {arXiv:2111.03606 [gr-qc]}
  \BibitemShut {NoStop}%
\bibitem [{\citenamefont {Buonanno}\ and\ \citenamefont
  {Damour}(1999)}]{Buonanno:1998gg}%
  \BibitemOpen
  \bibfield  {author} {\bibinfo {author} {\bibfnamefont {A.}~\bibnamefont
  {Buonanno}}\ and\ \bibinfo {author} {\bibfnamefont {T.}~\bibnamefont
  {Damour}},\ }\href {\doibase 10.1103/PhysRevD.59.084006} {\bibfield
  {journal} {\bibinfo  {journal} {Phys. Rev.}\ }\textbf {\bibinfo {volume}
  {D59}},\ \bibinfo {pages} {084006} (\bibinfo {year} {1999})},\ \Eprint
  {http://arxiv.org/abs/gr-qc/9811091} {arXiv:gr-qc/9811091} \BibitemShut
  {NoStop}%
%%CITATION = GR-QC/9811091;%%
\bibitem [{\citenamefont {Buonanno}\ and\ \citenamefont
  {Damour}(2000)}]{Buonanno:2000ef}%
  \BibitemOpen
  \bibfield  {author} {\bibinfo {author} {\bibfnamefont {A.}~\bibnamefont
  {Buonanno}}\ and\ \bibinfo {author} {\bibfnamefont {T.}~\bibnamefont
  {Damour}},\ }\href {\doibase 10.1103/PhysRevD.62.064015} {\bibfield
  {journal} {\bibinfo  {journal} {Phys. Rev.}\ }\textbf {\bibinfo {volume}
  {D62}},\ \bibinfo {pages} {064015} (\bibinfo {year} {2000})},\ \Eprint
  {http://arxiv.org/abs/gr-qc/0001013} {arXiv:gr-qc/0001013} \BibitemShut
  {NoStop}%
%%CITATION = GR-QC/0001013;%%
\bibitem [{\citenamefont {Damour}\ \emph {et~al.}(2000)\citenamefont {Damour},
  \citenamefont {Jaranowski},\ and\ \citenamefont {Schaefer}}]{Damour:2000we}%
  \BibitemOpen
  \bibfield  {author} {\bibinfo {author} {\bibfnamefont {T.}~\bibnamefont
  {Damour}}, \bibinfo {author} {\bibfnamefont {P.}~\bibnamefont {Jaranowski}},
  \ and\ \bibinfo {author} {\bibfnamefont {G.}~\bibnamefont {Schaefer}},\
  }\href {\doibase 10.1103/PhysRevD.62.084011} {\bibfield  {journal} {\bibinfo
  {journal} {Phys. Rev.}\ }\textbf {\bibinfo {volume} {D62}},\ \bibinfo {pages}
  {084011} (\bibinfo {year} {2000})},\ \Eprint
  {http://arxiv.org/abs/gr-qc/0005034} {arXiv:gr-qc/0005034 [gr-qc]}
  \BibitemShut {NoStop}%
%%CITATION = GR-QC/0005034;%%
\bibitem [{\citenamefont {Damour}(2001)}]{Damour:2001tu}%
  \BibitemOpen
  \bibfield  {author} {\bibinfo {author} {\bibfnamefont {T.}~\bibnamefont
  {Damour}},\ }\href {\doibase 10.1103/PhysRevD.64.124013} {\bibfield
  {journal} {\bibinfo  {journal} {Phys. Rev.}\ }\textbf {\bibinfo {volume}
  {D64}},\ \bibinfo {pages} {124013} (\bibinfo {year} {2001})},\ \Eprint
  {http://arxiv.org/abs/gr-qc/0103018} {arXiv:gr-qc/0103018} \BibitemShut
  {NoStop}%
%%CITATION = GR-QC/0103018;%%
\bibitem [{\citenamefont {Damour}\ \emph {et~al.}(2015)\citenamefont {Damour},
  \citenamefont {Jaranowski},\ and\ \citenamefont
  {Sch\"afer}}]{Damour:2015isa}%
  \BibitemOpen
  \bibfield  {author} {\bibinfo {author} {\bibfnamefont {T.}~\bibnamefont
  {Damour}}, \bibinfo {author} {\bibfnamefont {P.}~\bibnamefont {Jaranowski}},
  \ and\ \bibinfo {author} {\bibfnamefont {G.}~\bibnamefont {Sch\"afer}},\
  }\href {\doibase 10.1103/PhysRevD.91.084024} {\bibfield  {journal} {\bibinfo
  {journal} {Phys. Rev. D}\ }\textbf {\bibinfo {volume} {91}},\ \bibinfo
  {pages} {084024} (\bibinfo {year} {2015})},\ \Eprint
  {http://arxiv.org/abs/1502.07245} {arXiv:1502.07245 [gr-qc]} \BibitemShut
  {NoStop}%
\bibitem [{\citenamefont {Gamba}\ \emph {et~al.}(2024)\citenamefont {Gamba},
  \citenamefont {Chiaramello},\ and\ \citenamefont {Neogi}}]{Gamba:2024cvy}%
  \BibitemOpen
  \bibfield  {author} {\bibinfo {author} {\bibfnamefont {R.}~\bibnamefont
  {Gamba}}, \bibinfo {author} {\bibfnamefont {D.}~\bibnamefont {Chiaramello}},
  \ and\ \bibinfo {author} {\bibfnamefont {S.}~\bibnamefont {Neogi}},\ }\href
  {\doibase 10.1103/PhysRevD.110.024031} {\bibfield  {journal} {\bibinfo
  {journal} {Phys. Rev. D}\ }\textbf {\bibinfo {volume} {110}},\ \bibinfo
  {pages} {024031} (\bibinfo {year} {2024})},\ \Eprint
  {http://arxiv.org/abs/2404.15408} {arXiv:2404.15408 [gr-qc]} \BibitemShut
  {NoStop}%
\bibitem [{\citenamefont {Albanesi}\ \emph
  {et~al.}(2025{\natexlab{a}})\citenamefont {Albanesi}, \citenamefont {Gamba},
  \citenamefont {Bernuzzi}, \citenamefont {Fontbut\'e}, \citenamefont
  {Gonzalez},\ and\ \citenamefont {Nagar}}]{Albanesi:2025txj}%
  \BibitemOpen
  \bibfield  {author} {\bibinfo {author} {\bibfnamefont {S.}~\bibnamefont
  {Albanesi}}, \bibinfo {author} {\bibfnamefont {R.}~\bibnamefont {Gamba}},
  \bibinfo {author} {\bibfnamefont {S.}~\bibnamefont {Bernuzzi}}, \bibinfo
  {author} {\bibfnamefont {J.}~\bibnamefont {Fontbut\'e}}, \bibinfo {author}
  {\bibfnamefont {A.}~\bibnamefont {Gonzalez}}, \ and\ \bibinfo {author}
  {\bibfnamefont {A.}~\bibnamefont {Nagar}},\ }\href@noop {} {\  (\bibinfo
  {year} {2025}{\natexlab{a}})},\ \Eprint {http://arxiv.org/abs/2503.14580}
  {arXiv:2503.14580 [gr-qc]} \BibitemShut {NoStop}%
\bibitem [{\citenamefont {Gamba}\ \emph {et~al.}(2025)\citenamefont {Gamba},
  \citenamefont {Lange}, \citenamefont {Chiaramello}, \citenamefont {Tissino},\
  and\ \citenamefont {Tibrewal}}]{Gamba:2025qfg}%
  \BibitemOpen
  \bibfield  {author} {\bibinfo {author} {\bibfnamefont {R.}~\bibnamefont
  {Gamba}}, \bibinfo {author} {\bibfnamefont {J.}~\bibnamefont {Lange}},
  \bibinfo {author} {\bibfnamefont {D.}~\bibnamefont {Chiaramello}}, \bibinfo
  {author} {\bibfnamefont {J.}~\bibnamefont {Tissino}}, \ and\ \bibinfo
  {author} {\bibfnamefont {S.}~\bibnamefont {Tibrewal}},\ }\href@noop {} {\
  (\bibinfo {year} {2025})},\ \Eprint {http://arxiv.org/abs/2505.21612}
  {arXiv:2505.21612 [gr-qc]} \BibitemShut {NoStop}%
\bibitem [{\citenamefont {Jan}\ \emph {et~al.}(2025)\citenamefont {Jan},
  \citenamefont {Tsao}, \citenamefont {O'Shaughnessy}, \citenamefont
  {Shoemaker},\ and\ \citenamefont {Laguna}}]{Jan:2025fps}%
  \BibitemOpen
  \bibfield  {author} {\bibinfo {author} {\bibfnamefont {A.}~\bibnamefont
  {Jan}}, \bibinfo {author} {\bibfnamefont {B.-J.}\ \bibnamefont {Tsao}},
  \bibinfo {author} {\bibfnamefont {R.}~\bibnamefont {O'Shaughnessy}}, \bibinfo
  {author} {\bibfnamefont {D.}~\bibnamefont {Shoemaker}}, \ and\ \bibinfo
  {author} {\bibfnamefont {P.}~\bibnamefont {Laguna}},\ }\href@noop {} {\
  (\bibinfo {year} {2025})},\ \Eprint {http://arxiv.org/abs/2508.12460}
  {arXiv:2508.12460 [gr-qc]} \BibitemShut {NoStop}%
\bibitem [{\citenamefont {Gamba}\ \emph {et~al.}(2023)\citenamefont {Gamba},
  \citenamefont {Breschi}, \citenamefont {Carullo}, \citenamefont {Albanesi},
  \citenamefont {Rettegno}, \citenamefont {Bernuzzi},\ and\ \citenamefont
  {Nagar}}]{Gamba:2021gap}%
  \BibitemOpen
  \bibfield  {author} {\bibinfo {author} {\bibfnamefont {R.}~\bibnamefont
  {Gamba}}, \bibinfo {author} {\bibfnamefont {M.}~\bibnamefont {Breschi}},
  \bibinfo {author} {\bibfnamefont {G.}~\bibnamefont {Carullo}}, \bibinfo
  {author} {\bibfnamefont {S.}~\bibnamefont {Albanesi}}, \bibinfo {author}
  {\bibfnamefont {P.}~\bibnamefont {Rettegno}}, \bibinfo {author}
  {\bibfnamefont {S.}~\bibnamefont {Bernuzzi}}, \ and\ \bibinfo {author}
  {\bibfnamefont {A.}~\bibnamefont {Nagar}},\ }\href {\doibase
  10.1038/s41550-022-01813-w} {\bibfield  {journal} {\bibinfo  {journal}
  {Nature Astron.}\ }\textbf {\bibinfo {volume} {7}},\ \bibinfo {pages} {11}
  (\bibinfo {year} {2023})},\ \Eprint {http://arxiv.org/abs/2106.05575}
  {arXiv:2106.05575 [gr-qc]} \BibitemShut {NoStop}%
\bibitem [{\citenamefont {Henshaw}\ \emph {et~al.}(2025)\citenamefont
  {Henshaw}, \citenamefont {Lange}, \citenamefont {Lott}, \citenamefont
  {O'Shaughnessy},\ and\ \citenamefont {Cadonati}}]{Henshaw:2025arb}%
  \BibitemOpen
  \bibfield  {author} {\bibinfo {author} {\bibfnamefont {C.}~\bibnamefont
  {Henshaw}}, \bibinfo {author} {\bibfnamefont {J.}~\bibnamefont {Lange}},
  \bibinfo {author} {\bibfnamefont {P.}~\bibnamefont {Lott}}, \bibinfo {author}
  {\bibfnamefont {R.}~\bibnamefont {O'Shaughnessy}}, \ and\ \bibinfo {author}
  {\bibfnamefont {L.}~\bibnamefont {Cadonati}},\ }\href@noop {} {\  (\bibinfo
  {year} {2025})},\ \Eprint {http://arxiv.org/abs/2507.01156} {arXiv:2507.01156
  [gr-qc]} \BibitemShut {NoStop}%
\bibitem [{\citenamefont {Nagar}\ \emph {et~al.}(2025)\citenamefont {Nagar},
  \citenamefont {Chiaramello}, \citenamefont {Gamba}, \citenamefont {Albanesi},
  \citenamefont {Bernuzzi}, \citenamefont {Fantini}, \citenamefont {Panzeri},\
  and\ \citenamefont {Rettegno}}]{Nagar:2024oyk}%
  \BibitemOpen
  \bibfield  {author} {\bibinfo {author} {\bibfnamefont {A.}~\bibnamefont
  {Nagar}}, \bibinfo {author} {\bibfnamefont {D.}~\bibnamefont {Chiaramello}},
  \bibinfo {author} {\bibfnamefont {R.}~\bibnamefont {Gamba}}, \bibinfo
  {author} {\bibfnamefont {S.}~\bibnamefont {Albanesi}}, \bibinfo {author}
  {\bibfnamefont {S.}~\bibnamefont {Bernuzzi}}, \bibinfo {author}
  {\bibfnamefont {V.}~\bibnamefont {Fantini}}, \bibinfo {author} {\bibfnamefont
  {M.}~\bibnamefont {Panzeri}}, \ and\ \bibinfo {author} {\bibfnamefont
  {P.}~\bibnamefont {Rettegno}},\ }\href {\doibase 10.1103/PhysRevD.111.064050}
  {\bibfield  {journal} {\bibinfo  {journal} {Phys. Rev. D}\ }\textbf {\bibinfo
  {volume} {111}},\ \bibinfo {pages} {064050} (\bibinfo {year} {2025})},\
  \Eprint {http://arxiv.org/abs/2407.04762} {arXiv:2407.04762 [gr-qc]}
  \BibitemShut {NoStop}%
\bibitem [{\citenamefont {Pompili}\ \emph {et~al.}(2023)\citenamefont {Pompili}
  \emph {et~al.}}]{Pompili:2023tna}%
  \BibitemOpen
  \bibfield  {author} {\bibinfo {author} {\bibfnamefont {L.}~\bibnamefont
  {Pompili}} \emph {et~al.},\ }\href {\doibase 10.1103/PhysRevD.108.124035}
  {\bibfield  {journal} {\bibinfo  {journal} {Phys. Rev. D}\ }\textbf {\bibinfo
  {volume} {108}},\ \bibinfo {pages} {124035} (\bibinfo {year} {2023})},\
  \Eprint {http://arxiv.org/abs/2303.18039} {arXiv:2303.18039 [gr-qc]}
  \BibitemShut {NoStop}%
\bibitem [{\citenamefont {Estell{\'e}s}\ \emph {et~al.}(2025)\citenamefont
  {Estell{\'e}s}, \citenamefont {Buonanno}, \citenamefont {Enficiaud},
  \citenamefont {Foo},\ and\ \citenamefont {Pompili}}]{Estelles:2025zah}%
  \BibitemOpen
  \bibfield  {author} {\bibinfo {author} {\bibfnamefont {H.}~\bibnamefont
  {Estell{\'e}s}}, \bibinfo {author} {\bibfnamefont {A.}~\bibnamefont
  {Buonanno}}, \bibinfo {author} {\bibfnamefont {R.}~\bibnamefont {Enficiaud}},
  \bibinfo {author} {\bibfnamefont {C.}~\bibnamefont {Foo}}, \ and\ \bibinfo
  {author} {\bibfnamefont {L.}~\bibnamefont {Pompili}},\ }\href@noop {} {\
  (\bibinfo {year} {2025})},\ \Eprint {http://arxiv.org/abs/2506.19911}
  {arXiv:2506.19911 [gr-qc]} \BibitemShut {NoStop}%
\bibitem [{\citenamefont {Gamboa}\ \emph {et~al.}(2024)\citenamefont {Gamboa}
  \emph {et~al.}}]{Gamboa:2024hli}%
  \BibitemOpen
  \bibfield  {author} {\bibinfo {author} {\bibfnamefont {A.}~\bibnamefont
  {Gamboa}} \emph {et~al.},\ }\href@noop {} {\  (\bibinfo {year} {2024})},\
  \Eprint {http://arxiv.org/abs/2412.12823} {arXiv:2412.12823 [gr-qc]}
  \BibitemShut {NoStop}%
\bibitem [{\citenamefont {Davis}\ \emph {et~al.}(1971)\citenamefont {Davis},
  \citenamefont {Ruffini}, \citenamefont {Press},\ and\ \citenamefont
  {Price}}]{Davis:1971gg}%
  \BibitemOpen
  \bibfield  {author} {\bibinfo {author} {\bibfnamefont {M.}~\bibnamefont
  {Davis}}, \bibinfo {author} {\bibfnamefont {R.}~\bibnamefont {Ruffini}},
  \bibinfo {author} {\bibfnamefont {W.~H.}\ \bibnamefont {Press}}, \ and\
  \bibinfo {author} {\bibfnamefont {R.~H.}\ \bibnamefont {Price}},\ }\href
  {\doibase 10.1103/PhysRevLett.27.1466} {\bibfield  {journal} {\bibinfo
  {journal} {Phys. Rev. Lett.}\ }\textbf {\bibinfo {volume} {27}},\ \bibinfo
  {pages} {1466} (\bibinfo {year} {1971})}\BibitemShut {NoStop}%
\bibitem [{\citenamefont {Davis}\ \emph {et~al.}(1972)\citenamefont {Davis},
  \citenamefont {Ruffini},\ and\ \citenamefont {Tiomno}}]{Davis:1972ud}%
  \BibitemOpen
  \bibfield  {author} {\bibinfo {author} {\bibfnamefont {M.}~\bibnamefont
  {Davis}}, \bibinfo {author} {\bibfnamefont {R.}~\bibnamefont {Ruffini}}, \
  and\ \bibinfo {author} {\bibfnamefont {J.}~\bibnamefont {Tiomno}},\ }\href
  {\doibase 10.1103/PhysRevD.5.2932} {\bibfield  {journal} {\bibinfo  {journal}
  {Phys. Rev. D}\ }\textbf {\bibinfo {volume} {5}},\ \bibinfo {pages} {2932}
  (\bibinfo {year} {1972})}\BibitemShut {NoStop}%
\bibitem [{\citenamefont {Lousto}\ and\ \citenamefont
  {Price}(1997{\natexlab{a}})}]{Lousto:1996sx}%
  \BibitemOpen
  \bibfield  {author} {\bibinfo {author} {\bibfnamefont {C.~O.}\ \bibnamefont
  {Lousto}}\ and\ \bibinfo {author} {\bibfnamefont {R.~H.}\ \bibnamefont
  {Price}},\ }\href {\doibase 10.1103/PhysRevD.55.2124} {\bibfield  {journal}
  {\bibinfo  {journal} {Phys.Rev.}\ }\textbf {\bibinfo {volume} {D55}},\
  \bibinfo {pages} {2124} (\bibinfo {year} {1997}{\natexlab{a}})},\ \Eprint
  {http://arxiv.org/abs/gr-qc/9609012} {arXiv:gr-qc/9609012 [gr-qc]}
  \BibitemShut {NoStop}%
%%CITATION = GR-QC/9609012;%%
\bibitem [{\citenamefont {Lousto}\ and\ \citenamefont
  {Price}(1997{\natexlab{b}})}]{Lousto:1997wf}%
  \BibitemOpen
  \bibfield  {author} {\bibinfo {author} {\bibfnamefont {C.~O.}\ \bibnamefont
  {Lousto}}\ and\ \bibinfo {author} {\bibfnamefont {R.~H.}\ \bibnamefont
  {Price}},\ }\href {\doibase 10.1103/PhysRevD.56.6439} {\bibfield  {journal}
  {\bibinfo  {journal} {Phys. Rev.}\ }\textbf {\bibinfo {volume} {D56}},\
  \bibinfo {pages} {6439} (\bibinfo {year} {1997}{\natexlab{b}})},\ \Eprint
  {http://arxiv.org/abs/gr-qc/9705071} {arXiv:gr-qc/9705071} \BibitemShut
  {NoStop}%
%%CITATION = GR-QC/9705071;%%
\bibitem [{\citenamefont {Lousto}\ and\ \citenamefont
  {Price}(1998)}]{Lousto:1997ge}%
  \BibitemOpen
  \bibfield  {author} {\bibinfo {author} {\bibfnamefont {C.~O.}\ \bibnamefont
  {Lousto}}\ and\ \bibinfo {author} {\bibfnamefont {R.~H.}\ \bibnamefont
  {Price}},\ }\href {\doibase 10.1103/PhysRevD.57.1073} {\bibfield  {journal}
  {\bibinfo  {journal} {Phys. Rev. D}\ }\textbf {\bibinfo {volume} {57}},\
  \bibinfo {pages} {1073} (\bibinfo {year} {1998})},\ \Eprint
  {http://arxiv.org/abs/gr-qc/9708022} {arXiv:gr-qc/9708022} \BibitemShut
  {NoStop}%
\bibitem [{\citenamefont {Martel}\ and\ \citenamefont
  {Poisson}(2002)}]{Martel:2001yf}%
  \BibitemOpen
  \bibfield  {author} {\bibinfo {author} {\bibfnamefont {K.}~\bibnamefont
  {Martel}}\ and\ \bibinfo {author} {\bibfnamefont {E.}~\bibnamefont
  {Poisson}},\ }\href {\doibase 10.1103/PhysRevD.66.084001} {\bibfield
  {journal} {\bibinfo  {journal} {Phys. Rev.}\ }\textbf {\bibinfo {volume}
  {D66}},\ \bibinfo {pages} {084001} (\bibinfo {year} {2002})},\ \Eprint
  {http://arxiv.org/abs/gr-qc/0107104} {arXiv:gr-qc/0107104} \BibitemShut
  {NoStop}%
%%CITATION = GR-QC/0107104;%%
\bibitem [{\citenamefont {Lousto}\ and\ \citenamefont
  {Healy}(2020)}]{Lousto:2020tnb}%
  \BibitemOpen
  \bibfield  {author} {\bibinfo {author} {\bibfnamefont {C.~O.}\ \bibnamefont
  {Lousto}}\ and\ \bibinfo {author} {\bibfnamefont {J.}~\bibnamefont {Healy}},\
  }\href {\doibase 10.1103/PhysRevLett.125.191102} {\bibfield  {journal}
  {\bibinfo  {journal} {Phys. Rev. Lett.}\ }\textbf {\bibinfo {volume} {125}},\
  \bibinfo {pages} {191102} (\bibinfo {year} {2020})},\ \Eprint
  {http://arxiv.org/abs/2006.04818} {arXiv:2006.04818 [gr-qc]} \BibitemShut
  {NoStop}%
\bibitem [{\citenamefont {Lousto}\ and\ \citenamefont
  {Healy}(2023)}]{Lousto:2022hoq}%
  \BibitemOpen
  \bibfield  {author} {\bibinfo {author} {\bibfnamefont {C.~O.}\ \bibnamefont
  {Lousto}}\ and\ \bibinfo {author} {\bibfnamefont {J.}~\bibnamefont {Healy}},\
  }\href {\doibase 10.1088/1361-6382/acc7ef} {\bibfield  {journal} {\bibinfo
  {journal} {Class. Quant. Grav.}\ }\textbf {\bibinfo {volume} {40}},\ \bibinfo
  {pages} {09LT01} (\bibinfo {year} {2023})},\ \Eprint
  {http://arxiv.org/abs/2203.08831} {arXiv:2203.08831 [gr-qc]} \BibitemShut
  {NoStop}%
\bibitem [{\citenamefont {Nagar}\ \emph {et~al.}(2007)\citenamefont {Nagar},
  \citenamefont {Damour},\ and\ \citenamefont {Tartaglia}}]{Nagar:2006xv}%
  \BibitemOpen
  \bibfield  {author} {\bibinfo {author} {\bibfnamefont {A.}~\bibnamefont
  {Nagar}}, \bibinfo {author} {\bibfnamefont {T.}~\bibnamefont {Damour}}, \
  and\ \bibinfo {author} {\bibfnamefont {A.}~\bibnamefont {Tartaglia}},\ }\href
  {\doibase 10.1088/0264-9381/24/12/S08} {\bibfield  {journal} {\bibinfo
  {journal} {Class. Quant. Grav.}\ }\textbf {\bibinfo {volume} {24}},\ \bibinfo
  {pages} {S109} (\bibinfo {year} {2007})},\ \Eprint
  {http://arxiv.org/abs/gr-qc/0612096} {arXiv:gr-qc/0612096} \BibitemShut
  {NoStop}%
%%CITATION = GR-QC/0612096;%%
\bibitem [{\citenamefont {Damour}\ and\ \citenamefont
  {Nagar}(2007)}]{Damour:2007xr}%
  \BibitemOpen
  \bibfield  {author} {\bibinfo {author} {\bibfnamefont {T.}~\bibnamefont
  {Damour}}\ and\ \bibinfo {author} {\bibfnamefont {A.}~\bibnamefont {Nagar}},\
  }\href {\doibase 10.1103/PhysRevD.76.064028} {\bibfield  {journal} {\bibinfo
  {journal} {Phys. Rev.}\ }\textbf {\bibinfo {volume} {D76}},\ \bibinfo {pages}
  {064028} (\bibinfo {year} {2007})},\ \Eprint {http://arxiv.org/abs/0705.2519}
  {arXiv:0705.2519 [gr-qc]} \BibitemShut {NoStop}%
%%CITATION = 0705.2519;%%
\bibitem [{\citenamefont {Ruoff}\ \emph {et~al.}(2001)\citenamefont {Ruoff},
  \citenamefont {Laguna},\ and\ \citenamefont {Pullin}}]{Ruoff:2000et}%
  \BibitemOpen
  \bibfield  {author} {\bibinfo {author} {\bibfnamefont {J.}~\bibnamefont
  {Ruoff}}, \bibinfo {author} {\bibfnamefont {P.}~\bibnamefont {Laguna}}, \
  and\ \bibinfo {author} {\bibfnamefont {J.}~\bibnamefont {Pullin}},\ }\href
  {\doibase 10.1103/PhysRevD.63.064019} {\bibfield  {journal} {\bibinfo
  {journal} {Phys. Rev.}\ }\textbf {\bibinfo {volume} {D63}},\ \bibinfo {pages}
  {064019} (\bibinfo {year} {2001})},\ \Eprint
  {http://arxiv.org/abs/gr-qc/0005002} {arXiv:gr-qc/0005002} \BibitemShut
  {NoStop}%
%%CITATION = GR-QC/0005002;%%
\bibitem [{\citenamefont {Nagar}\ \emph {et~al.}(2004)\citenamefont {Nagar},
  \citenamefont {Diaz}, \citenamefont {Pons},\ and\ \citenamefont
  {Font}}]{Nagar:2004ns}%
  \BibitemOpen
  \bibfield  {author} {\bibinfo {author} {\bibfnamefont {A.}~\bibnamefont
  {Nagar}}, \bibinfo {author} {\bibfnamefont {G.}~\bibnamefont {Diaz}},
  \bibinfo {author} {\bibfnamefont {J.~A.}\ \bibnamefont {Pons}}, \ and\
  \bibinfo {author} {\bibfnamefont {J.~A.}\ \bibnamefont {Font}},\ }\href@noop
  {} {\bibfield  {journal} {\bibinfo  {journal} {Phys. Rev.}\ }\textbf
  {\bibinfo {volume} {D69}},\ \bibinfo {pages} {124028} (\bibinfo {year}
  {2004})},\ \Eprint {http://arxiv.org/abs/gr-qc/0403077} {gr-qc/0403077}
  \BibitemShut {NoStop}%
%%CITATION = GR-QC/0403077;%%
\bibitem [{\citenamefont {Bernuzzi}\ and\ \citenamefont
  {Nagar}(2010)}]{Bernuzzi:2010ty}%
  \BibitemOpen
  \bibfield  {author} {\bibinfo {author} {\bibfnamefont {S.}~\bibnamefont
  {Bernuzzi}}\ and\ \bibinfo {author} {\bibfnamefont {A.}~\bibnamefont
  {Nagar}},\ }\href {\doibase 10.1103/PhysRevD.81.084056} {\bibfield  {journal}
  {\bibinfo  {journal} {Phys. Rev.}\ }\textbf {\bibinfo {volume} {D81}},\
  \bibinfo {pages} {084056} (\bibinfo {year} {2010})},\ \Eprint
  {http://arxiv.org/abs/1003.0597} {arXiv:1003.0597 [gr-qc]} \BibitemShut
  {NoStop}%
%%CITATION = 1003.0597;%%
\bibitem [{\citenamefont {Bernuzzi}\ \emph
  {et~al.}(2011{\natexlab{a}})\citenamefont {Bernuzzi}, \citenamefont {Nagar},\
  and\ \citenamefont {Zenginoglu}}]{Bernuzzi:2011aj}%
  \BibitemOpen
  \bibfield  {author} {\bibinfo {author} {\bibfnamefont {S.}~\bibnamefont
  {Bernuzzi}}, \bibinfo {author} {\bibfnamefont {A.}~\bibnamefont {Nagar}}, \
  and\ \bibinfo {author} {\bibfnamefont {A.}~\bibnamefont {Zenginoglu}},\
  }\href {\doibase 10.1103/PhysRevD.84.084026} {\bibfield  {journal} {\bibinfo
  {journal} {Phys.Rev.}\ }\textbf {\bibinfo {volume} {D84}},\ \bibinfo {pages}
  {084026} (\bibinfo {year} {2011}{\natexlab{a}})},\ \Eprint
  {http://arxiv.org/abs/1107.5402} {arXiv:1107.5402 [gr-qc]} \BibitemShut
  {NoStop}%
%%CITATION = ARXIV:1107.5402;%%
\bibitem [{\citenamefont {Bernuzzi}\ \emph {et~al.}(2012)\citenamefont
  {Bernuzzi}, \citenamefont {Nagar},\ and\ \citenamefont
  {Zenginoglu}}]{Bernuzzi:2012ku}%
  \BibitemOpen
  \bibfield  {author} {\bibinfo {author} {\bibfnamefont {S.}~\bibnamefont
  {Bernuzzi}}, \bibinfo {author} {\bibfnamefont {A.}~\bibnamefont {Nagar}}, \
  and\ \bibinfo {author} {\bibfnamefont {A.}~\bibnamefont {Zenginoglu}},\
  }\href {\doibase 10.1103/PhysRevD.86.104038} {\bibfield  {journal} {\bibinfo
  {journal} {Phys.Rev.}\ }\textbf {\bibinfo {volume} {D86}},\ \bibinfo {pages}
  {104038} (\bibinfo {year} {2012})},\ \Eprint {http://arxiv.org/abs/1207.0769}
  {arXiv:1207.0769 [gr-qc]} \BibitemShut {NoStop}%
%%CITATION = ARXIV:1207.0769;%%
\bibitem [{\citenamefont {Barausse}\ \emph {et~al.}(2012)\citenamefont
  {Barausse}, \citenamefont {Buonanno}, \citenamefont {Hughes}, \citenamefont
  {Khanna}, \citenamefont {O'Sullivan} \emph {et~al.}}]{Barausse:2011kb}%
  \BibitemOpen
  \bibfield  {author} {\bibinfo {author} {\bibfnamefont {E.}~\bibnamefont
  {Barausse}}, \bibinfo {author} {\bibfnamefont {A.}~\bibnamefont {Buonanno}},
  \bibinfo {author} {\bibfnamefont {S.~A.}\ \bibnamefont {Hughes}}, \bibinfo
  {author} {\bibfnamefont {G.}~\bibnamefont {Khanna}}, \bibinfo {author}
  {\bibfnamefont {S.}~\bibnamefont {O'Sullivan}},  \emph {et~al.},\ }\href
  {\doibase 10.1103/PhysRevD.85.024046} {\bibfield  {journal} {\bibinfo
  {journal} {Phys.Rev.}\ }\textbf {\bibinfo {volume} {D85}},\ \bibinfo {pages}
  {024046} (\bibinfo {year} {2012})},\ \Eprint {http://arxiv.org/abs/1110.3081}
  {arXiv:1110.3081 [gr-qc]} \BibitemShut {NoStop}%
%%CITATION = ARXIV:1110.3081;%%
\bibitem [{\citenamefont {Harms}\ \emph {et~al.}(2014)\citenamefont {Harms},
  \citenamefont {Bernuzzi}, \citenamefont {Nagar},\ and\ \citenamefont
  {Zenginoglu}}]{Harms:2014dqa}%
  \BibitemOpen
  \bibfield  {author} {\bibinfo {author} {\bibfnamefont {E.}~\bibnamefont
  {Harms}}, \bibinfo {author} {\bibfnamefont {S.}~\bibnamefont {Bernuzzi}},
  \bibinfo {author} {\bibfnamefont {A.}~\bibnamefont {Nagar}}, \ and\ \bibinfo
  {author} {\bibfnamefont {A.}~\bibnamefont {Zenginoglu}},\ }\href {\doibase
  10.1088/0264-9381/31/24/245004} {\bibfield  {journal} {\bibinfo  {journal}
  {Class.Quant.Grav.}\ }\textbf {\bibinfo {volume} {31}},\ \bibinfo {pages}
  {245004} (\bibinfo {year} {2014})},\ \Eprint {http://arxiv.org/abs/1406.5983}
  {arXiv:1406.5983 [gr-qc]} \BibitemShut {NoStop}%
%%CITATION = ARXIV:1406.5983;%%
\bibitem [{\citenamefont {Taracchini}\ \emph {et~al.}(2013)\citenamefont
  {Taracchini}, \citenamefont {Buonanno}, \citenamefont {Hughes},\ and\
  \citenamefont {Khanna}}]{Taracchini:2013wfa}%
  \BibitemOpen
  \bibfield  {author} {\bibinfo {author} {\bibfnamefont {A.}~\bibnamefont
  {Taracchini}}, \bibinfo {author} {\bibfnamefont {A.}~\bibnamefont
  {Buonanno}}, \bibinfo {author} {\bibfnamefont {S.~A.}\ \bibnamefont
  {Hughes}}, \ and\ \bibinfo {author} {\bibfnamefont {G.}~\bibnamefont
  {Khanna}},\ }\href {\doibase 10.1103/PhysRevD.88.044001} {\bibfield
  {journal} {\bibinfo  {journal} {Phys.Rev.}\ }\textbf {\bibinfo {volume}
  {D88}},\ \bibinfo {pages} {044001} (\bibinfo {year} {2013})},\ \Eprint
  {http://arxiv.org/abs/1305.2184} {arXiv:1305.2184 [gr-qc]} \BibitemShut
  {NoStop}%
%%CITATION = ARXIV:1305.2184;%%
\bibitem [{\citenamefont {Taracchini}\ \emph {et~al.}(2014)\citenamefont
  {Taracchini}, \citenamefont {Buonanno}, \citenamefont {Khanna},\ and\
  \citenamefont {Hughes}}]{Taracchini:2014zpa}%
  \BibitemOpen
  \bibfield  {author} {\bibinfo {author} {\bibfnamefont {A.}~\bibnamefont
  {Taracchini}}, \bibinfo {author} {\bibfnamefont {A.}~\bibnamefont
  {Buonanno}}, \bibinfo {author} {\bibfnamefont {G.}~\bibnamefont {Khanna}}, \
  and\ \bibinfo {author} {\bibfnamefont {S.~A.}\ \bibnamefont {Hughes}},\
  }\href {\doibase 10.1103/PhysRevD.90.084025} {\bibfield  {journal} {\bibinfo
  {journal} {Phys.Rev.}\ }\textbf {\bibinfo {volume} {D90}},\ \bibinfo {pages}
  {084025} (\bibinfo {year} {2014})},\ \Eprint {http://arxiv.org/abs/1404.1819}
  {arXiv:1404.1819 [gr-qc]} \BibitemShut {NoStop}%
%%CITATION = ARXIV:1404.1819;%%
\bibitem [{\citenamefont {Nagar}\ \emph {et~al.}(2014)\citenamefont {Nagar},
  \citenamefont {Harms}, \citenamefont {Bernuzzi},\ and\ \citenamefont
  {Zenginoğlu}}]{Nagar:2014kha}%
  \BibitemOpen
  \bibfield  {author} {\bibinfo {author} {\bibfnamefont {A.}~\bibnamefont
  {Nagar}}, \bibinfo {author} {\bibfnamefont {E.}~\bibnamefont {Harms}},
  \bibinfo {author} {\bibfnamefont {S.}~\bibnamefont {Bernuzzi}}, \ and\
  \bibinfo {author} {\bibfnamefont {A.}~\bibnamefont {Zenginoğlu}},\ }\href
  {\doibase 10.1103/PhysRevD.90.124086} {\bibfield  {journal} {\bibinfo
  {journal} {Phys. Rev.}\ }\textbf {\bibinfo {volume} {D90}},\ \bibinfo {pages}
  {124086} (\bibinfo {year} {2014})},\ \Eprint {http://arxiv.org/abs/1407.5033}
  {arXiv:1407.5033 [gr-qc]} \BibitemShut {NoStop}%
%%CITATION = ARXIV:1407.5033;%%
\bibitem [{\citenamefont {Albanesi}\ \emph {et~al.}(2021)\citenamefont
  {Albanesi}, \citenamefont {Nagar},\ and\ \citenamefont
  {Bernuzzi}}]{Albanesi:2021rby}%
  \BibitemOpen
  \bibfield  {author} {\bibinfo {author} {\bibfnamefont {S.}~\bibnamefont
  {Albanesi}}, \bibinfo {author} {\bibfnamefont {A.}~\bibnamefont {Nagar}}, \
  and\ \bibinfo {author} {\bibfnamefont {S.}~\bibnamefont {Bernuzzi}},\ }\href
  {\doibase 10.1103/PhysRevD.104.024067} {\bibfield  {journal} {\bibinfo
  {journal} {Phys. Rev. D}\ }\textbf {\bibinfo {volume} {104}},\ \bibinfo
  {pages} {024067} (\bibinfo {year} {2021})},\ \Eprint
  {http://arxiv.org/abs/2104.10559} {arXiv:2104.10559 [gr-qc]} \BibitemShut
  {NoStop}%
\bibitem [{\citenamefont {Albanesi}\ \emph {et~al.}(2023)\citenamefont
  {Albanesi}, \citenamefont {Bernuzzi}, \citenamefont {Damour}, \citenamefont
  {Nagar},\ and\ \citenamefont {Placidi}}]{Albanesi:2023bgi}%
  \BibitemOpen
  \bibfield  {author} {\bibinfo {author} {\bibfnamefont {S.}~\bibnamefont
  {Albanesi}}, \bibinfo {author} {\bibfnamefont {S.}~\bibnamefont {Bernuzzi}},
  \bibinfo {author} {\bibfnamefont {T.}~\bibnamefont {Damour}}, \bibinfo
  {author} {\bibfnamefont {A.}~\bibnamefont {Nagar}}, \ and\ \bibinfo {author}
  {\bibfnamefont {A.}~\bibnamefont {Placidi}},\ }\href {\doibase
  10.1103/PhysRevD.108.084037} {\bibfield  {journal} {\bibinfo  {journal}
  {Phys. Rev. D}\ }\textbf {\bibinfo {volume} {108}},\ \bibinfo {pages}
  {084037} (\bibinfo {year} {2023})},\ \Eprint
  {http://arxiv.org/abs/2305.19336} {arXiv:2305.19336 [gr-qc]} \BibitemShut
  {NoStop}%
\bibitem [{\citenamefont {Faggioli}\ \emph
  {et~al.}(2025{\natexlab{a}})\citenamefont {Faggioli}, \citenamefont {van~de
  Meent}, \citenamefont {Buonanno}, \citenamefont {Gamboa}, \citenamefont
  {Khalil},\ and\ \citenamefont {Khanna}}]{Faggioli:2024ugn}%
  \BibitemOpen
  \bibfield  {author} {\bibinfo {author} {\bibfnamefont {G.}~\bibnamefont
  {Faggioli}}, \bibinfo {author} {\bibfnamefont {M.}~\bibnamefont {van~de
  Meent}}, \bibinfo {author} {\bibfnamefont {A.}~\bibnamefont {Buonanno}},
  \bibinfo {author} {\bibfnamefont {A.}~\bibnamefont {Gamboa}}, \bibinfo
  {author} {\bibfnamefont {M.}~\bibnamefont {Khalil}}, \ and\ \bibinfo {author}
  {\bibfnamefont {G.}~\bibnamefont {Khanna}},\ }\href {\doibase
  10.1103/PhysRevD.111.044036} {\bibfield  {journal} {\bibinfo  {journal}
  {Phys. Rev. D}\ }\textbf {\bibinfo {volume} {111}},\ \bibinfo {pages}
  {044036} (\bibinfo {year} {2025}{\natexlab{a}})},\ \Eprint
  {http://arxiv.org/abs/2405.19006} {arXiv:2405.19006 [gr-qc]} \BibitemShut
  {NoStop}%
\bibitem [{\citenamefont {Faggioli}\ \emph
  {et~al.}(2025{\natexlab{b}})\citenamefont {Faggioli}, \citenamefont {van~de
  Meent}, \citenamefont {Buonanno},\ and\ \citenamefont
  {Khanna}}]{Faggioli:2025hff}%
  \BibitemOpen
  \bibfield  {author} {\bibinfo {author} {\bibfnamefont {G.}~\bibnamefont
  {Faggioli}}, \bibinfo {author} {\bibfnamefont {M.}~\bibnamefont {van~de
  Meent}}, \bibinfo {author} {\bibfnamefont {A.}~\bibnamefont {Buonanno}}, \
  and\ \bibinfo {author} {\bibfnamefont {G.}~\bibnamefont {Khanna}},\
  }\href@noop {} {\  (\bibinfo {year} {2025}{\natexlab{b}})},\ \Eprint
  {http://arxiv.org/abs/2507.05870} {arXiv:2507.05870 [gr-qc]} \BibitemShut
  {NoStop}%
\bibitem [{\citenamefont {Becker}\ and\ \citenamefont
  {Hughes}(2025)}]{Becker:2024xdi}%
  \BibitemOpen
  \bibfield  {author} {\bibinfo {author} {\bibfnamefont {D.~R.}\ \bibnamefont
  {Becker}}\ and\ \bibinfo {author} {\bibfnamefont {S.~A.}\ \bibnamefont
  {Hughes}},\ }\href {\doibase 10.1103/PhysRevD.111.064003} {\bibfield
  {journal} {\bibinfo  {journal} {Phys. Rev. D}\ }\textbf {\bibinfo {volume}
  {111}},\ \bibinfo {pages} {064003} (\bibinfo {year} {2025})},\ \Eprint
  {http://arxiv.org/abs/2410.09160} {arXiv:2410.09160 [gr-qc]} \BibitemShut
  {NoStop}%
\bibitem [{\citenamefont {Ori}\ and\ \citenamefont
  {Thorne}(2000)}]{Ori:2000zn}%
  \BibitemOpen
  \bibfield  {author} {\bibinfo {author} {\bibfnamefont {A.}~\bibnamefont
  {Ori}}\ and\ \bibinfo {author} {\bibfnamefont {K.~S.}\ \bibnamefont
  {Thorne}},\ }\href {\doibase 10.1103/PhysRevD.62.124022} {\bibfield
  {journal} {\bibinfo  {journal} {Phys.Rev.}\ }\textbf {\bibinfo {volume}
  {D62}},\ \bibinfo {pages} {124022} (\bibinfo {year} {2000})},\ \Eprint
  {http://arxiv.org/abs/gr-qc/0003032} {arXiv:gr-qc/0003032 [gr-qc]}
  \BibitemShut {NoStop}%
%%CITATION = GR-QC/0003032;%%
\bibitem [{\citenamefont {Sundararajan}(2008)}]{Sundararajan:2008bw}%
  \BibitemOpen
  \bibfield  {author} {\bibinfo {author} {\bibfnamefont {P.~A.}\ \bibnamefont
  {Sundararajan}},\ }\href {\doibase 10.1103/PhysRevD.77.124050} {\bibfield
  {journal} {\bibinfo  {journal} {Phys. Rev.}\ }\textbf {\bibinfo {volume}
  {D77}},\ \bibinfo {pages} {124050} (\bibinfo {year} {2008})},\ \Eprint
  {http://arxiv.org/abs/0803.4482} {arXiv:0803.4482 [gr-qc]} \BibitemShut
  {NoStop}%
%%CITATION = 0803.4482;%%
\bibitem [{\citenamefont {Apte}\ and\ \citenamefont
  {Hughes}(2019)}]{Apte:2019txp}%
  \BibitemOpen
  \bibfield  {author} {\bibinfo {author} {\bibfnamefont {A.}~\bibnamefont
  {Apte}}\ and\ \bibinfo {author} {\bibfnamefont {S.~A.}\ \bibnamefont
  {Hughes}},\ }\href {\doibase 10.1103/PhysRevD.100.084031} {\bibfield
  {journal} {\bibinfo  {journal} {Phys. Rev. D}\ }\textbf {\bibinfo {volume}
  {100}},\ \bibinfo {pages} {084031} (\bibinfo {year} {2019})},\ \Eprint
  {http://arxiv.org/abs/1901.05901} {arXiv:1901.05901 [gr-qc]} \BibitemShut
  {NoStop}%
\bibitem [{\citenamefont {Lim}\ \emph {et~al.}(2019)\citenamefont {Lim},
  \citenamefont {Khanna}, \citenamefont {Apte},\ and\ \citenamefont
  {Hughes}}]{Lim:2019xrb}%
  \BibitemOpen
  \bibfield  {author} {\bibinfo {author} {\bibfnamefont {H.}~\bibnamefont
  {Lim}}, \bibinfo {author} {\bibfnamefont {G.}~\bibnamefont {Khanna}},
  \bibinfo {author} {\bibfnamefont {A.}~\bibnamefont {Apte}}, \ and\ \bibinfo
  {author} {\bibfnamefont {S.~A.}\ \bibnamefont {Hughes}},\ }\href {\doibase
  10.1103/PhysRevD.100.084032} {\bibfield  {journal} {\bibinfo  {journal}
  {Phys. Rev. D}\ }\textbf {\bibinfo {volume} {100}},\ \bibinfo {pages}
  {084032} (\bibinfo {year} {2019})},\ \Eprint
  {http://arxiv.org/abs/1901.05902} {arXiv:1901.05902 [gr-qc]} \BibitemShut
  {NoStop}%
\bibitem [{\citenamefont {Khalil}\ \emph {et~al.}(2023)\citenamefont {Khalil},
  \citenamefont {Buonanno}, \citenamefont {Estelles}, \citenamefont {Mihaylov},
  \citenamefont {Ossokine}, \citenamefont {Pompili},\ and\ \citenamefont
  {Ramos-Buades}}]{Khalil:2023kep}%
  \BibitemOpen
  \bibfield  {author} {\bibinfo {author} {\bibfnamefont {M.}~\bibnamefont
  {Khalil}}, \bibinfo {author} {\bibfnamefont {A.}~\bibnamefont {Buonanno}},
  \bibinfo {author} {\bibfnamefont {H.}~\bibnamefont {Estelles}}, \bibinfo
  {author} {\bibfnamefont {D.~P.}\ \bibnamefont {Mihaylov}}, \bibinfo {author}
  {\bibfnamefont {S.}~\bibnamefont {Ossokine}}, \bibinfo {author}
  {\bibfnamefont {L.}~\bibnamefont {Pompili}}, \ and\ \bibinfo {author}
  {\bibfnamefont {A.}~\bibnamefont {Ramos-Buades}},\ }\href {\doibase
  10.1103/PhysRevD.108.124036} {\bibfield  {journal} {\bibinfo  {journal}
  {Phys. Rev. D}\ }\textbf {\bibinfo {volume} {108}},\ \bibinfo {pages}
  {124036} (\bibinfo {year} {2023})},\ \Eprint
  {http://arxiv.org/abs/2303.18143} {arXiv:2303.18143 [gr-qc]} \BibitemShut
  {NoStop}%
\bibitem [{\citenamefont {Gamba}\ \emph {et~al.}(2022)\citenamefont {Gamba},
  \citenamefont {Ak\c{c}ay}, \citenamefont {Bernuzzi},\ and\ \citenamefont
  {Williams}}]{Gamba:2021ydi}%
  \BibitemOpen
  \bibfield  {author} {\bibinfo {author} {\bibfnamefont {R.}~\bibnamefont
  {Gamba}}, \bibinfo {author} {\bibfnamefont {S.}~\bibnamefont {Ak\c{c}ay}},
  \bibinfo {author} {\bibfnamefont {S.}~\bibnamefont {Bernuzzi}}, \ and\
  \bibinfo {author} {\bibfnamefont {J.}~\bibnamefont {Williams}},\ }\href
  {\doibase 10.1103/PhysRevD.106.024020} {\bibfield  {journal} {\bibinfo
  {journal} {Phys. Rev. D}\ }\textbf {\bibinfo {volume} {106}},\ \bibinfo
  {pages} {024020} (\bibinfo {year} {2022})},\ \Eprint
  {http://arxiv.org/abs/2111.03675} {arXiv:2111.03675 [gr-qc]} \BibitemShut
  {NoStop}%
\bibitem [{\citenamefont {Buonanno}\ \emph {et~al.}(2006)\citenamefont
  {Buonanno}, \citenamefont {Chen},\ and\ \citenamefont
  {Damour}}]{Buonanno:2005xu}%
  \BibitemOpen
  \bibfield  {author} {\bibinfo {author} {\bibfnamefont {A.}~\bibnamefont
  {Buonanno}}, \bibinfo {author} {\bibfnamefont {Y.}~\bibnamefont {Chen}}, \
  and\ \bibinfo {author} {\bibfnamefont {T.}~\bibnamefont {Damour}},\ }\href
  {\doibase 10.1103/PhysRevD.74.104005} {\bibfield  {journal} {\bibinfo
  {journal} {Phys. Rev.}\ }\textbf {\bibinfo {volume} {D74}},\ \bibinfo {pages}
  {104005} (\bibinfo {year} {2006})},\ \Eprint
  {http://arxiv.org/abs/gr-qc/0508067} {arXiv:gr-qc/0508067} \BibitemShut
  {NoStop}%
%%CITATION = GR-QC/0508067;%%
\bibitem [{\citenamefont {Balmelli}\ and\ \citenamefont
  {Damour}(2015)}]{Balmelli:2015zsa}%
  \BibitemOpen
  \bibfield  {author} {\bibinfo {author} {\bibfnamefont {S.}~\bibnamefont
  {Balmelli}}\ and\ \bibinfo {author} {\bibfnamefont {T.}~\bibnamefont
  {Damour}},\ }\href {\doibase 10.1103/PhysRevD.92.124022} {\bibfield
  {journal} {\bibinfo  {journal} {Phys. Rev.}\ }\textbf {\bibinfo {volume}
  {D92}},\ \bibinfo {pages} {124022} (\bibinfo {year} {2015})},\ \Eprint
  {http://arxiv.org/abs/1509.08135} {arXiv:1509.08135 [gr-qc]} \BibitemShut
  {NoStop}%
%%CITATION = ARXIV:1509.08135;%%
\bibitem [{\citenamefont {Damour}\ and\ \citenamefont
  {Nagar}(2014)}]{Damour:2014sva}%
  \BibitemOpen
  \bibfield  {author} {\bibinfo {author} {\bibfnamefont {T.}~\bibnamefont
  {Damour}}\ and\ \bibinfo {author} {\bibfnamefont {A.}~\bibnamefont {Nagar}},\
  }\href {\doibase 10.1103/PhysRevD.90.044018} {\bibfield  {journal} {\bibinfo
  {journal} {Phys.Rev.}\ }\textbf {\bibinfo {volume} {D90}},\ \bibinfo {pages}
  {044018} (\bibinfo {year} {2014})},\ \Eprint {http://arxiv.org/abs/1406.6913}
  {arXiv:1406.6913 [gr-qc]} \BibitemShut {NoStop}%
%%CITATION = ARXIV:1406.6913;%%
\bibitem [{\citenamefont {Damour}\ \emph {et~al.}(1998)\citenamefont {Damour},
  \citenamefont {Iyer},\ and\ \citenamefont {Sathyaprakash}}]{Damour:1997ub}%
  \BibitemOpen
  \bibfield  {author} {\bibinfo {author} {\bibfnamefont {T.}~\bibnamefont
  {Damour}}, \bibinfo {author} {\bibfnamefont {B.~R.}\ \bibnamefont {Iyer}}, \
  and\ \bibinfo {author} {\bibfnamefont {B.~S.}\ \bibnamefont
  {Sathyaprakash}},\ }\href {\doibase 10.1103/PhysRevD.57.885} {\bibfield
  {journal} {\bibinfo  {journal} {Phys. Rev.}\ }\textbf {\bibinfo {volume}
  {D57}},\ \bibinfo {pages} {885} (\bibinfo {year} {1998})},\ \Eprint
  {http://arxiv.org/abs/gr-qc/9708034} {arXiv:gr-qc/9708034 [gr-qc]}
  \BibitemShut {NoStop}%
%%CITATION = GR-QC/9708034;%%
\bibitem [{\citenamefont {Damour}\ \emph {et~al.}(2009)\citenamefont {Damour},
  \citenamefont {Iyer},\ and\ \citenamefont {Nagar}}]{Damour:2008gu}%
  \BibitemOpen
  \bibfield  {author} {\bibinfo {author} {\bibfnamefont {T.}~\bibnamefont
  {Damour}}, \bibinfo {author} {\bibfnamefont {B.~R.}\ \bibnamefont {Iyer}}, \
  and\ \bibinfo {author} {\bibfnamefont {A.}~\bibnamefont {Nagar}},\ }\href
  {\doibase 10.1103/PhysRevD.79.064004} {\bibfield  {journal} {\bibinfo
  {journal} {Phys. Rev.}\ }\textbf {\bibinfo {volume} {D79}},\ \bibinfo {pages}
  {064004} (\bibinfo {year} {2009})},\ \Eprint {http://arxiv.org/abs/0811.2069}
  {arXiv:0811.2069 [gr-qc]} \BibitemShut {NoStop}%
%%CITATION = 0811.2069;%%
\bibitem [{\citenamefont {Fujita}(2015)}]{Fujita:2014eta}%
  \BibitemOpen
  \bibfield  {author} {\bibinfo {author} {\bibfnamefont {R.}~\bibnamefont
  {Fujita}},\ }\href {\doibase 10.1093/ptep/ptv012} {\bibfield  {journal}
  {\bibinfo  {journal} {PTEP}\ }\textbf {\bibinfo {volume} {2015}},\ \bibinfo
  {pages} {033E01} (\bibinfo {year} {2015})},\ \Eprint
  {http://arxiv.org/abs/1412.5689} {arXiv:1412.5689 [gr-qc]} \BibitemShut
  {NoStop}%
%%CITATION = ARXIV:1412.5689;%%
\bibitem [{\citenamefont {Albertini}\ \emph {et~al.}(2024)\citenamefont
  {Albertini}, \citenamefont {Gamba}, \citenamefont {Nagar},\ and\
  \citenamefont {Bernuzzi}}]{Albertini:2023aol}%
  \BibitemOpen
  \bibfield  {author} {\bibinfo {author} {\bibfnamefont {A.}~\bibnamefont
  {Albertini}}, \bibinfo {author} {\bibfnamefont {R.}~\bibnamefont {Gamba}},
  \bibinfo {author} {\bibfnamefont {A.}~\bibnamefont {Nagar}}, \ and\ \bibinfo
  {author} {\bibfnamefont {S.}~\bibnamefont {Bernuzzi}},\ }\href {\doibase
  10.1103/PhysRevD.109.044022} {\bibfield  {journal} {\bibinfo  {journal}
  {Phys. Rev. D}\ }\textbf {\bibinfo {volume} {109}},\ \bibinfo {pages}
  {044022} (\bibinfo {year} {2024})},\ \Eprint
  {http://arxiv.org/abs/2310.13578} {arXiv:2310.13578 [gr-qc]} \BibitemShut
  {NoStop}%
\bibitem [{\citenamefont {Warburton}\ \emph {et~al.}(2021)\citenamefont
  {Warburton}, \citenamefont {Pound}, \citenamefont {Wardell}, \citenamefont
  {Miller},\ and\ \citenamefont {Durkan}}]{Warburton:2021kwk}%
  \BibitemOpen
  \bibfield  {author} {\bibinfo {author} {\bibfnamefont {N.}~\bibnamefont
  {Warburton}}, \bibinfo {author} {\bibfnamefont {A.}~\bibnamefont {Pound}},
  \bibinfo {author} {\bibfnamefont {B.}~\bibnamefont {Wardell}}, \bibinfo
  {author} {\bibfnamefont {J.}~\bibnamefont {Miller}}, \ and\ \bibinfo {author}
  {\bibfnamefont {L.}~\bibnamefont {Durkan}},\ }\href {\doibase
  10.1103/PhysRevLett.127.151102} {\bibfield  {journal} {\bibinfo  {journal}
  {Phys. Rev. Lett.}\ }\textbf {\bibinfo {volume} {127}},\ \bibinfo {pages}
  {151102} (\bibinfo {year} {2021})},\ \Eprint
  {http://arxiv.org/abs/2107.01298} {arXiv:2107.01298 [gr-qc]} \BibitemShut
  {NoStop}%
\bibitem [{\citenamefont {Damour}\ \emph {et~al.}(2013)\citenamefont {Damour},
  \citenamefont {Nagar},\ and\ \citenamefont {Bernuzzi}}]{Damour:2012ky}%
  \BibitemOpen
  \bibfield  {author} {\bibinfo {author} {\bibfnamefont {T.}~\bibnamefont
  {Damour}}, \bibinfo {author} {\bibfnamefont {A.}~\bibnamefont {Nagar}}, \
  and\ \bibinfo {author} {\bibfnamefont {S.}~\bibnamefont {Bernuzzi}},\ }\href
  {\doibase 10.1103/PhysRevD.87.084035} {\bibfield  {journal} {\bibinfo
  {journal} {Phys.Rev.}\ }\textbf {\bibinfo {volume} {D87}},\ \bibinfo {pages}
  {084035} (\bibinfo {year} {2013})},\ \Eprint {http://arxiv.org/abs/1212.4357}
  {arXiv:1212.4357 [gr-qc]} \BibitemShut {NoStop}%
%%CITATION = ARXIV:1212.4357;%%
\bibitem [{\citenamefont {Nagar}\ and\ \citenamefont
  {Rettegno}(2019)}]{Nagar:2018gnk}%
  \BibitemOpen
  \bibfield  {author} {\bibinfo {author} {\bibfnamefont {A.}~\bibnamefont
  {Nagar}}\ and\ \bibinfo {author} {\bibfnamefont {P.}~\bibnamefont
  {Rettegno}},\ }\href {\doibase 10.1103/PhysRevD.99.021501} {\bibfield
  {journal} {\bibinfo  {journal} {Phys. Rev.}\ }\textbf {\bibinfo {volume}
  {D99}},\ \bibinfo {pages} {021501} (\bibinfo {year} {2019})},\ \Eprint
  {http://arxiv.org/abs/1805.03891} {arXiv:1805.03891 [gr-qc]} \BibitemShut
  {NoStop}%
%%CITATION = ARXIV:1805.03891;%%
\bibitem [{\citenamefont {Pan}\ \emph {et~al.}(2014)\citenamefont {Pan},
  \citenamefont {Buonanno}, \citenamefont {Taracchini}, \citenamefont {Kidder},
  \citenamefont {Mroue} \emph {et~al.}}]{Pan:2013rra}%
  \BibitemOpen
  \bibfield  {author} {\bibinfo {author} {\bibfnamefont {Y.}~\bibnamefont
  {Pan}}, \bibinfo {author} {\bibfnamefont {A.}~\bibnamefont {Buonanno}},
  \bibinfo {author} {\bibfnamefont {A.}~\bibnamefont {Taracchini}}, \bibinfo
  {author} {\bibfnamefont {L.~E.}\ \bibnamefont {Kidder}}, \bibinfo {author}
  {\bibfnamefont {A.~H.}\ \bibnamefont {Mroue}},  \emph {et~al.},\ }\href
  {\doibase 10.1103/PhysRevD.89.084006} {\bibfield  {journal} {\bibinfo
  {journal} {Phys.Rev.}\ }\textbf {\bibinfo {volume} {D89}},\ \bibinfo {pages}
  {084006} (\bibinfo {year} {2014})},\ \Eprint {http://arxiv.org/abs/1307.6232}
  {arXiv:1307.6232 [gr-qc]} \BibitemShut {NoStop}%
%%CITATION = ARXIV:1307.6232;%%
\bibitem [{\citenamefont {Stein}\ and\ \citenamefont
  {Warburton}(2020)}]{Stein:2019buj}%
  \BibitemOpen
  \bibfield  {author} {\bibinfo {author} {\bibfnamefont {L.~C.}\ \bibnamefont
  {Stein}}\ and\ \bibinfo {author} {\bibfnamefont {N.}~\bibnamefont
  {Warburton}},\ }\href {\doibase 10.1103/PhysRevD.101.064007} {\bibfield
  {journal} {\bibinfo  {journal} {Phys. Rev. D}\ }\textbf {\bibinfo {volume}
  {101}},\ \bibinfo {pages} {064007} (\bibinfo {year} {2020})},\ \Eprint
  {http://arxiv.org/abs/1912.07609} {arXiv:1912.07609 [gr-qc]} \BibitemShut
  {NoStop}%
\bibitem [{\citenamefont {Chandrasekhar}(1985)}]{Chandrasekhar:1985kt}%
  \BibitemOpen
  \bibfield  {author} {\bibinfo {author} {\bibfnamefont {S.}~\bibnamefont
  {Chandrasekhar}},\ }\href@noop {} {\emph {\bibinfo {title} {{Oxford, UK:
  Clarendon (1992) 646 p., OXFORD, UK: CLARENDON (1985) 646 P.}}}}\ (\bibinfo
  {year} {1985})\BibitemShut {NoStop}%
%%CITATION = INSPIRE-224457;%%
\bibitem [{\citenamefont {Gralla}\ \emph {et~al.}(2016)\citenamefont {Gralla},
  \citenamefont {Hughes},\ and\ \citenamefont {Warburton}}]{Gralla:2016qfw}%
  \BibitemOpen
  \bibfield  {author} {\bibinfo {author} {\bibfnamefont {S.~E.}\ \bibnamefont
  {Gralla}}, \bibinfo {author} {\bibfnamefont {S.~A.}\ \bibnamefont {Hughes}},
  \ and\ \bibinfo {author} {\bibfnamefont {N.}~\bibnamefont {Warburton}},\
  }\href {\doibase 10.1088/0264-9381/33/15/155002} {\bibfield  {journal}
  {\bibinfo  {journal} {Class. Quant. Grav.}\ }\textbf {\bibinfo {volume}
  {33}},\ \bibinfo {pages} {155002} (\bibinfo {year} {2016})},\ \bibinfo {note}
  {[Erratum: Class.Quant.Grav. 37, 109501 (2020)]},\ \Eprint
  {http://arxiv.org/abs/1603.01221} {arXiv:1603.01221 [gr-qc]} \BibitemShut
  {NoStop}%
\bibitem [{\citenamefont {Teukolsky}(1973)}]{Teukolsky:1973ha}%
  \BibitemOpen
  \bibfield  {author} {\bibinfo {author} {\bibfnamefont {S.~A.}\ \bibnamefont
  {Teukolsky}},\ }\href {\doibase 10.1086/152444} {\bibfield  {journal}
  {\bibinfo  {journal} {Astrophys. J.}\ }\textbf {\bibinfo {volume} {185}},\
  \bibinfo {pages} {635} (\bibinfo {year} {1973})}\BibitemShut {NoStop}%
%%CITATION = ASJOA,185,635;%%
\bibitem [{\citenamefont {Zengino{\u g}lu}(2008)}]{Zenginoglu:2007jw}%
  \BibitemOpen
  \bibfield  {author} {\bibinfo {author} {\bibfnamefont {A.}~\bibnamefont
  {Zengino{\u g}lu}},\ }\href {\doibase 10.1088/0264-9381/25/14/145002}
  {\bibfield  {journal} {\bibinfo  {journal} {Class. Quant. Grav.}\ }\textbf
  {\bibinfo {volume} {25}},\ \bibinfo {pages} {145002} (\bibinfo {year}
  {2008})},\ \Eprint {http://arxiv.org/abs/0712.4333} {arXiv:0712.4333 [gr-qc]}
  \BibitemShut {NoStop}%
%%CITATION = 0712.4333;%%
\bibitem [{\citenamefont {Zengino{\u g}lu}\ and\ \citenamefont
  {Tiglio}(2009)}]{Zenginoglu:2009hd}%
  \BibitemOpen
  \bibfield  {author} {\bibinfo {author} {\bibfnamefont {A.}~\bibnamefont
  {Zengino{\u g}lu}}\ and\ \bibinfo {author} {\bibfnamefont {M.}~\bibnamefont
  {Tiglio}},\ }\href {\doibase 10.1103/PhysRevD.80.024044} {\bibfield
  {journal} {\bibinfo  {journal} {Phys. Rev.}\ }\textbf {\bibinfo {volume}
  {D80}},\ \bibinfo {pages} {024044} (\bibinfo {year} {2009})},\ \Eprint
  {http://arxiv.org/abs/0906.3342} {arXiv:0906.3342 [gr-qc]} \BibitemShut
  {NoStop}%
%%CITATION = 0906.3342;%%
\bibitem [{\citenamefont {Zengino{\u g}lu}(2011)}]{Zenginoglu:2010cq}%
  \BibitemOpen
  \bibfield  {author} {\bibinfo {author} {\bibfnamefont {A.}~\bibnamefont
  {Zengino{\u g}lu}},\ }\href {\doibase 10.1016/j.jcp.2010.12.016} {\bibfield
  {journal} {\bibinfo  {journal} {J.Comput.Phys.}\ }\textbf {\bibinfo {volume}
  {230}},\ \bibinfo {pages} {2286} (\bibinfo {year} {2011})},\ \Eprint
  {http://arxiv.org/abs/1008.3809} {arXiv:1008.3809 [math.NA]} \BibitemShut
  {NoStop}%
%%CITATION = ARXIV:1008.3809;%%
\bibitem [{\citenamefont {Bernuzzi}\ \emph
  {et~al.}(2011{\natexlab{b}})\citenamefont {Bernuzzi}, \citenamefont {Nagar},\
  and\ \citenamefont {Zenginoglu}}]{Bernuzzi:2010xj}%
  \BibitemOpen
  \bibfield  {author} {\bibinfo {author} {\bibfnamefont {S.}~\bibnamefont
  {Bernuzzi}}, \bibinfo {author} {\bibfnamefont {A.}~\bibnamefont {Nagar}}, \
  and\ \bibinfo {author} {\bibfnamefont {A.}~\bibnamefont {Zenginoglu}},\
  }\href {\doibase 10.1103/PhysRevD.83.064010} {\bibfield  {journal} {\bibinfo
  {journal} {Phys.Rev.}\ }\textbf {\bibinfo {volume} {D83}},\ \bibinfo {pages}
  {064010} (\bibinfo {year} {2011}{\natexlab{b}})},\ \Eprint
  {http://arxiv.org/abs/1012.2456} {arXiv:1012.2456 [gr-qc]} \BibitemShut
  {NoStop}%
\bibitem [{\citenamefont {Nagar}(2013)}]{Nagar:2013sga}%
  \BibitemOpen
  \bibfield  {author} {\bibinfo {author} {\bibfnamefont {A.}~\bibnamefont
  {Nagar}},\ }\href@noop {} {\bibfield  {journal} {\bibinfo  {journal}
  {Phys.Rev.}\ }\textbf {\bibinfo {volume} {D88}},\ \bibinfo {pages} {121501}
  (\bibinfo {year} {2013})},\ \Eprint {http://arxiv.org/abs/1306.6299}
  {arXiv:1306.6299 [gr-qc]} \BibitemShut {NoStop}%
%%CITATION = ARXIV:1306.6299;%%
\bibitem [{\citenamefont {Regge}\ and\ \citenamefont
  {Wheeler}(1957)}]{Regge:1957td}%
  \BibitemOpen
  \bibfield  {author} {\bibinfo {author} {\bibfnamefont {T.}~\bibnamefont
  {Regge}}\ and\ \bibinfo {author} {\bibfnamefont {J.~A.}\ \bibnamefont
  {Wheeler}},\ }\href@noop {} {\bibfield  {journal} {\bibinfo  {journal} {Phys.
  Rev.}\ }\textbf {\bibinfo {volume} {108}},\ \bibinfo {pages} {1063} (\bibinfo
  {year} {1957})}\BibitemShut {NoStop}%
%%CITATION = PHRVA,108,1063;%%
\bibitem [{\citenamefont {Zerilli}(1970{\natexlab{a}})}]{Zerilli:1970se}%
  \BibitemOpen
  \bibfield  {author} {\bibinfo {author} {\bibfnamefont {F.~J.}\ \bibnamefont
  {Zerilli}},\ }\href {\doibase 10.1103/PhysRevLett.24.737} {\bibfield
  {journal} {\bibinfo  {journal} {Phys. Rev. Lett.}\ }\textbf {\bibinfo
  {volume} {24}},\ \bibinfo {pages} {737} (\bibinfo {year}
  {1970}{\natexlab{a}})}\BibitemShut {NoStop}%
%%CITATION = PRLTA,24,737;%%
\bibitem [{\citenamefont {Zerilli}(1970{\natexlab{b}})}]{Zerilli:1970wzz}%
  \BibitemOpen
  \bibfield  {author} {\bibinfo {author} {\bibfnamefont {F.~J.}\ \bibnamefont
  {Zerilli}},\ }\href {\doibase 10.1103/PhysRevD.2.2141} {\bibfield  {journal}
  {\bibinfo  {journal} {Phys. Rev. D}\ }\textbf {\bibinfo {volume} {2}},\
  \bibinfo {pages} {2141} (\bibinfo {year} {1970}{\natexlab{b}})}\BibitemShut
  {NoStop}%
\bibitem [{\citenamefont {Nagar}\ and\ \citenamefont
  {Rezzolla}(2005)}]{Nagar:2005ea}%
  \BibitemOpen
  \bibfield  {author} {\bibinfo {author} {\bibfnamefont {A.}~\bibnamefont
  {Nagar}}\ and\ \bibinfo {author} {\bibfnamefont {L.}~\bibnamefont
  {Rezzolla}},\ }\href {\doibase 10.1088/0264-9381/22/16/R01} {\bibfield
  {journal} {\bibinfo  {journal} {Class. Quant. Grav.}\ }\textbf {\bibinfo
  {volume} {22}},\ \bibinfo {pages} {R167} (\bibinfo {year} {2005})},\ \Eprint
  {http://arxiv.org/abs/gr-qc/0502064} {arXiv:gr-qc/0502064} \BibitemShut
  {NoStop}%
%%CITATION = GR-QC/0502064;%%
\bibitem [{\citenamefont {Martel}\ and\ \citenamefont
  {Poisson}(2005)}]{Martel:2005ir}%
  \BibitemOpen
  \bibfield  {author} {\bibinfo {author} {\bibfnamefont {K.}~\bibnamefont
  {Martel}}\ and\ \bibinfo {author} {\bibfnamefont {E.}~\bibnamefont
  {Poisson}},\ }\href {\doibase 10.1103/PhysRevD.71.104003} {\bibfield
  {journal} {\bibinfo  {journal} {Physical Review D (Particles, Fields,
  Gravitation, and Cosmology)}\ }\textbf {\bibinfo {volume} {71}},\ \bibinfo
  {eid} {104003} (\bibinfo {year} {2005})}\BibitemShut {NoStop}%
\bibitem [{\citenamefont {Boyle}\ \emph {et~al.}(2014)\citenamefont {Boyle},
  \citenamefont {Kidder}, \citenamefont {Ossokine},\ and\ \citenamefont
  {Pfeiffer}}]{Boyle:2014ioa}%
  \BibitemOpen
  \bibfield  {author} {\bibinfo {author} {\bibfnamefont {M.}~\bibnamefont
  {Boyle}}, \bibinfo {author} {\bibfnamefont {L.~E.}\ \bibnamefont {Kidder}},
  \bibinfo {author} {\bibfnamefont {S.}~\bibnamefont {Ossokine}}, \ and\
  \bibinfo {author} {\bibfnamefont {H.~P.}\ \bibnamefont {Pfeiffer}},\
  }\href@noop {} {\  (\bibinfo {year} {2014})},\ \Eprint
  {http://arxiv.org/abs/1409.4431} {arXiv:1409.4431 [gr-qc]} \BibitemShut
  {NoStop}%
\bibitem [{\citenamefont {De~Amicis}\ \emph
  {et~al.}(2024{\natexlab{a}})\citenamefont {De~Amicis} \emph
  {et~al.}}]{DeAmicis:2024eoy}%
  \BibitemOpen
  \bibfield  {author} {\bibinfo {author} {\bibfnamefont {M.}~\bibnamefont
  {De~Amicis}} \emph {et~al.},\ }\href@noop {} {\  (\bibinfo {year}
  {2024}{\natexlab{a}})},\ \Eprint {http://arxiv.org/abs/2412.06887}
  {arXiv:2412.06887 [gr-qc]} \BibitemShut {NoStop}%
\bibitem [{\citenamefont {Kidder}(1995)}]{Kidder:1995zr}%
  \BibitemOpen
  \bibfield  {author} {\bibinfo {author} {\bibfnamefont {L.~E.}\ \bibnamefont
  {Kidder}},\ }\href {\doibase 10.1103/PhysRevD.52.821} {\bibfield  {journal}
  {\bibinfo  {journal} {Phys.Rev.}\ }\textbf {\bibinfo {volume} {D52}},\
  \bibinfo {pages} {821} (\bibinfo {year} {1995})},\ \Eprint
  {http://arxiv.org/abs/gr-qc/9506022} {arXiv:gr-qc/9506022 [gr-qc]}
  \BibitemShut {NoStop}%
%%CITATION = GR-QC/9506022;%%
\bibitem [{\citenamefont {Akcay}\ \emph {et~al.}(2021)\citenamefont {Akcay},
  \citenamefont {Gamba},\ and\ \citenamefont {Bernuzzi}}]{Akcay:2020qrj}%
  \BibitemOpen
  \bibfield  {author} {\bibinfo {author} {\bibfnamefont {S.}~\bibnamefont
  {Akcay}}, \bibinfo {author} {\bibfnamefont {R.}~\bibnamefont {Gamba}}, \ and\
  \bibinfo {author} {\bibfnamefont {S.}~\bibnamefont {Bernuzzi}},\ }\href
  {\doibase 10.1103/PhysRevD.103.024014} {\bibfield  {journal} {\bibinfo
  {journal} {Phys. Rev. D}\ }\textbf {\bibinfo {volume} {103}},\ \bibinfo
  {pages} {024014} (\bibinfo {year} {2021})},\ \Eprint
  {http://arxiv.org/abs/2005.05338} {arXiv:2005.05338 [gr-qc]} \BibitemShut
  {NoStop}%
\bibitem [{\citenamefont {Ramos-Buades}\ \emph {et~al.}(2023)\citenamefont
  {Ramos-Buades}, \citenamefont {Buonanno}, \citenamefont {Estell\'es},
  \citenamefont {Khalil}, \citenamefont {Mihaylov}, \citenamefont {Ossokine},
  \citenamefont {Pompili},\ and\ \citenamefont
  {Shiferaw}}]{Ramos-Buades:2023ehm}%
  \BibitemOpen
  \bibfield  {author} {\bibinfo {author} {\bibfnamefont {A.}~\bibnamefont
  {Ramos-Buades}}, \bibinfo {author} {\bibfnamefont {A.}~\bibnamefont
  {Buonanno}}, \bibinfo {author} {\bibfnamefont {H.}~\bibnamefont
  {Estell\'es}}, \bibinfo {author} {\bibfnamefont {M.}~\bibnamefont {Khalil}},
  \bibinfo {author} {\bibfnamefont {D.~P.}\ \bibnamefont {Mihaylov}}, \bibinfo
  {author} {\bibfnamefont {S.}~\bibnamefont {Ossokine}}, \bibinfo {author}
  {\bibfnamefont {L.}~\bibnamefont {Pompili}}, \ and\ \bibinfo {author}
  {\bibfnamefont {M.}~\bibnamefont {Shiferaw}},\ }\href {\doibase
  10.1103/PhysRevD.108.124037} {\bibfield  {journal} {\bibinfo  {journal}
  {Phys. Rev. D}\ }\textbf {\bibinfo {volume} {108}},\ \bibinfo {pages}
  {124037} (\bibinfo {year} {2023})},\ \Eprint
  {http://arxiv.org/abs/2303.18046} {arXiv:2303.18046 [gr-qc]} \BibitemShut
  {NoStop}%
\bibitem [{\citenamefont {Hamilton}\ \emph {et~al.}(2025)\citenamefont
  {Hamilton} \emph {et~al.}}]{Hamilton:2025xru}%
  \BibitemOpen
  \bibfield  {author} {\bibinfo {author} {\bibfnamefont {E.}~\bibnamefont
  {Hamilton}} \emph {et~al.},\ }\href@noop {} {\  (\bibinfo {year} {2025})},\
  \Eprint {http://arxiv.org/abs/2507.02604} {arXiv:2507.02604 [gr-qc]}
  \BibitemShut {NoStop}%
\bibitem [{\citenamefont {Boyle}\ \emph {et~al.}(2011)\citenamefont {Boyle},
  \citenamefont {Owen},\ and\ \citenamefont {Pfeiffer}}]{Boyle:2011gg}%
  \BibitemOpen
  \bibfield  {author} {\bibinfo {author} {\bibfnamefont {M.}~\bibnamefont
  {Boyle}}, \bibinfo {author} {\bibfnamefont {R.}~\bibnamefont {Owen}}, \ and\
  \bibinfo {author} {\bibfnamefont {H.~P.}\ \bibnamefont {Pfeiffer}},\ }\href
  {\doibase 10.1103/PhysRevD.84.124011} {\bibfield  {journal} {\bibinfo
  {journal} {Phys. Rev.}\ }\textbf {\bibinfo {volume} {D84}},\ \bibinfo {pages}
  {124011} (\bibinfo {year} {2011})},\ \Eprint {http://arxiv.org/abs/1110.2965}
  {arXiv:1110.2965 [gr-qc]} \BibitemShut {NoStop}%
%%CITATION = ARXIV:1110.2965;%%
\bibitem [{\citenamefont {Buonanno}\ \emph {et~al.}(2003)\citenamefont
  {Buonanno}, \citenamefont {Chen},\ and\ \citenamefont
  {Vallisneri}}]{Buonanno:2002fy}%
  \BibitemOpen
  \bibfield  {author} {\bibinfo {author} {\bibfnamefont {A.}~\bibnamefont
  {Buonanno}}, \bibinfo {author} {\bibfnamefont {Y.-b.}\ \bibnamefont {Chen}},
  \ and\ \bibinfo {author} {\bibfnamefont {M.}~\bibnamefont {Vallisneri}},\
  }\href {\doibase 10.1103/PhysRevD.67.104025, 10.1103/PhysRevD.74.029904}
  {\bibfield  {journal} {\bibinfo  {journal} {Phys. Rev.}\ }\textbf {\bibinfo
  {volume} {D67}},\ \bibinfo {pages} {104025} (\bibinfo {year} {2003})},\
  \bibinfo {note} {[Erratum: Phys. Rev.D74,029904(2006)]},\ \Eprint
  {http://arxiv.org/abs/gr-qc/0211087} {arXiv:gr-qc/0211087 [gr-qc]}
  \BibitemShut {NoStop}%
%%CITATION = GR-QC/0211087;%%
\bibitem [{\citenamefont {Schmidt}\ \emph {et~al.}(2011)\citenamefont
  {Schmidt}, \citenamefont {Hannam}, \citenamefont {Husa},\ and\ \citenamefont
  {Ajith}}]{Schmidt:2010it}%
  \BibitemOpen
  \bibfield  {author} {\bibinfo {author} {\bibfnamefont {P.}~\bibnamefont
  {Schmidt}}, \bibinfo {author} {\bibfnamefont {M.}~\bibnamefont {Hannam}},
  \bibinfo {author} {\bibfnamefont {S.}~\bibnamefont {Husa}}, \ and\ \bibinfo
  {author} {\bibfnamefont {P.}~\bibnamefont {Ajith}},\ }\href {\doibase
  10.1103/PhysRevD.84.024046} {\bibfield  {journal} {\bibinfo  {journal} {Phys.
  Rev.}\ }\textbf {\bibinfo {volume} {D84}},\ \bibinfo {pages} {024046}
  (\bibinfo {year} {2011})},\ \Eprint {http://arxiv.org/abs/1012.2879}
  {arXiv:1012.2879 [gr-qc]} \BibitemShut {NoStop}%
%%CITATION = ARXIV:1012.2879;%%
\bibitem [{\citenamefont {Schmidt}\ \emph {et~al.}(2012)\citenamefont
  {Schmidt}, \citenamefont {Hannam},\ and\ \citenamefont
  {Husa}}]{Schmidt:2012rh}%
  \BibitemOpen
  \bibfield  {author} {\bibinfo {author} {\bibfnamefont {P.}~\bibnamefont
  {Schmidt}}, \bibinfo {author} {\bibfnamefont {M.}~\bibnamefont {Hannam}}, \
  and\ \bibinfo {author} {\bibfnamefont {S.}~\bibnamefont {Husa}},\ }\href
  {\doibase 10.1103/PhysRevD.86.104063} {\bibfield  {journal} {\bibinfo
  {journal} {Phys. Rev.}\ }\textbf {\bibinfo {volume} {D86}},\ \bibinfo {pages}
  {104063} (\bibinfo {year} {2012})},\ \Eprint {http://arxiv.org/abs/1207.3088}
  {arXiv:1207.3088 [gr-qc]} \BibitemShut {NoStop}%
%%CITATION = ARXIV:1207.3088;%%
\bibitem [{\citenamefont {O'Shaughnessy}\ \emph {et~al.}(2011)\citenamefont
  {O'Shaughnessy}, \citenamefont {Vaishnav}, \citenamefont {Healy},
  \citenamefont {Meeks},\ and\ \citenamefont
  {Shoemaker}}]{OShaughnessy:2011pmr}%
  \BibitemOpen
  \bibfield  {author} {\bibinfo {author} {\bibfnamefont {R.}~\bibnamefont
  {O'Shaughnessy}}, \bibinfo {author} {\bibfnamefont {B.}~\bibnamefont
  {Vaishnav}}, \bibinfo {author} {\bibfnamefont {J.}~\bibnamefont {Healy}},
  \bibinfo {author} {\bibfnamefont {Z.}~\bibnamefont {Meeks}}, \ and\ \bibinfo
  {author} {\bibfnamefont {D.}~\bibnamefont {Shoemaker}},\ }\href {\doibase
  10.1103/PhysRevD.84.124002} {\bibfield  {journal} {\bibinfo  {journal} {Phys.
  Rev.}\ }\textbf {\bibinfo {volume} {D84}},\ \bibinfo {pages} {124002}
  (\bibinfo {year} {2011})},\ \Eprint {http://arxiv.org/abs/1109.5224}
  {arXiv:1109.5224 [gr-qc]} \BibitemShut {NoStop}%
%%CITATION = ARXIV:1109.5224;%%
\bibitem [{\citenamefont {Baiotti}\ \emph {et~al.}(2011)\citenamefont
  {Baiotti}, \citenamefont {Damour}, \citenamefont {Giacomazzo}, \citenamefont
  {Nagar},\ and\ \citenamefont {Rezzolla}}]{Baiotti:2011am}%
  \BibitemOpen
  \bibfield  {author} {\bibinfo {author} {\bibfnamefont {L.}~\bibnamefont
  {Baiotti}}, \bibinfo {author} {\bibfnamefont {T.}~\bibnamefont {Damour}},
  \bibinfo {author} {\bibfnamefont {B.}~\bibnamefont {Giacomazzo}}, \bibinfo
  {author} {\bibfnamefont {A.}~\bibnamefont {Nagar}}, \ and\ \bibinfo {author}
  {\bibfnamefont {L.}~\bibnamefont {Rezzolla}},\ }\href {\doibase
  10.1103/PhysRevD.84.024017} {\bibfield  {journal} {\bibinfo  {journal} {Phys.
  Rev.}\ }\textbf {\bibinfo {volume} {D84}},\ \bibinfo {pages} {024017}
  (\bibinfo {year} {2011})},\ \Eprint {http://arxiv.org/abs/1103.3874}
  {arXiv:1103.3874 [gr-qc]} \BibitemShut {NoStop}%
%%CITATION = 1103.3874;%%
\bibitem [{\citenamefont {De~Amicis}\ \emph
  {et~al.}(2024{\natexlab{b}})\citenamefont {De~Amicis}, \citenamefont
  {Albanesi},\ and\ \citenamefont {Carullo}}]{DeAmicis:2024not}%
  \BibitemOpen
  \bibfield  {author} {\bibinfo {author} {\bibfnamefont {M.}~\bibnamefont
  {De~Amicis}}, \bibinfo {author} {\bibfnamefont {S.}~\bibnamefont {Albanesi}},
  \ and\ \bibinfo {author} {\bibfnamefont {G.}~\bibnamefont {Carullo}},\ }\href
  {\doibase 10.1103/PhysRevD.110.104005} {\bibfield  {journal} {\bibinfo
  {journal} {Phys. Rev. D}\ }\textbf {\bibinfo {volume} {110}},\ \bibinfo
  {pages} {104005} (\bibinfo {year} {2024}{\natexlab{b}})},\ \Eprint
  {http://arxiv.org/abs/2406.17018} {arXiv:2406.17018 [gr-qc]} \BibitemShut
  {NoStop}%
\bibitem [{\citenamefont {Hughes}\ \emph {et~al.}(2019)\citenamefont {Hughes},
  \citenamefont {Apte}, \citenamefont {Khanna},\ and\ \citenamefont
  {Lim}}]{Hughes:2019zmt}%
  \BibitemOpen
  \bibfield  {author} {\bibinfo {author} {\bibfnamefont {S.~A.}\ \bibnamefont
  {Hughes}}, \bibinfo {author} {\bibfnamefont {A.}~\bibnamefont {Apte}},
  \bibinfo {author} {\bibfnamefont {G.}~\bibnamefont {Khanna}}, \ and\ \bibinfo
  {author} {\bibfnamefont {H.}~\bibnamefont {Lim}},\ }\href {\doibase
  10.1103/PhysRevLett.123.161101} {\bibfield  {journal} {\bibinfo  {journal}
  {Phys. Rev. Lett.}\ }\textbf {\bibinfo {volume} {123}},\ \bibinfo {pages}
  {161101} (\bibinfo {year} {2019})},\ \Eprint
  {http://arxiv.org/abs/1901.05900} {arXiv:1901.05900 [gr-qc]} \BibitemShut
  {NoStop}%
\bibitem [{\citenamefont {Lim}\ \emph {et~al.}(2022)\citenamefont {Lim},
  \citenamefont {Hughes},\ and\ \citenamefont {Khanna}}]{Lim:2022veo}%
  \BibitemOpen
  \bibfield  {author} {\bibinfo {author} {\bibfnamefont {H.}~\bibnamefont
  {Lim}}, \bibinfo {author} {\bibfnamefont {S.~A.}\ \bibnamefont {Hughes}}, \
  and\ \bibinfo {author} {\bibfnamefont {G.}~\bibnamefont {Khanna}},\ }\href
  {\doibase 10.1103/PhysRevD.105.124030} {\bibfield  {journal} {\bibinfo
  {journal} {Phys. Rev. D}\ }\textbf {\bibinfo {volume} {105}},\ \bibinfo
  {pages} {124030} (\bibinfo {year} {2022})},\ \Eprint
  {http://arxiv.org/abs/2204.06007} {arXiv:2204.06007 [gr-qc]} \BibitemShut
  {NoStop}%
\bibitem [{\citenamefont {Pan}\ \emph {et~al.}(2010)\citenamefont {Pan},
  \citenamefont {Buonanno}, \citenamefont {Buchman}, \citenamefont {Chu},
  \citenamefont {Kidder} \emph {et~al.}}]{Pan:2009wj}%
  \BibitemOpen
  \bibfield  {author} {\bibinfo {author} {\bibfnamefont {Y.}~\bibnamefont
  {Pan}}, \bibinfo {author} {\bibfnamefont {A.}~\bibnamefont {Buonanno}},
  \bibinfo {author} {\bibfnamefont {L.~T.}\ \bibnamefont {Buchman}}, \bibinfo
  {author} {\bibfnamefont {T.}~\bibnamefont {Chu}}, \bibinfo {author}
  {\bibfnamefont {L.~E.}\ \bibnamefont {Kidder}},  \emph {et~al.},\ }\href
  {\doibase 10.1103/PhysRevD.81.084041} {\bibfield  {journal} {\bibinfo
  {journal} {Phys.Rev.}\ }\textbf {\bibinfo {volume} {D81}},\ \bibinfo {pages}
  {084041} (\bibinfo {year} {2010})},\ \Eprint {http://arxiv.org/abs/0912.3466}
  {arXiv:0912.3466 [gr-qc]} \BibitemShut {NoStop}%
\bibitem [{\citenamefont {Wigner}(1960)}]{WignerGroups}%
  \BibitemOpen
  \bibfield  {author} {\bibinfo {author} {\bibfnamefont {E.}~\bibnamefont
  {Wigner}},\ }\href {\doibase 10.1017/s0008439500025364} {\emph {\bibinfo
  {title} {Group Theory and its Application to the Quantum-Mechanics of Atomic
  Spectra}}}\ (\bibinfo  {publisher} {Canadian Mathematical Society},\ \bibinfo
  {year} {1960})\ p.\ \bibinfo {pages} {95–95}\BibitemShut {NoStop}%
\bibitem [{\citenamefont {Brown}\ \emph {et~al.}(2007)\citenamefont {Brown},
  \citenamefont {Fairhurst}, \citenamefont {Krishnan}, \citenamefont {Mercer},
  \citenamefont {Kopparapu}, \citenamefont {Santamaria},\ and\ \citenamefont
  {Whelan}}]{Brown:2007jx}%
  \BibitemOpen
  \bibfield  {author} {\bibinfo {author} {\bibfnamefont {D.}~\bibnamefont
  {Brown}}, \bibinfo {author} {\bibfnamefont {S.}~\bibnamefont {Fairhurst}},
  \bibinfo {author} {\bibfnamefont {B.}~\bibnamefont {Krishnan}}, \bibinfo
  {author} {\bibfnamefont {R.~A.}\ \bibnamefont {Mercer}}, \bibinfo {author}
  {\bibfnamefont {R.~K.}\ \bibnamefont {Kopparapu}}, \bibinfo {author}
  {\bibfnamefont {L.}~\bibnamefont {Santamaria}}, \ and\ \bibinfo {author}
  {\bibfnamefont {J.~T.}\ \bibnamefont {Whelan}},\ }\href@noop {} {\  (\bibinfo
  {year} {2007})},\ \Eprint {http://arxiv.org/abs/0709.0093} {arXiv:0709.0093
  [gr-qc]} \BibitemShut {NoStop}%
\bibitem [{\citenamefont {Nagar}\ \emph {et~al.}(2024)\citenamefont {Nagar},
  \citenamefont {Gamba}, \citenamefont {Rettegno}, \citenamefont {Fantini},\
  and\ \citenamefont {Bernuzzi}}]{Nagar:2024dzj}%
  \BibitemOpen
  \bibfield  {author} {\bibinfo {author} {\bibfnamefont {A.}~\bibnamefont
  {Nagar}}, \bibinfo {author} {\bibfnamefont {R.}~\bibnamefont {Gamba}},
  \bibinfo {author} {\bibfnamefont {P.}~\bibnamefont {Rettegno}}, \bibinfo
  {author} {\bibfnamefont {V.}~\bibnamefont {Fantini}}, \ and\ \bibinfo
  {author} {\bibfnamefont {S.}~\bibnamefont {Bernuzzi}},\ }\href {\doibase
  10.1103/PhysRevD.110.084001} {\bibfield  {journal} {\bibinfo  {journal}
  {Phys. Rev. D}\ }\textbf {\bibinfo {volume} {110}},\ \bibinfo {pages}
  {084001} (\bibinfo {year} {2024})},\ \Eprint
  {http://arxiv.org/abs/2404.05288} {arXiv:2404.05288 [gr-qc]} \BibitemShut
  {NoStop}%
\bibitem [{\citenamefont {Damour}\ \emph {et~al.}(2025)\citenamefont {Damour},
  \citenamefont {Nagar}, \citenamefont {Placidi},\ and\ \citenamefont
  {Rettegno}}]{Damour:2025uka}%
  \BibitemOpen
  \bibfield  {author} {\bibinfo {author} {\bibfnamefont {T.}~\bibnamefont
  {Damour}}, \bibinfo {author} {\bibfnamefont {A.}~\bibnamefont {Nagar}},
  \bibinfo {author} {\bibfnamefont {A.}~\bibnamefont {Placidi}}, \ and\
  \bibinfo {author} {\bibfnamefont {P.}~\bibnamefont {Rettegno}},\ }\href@noop
  {} {\  (\bibinfo {year} {2025})},\ \Eprint {http://arxiv.org/abs/2503.05487}
  {arXiv:2503.05487 [gr-qc]} \BibitemShut {NoStop}%
\bibitem [{\citenamefont {Albanesi}\ \emph
  {et~al.}(2025{\natexlab{b}})\citenamefont {Albanesi}, \citenamefont {Nagar},\
  and\ \citenamefont {Bernuzzi}}]{Albanesi:2025prep}%
  \BibitemOpen
  \bibfield  {author} {\bibinfo {author} {\bibfnamefont {S.}~\bibnamefont
  {Albanesi}}, \bibinfo {author} {\bibfnamefont {A.}~\bibnamefont {Nagar}}, \
  and\ \bibinfo {author} {\bibfnamefont {S.}~\bibnamefont {Bernuzzi}},\
  }\href@noop {} {\  (\bibinfo {year} {2025}{\natexlab{b}})},\ \Eprint
  {http://arxiv.org/abs/in preparation} {in preparation} \BibitemShut {NoStop}%
\end{thebibliography}%

\end{document}